\documentclass[11pt,a4paper,onecolumn]{article}
\usepackage[margin=22mm]{geometry}
\usepackage[T1]{fontenc}
\usepackage{lmodern}
\usepackage{amsmath,amssymb,bm,graphicx}
\usepackage[monochrome]{xcolor}
\usepackage[numbers,sort&compress]{natbib}
\usepackage{authblk}
\usepackage{tocloft}
\usepackage{microtype}
\usepackage[font=small,labelfont=bf]{caption}

\usepackage[hidelinks]{hyperref}

\renewcommand{\thesection}{\Roman{section}}
\renewcommand{\thesubsection}{\Alph{subsection}}

\makeatletter
\def\p@subsection{\thesection\,}
\def\p@subsubsection{\thesection\,\thesubsection\,}
\makeatother

\newcommand{\vct}[1]{\bm{\mathrm{#1}}}
\newcommand{\mean}[1]{\langle #1\rangle}
\newcommand{\tr}{\operatorname{Tr}}
\newcommand{\dd}{\mathrm d}

\hypersetup{pdftitle={Quantum transport of vortex electrons in bending magnets and circular accelerators},pdfauthor={D. Karlovets; I. Pavlov; A. Murtazin; D. Grosman; X. Chen},pdfsubject={Orbital angular momentum transport in bending magnets and circular accelerators}}
\def\CombinedDocument{1}
\makeatletter
\DeclareRobustCommand{\CrossDocumentLink}[3]{%
  \begingroup\let\HyperDestNameFilter\@firstofone
  \ifdefined\CombinedDocument
    \hyperlink{#2}{#3}%
  \else
    \hyper@linkfile{#3}{#1}{#2}%
  \fi
  \endgroup}
\makeatother
\expandafter\def\csname SMmovie@1\endcsname{Movie1_Gaussian_lab.mp4}
\expandafter\def\csname SMmovie@2\endcsname{Movie2_Vortex_comoving.mp4}
\expandafter\def\csname SMmovie@3\endcsname{Movie3_Vortex_lab.mp4}
\expandafter\def\csname SMmovie@4\endcsname{Movie4_Correlations.mp4}
\expandafter\def\csname SMmovie@5\endcsname{Movie5_Ring.mp4}
\DeclareRobustCommand{\MovieIndex}[1]{\hyperlink{movie:\csname SMmovie@#1\endcsname}{#1}}
\DeclareRobustCommand{\MovieNumber}[1]{\hyperlink{movie:\csname SMmovie@#1\endcsname}{Movie~#1}}

\DeclareRobustCommand{\SMref}[1]{%
  \ifdefined\CombinedDocument
    \CrossDocumentLink{arXiv_Supplementary.pdf}{SM.\getrefbykeydefault{SM:#1}{anchor}{}}{\getrefnumber{SM:#1}}%
  \else
    \ifcsname SMnumber@#1\endcsname
      \CrossDocumentLink{arXiv_Supplementary.pdf}{\csname SMtarget@#1\endcsname}{\csname SMnumber@#1\endcsname}%
    \else\PackageError{Crossrefs}{Undefined reference #1}{Rebuild cross-reference files.}\fi
  \fi}
\expandafter\def\csname SMnumber@eq:B\endcsname{S43}
\expandafter\def\csname SMnumber@eq:Horb\endcsname{S38}
\expandafter\def\csname SMnumber@eq:LG\endcsname{S1}
\expandafter\def\csname SMnumber@eq:LGcharacteristic\endcsname{S51}
\expandafter\def\csname SMnumber@eq:LGfourth\endcsname{S53}
\expandafter\def\csname SMnumber@eq:LH\endcsname{S49}
\expandafter\def\csname SMnumber@eq:OAM_covariance\endcsname{S4}
\expandafter\def\csname SMnumber@eq:OAMmatrix\endcsname{S54}
\expandafter\def\csname SMnumber@eq:SMcovariant\endcsname{S32}
\expandafter\def\csname SMnumber@eq:SMcrossing\endcsname{S34}
\expandafter\def\csname SMnumber@eq:SMentry\endcsname{S35}
\expandafter\def\csname SMnumber@eq:SMentry_acceleration\endcsname{S20}
\expandafter\def\csname SMnumber@eq:SMentry_momentum\endcsname{S21}
\expandafter\def\csname SMnumber@eq:SMentry_size_bounds\endcsname{S22}
\expandafter\def\csname SMnumber@eq:SMexpansion\endcsname{S18}
\expandafter\def\csname SMnumber@eq:SMfaces\endcsname{S36}
\expandafter\def\csname SMnumber@eq:SMfield_rate\endcsname{S13}
\expandafter\def\csname SMnumber@eq:SMincident_momentum\endcsname{S19}
\expandafter\def\csname SMnumber@eq:SMlabroot\endcsname{S9}
\expandafter\def\csname SMnumber@eq:SMlocal_energy\endcsname{S10}
\expandafter\def\csname SMnumber@eq:SMlorentz\endcsname{S33}
\expandafter\def\csname SMnumber@eq:SMmetric\endcsname{S8}
\expandafter\def\csname SMnumber@eq:SMpotential\endcsname{S14}
\expandafter\def\csname SMnumber@eq:SMpotential_gradient\endcsname{S15}
\expandafter\def\csname SMnumber@eq:SMpsi\endcsname{S84}
\expandafter\def\csname SMnumber@eq:SMpsiMom\endcsname{S85}
\expandafter\def\csname SMnumber@eq:SMroot\endcsname{S11}
\expandafter\def\csname SMnumber@eq:SMtorques\endcsname{S30}
\expandafter\def\csname SMnumber@eq:Sode\endcsname{S26}
\expandafter\def\csname SMnumber@eq:V\endcsname{S56}
\expandafter\def\csname SMnumber@eq:angleH\endcsname{S39}
\expandafter\def\csname SMnumber@eq:axisymmetric_lens\endcsname{S116}
\expandafter\def\csname SMnumber@eq:bulkmap\endcsname{S41}
\expandafter\def\csname SMnumber@eq:circlemap\endcsname{S57}
\expandafter\def\csname SMnumber@eq:circlevariance\endcsname{S58}
\expandafter\def\csname SMnumber@eq:covariance_definition\endcsname{S2}
\expandafter\def\csname SMnumber@eq:covariance_ellipse\endcsname{S45}
\expandafter\def\csname SMnumber@eq:covariance_transport\endcsname{S27}
\expandafter\def\csname SMnumber@eq:density_surface\endcsname{S125}
\expandafter\def\csname SMnumber@eq:detector\endcsname{S7}
\expandafter\def\csname SMnumber@eq:effective_length\endcsname{S99}
\expandafter\def\csname SMnumber@eq:entrance_field_axis_mean\endcsname{S37}
\expandafter\def\csname SMnumber@eq:exit_H_removal\endcsname{S89}
\expandafter\def\csname SMnumber@eq:fringe_field\endcsname{S93}
\expandafter\def\csname SMnumber@eq:fringe_integral\endcsname{S91}
\expandafter\def\csname SMnumber@eq:fringe_orbit\endcsname{S94}
\expandafter\def\csname SMnumber@eq:fringe_transfer\endcsname{S95}
\expandafter\def\csname SMnumber@eq:halfoperator\endcsname{S60}
\expandafter\def\csname SMnumber@eq:halfvariance\endcsname{S61}
\expandafter\def\csname SMnumber@eq:input\endcsname{S3}
\expandafter\def\csname SMnumber@eq:kinetic_canonical_oam\endcsname{S31}
\expandafter\def\csname SMnumber@eq:kinetic_transfer\endcsname{S44}
\expandafter\def\csname SMnumber@eq:labH2\endcsname{S23}
\expandafter\def\csname SMnumber@eq:linac_validity\endcsname{S119}
\expandafter\def\csname SMnumber@eq:local_crossing\endcsname{S100}
\expandafter\def\csname SMnumber@eq:local_kinetic_momentum\endcsname{S88}
\expandafter\def\csname SMnumber@eq:logistic_profile\endcsname{S92}
\expandafter\def\csname SMnumber@eq:map\endcsname{S42}
\expandafter\def\csname SMnumber@eq:matchedwidth\endcsname{S63}
\expandafter\def\csname SMnumber@eq:matrix_uncertainty\endcsname{S46}
\expandafter\def\csname SMnumber@eq:mean_trace\endcsname{S55}
\expandafter\def\csname SMnumber@eq:mixture\endcsname{S6}
\expandafter\def\csname SMnumber@eq:movie_exit_mean\endcsname{S90}
\expandafter\def\csname SMnumber@eq:movie_lab_density\endcsname{S123}
\expandafter\def\csname SMnumber@eq:movie_lab_events\endcsname{S122}
\expandafter\def\csname SMnumber@eq:movie_rest_current\endcsname{S121}
\expandafter\def\csname SMnumber@eq:movie_times\endcsname{S124}
\expandafter\def\csname SMnumber@eq:movie_two_modes\endcsname{S87}
\expandafter\def\csname SMnumber@eq:offwaist_correlation\endcsname{S104}
\expandafter\def\csname SMnumber@eq:offwaist_mean\endcsname{S106}
\expandafter\def\csname SMnumber@eq:offwaist_radius\endcsname{S105}
\expandafter\def\csname SMnumber@eq:offwaist_state\endcsname{S102}
\expandafter\def\csname SMnumber@eq:offwaist_transfer\endcsname{S103}
\expandafter\def\csname SMnumber@eq:opposite_bends\endcsname{S120}
\expandafter\def\csname SMnumber@eq:projection\endcsname{S86}
\expandafter\def\csname SMnumber@eq:psi_blocks\endcsname{S78}
\expandafter\def\csname SMnumber@eq:psi_classical_action\endcsname{S76}
\expandafter\def\csname SMnumber@eq:psi_diffraction_times\endcsname{S74}
\expandafter\def\csname SMnumber@eq:psi_free_spreading\endcsname{S82}
\expandafter\def\csname SMnumber@eq:psi_gaussian_coefficients\endcsname{S81}
\expandafter\def\csname SMnumber@eq:psi_gaussian_integral\endcsname{S80}
\expandafter\def\csname SMnumber@eq:psi_initial\endcsname{S73}
\expandafter\def\csname SMnumber@eq:psi_polynomial_coefficients\endcsname{S83}
\expandafter\def\csname SMnumber@eq:psi_propagator\endcsname{S79}
\expandafter\def\csname SMnumber@eq:psi_stages\endcsname{S75}
\expandafter\def\csname SMnumber@eq:psi_transfer\endcsname{S77}
\expandafter\def\csname SMnumber@eq:reslice\endcsname{S24}
\expandafter\def\csname SMnumber@eq:rigidity\endcsname{S107}
\expandafter\def\csname SMnumber@eq:ring_cell\endcsname{S113}
\expandafter\def\csname SMnumber@eq:ring_return_variance\endcsname{S115}
\expandafter\def\csname SMnumber@eq:ring_secular_shifts\endcsname{S114}
\expandafter\def\csname SMnumber@eq:ring_transfer\endcsname{S112}
\expandafter\def\csname SMnumber@eq:sectorode\endcsname{S40}
\expandafter\def\csname SMnumber@eq:shearweak\endcsname{S64}
\expandafter\def\csname SMnumber@eq:size_box_bound\endcsname{S110}
\expandafter\def\csname SMnumber@eq:size_errors\endcsname{S111}
\expandafter\def\csname SMnumber@eq:size_small_bend\endcsname{S108}
\expandafter\def\csname SMnumber@eq:sizetilt\endcsname{S62}
\expandafter\def\csname SMnumber@eq:spatial_validity\endcsname{S97}
\expandafter\def\csname SMnumber@eq:two_incidence_angles\endcsname{S98}
\expandafter\def\csname SMnumber@eq:validity_ratios\endcsname{S96}
\expandafter\def\csname SMnumber@eq:weak_aspect_ratio\endcsname{S109}
\expandafter\def\csname SMnumber@eq:weakbalance\endcsname{S70}
\expandafter\def\csname SMnumber@eq:weakmean\endcsname{S66}
\expandafter\def\csname SMnumber@eq:weakmixed\endcsname{S68}
\expandafter\def\csname SMnumber@eq:weakprojection\endcsname{S67}
\expandafter\def\csname SMnumber@eq:weakwavefunction\endcsname{S65}
\expandafter\def\csname SMnumber@eq:weakwidth\endcsname{S72}
\expandafter\def\csname SMnumber@eq:weakxp\endcsname{S69}
\expandafter\def\csname SMnumber@eq:width_angle\endcsname{S5}
\expandafter\def\csname SMnumber@eq:window\endcsname{S101}
\expandafter\def\csname SMnumber@fig:acceleration\endcsname{S6}
\expandafter\def\csname SMnumber@fig:coordinate_frames\endcsname{S1}
\expandafter\def\csname SMnumber@fig:covariance_ellipses\endcsname{S2}
\expandafter\def\csname SMnumber@fig:facilitywidths\endcsname{S8}
\expandafter\def\csname SMnumber@fig:half_extended\endcsname{S11}
\expandafter\def\csname SMnumber@fig:linacvalidity\endcsname{S16}
\expandafter\def\csname SMnumber@fig:offwaist\endcsname{S7}
\expandafter\def\csname SMnumber@fig:packet_story\endcsname{S4}
\expandafter\def\csname SMnumber@fig:projection\endcsname{S3}
\expandafter\def\csname SMnumber@fig:ring\endcsname{S14}
\expandafter\def\csname SMnumber@fig:ringturns_extended\endcsname{S15}
\expandafter\def\csname SMnumber@fig:size_fixed_aspect\endcsname{S9}
\expandafter\def\csname SMnumber@fig:size_low\endcsname{S9}
\expandafter\def\csname SMnumber@fig:size_surfaces\endcsname{S10}
\expandafter\def\csname SMnumber@fig:sizetolerance\endcsname{S12}
\expandafter\def\csname SMnumber@fig:weak_oam_trace\endcsname{S5}
\expandafter\def\csname SMnumber@fig:windows_extended\endcsname{S13}
\expandafter\def\csname SMnumber@sec:analytic\endcsname{IV}
\expandafter\def\csname SMnumber@sec:boundary\endcsname{III\,A}
\expandafter\def\csname SMnumber@sec:circle\endcsname{IV\,C}
\expandafter\def\csname SMnumber@sec:covariance_ellipses\endcsname{III\,C}
\expandafter\def\csname SMnumber@sec:density_notes\endcsname{XII\,B}
\expandafter\def\csname SMnumber@sec:dipole\endcsname{III\,B}
\expandafter\def\csname SMnumber@sec:ellipse_example\endcsname{III\,C}
\expandafter\def\csname SMnumber@sec:entry_size_validity\endcsname{II\,D}
\expandafter\def\csname SMnumber@sec:facilities\endcsname{VIII}
\expandafter\def\csname SMnumber@sec:favorable\endcsname{VIII\,B}
\expandafter\def\csname SMnumber@sec:fixed_checks\endcsname{IX}
\expandafter\def\csname SMnumber@sec:fixed_recovery\endcsname{IX\,A}
\expandafter\def\csname SMnumber@sec:fringes\endcsname{VI}
\expandafter\def\csname SMnumber@sec:halfturn\endcsname{IV\,D}
\expandafter\def\csname SMnumber@sec:hamiltonian\endcsname{II}
\expandafter\def\csname SMnumber@sec:highenergy\endcsname{VIII\,B}
\expandafter\def\csname SMnumber@sec:incidence\endcsname{VI\,A}
\expandafter\def\csname SMnumber@sec:incident_state\endcsname{I\,A}
\expandafter\def\csname SMnumber@sec:lab_density_movies\endcsname{XII\,A}
\expandafter\def\csname SMnumber@sec:lab_expansion\endcsname{II\,E}
\expandafter\def\csname SMnumber@sec:lab_vortex_movie\endcsname{XII\,C}
\expandafter\def\csname SMnumber@sec:linac\endcsname{X}
\expandafter\def\csname SMnumber@sec:matched_search\endcsname{IX\,D}
\expandafter\def\csname SMnumber@sec:mean_equation\endcsname{II\,F}
\expandafter\def\csname SMnumber@sec:mean_oam\endcsname{IV\,A}
\expandafter\def\csname SMnumber@sec:measurement\endcsname{I\,C}
\expandafter\def\csname SMnumber@sec:moved_derivations\endcsname{II}
\expandafter\def\csname SMnumber@sec:moved_fw\endcsname{II\,A}
\expandafter\def\csname SMnumber@sec:moved_uniform\endcsname{III\,B}
\expandafter\def\csname SMnumber@sec:movie_acceleration\endcsname{V\,E}
\expandafter\def\csname SMnumber@sec:movie_comparison\endcsname{V\,D}
\expandafter\def\csname SMnumber@sec:movie_correlations\endcsname{XII\,D}
\expandafter\def\csname SMnumber@sec:movie_guide\endcsname{XII}
\expandafter\def\csname SMnumber@sec:movie_mirrorcle\endcsname{XII\,E}
\expandafter\def\csname SMnumber@sec:offwaist\endcsname{VII}
\expandafter\def\csname SMnumber@sec:packet_visualization\endcsname{XII\,B}
\expandafter\def\csname SMnumber@sec:potentials\endcsname{II\,C}
\expandafter\def\csname SMnumber@sec:precession\endcsname{II\,G}
\expandafter\def\csname SMnumber@sec:preparation\endcsname{I}
\expandafter\def\csname SMnumber@sec:preparation_guide\endcsname{I\,B}
\expandafter\def\csname SMnumber@sec:projection\endcsname{V\,B}
\expandafter\def\csname SMnumber@sec:projection_spectrum\endcsname{V\,B}
\expandafter\def\csname SMnumber@sec:quadratic_internal\endcsname{II\,D}
\expandafter\def\csname SMnumber@sec:recovery\endcsname{IX\,C}
\expandafter\def\csname SMnumber@sec:repeated_bends\endcsname{IX\,E}
\expandafter\def\csname SMnumber@sec:ring\endcsname{IX\,E}
\expandafter\def\csname SMnumber@sec:ring_focusing\endcsname{IX\,E}
\expandafter\def\csname SMnumber@sec:size_preparations\endcsname{VIII\,C}
\expandafter\def\csname SMnumber@sec:size_scan\endcsname{VIII\,C}
\expandafter\def\csname SMnumber@sec:size_weak\endcsname{VIII\,C}
\expandafter\def\csname SMnumber@sec:sizetolerance\endcsname{IX\,D}
\expandafter\def\csname SMnumber@sec:summary\endcsname{XI}
\expandafter\def\csname SMnumber@sec:supplemental_movies\endcsname{XII}
\expandafter\def\csname SMnumber@sec:transport\endcsname{III}
\expandafter\def\csname SMnumber@sec:variance\endcsname{IV\,B}
\expandafter\def\csname SMnumber@sec:wavefunction\endcsname{V\,A}
\expandafter\def\csname SMnumber@sec:wavefunction_spectrum\endcsname{V}
\expandafter\def\csname SMnumber@sec:weak_density\endcsname{V\,C}
\expandafter\def\csname SMnumber@sec:weak_mean\endcsname{IV\,E}
\expandafter\def\csname SMnumber@sec:weaklaw\endcsname{IV\,E}
\expandafter\def\csname SMnumber@sec:wide_preparation_scan\endcsname{IX\,D}
\expandafter\def\csname SMnumber@sec:window\endcsname{VI\,B}
\expandafter\def\csname SMnumber@sec:worked_example\endcsname{IV\,F}
\expandafter\def\csname SMnumber@tab:facilities\endcsname{SI}
\expandafter\def\csname SMnumber@tab:favorable\endcsname{SIII}
\expandafter\def\csname SMnumber@tab:favorable_windows\endcsname{SV}
\expandafter\def\csname SMnumber@tab:finite_bends\endcsname{SII}
\expandafter\def\csname SMnumber@tab:fixedrings\endcsname{SVII}
\expandafter\def\csname SMnumber@tab:highenergy_checks\endcsname{SIII}
\expandafter\def\csname SMnumber@tab:linacformal\endcsname{SVIII}
\expandafter\def\csname SMnumber@tab:matched_neighborhoods\endcsname{SVI}
\expandafter\def\csname SMnumber@tab:recovery\endcsname{SV}
\expandafter\def\csname SMnumber@tab:ringfringes\endcsname{SVII}
\expandafter\def\csname SMnumber@tab:size_additional\endcsname{SIV}
\expandafter\def\csname SMnumber@tab:size_geometries\endcsname{SI}
\expandafter\def\csname SMnumber@tab:sizetolerance\endcsname{SVI}
\expandafter\def\csname SMnumber@tab:weak_minima\endcsname{SIII}
\expandafter\def\csname SMtarget@eq:B\endcsname{SM.equation.3.43}
\expandafter\def\csname SMtarget@eq:Horb\endcsname{SM.equation.3.38}
\expandafter\def\csname SMtarget@eq:LG\endcsname{SM.equation.1.1}
\expandafter\def\csname SMtarget@eq:LGcharacteristic\endcsname{SM.equation.4.51}
\expandafter\def\csname SMtarget@eq:LGfourth\endcsname{SM.equation.4.53}
\expandafter\def\csname SMtarget@eq:LH\endcsname{SM.equation.4.49}
\expandafter\def\csname SMtarget@eq:OAM_covariance\endcsname{SM.equation.1.4}
\expandafter\def\csname SMtarget@eq:OAMmatrix\endcsname{SM.equation.4.54}
\expandafter\def\csname SMtarget@eq:SMcovariant\endcsname{SM.equation.2.32}
\expandafter\def\csname SMtarget@eq:SMcrossing\endcsname{SM.equation.3.34}
\expandafter\def\csname SMtarget@eq:SMentry\endcsname{SM.equation.3.35}
\expandafter\def\csname SMtarget@eq:SMentry_acceleration\endcsname{SM.equation.2.20}
\expandafter\def\csname SMtarget@eq:SMentry_momentum\endcsname{SM.equation.2.21}
\expandafter\def\csname SMtarget@eq:SMentry_size_bounds\endcsname{SM.equation.2.22}
\expandafter\def\csname SMtarget@eq:SMexpansion\endcsname{SM.equation.2.18}
\expandafter\def\csname SMtarget@eq:SMfaces\endcsname{SM.equation.3.36}
\expandafter\def\csname SMtarget@eq:SMfield_rate\endcsname{SM.equation.2.13}
\expandafter\def\csname SMtarget@eq:SMincident_momentum\endcsname{SM.equation.2.19}
\expandafter\def\csname SMtarget@eq:SMlabroot\endcsname{SM.equation.2.9}
\expandafter\def\csname SMtarget@eq:SMlocal_energy\endcsname{SM.equation.2.10}
\expandafter\def\csname SMtarget@eq:SMlorentz\endcsname{SM.equation.3.33}
\expandafter\def\csname SMtarget@eq:SMmetric\endcsname{SM.equation.2.8}
\expandafter\def\csname SMtarget@eq:SMpotential\endcsname{SM.equation.2.14}
\expandafter\def\csname SMtarget@eq:SMpotential_gradient\endcsname{SM.equation.2.15}
\expandafter\def\csname SMtarget@eq:SMpsi\endcsname{SM.equation.5.84}
\expandafter\def\csname SMtarget@eq:SMpsiMom\endcsname{SM.equation.5.85}
\expandafter\def\csname SMtarget@eq:SMroot\endcsname{SM.equation.2.11}
\expandafter\def\csname SMtarget@eq:SMtorques\endcsname{SM.equation.2.30}
\expandafter\def\csname SMtarget@eq:Sode\endcsname{SM.equation.2.26}
\expandafter\def\csname SMtarget@eq:V\endcsname{SM.equation.4.56}
\expandafter\def\csname SMtarget@eq:angleH\endcsname{SM.equation.3.39}
\expandafter\def\csname SMtarget@eq:axisymmetric_lens\endcsname{SM.equation.9.116}
\expandafter\def\csname SMtarget@eq:bulkmap\endcsname{SM.equation.3.41}
\expandafter\def\csname SMtarget@eq:circlemap\endcsname{SM.equation.4.57}
\expandafter\def\csname SMtarget@eq:circlevariance\endcsname{SM.equation.4.58}
\expandafter\def\csname SMtarget@eq:covariance_definition\endcsname{SM.equation.1.2}
\expandafter\def\csname SMtarget@eq:covariance_ellipse\endcsname{SM.equation.3.45}
\expandafter\def\csname SMtarget@eq:covariance_transport\endcsname{SM.equation.2.27}
\expandafter\def\csname SMtarget@eq:density_surface\endcsname{SM.equation.12.125}
\expandafter\def\csname SMtarget@eq:detector\endcsname{SM.equation.1.7}
\expandafter\def\csname SMtarget@eq:effective_length\endcsname{SM.equation.6.99}
\expandafter\def\csname SMtarget@eq:entrance_field_axis_mean\endcsname{SM.equation.3.37}
\expandafter\def\csname SMtarget@eq:exit_H_removal\endcsname{SM.equation.5.89}
\expandafter\def\csname SMtarget@eq:fringe_field\endcsname{SM.equation.6.93}
\expandafter\def\csname SMtarget@eq:fringe_integral\endcsname{SM.equation.6.91}
\expandafter\def\csname SMtarget@eq:fringe_orbit\endcsname{SM.equation.6.94}
\expandafter\def\csname SMtarget@eq:fringe_transfer\endcsname{SM.equation.6.95}
\expandafter\def\csname SMtarget@eq:halfoperator\endcsname{SM.equation.4.60}
\expandafter\def\csname SMtarget@eq:halfvariance\endcsname{SM.equation.4.61}
\expandafter\def\csname SMtarget@eq:input\endcsname{SM.equation.1.3}
\expandafter\def\csname SMtarget@eq:kinetic_canonical_oam\endcsname{SM.equation.2.31}
\expandafter\def\csname SMtarget@eq:kinetic_transfer\endcsname{SM.equation.3.44}
\expandafter\def\csname SMtarget@eq:labH2\endcsname{SM.equation.2.23}
\expandafter\def\csname SMtarget@eq:linac_validity\endcsname{SM.equation.10.119}
\expandafter\def\csname SMtarget@eq:local_crossing\endcsname{SM.equation.6.100}
\expandafter\def\csname SMtarget@eq:local_kinetic_momentum\endcsname{SM.equation.5.88}
\expandafter\def\csname SMtarget@eq:logistic_profile\endcsname{SM.equation.6.92}
\expandafter\def\csname SMtarget@eq:map\endcsname{SM.equation.3.42}
\expandafter\def\csname SMtarget@eq:matchedwidth\endcsname{SM.equation.4.63}
\expandafter\def\csname SMtarget@eq:matrix_uncertainty\endcsname{SM.equation.3.46}
\expandafter\def\csname SMtarget@eq:mean_trace\endcsname{SM.equation.4.55}
\expandafter\def\csname SMtarget@eq:mixture\endcsname{SM.equation.1.6}
\expandafter\def\csname SMtarget@eq:movie_exit_mean\endcsname{SM.equation.5.90}
\expandafter\def\csname SMtarget@eq:movie_lab_density\endcsname{SM.equation.12.123}
\expandafter\def\csname SMtarget@eq:movie_lab_events\endcsname{SM.equation.12.122}
\expandafter\def\csname SMtarget@eq:movie_rest_current\endcsname{SM.equation.12.121}
\expandafter\def\csname SMtarget@eq:movie_times\endcsname{SM.equation.12.124}
\expandafter\def\csname SMtarget@eq:movie_two_modes\endcsname{SM.equation.5.87}
\expandafter\def\csname SMtarget@eq:offwaist_correlation\endcsname{SM.equation.7.104}
\expandafter\def\csname SMtarget@eq:offwaist_mean\endcsname{SM.equation.7.106}
\expandafter\def\csname SMtarget@eq:offwaist_radius\endcsname{SM.equation.7.105}
\expandafter\def\csname SMtarget@eq:offwaist_state\endcsname{SM.equation.7.102}
\expandafter\def\csname SMtarget@eq:offwaist_transfer\endcsname{SM.equation.7.103}
\expandafter\def\csname SMtarget@eq:opposite_bends\endcsname{SM.equation.10.120}
\expandafter\def\csname SMtarget@eq:projection\endcsname{SM.equation.5.86}
\expandafter\def\csname SMtarget@eq:psi_blocks\endcsname{SM.equation.5.78}
\expandafter\def\csname SMtarget@eq:psi_classical_action\endcsname{SM.equation.5.76}
\expandafter\def\csname SMtarget@eq:psi_diffraction_times\endcsname{SM.equation.5.74}
\expandafter\def\csname SMtarget@eq:psi_free_spreading\endcsname{SM.equation.5.82}
\expandafter\def\csname SMtarget@eq:psi_gaussian_coefficients\endcsname{SM.equation.5.81}
\expandafter\def\csname SMtarget@eq:psi_gaussian_integral\endcsname{SM.equation.5.80}
\expandafter\def\csname SMtarget@eq:psi_initial\endcsname{SM.equation.5.73}
\expandafter\def\csname SMtarget@eq:psi_polynomial_coefficients\endcsname{SM.equation.5.83}
\expandafter\def\csname SMtarget@eq:psi_propagator\endcsname{SM.equation.5.79}
\expandafter\def\csname SMtarget@eq:psi_stages\endcsname{SM.equation.5.75}
\expandafter\def\csname SMtarget@eq:psi_transfer\endcsname{SM.equation.5.77}
\expandafter\def\csname SMtarget@eq:reslice\endcsname{SM.equation.2.24}
\expandafter\def\csname SMtarget@eq:rigidity\endcsname{SM.equation.8.107}
\expandafter\def\csname SMtarget@eq:ring_cell\endcsname{SM.equation.9.113}
\expandafter\def\csname SMtarget@eq:ring_return_variance\endcsname{SM.equation.9.115}
\expandafter\def\csname SMtarget@eq:ring_secular_shifts\endcsname{SM.equation.9.114}
\expandafter\def\csname SMtarget@eq:ring_transfer\endcsname{SM.equation.9.112}
\expandafter\def\csname SMtarget@eq:sectorode\endcsname{SM.equation.3.40}
\expandafter\def\csname SMtarget@eq:shearweak\endcsname{SM.equation.4.64}
\expandafter\def\csname SMtarget@eq:size_box_bound\endcsname{SM.equation.8.110}
\expandafter\def\csname SMtarget@eq:size_errors\endcsname{SM.equation.9.111}
\expandafter\def\csname SMtarget@eq:size_small_bend\endcsname{SM.equation.8.108}
\expandafter\def\csname SMtarget@eq:sizetilt\endcsname{SM.equation.4.62}
\expandafter\def\csname SMtarget@eq:spatial_validity\endcsname{SM.equation.6.97}
\expandafter\def\csname SMtarget@eq:two_incidence_angles\endcsname{SM.equation.6.98}
\expandafter\def\csname SMtarget@eq:validity_ratios\endcsname{SM.equation.6.96}
\expandafter\def\csname SMtarget@eq:weak_aspect_ratio\endcsname{SM.equation.8.109}
\expandafter\def\csname SMtarget@eq:weakbalance\endcsname{SM.equation.4.70}
\expandafter\def\csname SMtarget@eq:weakmean\endcsname{SM.equation.4.66}
\expandafter\def\csname SMtarget@eq:weakmixed\endcsname{SM.equation.4.68}
\expandafter\def\csname SMtarget@eq:weakprojection\endcsname{SM.equation.4.67}
\expandafter\def\csname SMtarget@eq:weakwavefunction\endcsname{SM.equation.4.65}
\expandafter\def\csname SMtarget@eq:weakwidth\endcsname{SM.equation.4.72}
\expandafter\def\csname SMtarget@eq:weakxp\endcsname{SM.equation.4.69}
\expandafter\def\csname SMtarget@eq:width_angle\endcsname{SM.equation.1.5}
\expandafter\def\csname SMtarget@eq:window\endcsname{SM.equation.6.101}
\expandafter\def\csname SMtarget@fig:acceleration\endcsname{SM.figure.caption.10}
\expandafter\def\csname SMtarget@fig:coordinate_frames\endcsname{SM.figure.caption.2}
\expandafter\def\csname SMtarget@fig:covariance_ellipses\endcsname{SM.figure.caption.4}
\expandafter\def\csname SMtarget@fig:facilitywidths\endcsname{SM.figure.caption.14}
\expandafter\def\csname SMtarget@fig:half_extended\endcsname{SM.figure.caption.24}
\expandafter\def\csname SMtarget@fig:linacvalidity\endcsname{SM.figure.caption.38}
\expandafter\def\csname SMtarget@fig:offwaist\endcsname{SM.figure.caption.11}
\expandafter\def\csname SMtarget@fig:packet_story\endcsname{SM.figure.caption.8}
\expandafter\def\csname SMtarget@fig:projection\endcsname{SM.figure.caption.7}
\expandafter\def\csname SMtarget@fig:ring\endcsname{SM.figure.caption.32}
\expandafter\def\csname SMtarget@fig:ringturns_extended\endcsname{SM.figure.caption.34}
\expandafter\def\csname SMtarget@fig:size_fixed_aspect\endcsname{SM.figure.caption.16}
\expandafter\def\csname SMtarget@fig:size_low\endcsname{SM.figure.caption.16}
\expandafter\def\csname SMtarget@fig:size_surfaces\endcsname{SM.figure.caption.17}
\expandafter\def\csname SMtarget@fig:sizetolerance\endcsname{SM.figure.caption.25}
\expandafter\def\csname SMtarget@fig:weak_oam_trace\endcsname{SM.figure.caption.9}
\expandafter\def\csname SMtarget@fig:windows_extended\endcsname{SM.figure.caption.28}
\expandafter\def\csname SMtarget@sec:analytic\endcsname{SM.section.4}
\expandafter\def\csname SMtarget@sec:boundary\endcsname{SM.subsection.3.1}
\expandafter\def\csname SMtarget@sec:circle\endcsname{SM.subsection.4.3}
\expandafter\def\csname SMtarget@sec:covariance_ellipses\endcsname{SM.subsection.3.3}
\expandafter\def\csname SMtarget@sec:density_notes\endcsname{SM.subsection.12.2}
\expandafter\def\csname SMtarget@sec:dipole\endcsname{SM.subsection.3.2}
\expandafter\def\csname SMtarget@sec:ellipse_example\endcsname{SM.subsection.3.3}
\expandafter\def\csname SMtarget@sec:entry_size_validity\endcsname{SM.section*.3}
\expandafter\def\csname SMtarget@sec:facilities\endcsname{SM.section.8}
\expandafter\def\csname SMtarget@sec:favorable\endcsname{SM.subsection.8.2}
\expandafter\def\csname SMtarget@sec:fixed_checks\endcsname{SM.section.9}
\expandafter\def\csname SMtarget@sec:fixed_recovery\endcsname{SM.figure.caption.24}
\expandafter\def\csname SMtarget@sec:fringes\endcsname{SM.section.6}
\expandafter\def\csname SMtarget@sec:halfturn\endcsname{SM.subsection.4.4}
\expandafter\def\csname SMtarget@sec:hamiltonian\endcsname{SM.section.2}
\expandafter\def\csname SMtarget@sec:highenergy\endcsname{SM.subsection.8.2}
\expandafter\def\csname SMtarget@sec:incidence\endcsname{SM.subsection.6.1}
\expandafter\def\csname SMtarget@sec:incident_state\endcsname{SM.subsection.1.1}
\expandafter\def\csname SMtarget@sec:lab_density_movies\endcsname{SM.subsection.12.1}
\expandafter\def\csname SMtarget@sec:lab_expansion\endcsname{SM.subsection.2.5}
\expandafter\def\csname SMtarget@sec:lab_vortex_movie\endcsname{SM.subsection.12.3}
\expandafter\def\csname SMtarget@sec:linac\endcsname{SM.section.10}
\expandafter\def\csname SMtarget@sec:matched_search\endcsname{SM.section*.29}
\expandafter\def\csname SMtarget@sec:mean_equation\endcsname{SM.subsection.2.6}
\expandafter\def\csname SMtarget@sec:mean_oam\endcsname{SM.subsection.4.1}
\expandafter\def\csname SMtarget@sec:measurement\endcsname{SM.subsection.1.3}
\expandafter\def\csname SMtarget@sec:moved_derivations\endcsname{SM.section.2}
\expandafter\def\csname SMtarget@sec:moved_fw\endcsname{SM.subsection.2.1}
\expandafter\def\csname SMtarget@sec:moved_uniform\endcsname{SM.subsection.3.2}
\expandafter\def\csname SMtarget@sec:movie_acceleration\endcsname{SM.subsection.5.5}
\expandafter\def\csname SMtarget@sec:movie_comparison\endcsname{SM.subsection.5.4}
\expandafter\def\csname SMtarget@sec:movie_correlations\endcsname{SM.subsection.12.4}
\expandafter\def\csname SMtarget@sec:movie_guide\endcsname{SM.section*.41}
\expandafter\def\csname SMtarget@sec:movie_mirrorcle\endcsname{SM.subsection.12.5}
\expandafter\def\csname SMtarget@sec:offwaist\endcsname{SM.section.7}
\expandafter\def\csname SMtarget@sec:packet_visualization\endcsname{SM.subsection.12.2}
\expandafter\def\csname SMtarget@sec:potentials\endcsname{SM.subsection.2.3}
\expandafter\def\csname SMtarget@sec:precession\endcsname{SM.subsection.2.7}
\expandafter\def\csname SMtarget@sec:preparation\endcsname{SM.section.1}
\expandafter\def\csname SMtarget@sec:preparation_guide\endcsname{SM.subsection.1.2}
\expandafter\def\csname SMtarget@sec:projection\endcsname{SM.subsection.5.2}
\expandafter\def\csname SMtarget@sec:projection_spectrum\endcsname{SM.subsection.5.2}
\expandafter\def\csname SMtarget@sec:quadratic_internal\endcsname{SM.subsection.2.4}
\expandafter\def\csname SMtarget@sec:recovery\endcsname{SM.subsection.9.3}
\expandafter\def\csname SMtarget@sec:repeated_bends\endcsname{SM.subsection.9.5}
\expandafter\def\csname SMtarget@sec:ring\endcsname{SM.subsection.9.5}
\expandafter\def\csname SMtarget@sec:ring_focusing\endcsname{SM.section*.37}
\expandafter\def\csname SMtarget@sec:size_preparations\endcsname{SM.subsection.8.3}
\expandafter\def\csname SMtarget@sec:size_scan\endcsname{SM.subsection.8.3}
\expandafter\def\csname SMtarget@sec:size_weak\endcsname{SM.section*.18}
\expandafter\def\csname SMtarget@sec:sizetolerance\endcsname{SM.subsection.9.4}
\expandafter\def\csname SMtarget@sec:summary\endcsname{SM.section.11}
\expandafter\def\csname SMtarget@sec:supplemental_movies\endcsname{SM.section.12}
\expandafter\def\csname SMtarget@sec:transport\endcsname{SM.section.3}
\expandafter\def\csname SMtarget@sec:variance\endcsname{SM.subsection.4.2}
\expandafter\def\csname SMtarget@sec:wavefunction\endcsname{SM.subsection.5.1}
\expandafter\def\csname SMtarget@sec:wavefunction_spectrum\endcsname{SM.section.5}
\expandafter\def\csname SMtarget@sec:weak_density\endcsname{SM.subsection.5.3}
\expandafter\def\csname SMtarget@sec:weak_mean\endcsname{SM.section*.5}
\expandafter\def\csname SMtarget@sec:weaklaw\endcsname{SM.subsection.4.5}
\expandafter\def\csname SMtarget@sec:wide_preparation_scan\endcsname{SM.section*.31}
\expandafter\def\csname SMtarget@sec:window\endcsname{SM.subsection.6.2}
\expandafter\def\csname SMtarget@sec:worked_example\endcsname{SM.subsection.4.6}
\expandafter\def\csname SMtarget@tab:facilities\endcsname{SM.table.caption.12}
\expandafter\def\csname SMtarget@tab:favorable\endcsname{SM.table.caption.15}
\expandafter\def\csname SMtarget@tab:favorable_windows\endcsname{SM.table.caption.26}
\expandafter\def\csname SMtarget@tab:finite_bends\endcsname{SM.table.caption.13}
\expandafter\def\csname SMtarget@tab:fixedrings\endcsname{SM.table.caption.33}
\expandafter\def\csname SMtarget@tab:highenergy_checks\endcsname{SM.table.caption.15}
\expandafter\def\csname SMtarget@tab:linacformal\endcsname{SM.table.caption.39}
\expandafter\def\csname SMtarget@tab:matched_neighborhoods\endcsname{SM.table.caption.30}
\expandafter\def\csname SMtarget@tab:recovery\endcsname{SM.table.caption.26}
\expandafter\def\csname SMtarget@tab:ringfringes\endcsname{SM.table.caption.33}
\expandafter\def\csname SMtarget@tab:size_additional\endcsname{SM.table.caption.21}
\expandafter\def\csname SMtarget@tab:size_geometries\endcsname{SM.table.caption.12}
\expandafter\def\csname SMtarget@tab:sizetolerance\endcsname{SM.table.caption.30}
\expandafter\def\csname SMtarget@tab:weak_minima\endcsname{SM.table.caption.15}
\expandafter\def\csname SMmovie@1\endcsname{Movie1_Gaussian_lab.mp4}
\expandafter\def\csname SMmovie@2\endcsname{Movie2_Vortex_comoving.mp4}
\expandafter\def\csname SMmovie@3\endcsname{Movie3_Vortex_lab.mp4}
\expandafter\def\csname SMmovie@4\endcsname{Movie4_Correlations.mp4}
\expandafter\def\csname SMmovie@5\endcsname{Movie5_Ring.mp4}
\DeclareRobustCommand{\SMmovie}[1]{\CrossDocumentLink{arXiv_Supplementary.pdf}{SM.movie:\csname SMmovie@#1\endcsname}{Movie~#1}}

\ifdefined\CrossDocumentLink\else
\makeatletter
\DeclareRobustCommand{\CrossDocumentLink}[3]{%
  \begingroup\let\HyperDestNameFilter\@firstofone
  \ifdefined\CombinedDocument
    \hyperlink{#2}{#3}%
  \else
    \hyper@linkfile{#3}{#1}{#2}%
  \fi
  \endgroup}
\makeatother
\expandafter\def\csname SMmovie@1\endcsname{Movie1_Gaussian_lab.mp4}
\expandafter\def\csname SMmovie@2\endcsname{Movie2_Vortex_comoving.mp4}
\expandafter\def\csname SMmovie@3\endcsname{Movie3_Vortex_lab.mp4}
\expandafter\def\csname SMmovie@4\endcsname{Movie4_Correlations.mp4}
\expandafter\def\csname SMmovie@5\endcsname{Movie5_Ring.mp4}
\DeclareRobustCommand{\MovieIndex}[1]{\hyperlink{movie:\csname SMmovie@#1\endcsname}{#1}}
\DeclareRobustCommand{\MovieNumber}[1]{\hyperlink{movie:\csname SMmovie@#1\endcsname}{Movie~#1}}
\fi
\DeclareRobustCommand{\Lref}[1]{%
  \ifdefined\CombinedDocument
    \CrossDocumentLink{arXiv_Article.pdf}{\getrefbykeydefault{#1}{anchor}{}}{\getrefnumber{#1}}%
  \else
    \ifcsname Papernumber@#1\endcsname
      \CrossDocumentLink{arXiv_Article.pdf}{\csname Papertarget@#1\endcsname}{\csname Papernumber@#1\endcsname}%
    \else\PackageError{Crossrefs}{Undefined reference #1}{Rebuild cross-reference files.}\fi
  \fi}
\expandafter\def\csname Papernumber@eq:H\endcsname{1}
\expandafter\def\csname Papernumber@eq:W\endcsname{7}
\expandafter\def\csname Papernumber@eq:bmt\endcsname{3}
\expandafter\def\csname Papernumber@eq:cov\endcsname{2}
\expandafter\def\csname Papernumber@eq:entry\endcsname{4}
\expandafter\def\csname Papernumber@eq:half\endcsname{6}
\expandafter\def\csname Papernumber@eq:halfwidth\endcsname{8}
\expandafter\def\csname Papernumber@eq:mean\endcsname{5}
\expandafter\def\csname Papernumber@eq:shape\endcsname{9}
\expandafter\def\csname Papernumber@eq:tiltwidth\endcsname{10}
\expandafter\def\csname Papernumber@fig:boundary_comparison\endcsname{5}
\expandafter\def\csname Papernumber@fig:matching\endcsname{1}
\expandafter\def\csname Papernumber@fig:mechanism\endcsname{4}
\expandafter\def\csname Papernumber@fig:packet_transport\endcsname{2}
\expandafter\def\csname Papernumber@fig:size_scan\endcsname{3}
\expandafter\def\csname Papernumber@fig:windows\endcsname{3}
\expandafter\def\csname Papernumber@sec:boundary_dynamics\endcsname{VIII}
\expandafter\def\csname Papernumber@sec:repeated_turns\endcsname{VII}
\expandafter\def\csname Papertarget@eq:H\endcsname{equation.2.1}
\expandafter\def\csname Papertarget@eq:W\endcsname{equation.4.7}
\expandafter\def\csname Papertarget@eq:bmt\endcsname{equation.2.3}
\expandafter\def\csname Papertarget@eq:cov\endcsname{equation.2.2}
\expandafter\def\csname Papertarget@eq:entry\endcsname{equation.3.4}
\expandafter\def\csname Papertarget@eq:half\endcsname{equation.4.6}
\expandafter\def\csname Papertarget@eq:halfwidth\endcsname{equation.4.8}
\expandafter\def\csname Papertarget@eq:mean\endcsname{equation.4.5}
\expandafter\def\csname Papertarget@eq:shape\endcsname{equation.5.9}
\expandafter\def\csname Papertarget@eq:tiltwidth\endcsname{equation.5.10}
\expandafter\def\csname Papertarget@fig:boundary_comparison\endcsname{figure.caption.5}
\expandafter\def\csname Papertarget@fig:matching\endcsname{figure.caption.1}
\expandafter\def\csname Papertarget@fig:mechanism\endcsname{figure.caption.4}
\expandafter\def\csname Papertarget@fig:packet_transport\endcsname{figure.caption.2}
\expandafter\def\csname Papertarget@fig:size_scan\endcsname{figure.caption.3}
\expandafter\def\csname Papertarget@fig:windows\endcsname{figure.caption.3}
\expandafter\def\csname Papertarget@sec:boundary_dynamics\endcsname{section.8}
\expandafter\def\csname Papertarget@sec:repeated_turns\endcsname{section.7}

\usepackage{needspace}
\newenvironment{ruledtabular}{\par\smallskip\hrule\smallskip\centering}{\par\smallskip\hrule\smallskip}
\newcommand{\MovieLink}[2]{\hyperlink{movie:#1}{#2}}
\makeatletter
\let\CombinedSavedMakeTitle\maketitle
\let\CombinedSavedInternalMakeTitle\@maketitle
\let\CombinedSavedAuthor\author
\let\CombinedSavedTitle\title
\let\CombinedSavedDate\date
\let\CombinedSavedThanks\thanks
\let\CombinedSavedAnd\and
\let\CombinedSavedAuthorBody\@author
\let\CombinedSavedAffilsep\AB@affilsep
\let\CombinedSavedBlkAnd\AB@blk@and
\newcommand{\ResetSupplementTitle}{%
  \let\maketitle\CombinedSavedMakeTitle
  \let\@maketitle\CombinedSavedInternalMakeTitle
  \let\author\CombinedSavedAuthor\let\title\CombinedSavedTitle
  \let\date\CombinedSavedDate\let\thanks\CombinedSavedThanks
  \let\and\CombinedSavedAnd\let\@author\CombinedSavedAuthorBody
  \let\AB@affilsep\CombinedSavedAffilsep\let\AB@blk@and\CombinedSavedBlkAnd
  \gdef\AB@authors{}\gdef\AB@authlist{}\gdef\AB@affillist{}
  \gdef\AB@las{}\gdef\AB@lasx{}\gdef\AB@as{}\gdef\AB@au@str{}
  \gdef\@thanks{}\setcounter{authors}{0}\setcounter{affil}{0}\newaffiltrue}
\def\CurrentContentsFile{maintoc}
\let\CombinedAddToContents\addtocontents
\renewcommand{\addtocontents}[2]{%
  \def\tempa{#1}\def\tempb{toc}%
  \ifx\tempa\tempb\CombinedAddToContents{\CurrentContentsFile}{#2}%
  \else\CombinedAddToContents{#1}{#2}\fi}
\let\CombinedStartToc\@starttoc
\renewcommand{\@starttoc}[1]{%
  \def\tempa{#1}\def\tempb{toc}%
  \ifx\tempa\tempb\expandafter\CombinedStartToc\expandafter{\CurrentContentsFile}%
  \else\CombinedStartToc{#1}\fi}
\makeatother

\begin{document}

\title{\texorpdfstring{Quantum transport of vortex electrons\\in bending magnets and circular accelerators}{Quantum transport of vortex electrons in bending magnets and circular accelerators}}
\author[1,2]{D. Karlovets\thanks{\href{mailto:d.karlovets@gmail.com}{d.karlovets@gmail.com}}}
\author[1,2]{I. Pavlov\thanks{\href{mailto:ilia.pavlov23@yandex.ru}{ilia.pavlov23@yandex.ru}}}
\author[1,3,4]{A. Murtazin}
\author[1]{D. Grosman}
\author[2]{X. Chen}
\affil[1]{School of Physics and Engineering, ITMO University, St. Petersburg 197101, Russia}
\affil[2]{Institute of Modern Physics, Chinese Academy of Sciences, Lanzhou 730000, China}
\affil[3]{Petersburg Nuclear Physics Institute of NRC ``Kurchatov Institute'', Gatchina 188300, Russia}
\affil[4]{St. Petersburg Department of Steklov Mathematical Institute of Russian Academy of Sciences, St. Petersburg 191023, Russia}
\date{\today}

\maketitle
\begin{abstract}
{Controlling the sizes and phase structure of single-particle wave packets in accelerators remains an experimental challenge and requires understanding how bending magnets transform these states.} {An important example is transporting relativistic vortex electrons through the magnets to a collision point while keeping a narrow orbital-angular-momentum (OAM) distribution.} {We derive an effective Hamiltonian that accounts for the packet acceleration, employ it to obtain a generalized BMT-like equation, match the electron state at both magnet boundaries, and calculate the mean OAM and its variance.} {A half turn can reverse the mean OAM relative to the momentum while broadening its distribution, so recovering the mean does not ensure state recovery.} {We identify windows in packet sizes and bend angles for a small OAM spread and show that weak bends can produce narrow distributions with the mean OAM tilted relative to the momentum, making these states candidates for spatiotemporal vortices.} {For a few MeV electrons, we propose to use half-turn magnets that leave the mean OAM nearly antiparallel to the momentum while keeping its distribution narrow.} {For a circular orbit in a uniform field,} the OAM variance after the exit grows as $N^2$ over $N$ turns, favoring linear accelerators, such as LINAC at JINR, for first experiments in MeV range.
\end{abstract}

\section{Introduction}
{Electron quantum states have been coherently shaped and reconstructed in microscopy \cite{Feist2015,Priebe2017,Gaida2024,Fang2026}, while beam coherence has been characterized in MeV electron diffraction \cite{Zhu2015,Hennicke2026}. Extending this control to accelerators requires understanding how magnets transform the sizes, phase structure, and internal correlations of particle wave packets. Controlled preparation and transport of such states remain an experimental challenge, with vortex electrons carrying orbital angular momentum (OAM) providing a topical example.} {Their manipulation in electron microscopes up to 300~keV} \cite{Uchida,Verbeeck,McMorran} has motivated proposals for particle collisions \cite{Ivanov2011,IvanovSerbo2011,IvanovHelicity2012,Serbo2015,Ivanov2016,Bliokh2017,Karlovets2017,KarlovetsSerbo2020,Bu2021,Ababekri2024,Ivanov2022,AnHIAF2025}, spin and parity measurements \cite{IvanovSpin2020,IvanovSpinPRD2020}, nuclear probes \cite{AfanasevRecoil2021,LuNuclear2025}, and gluon tomography in deep-inelastic scattering \cite{KouChen2026}. {Twisted quark states have been applied to parton distributions~\cite{AnikinChen2025}.} Laser interactions \cite{Karlovets2012,BuVortex2024,AbabekriPhoton2024} and generalized measurements \cite{KarlovetsMeasurements2023} offer routes to prepare high-energy vortex particles, while the superkick effect offers a diagnostic tool \cite{LiSuperkick2024}. Acceleration to relativistic energies and transport must approximately preserve the prepared state \cite{Karlovets2012,Karlovets2021,SilenkoParaxial2019,Zou2021,SizykhSolenoid2024,MengGouy2025,Meng2025,Murtazin2026,FilinaBaturin2026}.

\begin{figure}[tbp]
\centering
\includegraphics[width=0.60\textwidth]{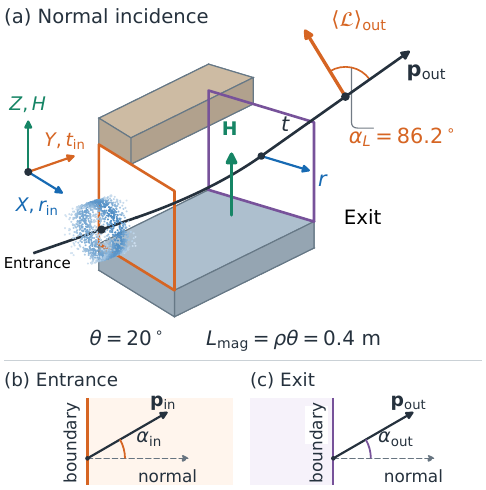}
\caption{(a) Normal incidence at a dipole with orbit bend $\theta$. (b,c) Laboratory in-plane incidence angles $\alpha_{\rm in,out}$ to the normals to the boundaries. {The OAM tilt $\alpha_L$ compares the outgoing rest-frame mean $\mean{\hat{\bm{\mathcal L}}}_{\rm out}$ with the orbit tangent, and $\rho$ is the orbit radius.} Axes $r,t,H$ point radially outward, along the orbit, and along the field. BNL ATF geometry \cite{BNL}: $K=25$~MeV, $H=74.2$~mT, $\theta=20^\circ$, effective dipole length $L_{\rm mag}=0.4$~m; $\ell=1000$, $R_\perp=1~\mu$m, $R_\parallel=0.3~\mu$m.}
\label{fig:matching}
\end{figure}

While an axisymmetric lens preserves the OAM \cite{Karlovets2021}, a bending magnet breaks {this} symmetry as it turns the trajectory. Earlier studies {address} wave-function {evolution in magnetic fields} \cite{Gallatin,BliokhPRX2012}, mean-OAM precession \cite{Silenko2017,Silenko2018,Karlovets2026}, higher multipole effects \cite{Silenko2019,KarlovetsZhevlakov2019}, and {invariants} in periodic {round} lattices \cite{BaturinTransport2026}. {Here we follow a relativistic vortex packet through a finite bending magnet to determine its outgoing free state, accounting for packet acceleration, internal evolution, entrance and exit matching, and realistic fringe fields. Its OAM distribution can broaden even when the mean recovers. We identify three experimentally distinct regimes:} \textit{(i)} {strong OAM broadening in standard bends}, \textit{(ii)} {narrow distributions with a tilted mean OAM in weak bends, which are candidates for spatiotemporal vortex states} \cite{BliokhNori2012}, and \textit{(iii)} orbital-helicity flips near a half turn. {The angular windows near a half turn are narrow, and recovery of the full vortex state remains experimentally demanding.}

\section{State and transport}
{We use SI units.} An electron of charge $-e$ and mass $m$ enters a dipole of laboratory field $\vct H=H\vct e_Z$ with sharp boundaries. For kinetic energy $K$, its center follows a circular orbit with $\gamma=1+K/(mc^2)$, speed $v=c\sqrt{1-\gamma^{-2}}$, and radius $\rho=\gamma mv/(eH)$. Initially, laboratory axes $X,Y,Z$ and comoving axes $x,y,z$ point radially outward, along the incident momentum, and along the field. The incident state is a Laguerre--Gaussian (LG) packet without radial nodes with a wave function in the comoving frame
\[
\psi_0\propto(x-i\operatorname{sgn}(\ell)z)^{|\ell|}
\exp\!\left[-\frac{x^2+z^2}{2w^2}-\frac{y^2}{4R_\parallel^2}\right],
\]
an OAM projection $\ell\hbar=0,\pm \hbar, \pm 2\hbar,...$ along $y$, $w=R_\perp/\sqrt{|\ell|+1}$, rest frame rms transverse radius $R_\perp$, and rms length $R_\parallel$. {Packet spreading allows sizes from 1~nm to 10~$\mu$m and larger~\cite{Karlovets2021}.}

We connect successive instantaneous rest frames by nonrotating Fermi--Walker (FW) axes $x,y,z$ \cite{Rindler,OST,Roura} and use the center's proper time $\tau$. The orbital axes $r,t,H$ (Fig.~\ref{fig:matching}) rotate relative to FW axes (SM, Sec.~\SMref{sec:dipole}), and their position components $(r_r,r_t,r_H)$ initially equal $(x,y,z)$. Centered~operators
\[
\delta\hat{\vct r}=\hat{\vct r}-\mean{\hat{\vct r}},\qquad
\delta\hat{\vct\pi}=\hat{\vct\pi}-\mean{\hat{\vct\pi}}
\]
define intrinsic kinetic OAM
\[
\hat{\bm{\mathcal L}}=\delta\hat{\vct r}\times\delta\hat{\vct\pi}.
\] Neglecting spin and expanding the transformed Hamiltonian to second order in internal position and momentum yields
\begin{equation}
\hat{\mathcal H}_2^{(\mathrm{FW})}=\frac{(\delta\hat{\vct\pi})^2}{2m}-\frac{m}{2c^2}(\vct a'\!\cdot\delta\hat{\vct r})^2,
\quad \vct A=\tfrac12\vct H'\times\delta\hat{\vct r},\label{eq:H}
\end{equation}
where $\delta\hat{\vct\pi}=\delta\hat{\vct p}+e\vct A$, $\delta\hat{\vct p}$ is centered canonical momentum, $\vct A$ the vector potential, and $\vct H'$ and $\vct a'$ are the rest-frame field and acceleration. The derivation and bounds are given in Supplemental Material (SM) \cite{SM}, Sec.~\SMref{sec:hamiltonian}. The acceleration term is absent from the internal approximations of Refs.~\cite{Silenko2017,Silenko2018,Karlovets2026}, although their laboratory relativistic dynamics contains its origin. Eq.~\eqref{eq:H} permits relativistic center motion but requires
\[
\sqrt{\mean{(\delta\hat{\vct\pi})^2}}\ll mc,\qquad
\sqrt{\mean{(\vct a'\cdot\delta\hat{\vct r})^2}}\ll c^2,
\]
and small packet sizes compared with field-variation scales.

In FW axes, the centered canonical variables
\[
\delta\hat Z=(\delta\hat x,\delta\hat y,\delta\hat z,\delta\hat p_x,\delta\hat p_y,\delta\hat p_z)^T
\]
have {Weyl-ordered} covariance
\[
\Sigma_{AB}=\frac12\mean{\{\delta\hat Z_A,\delta\hat Z_B\}},\qquad A,B=1,\ldots,6,
\]
and $\{\hat A,\hat B\}=\hat A\hat B+\hat B\hat A$. {Hamiltonian~\eqref{eq:H} gives the linear Heisenberg evolution $\delta\hat Z(\tau)=S(\tau)\delta\hat Z_0$, whose symplectic transfer $S$ preserves canonical commutators.} {This quadratic evolution follows the trajectory-coherent approach of Bagrov et al.}~{\cite{BagrovBelovMaslov1989,BagrovBelovTrifonov1993,BagrovBelovKondrateva1994,BagrovBelovTrifonov1996,BagrovBelovTrifonov2005,BagrovBelovTrifonov2006}}. {Different transfer matrices $S$ describe entrance matching, propagation inside the magnet, and exit matching
[SM, Eqs.~(\SMref{eq:SMfaces}) and (\SMref{eq:map})].} {In each case,}
\begin{equation}
 \Sigma(\tau)=S(\tau)\Sigma_0S^T(\tau),
 \label{eq:cov}
\end{equation}
where $0$ denotes the incident state.

\begin{samepage}
The rest-frame mean kinetic OAM obeys an analogue of the Bargmann--Michel--Telegdi equation with two additional torques (cf. Refs.~\cite{BMT,ManePolarized2005,Silenko2017,Silenko2018,Karlovets2026}):
\begin{equation}
 \frac{\dd\mean{\hat{\bm{\mathcal L}}}}{\dd\tau}
 =\frac{e}{2m}\vct H'\times\mean{\hat{\bm{\mathcal L}}}
  +\vct K_\pi+\vct K_a.
 \label{eq:bmt}
\end{equation}
\end{samepage}
Its covariant form is given in SM, Eq.~(\SMref{eq:SMcovariant}). The gauge-invariant torques $\vct K_\pi$ and $\vct K_a$ depend on the second moments (SM, Eq.~(\SMref{eq:SMtorques})). Eq.~\eqref{eq:bmt} must be solved together with the covariance evolution, Eq.~\eqref{eq:cov}, using the incident moments. {Importantly, the entrance matching changes the in-field moments -- that is, initial conditions for Eq.~\eqref{eq:bmt} --  from those of the free LG state.}

\section{Matching at the entrance}
The packet front enters the field before its rear. In the inertial rest frame at entry, a point at coordinate $y$ crosses the sharp boundary at $t'_\times=-y/v$. Continuing each crossing to $t'=0$, when the packet center enters, defines one initial in-field wave function. With $E'_x=vH'$ and $H'=\gamma H$, the operator Lorentz force gives
\[
\int_{-y/v}^0(-eE'_x)\dd t'=-eH'y.
\] This position-dependent momentum change we call a \textit{momentum shear}. The magnetic force contributes at second order over the crossing interval. Promoting the first-order relations to centered operators yields
\[
 \delta\hat\pi_x^+=\delta\hat p_x^--eH'\delta\hat y,\qquad
 \delta\hat\pi_y^+=\delta\hat p_y^-.
\]
The kinetic momentum is continuous at the boundary, and the shifts above arise from integrating the electric force between each crossing and the common time $t'=0$. Substituting these shifts and the symmetric-gauge potential $\vct A=\tfrac12H'(-\delta\hat y,\delta\hat x,0)$ from Eq.~\eqref{eq:H} into $\delta\hat{\vct p}^{+}=\delta\hat{\vct\pi}^{+}-e\vct A$ gives
\begin{equation}
 \delta\hat p_x^+=\delta\hat p_x^--\tfrac12eH'\delta\hat y,\qquad
 \delta\hat p_y^+=\delta\hat p_y^--\tfrac12eH'\delta\hat x,
 \label{eq:entry}
\end{equation}
with $\delta\hat p_z$ unchanged. Superscripts $-$ and $+$ denote incident and in-field states. {The crossing time} $R_\parallel/v$ is only about 1~fs in Fig.~\ref{fig:matching}, yet the mean OAM component along the field becomes
\[
\mean{\hat{\mathcal L}_H}_{\rm in}=eH'R_\parallel^2\simeq510\hbar.
\] {Internal propagation changes the packet sizes and correlations on which the inverse exit shear acts (Sec.~\ref{sec:boundary_dynamics}). \SMmovie{4} follows this evolution.} %

\begin{figure}[tbp]
\centering
\includegraphics[width=\textwidth]{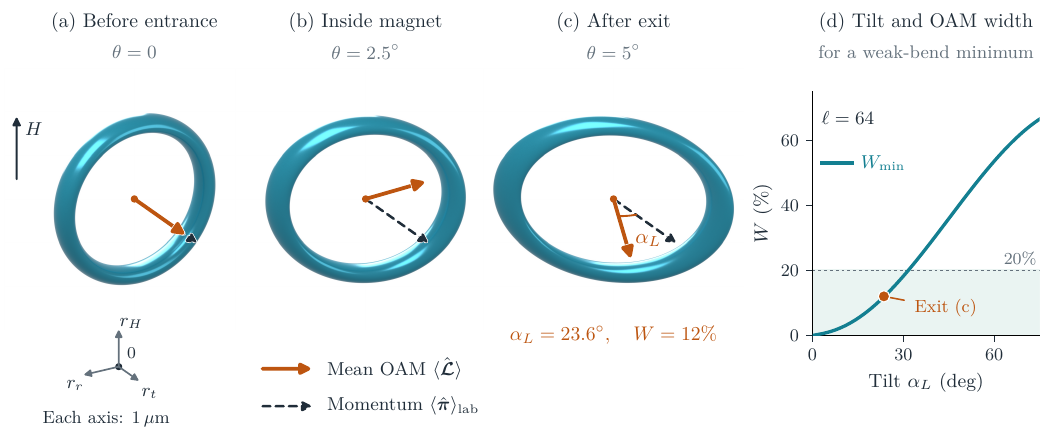}
\caption{{Probability-density evolution and OAM width for one LG packet.
(a)--(c) Isosurfaces $|\psi(\vct r,\tau)|^2=0.2\max_{\vct r}|\psi(\vct r,\tau)|^2$,
calculated from SM, Eq.~(\SMref{eq:SMpsi}), in the packet center's instantaneous rest frame
at proper time $\tau$: before entrance, inside the magnet at $\theta=2.5^\circ$,
and just after exit at $\theta=5^\circ$. Solid and dashed arrows give the directions of
$\mean{\hat{\bm{\mathcal L}}}$ and $\mean{\hat{\vct\pi}}_{\rm lab}$, respectively.
The arc in (c) marks $\alpha_L=\arccos[\mean{\hat{\mathcal L}_t}_{\rm out}/|\mean{\hat{\bm{\mathcal L}}}_{\rm out}|]$.
(d) Weak-bend minimum $W_{\min}$, Eq.~\eqref{eq:tiltwidth}, for $\gamma\gg1$,
$\theta\sim\gamma^{-1}$, $\ell=64$, and the size ratio in Eq.~\eqref{eq:shape}.
The point gives the exit state in (c), $\alpha_L=23.6^\circ$ and $W=12\%$
from Eq.~\eqref{eq:W}; the shaded band marks $W<20\%$.
Parameters: $K=4.6$~MeV~\cite{IMP2024}, $H=5$~mT, sharp boundaries, $\ell=64$,
$R_\perp=2~\mu$m, $R_\parallel=0.2~\mu$m.}}
\label{fig:packet_transport}
\end{figure}

\section{Mean OAM and its width}
After a dipole, the mean kinetic OAM becomes (SM, Sec.~\SMref{sec:mean_oam}):
\begin{align}
 \mean{\hat{\mathcal L}_r}_{\rm out}&=\tfrac12\gamma\ell\hbar\sin\theta,\label{eq:mean}\\
 \mean{\hat{\mathcal L}_t}_{\rm out}&=\ell\hbar(\cos\theta+\tfrac12\theta\sin\theta).\nonumber
\end{align}
The axes and OAM tilt $\alpha_L$ are shown in Fig.~\ref{fig:matching}. For our incoming LG state, the radial mean OAM vanishes in $0<\theta<2\pi$ \textit{only at} $\theta=\pi$. At that \textit{half turn},

\begin{equation}
 \begin{gathered}
 \mean{\hat{\mathcal L}_H}_{\theta=\pi}=\frac{2\gamma}{eH}
 [\mean{(\delta\hat p_y)^2}_0-\mean{(\delta\hat p_x)^2}_0],\\
 \mean{\hat{\mathcal L}_H}_{\theta=\pi}=0\quad\Longleftrightarrow\quad
 \frac{R_\perp}{R_\parallel}=\sqrt2(|\ell|+1),
 \end{gathered}
 \label{eq:half}
\end{equation}
or $\mean{(\delta\hat p_x)^2}_0=\mean{(\delta\hat p_y)^2}_0=\mean{(\delta\hat p_z)^2}_0$.
The mean tangential projection or \textit{orbital helicity} \cite{IvanovSerbo2011,IvanovHelicity2012} flips at $\theta=\pi$:
\[
\mean{\hat{\mathcal L}_t}_{\rm in}=\ell\hbar
\quad\longrightarrow\quad
\mean{\hat{\mathcal L}_t}_{\rm out}=-\ell\hbar.
\] %

For a nonzero mean OAM, we define its direction $\vct n=\mean{\hat{\bm{\mathcal L}}}/|\mean{\hat{\bm{\mathcal L}}}|$ and the relative rms OAM width
\begin{equation}
 \begin{aligned}
 W&=\frac{\Delta\mathcal L_\parallel}{|\mean{\hat{\bm{\mathcal L}}}|},\
 (\Delta\mathcal L_\parallel)^2
 =\mean{\big[\vct n\cdot(\hat{\bm{\mathcal L}}-\mean{\hat{\bm{\mathcal L}}})\big]^2}.
 \end{aligned}
 \label{eq:W}
\end{equation}
Here $\Delta\mathcal L_\parallel$ is the rms spread along the mean-OAM direction. The incident packet has $W=0${, and we seek $W<20\%$ after transport}. At $\theta=\pi$, with the packet sizes satisfying Eq.~\eqref{eq:half}, the mean OAM is collinear with the momentum, but the OAM projection along $\vct n$ still contains fluctuating products of momenta (SM, Sec.~\SMref{sec:halfturn}). With the magnetic length $l_H=\sqrt{\hbar/(eH)}$, {this gives}
\begin{equation}
W^2_{\theta=\pi}=\left(\frac{l_H}{R_\perp}\right)^4\frac{(|\ell|+1)^3}{\ell^2}\left(\gamma^2(|\ell|+1)+\frac{\pi^2}{8}(|\ell|+2)\right).\label{eq:halfwidth}
\end{equation}
At the fixed size ratio in Eq.~\eqref{eq:half}, increasing both sizes reduces the momentum uncertainties, giving $W\propto R_\perp^{-2}$. For $|\ell|\gg1$ and $\gamma\gg1$, {we get} $W\simeq\gamma|\ell|(l_H/R_\perp)^2$. A narrow distribution therefore requires $R_\perp\gg\sqrt{\gamma|\ell|}\,l_H$, subject to the Hamiltonian validity bounds (SM, Sec.~\SMref{sec:entry_size_validity}), field uniformity, and {the magnet aperture}. 

For example, at 1~MeV and 5~mT, $\ell=64$ and $R_\perp=3~\mu$m give $R_\parallel=32.6$~nm and {a mean OAM antiparallel to the outgoing momentum} but $W=305\%$. For $\ell=10$, correlated sizes $R_\perp=(10\pm7)~\mu$m and $R_\parallel$ from Eq.~\eqref{eq:half} give $W=1.8$--$56\%$, with $W=5.1\%$ at $R_\perp=10~\mu$m {(Fig.~\ref{fig:mechanism}(a))}. The $W=20\%$ threshold occurs at about $R_\perp=5~\mu$m, $R_\parallel=0.32~\mu$m. Here $l_H=0.36~\mu$m.

\begin{figure}[tbp]
\includegraphics[width=\textwidth]{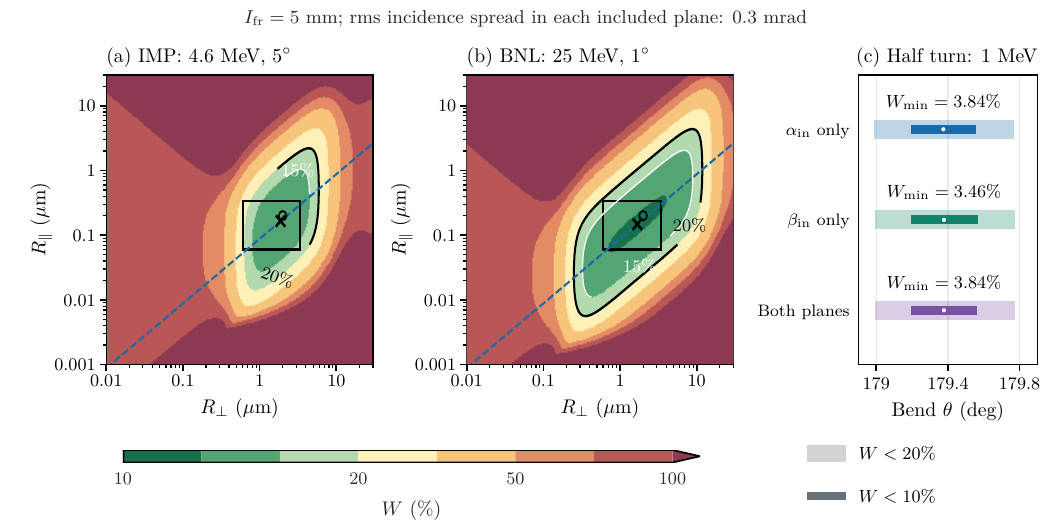}
\caption{Packet-size and angular windows of $W$, Eq.~\eqref{eq:W}, for logistic fringe integrals $I_{\rm fr}=5$~mm. Included incidence angles are independent zero-mean Gaussians with rms width 0.3~mrad each: $\alpha_{\rm in}$ lies in $(X,Y)$, while $\beta_{\rm in}$ tilts toward $Z$ [SM, Eq.~(\SMref{eq:two_incidence_angles})]. (a,b) Both angles; $\ell=64$. Proposed IMP magnet: 4.6~MeV, 5~mT, $5^\circ$, $L_{\rm mag}\simeq30$~cm. Legacy BNL model~\cite{BNL}: 25~MeV, 6.19~mT, $1^\circ$, $L_{\rm mag}=0.24$~m. Dashed lines: Eq.~\eqref{eq:shape}. Rectangles: independent $\pm70\%$ size changes about the open circles at $(2,0.2)~\mu$m; crosses: minima. {White and black contours mark 15\% and 20\%, respectively.} (c) Incidence-plane comparison near a half turn: 1~MeV, 5~mT, $\ell=10$, $(R_\perp,R_\parallel)=(10,1)~\mu$m. {Broad light and narrow dark segments: $W<20\%$ and $W<10\%$, respectively;} dots: minima. All OAM moments refer to the outgoing rest frame for zero incidence angles, with each packet centered at exit.}
\label{fig:windows}\label{fig:size_scan}
\end{figure}

\section{Weak bends}
For $\gamma\gg1$, an orbit bend $\theta\sim\gamma^{-1}\ll1$ can appreciably turn the mean OAM. The rest-frame field $H'=\gamma H$ acts for proper time $\tau=m\theta/(eH)$, giving a cyclotron angle $eH'\tau/m=\gamma\theta$. To leading order, transport gives $\mean{\hat{\bm{\mathcal L}}}_{\rm out}\simeq\ell\hbar(\gamma\theta/2,1,0)$ in the radial, tangential, and field directions. For $\ell>0$, $\alpha_L\simeq\arctan(\gamma\theta/2)$, while for $\ell<0$, the mean points oppositely and $\alpha_L\simeq\pi-\arctan(\gamma\theta/2)$. We seek packet sizes that keep the distribution narrow about this tilted direction, providing a candidate for a spatiotemporal vortex \cite{BliokhNori2012}. The leading weak-bend variance is minimized at (cf. Eq.~\eqref{eq:half})
\begin{equation}
\frac{R_\perp}{R_\parallel}=\sqrt{2(|\ell|+1)}{.}\label{eq:shape}
\end{equation}
At fixed transverse radius, increasing $R_\parallel$ enlarges longitudinal position fluctuations but reduces momentum ones. Their contributions to the OAM variance balance at this ratio. {This gives a lower bound relating the width to the tilt within this weak-bend LG family,}
\begin{equation}
W_{\min}^2=\tfrac12\sin^4\alpha_L+\frac{|\ell|+2}{2\ell^2}\sin^2\alpha_L,\label{eq:tiltwidth}
\end{equation}
see SM, Sec.~\SMref{sec:weaklaw}. For large $|\ell|$, $W_{\min}\simeq\sin^2\alpha_L/\sqrt2$: a width below $20\%$ \textit{requires the mean OAM to lie within about $32^\circ$} of either the parallel direction, $\ell>0$, or the antiparallel one, $\ell<0$. As $\alpha_L\to90^\circ$, $W_{\min}\to1/\sqrt2\simeq71\%$, and changing the packet sizes cannot make this transverse-OAM limit narrow. {The mean-OAM direction, the OAM width $W$, and the phase singularities characterize the state. The sheared envelope and phase-zero lines nearly parallel to the orbit tangent distinguish the OAM tilt from a rigid vortex rotation, consistent with L.~Ji's group~\cite{JiPrivate}. \SMmovie{2} shows singularity splitting analogous to astigmatic splitting~\cite{Dennis2006} (SM, Sec.~\SMref{sec:wavefunction}). The packet evolves unitarily, so the OAM broadening describes mode redistribution without decoherence.}

{For the packet in Fig.~\ref{fig:packet_transport}, the outgoing width lies close to the minimum in Eq.~\eqref{eq:tiltwidth}.
Fig.~\ref{fig:windows} compares our proposed IMP magnet \cite{IMP2024}
with a weak bend based on the legacy BNL ATF specification \cite{BNL,BNLcurrent}.
Both give $\gamma\theta\simeq0.87$ and $\alpha_L\simeq23.6^\circ$.
Eq.~\eqref{eq:tiltwidth} predicts $W_{\min}\simeq11.8\%$, close to
$12\%$ and $11.9\%$ from transport including fringes and incidence-angle spread
(SM, Secs.~\SMref{sec:fringes}, \SMref{sec:size_scan}).}

{For $(R_\perp,R_\parallel)=(2,0.2)~\mu$m, independent
$\pm70\%$ size changes give $W=12$--$27.5\%$ for IMP and
$11.9$--$19.1\%$ for BNL. The BNL rectangle stays below
$20\%$, while an IMP rectangle wholly below $20\%$ is
$R_\perp=1$--$5~\mu$m, $R_\parallel=0.1$--$0.3~\mu$m
(SM, Sec.~\SMref{sec:size_weak}).
The magnetic length sets the scale,
$R_\perp\sim l_H\sqrt{|\ell|+1}\sim3~\mu$m.}

\section{Fringes and recovery windows}
A gradual field transition changes the cancellation of the entry and exit momentum shifts. We average over independent 0.3-mrad rms incidence angles in both planes, integrating the orbit and internal deviations through a field satisfying Maxwell's equations in the fringe region. Each packet's intrinsic OAM is evaluated at exit in the outgoing rest frame for zero incidence angles (SM, Sec.~\SMref{sec:fringes}). The fringe integral
\[
I_{\rm fr}=\int f(1-f)\dd s
\]
measures the transition length, where $s$ is a distance along the normal to the boundary and $f=H_Z(s,0)/H$. For the profile
\[
f(s)=[1+\exp(-s/I_{\rm fr})]^{-1},
\]
the 10--90\% field-transition width is $2\ln9\,I_{\rm fr}$.

{Realistic smooth transitions can greatly broaden the OAM spectrum.
For a $180^\circ$ bend at 3~MeV and 0.1~T, with
$\ell=100$ and $(R_\perp,R_\parallel)=(10,2)~\mu$m,
a 1-mm transition (10--90\%) raises $W$ from $1.01\%$ for sharp boundaries
to $21.9\%$. Both profiles nevertheless admit narrow distributions
at different sizes (SM, Sec.~\SMref{sec:fixed_recovery}).}

{Varying the bend angle changes exit correlations and can restore a narrow distribution.} With both angular spreads, Fig.~\ref{fig:windows}(c) gives a 13.6-mrad interval with $W<20\%$, {while including $\beta_{\rm in}$ shifts the in-plane interval by only 0.06~mrad.} {At the fixed angle optimal for $(R_\perp,R_\parallel)=(10,1)~\mu$m,} independent sizes $R_\perp=10$--$16~\mu$m and $R_\parallel=0.7$--$1.3~\mu$m give $W=3.39$--$7.11\%$, {with the mean OAM within $2.85^\circ$ of the direction opposite to the outgoing momentum.} For $I_{\rm fr}=0.5$--1~mm, some off-plane trajectories leave the region where our local fringe-field model applies. Restricting the angular spread to the bending plane gives minima $W\simeq3.1\%$ for both profiles at 1~MeV (SM, Sec.~\SMref{sec:fringes}).

\section{Experiments and repeated bends}\label{sec:repeated_turns}
{We scan both sizes independently at $\ell=64$ and fixed energy, bend radius, angle, fringe profile, and incidence spread, transporting moments through entrance, magnet, and exit. Minima are $W=29$--$66\%$ for JINR's $28.5^\circ$ photo-injector bend (1--5~MeV), and $70$--$74\%$ for LINAC ($35^\circ$), BNL ($20^\circ$), MESA ($45^\circ$), and ELBE-inspired ($21^\circ$) bends. Sizes span 1~nm--10~$\mu$m (30~$\mu$m for MESA and ELBE) (SM, Sec.~\SMref{sec:size_scan}). A 3-m near-$180^\circ$ magnet (1~MeV, 5~mT) reaches $W\simeq3.1\%$ for $\ell=10$, $(R_\perp,R_\parallel)=(10,1)~\mu$m, $I_{\rm fr}=1$~mm, and 0.3-mrad planar incidence spread (SM, Table~\SMref{tab:recovery}).}

For a circular orbit in a uniform field, formal exit matching after
$N\geq1$ integer turns, $\theta=2\pi N$, gives the outgoing state.
The outgoing tangent and mean OAM recover their incident components in
the orbital axes [Eq.~\eqref{eq:mean}].
With $(0)$ denoting incident components, SM, Eq.~(\SMref{eq:map}), gives
\[
 \begin{aligned}
 \delta\hat r_r&=\delta\hat r_{r}^{(0)},
 &\delta\hat p_r&=\delta\hat p_{r}^{(0)},\\
 \delta\hat r_H&=\delta\hat r_{H}^{(0)}+
       \frac{2\pi N}{eH}\delta\hat p_{H}^{(0)},
 &\delta\hat p_H&=\delta\hat p_{H}^{(0)}.
 \end{aligned}
\]
{The second row describes free packet spreading along the field.} Inserting it into
$\hat{\mathcal L}_t=\delta\hat r_H\delta\hat p_r-\delta\hat r_r\delta\hat p_H$ yields
\[
 \hat{\mathcal L}_t=\hat{\mathcal L}_{t}^{(0)}
       +\frac{2\pi N}{eH}\delta\hat p_{r}^{(0)}\delta\hat p_{H}^{(0)}.
\]
For the incident LG state,
$\hat{\mathcal L}_{t}^{(0)}\psi_0=\ell\hbar\psi_0$ and
$\mean{\delta\hat p_{r}^{(0)}\delta\hat p_{H}^{(0)}}=0$.
The added term has zero mean and zero covariance with
$\hat{\mathcal L}_{t}^{(0)}$, so
\[
 (\Delta\mathcal L_t)^2=
 \left(\frac{2\pi N}{eH}\right)^2
 \mean{(\delta\hat p_{r}^{(0)})^2(\delta\hat p_{H}^{(0)})^2}.
\]
The mean OAM returns to its incident value, while free packet spreading
along the field adds the same momentum-product operator on every turn.
These contributions accumulate \textit{coherently}, giving an rms OAM
spread proportional to $N$ and the $N^2$ variance law
[SM, Eq.~(\SMref{eq:circlevariance}) and Sec.~\SMref{sec:circle}].

For four $90^\circ$ dipoles with sharp boundaries, equal drifts and
no focusing [SM, Fig.~\SMref{fig:ring}], the drifts conserve intrinsic OAM
but change the packet sizes and correlations entering the next magnet.
The variance along the mean has an oscillatory quadratic envelope for
$0<L_{\rm drift}<2\rho$. For $L_{\rm drift}>2\rho$, radial instability
gives exponential broadening within the quadratic model
[SM, Sec.~\SMref{sec:repeated_bends}]. At $K=1$~MeV and $H=5$~mT,
the 0.1- and 1-m examples are stable, with a nonmonotonic OAM variance
for the 1-m drifts. The 2-m drifts destabilize both the mean and rms spread
[SM, Fig.~\SMref{fig:ringturns_extended}].
The absolute OAM spread can grow while the relative width $W$ falls,
because the mean OAM changes both in magnitude and direction.
The growth of OAM fluctuations therefore depends on the lattice as well
as orbit closure. Aligning the final mean OAM with the momentum requires
additional lattice matching. A lattice of optical elements that preserves
both the mean OAM and its width remains to be found.

\clearpage
\section{Boundary dynamics}
\label{sec:boundary_dynamics}
{The OAM moments above follow from Heisenberg evolution without finding the wave function. To explain the boundary mechanism in the Schr\"odinger picture, we propagate the incident LG state $\psi_0$ [SM, Eq.~(\SMref{eq:LG})] with Eq.~\eqref{eq:H} and match both boundaries [SM, Eq.~(\SMref{eq:SMpsi})]. Fig.~\ref{fig:packet_transport}, SM, Fig.~\SMref{fig:packet_story}, and \SMmovie{2} show the resulting density $|\psi(\vct r,\tau)|^2$.}

\begin{figure}[tbp]
{\centering
\includegraphics[width=0.95\textwidth]{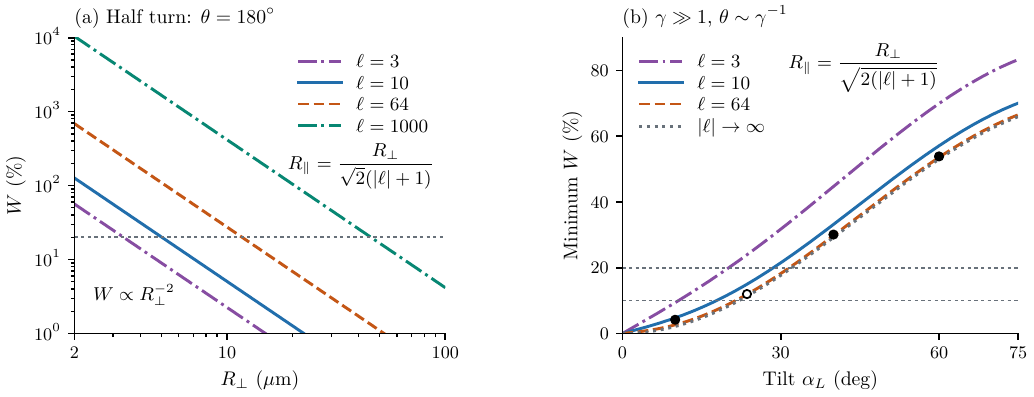}
\par}
\begingroup
\makeatletter\def\@captype{figure}\makeatother
\caption{{{Mean OAM reversal relative to the momentum and a small OAM spread require different conditions.} (a) Relative OAM width $W$, Eq.~\eqref{eq:W}, after a sharp half turn at 1~MeV and 5~mT. The sizes satisfy Eq.~\eqref{eq:half}, and the curves follow Eq.~\eqref{eq:halfwidth}. (b) Minimum $W$ at OAM tilt $\alpha_L$ for $\gamma\gg1$, $\theta\sim\gamma^{-1}$: Eqs.~\eqref{eq:shape} and \eqref{eq:tiltwidth}. Filled circles: finite-angle transport at $\theta=5^\circ$, sharp boundaries, $\ell=64$ (SM, Sec.~\SMref{sec:weaklaw}). Open circle: the IMP example in Fig.~\ref{fig:windows}(a), including fringes and incidence-angle spread. Dotted horizontal lines mark $20\%$ in (a) and $10\%$, $20\%$ in (b).}}
\label{fig:mechanism}
\endgroup
\end{figure}

\begin{figure}[tbp]
{\centering
\includegraphics[width=0.95\textwidth]{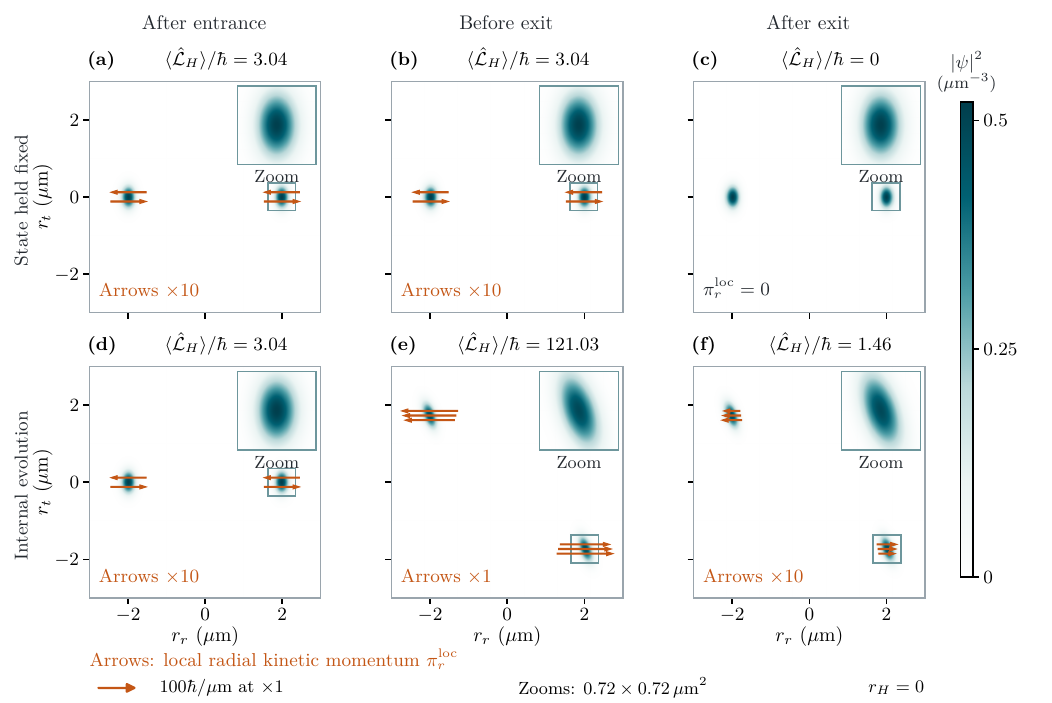}
\par}
\begingroup
\makeatletter\def\@captype{figure}\makeatother
\caption{Entrance and exit matching with the internal evolution
omitted [(a)--(c)] and retained [(d)--(f)], for the packet in
Fig.~\ref{fig:packet_transport}. The panels
[SM, Sec.~\SMref{sec:movie_comparison}] show the states just inside the
entrance, just inside the exit, and after leaving the magnet.
Color gives $|\psi(r_r,r_t,0;\tau)|^2$ in $\mu$m$^{-3}$, using the
same centered rest-frame coordinates as Fig.~\ref{fig:packet_transport}.
The arrows give the local radial kinetic momentum
$\pi_r^{\rm loc}=\operatorname{Re}[\psi^*{\delta\hat\pi_r}\psi]/|\psi|^2$
[SM, Eq.~(\SMref{eq:local_kinetic_momentum})], pointing toward positive
or negative $r_r$. {Arrow lengths use
$100\hbar/\mu$m per $1~\mu$m at $\times1$ and are magnified tenfold in (a,b,d,f), as marked.}
{Each inset magnifies the density in the outlined $0.72\times0.72~\mu\mathrm m^2$ window.}
With the quantum evolution inside the magnet taken into account, the outgoing density remains deformed and
$\mean{\hat{\mathcal L}_H}_{\rm out}\simeq1.46\hbar$.}
\label{fig:boundary_comparison}
\endgroup
\end{figure}

The front of the packet enters the magnetic field first and acquires a larger inward
change of radial momentum from the Lorentz force. Comparing all parts of the packet
at the center's entrance time in its rest frame gives the shear in
Eq.~\eqref{eq:entry}. In the symmetric gauge, the incoming wave
function is multiplied by a phase,
\[
 \psi^+(\vct r)=e^{-ieH'r_rr_t/(2\hbar)}\psi^-(\vct r).
\]
Applying the kinetic-momentum operator of Eq.~\eqref{eq:H} to
$\psi^+$ gives the entrance shear and hence
\[
 \mean{\delta\hat r_t\delta\hat\pi_r}_+
       =-eH'R_\parallel^2,\qquad
 \mean{\hat{\mathcal L}_H}_+=eH'R_\parallel^2
\]
right after entering the magnet. Thus, the position-dependent radial momentum produces the field-axis
OAM. The phase preserves $|\psi|^2$ at that common rest-frame time, and
its effect is visible in the momentum arrows of Fig.~\ref{fig:boundary_comparison}.
The probability density then deforms during propagation inside the magnet.

{The torques $\vct K_\pi$ and $\vct K_a$ in Eq.~\eqref{eq:bmt} couple the mean kinetic OAM to position--momentum correlations and spatial covariance [SM, Eq.~(\SMref{eq:SMtorques})], changing its direction and magnitude. Unlike point-particle spin precession~\cite{BMT}, the mean-OAM equation requires covariance evolution, Eq.~\eqref{eq:cov}. The results of the canonical-OAM dynamics from L.~Ji's group~\cite{JiPrivate} likewise depart from pure precession.}

{Fig.~\ref{fig:boundary_comparison} compares the same boundary phases with and without quantum dynamics of the packet inside the magnet [SM, Eq.~(\SMref{eq:movie_two_modes})].}
With the internal state held fixed, the inverse exit phase restores
$\psi_0$ and cancels both the shear and the induced field-axis OAM:
\[
 \psi_{\rm out}
 =e^{ieH'r_rr_t/(2\hbar)}e^{-ieH'r_rr_t/(2\hbar)}\psi_0=\psi_0.
\]

For the evolved packet, the exit shear acts on the larger longitudinal
extent and the correlations acquired inside the magnet. It changes the mean component
along the field as
\[
 \mean{\hat{\mathcal L}_H}_{\rm out}
 =\mean{\hat{\mathcal L}_H}_{\rm before\,exit}
       -eH'\mean{(\delta\hat r_t)^2}_{\rm exit}.
\]
For Fig.~\ref{fig:boundary_comparison}, the field-axis OAM grows from
$3.04\hbar$ at entry to $121.03\hbar$ before exit, while exit matching removes
$119.57\hbar$, leaving $1.46\hbar$.
The outgoing mean has radial component $27.9\hbar$ and makes
an angle $\alpha_L=23.6^\circ$ with the momentum. The OAM covariance
[SM, Eq.~(\SMref{eq:SMpsiMom})] gives $W=12\%$ from
Eq.~\eqref{eq:W}.
The evolution of the mean-OAM magnitude and $W$ is shown in
SM, Fig.~\SMref{fig:weak_oam_trace}.
During subsequent free propagation the packet spreads
[SM, Fig.~\SMref{fig:packet_story}(d)], while the intrinsic OAM is conserved.
{For the proposed IMP test, Figs.~\ref{fig:packet_transport} and \ref{fig:windows} specify the packet, magnet, and size/angle windows. Measuring $\alpha_L$ and $W$ tests OAM transport, while phase singularities require additional diagnostics. Small expansion parameters [SM, Eq.~(\SMref{eq:validity_ratios})] constrain local truncation, whereas accumulated errors in $\alpha_L$ and $W$ require further estimates. Approximate state conservation also depends on state preparation and lattice design.}

\section{Conclusions}
{Summarizing, quantum evolution of the packet inside a magnet changes packet correlations, so recovery of the mean OAM does not imply a narrow OAM distribution. The control of the incoming packet sizes for fixed magnet parameters yields narrow distributions near a half turn and after weak bends, but experimental tolerances remain demanding.} {Linear accelerators and beam lines with fewer bends, such as LINAC at JINR \cite{JINR,RSFproject}, offer the most direct route to ultrarelativistic vortex electrons.} {After a full turn in the uniform field, exit matching restores the incident mean OAM relative to the momentum. Magnetic sequences analogous to spin rotators and Siberian snakes \cite{ManePolarized2005,ManeSnakes2005,SilenkoTeryaevSnake2019} may compensate the $N^2$ OAM broadening, but a feasible lattice remains to be designed.}

\section*{Acknowledgments}
{We dedicate this work to the memory of Vladislav Bagrov and Andrey Trifonov, teachers of one of us, who greatly contributed to the development of the method used here.}
 We are grateful to L.\,Ji, P.\,Zhang, I.\,Ivanov, and L.\,Zou for the valuable comments, discussions, suggestions, as well as for hospitality. We are also grateful to M.\,Maximov, M.\,Nozdrin, A.\,Shchepkin, S.\, Baturin, K.\,Popov, A.\,Dyatlov, A.\,Silenko and I. Anikin for fruitful discussions and constructive criticism. Studies on dipole bends and linacs are supported by the Russian Science Foundation ({\href{https://rscf.ru/project/23-62-10026/}{23-62-10026}}). The work by DK on circular accelerators is supported by Chinese Academy of Sciences President's International Fellowship Initiative Grant No.2026PVA0222.

GPT-6 Astra was used to produce the supplemental movies and to cross-check selected formulas and the bibliography.

\clearpage
\ResetSupplementTitle
\renewcommand{\HyperDestNameFilter}[1]{SM.#1}
\setcounter{section}{0}\setcounter{subsection}{0}\setcounter{subsubsection}{0}
\setcounter{equation}{0}\setcounter{figure}{0}\setcounter{table}{0}\setcounter{footnote}{0}
\pagenumbering{arabic}
\renewcommand{\theequation}{S\arabic{equation}}
\renewcommand{\thefigure}{S\arabic{figure}}
\renewcommand{\thetable}{S\Roman{table}}
\setcounter{secnumdepth}{2}\setcounter{tocdepth}{1}
\def\CurrentContentsFile{smtoc}

\title{\texorpdfstring{Supplemental Material: Quantum transport of vortex electrons \\ in bending magnets and circular accelerators}{Supplemental Material: Quantum transport of vortex electrons in bending magnets and circular accelerators}}
\author[1,2]{D. Karlovets\thanks{\href{mailto:d.karlovets@gmail.com}{d.karlovets@gmail.com}}}
\author[1,2]{I. Pavlov\thanks{\href{mailto:ilia.pavlov23@yandex.ru}{ilia.pavlov23@yandex.ru}}}
\author[1,3,4]{A. Murtazin}
\author[1]{D. Grosman}
\author[2]{X. Chen}
\affil[1]{School of Physics and Engineering, ITMO University, St. Petersburg 197101, Russia}
\affil[2]{Institute of Modern Physics, Chinese Academy of Sciences, Lanzhou 730000, China}
\affil[3]{Petersburg Nuclear Physics Institute of NRC ``Kurchatov Institute'', Gatchina 188300, Russia}
\affil[4]{St. Petersburg Department of Steklov Mathematical Institute of Russian Academy of Sciences, St. Petersburg 191023, Russia}
\date{\today}
\maketitle

\raggedbottom

\hypertarget{def:kinematics}{}We consider an electron of charge $-e$ ($e>0$) and mass $m$, with laboratory kinetic energy $K$ and a Lorentz factor $\gamma=1+K/(mc^2)$, the speed $v=\beta c$, where $\beta=\sqrt{1-\gamma^{-2}}$. {We use SI units with $c$ and $\hbar$ explicit, we also denote the laboratory magnetic field strength by $\vct H$, measured in tesla. Primes denote the packet center's instantaneous rest frame, with $\vct H'$ the comoving magnetic field.} Packet sizes and quantum moments refer to one electron at a common time in its instantaneous rest frame. A superscript $(0)$ labels incident components before entrance to the magnet.

\hypertarget{def:operators}{}{Within a given frame, canonical and kinetic momenta are $\hat{\vct p}=-i\hbar\nabla$ and $\hat{\vct\pi}=\hat{\vct p}+e\vct A$. Their OAM about a fixed origin is $\hat{\vct r}\times\hat{\vct p}$ and $\hat{\vct r}\times\hat{\vct\pi}$, respectively. To describe internal packet dynamics, we use centered operators $\delta\hat A=\hat A-\mean{\hat A}$, centered canonical OAM $\hat{\vct L}=\delta\hat{\vct r}\times\delta\hat{\vct p}$, and centered intrinsic kinetic OAM $\hat{\bm{\mathcal L}}=\delta\hat{\vct r}\times\delta\hat{\vct\pi}$ [Eq.~\eqref{SM:eq:OAM_covariance}]. The latter is gauge invariant and is the OAM used below.}

\hypertarget{def:frames}{}{We use fixed laboratory axes $X,Y,Z$, nonrotating comoving Fermi--Walker (FW) axes $x,y,z$, and orbital axes $r,t,H$ along the outward radius, reference momentum, and magnetic field. The comoving bases coincide at entry and refer to events simultaneous in the packet center's instantaneous rest frame. Orbital position components are $(r_r,r_t,r_H)$, with axes fixed at their exit directions after the magnet. We use $\{\hat A,\hat B\}=\hat A\hat B+\hat B\hat A$.}

Greek indices run from $0$ to $3$ with metric $(+---)$. The dimensionless four-velocity is $u^\mu=\dot X^\mu/c$, {and a dot denotes differentiation with respect to the center's proper time $\tau$, except in the laboratory-time equations of Sec.~\ref{SM:sec:fringes}.}

\begingroup

\section*{Notation and abbreviations}
\setlength{\tabcolsep}{0pt}
\renewcommand{\arraystretch}{1.0}
\fontsize{9}{10.5}\selectfont
\noindent\begin{tabular}{@{}p{0.18\textwidth}@{\hspace{0.02\textwidth}}p{0.62\textwidth}@{\hspace{0.02\textwidth}}p{0.16\textwidth}@{}}
\textbf{Notation} & \textbf{Meaning} & \textbf{Definition}\\
\hline
\multicolumn{3}{@{}l}{\textit{State and orbital angular momentum}}\\*[2pt]
OAM, $\hat{\bm{\mathcal L}}$ & Orbital angular momentum. The observable used here is the intrinsic kinetic OAM about the packet center, $\delta\hat{\vct r}\times\delta\hat{\vct\pi}$. & Eq.~\eqref{SM:eq:OAM_covariance}\\
LG, $\psi_0$, $\ell$ & Laguerre--Gaussian incident state, with radial index zero and integer azimuthal index $\ell$. Its incident OAM along the motion is $\ell\hbar$. & Eq.~\eqref{SM:eq:LG}\\
$R_\perp$, $R_\parallel$ & Entrance rms transverse radius and longitudinal length in the packet rest frame. For the reference waist state, $R_\perp^2=\Sigma_{xx}+\Sigma_{zz}$ and $R_\parallel^2=\Sigma_{yy}$. & Eq.~\eqref{SM:eq:input}\\
$w$ & Transverse envelope parameter of the incident LG state, $w=R_\perp/\sqrt{|\ell|+1}$. & Eq.~\eqref{SM:eq:LG}\\
$\psi$ & Evolving rest-frame wave function, with probability density $|\psi|^2$. & Eq.~\eqref{SM:eq:psi_stages}\\
$\hat{\vct L}$ & Canonical OAM about the packet center, $\delta\hat{\vct r}\times\delta\hat{\vct p}$. & \hyperlink{def:operators}{Opening text}\\
$V$, $V_{\rm total}$ & Covariance matrix of the three kinetic-OAM components, and its value for an ensemble including the spread of the individual mean OAM vectors. & Eqs.~\eqref{SM:eq:OAM_covariance}, \eqref{SM:eq:mixture}\\
$\vct n$, $\Delta\mathcal L_\parallel$ & Unit vector along the mean OAM, and rms spread of its projection on this direction. & Eq.~\eqref{SM:eq:width_angle}\\
$W$ & Relative rms OAM width, $\Delta\mathcal L_\parallel/|\mean{\hat{\bm{\mathcal L}}}|$, defined for a nonzero mean OAM. & Eq.~\eqref{SM:eq:width_angle}\\
$\alpha_L$ & Angle between the mean OAM and the reference momentum. & Eq.~\eqref{SM:eq:width_angle}\\
$P_k$ & Probability of measuring the outgoing tangential OAM projection $k\hbar$, with integer $k$. & Eq.~\eqref{SM:eq:projection}\\[3pt]
\hline
\end{tabular}\par
\clearpage
\noindent\begin{tabular}{@{}p{0.18\textwidth}@{\hspace{0.02\textwidth}}p{0.62\textwidth}@{\hspace{0.02\textwidth}}p{0.16\textwidth}@{}}
\textbf{Notation} & \textbf{Meaning} & \textbf{Definition}\\
\hline
\multicolumn{3}{@{}l}{\textit{Reference orbit, frames, and fields}}\\*[2pt]
$\pi_{\rm ref}$, $\rho$ & Reference momentum magnitude and bending radius in the uniform field, $\rho=\pi_{\rm ref}/(eH)$. & Eq.~\eqref{SM:eq:two_incidence_angles}, \hyperlink{def:rho_compton}{Sec.~\ref*{SM:sec:entry_size_validity}}\\
$H$, $H'$ & Laboratory magnetic field in tesla and its rest-frame value. For the transverse uniform dipole, $H'=\gamma H$. & \hyperlink{def:kinematics}{Opening paragraph}, Fig.~\ref{SM:fig:coordinate_frames}\\
$\theta$, $L_{\rm mag}$ & Reference-orbit bend angle and effective magnetic length, $L_{\rm mag}=\rho\theta$. & Fig.~\ref{SM:fig:coordinate_frames}, Eq.~\eqref{SM:eq:effective_length}\\
FW, $(x,y,z)$ & Fermi--Walker axes carried by the packet center. They are locally nonrotating in its instantaneous rest frame. & Sec.~\ref{SM:sec:moved_fw}\\
$(X,Y,Z)$, $(r,t,H)$ & Fixed laboratory axes, and orbital axes along the radius, tangent, and magnetic field. The latter have position components $(r_r,r_t,r_H)$. & \hyperlink{def:frames}{Opening text}\\
$t_{\rm lab}$, $\tau$ & Laboratory time and packet-center proper time. At the center, $t_{\rm lab}=\gamma\tau$ for the constant-speed orbit. & Eq.~\eqref{SM:eq:SMmetric}, Sec.~\ref{SM:sec:lab_expansion}\\
$\vct a'$, $\mathcal N$ & Center's proper acceleration and clock-rate factor $\mathcal N=1+\vct a'\cdot\vct r/c^2$ in FW coordinates. & Eq.~\eqref{SM:eq:SMmetric}\\
$\vct A'$, $\Phi'$ & Vector and scalar potentials in the instantaneous rest frame, with $\vct A'=\vct H'\times\vct r/2$ in the symmetric gauge. & Sec.~\ref{SM:sec:potentials}, Eq.~\eqref{SM:eq:SMpotential}\\
$N$, $L_{\rm drift}$ & Number of completed turns and laboratory length of a field-free drift. & Eq.~\eqref{SM:eq:circlemap}, \hyperlink{def:drift_length}{Sec.~\ref*{SM:sec:dipole}}\\[3pt]
\multicolumn{3}{@{}l}{\textit{Quantum moments and their transport}}\\*[2pt]
$\hat{\vct p}$, $\hat{\vct\pi}$ & Canonical momentum $-i\hbar\nabla$ and kinetic momentum $\hat{\vct p}+e\vct A'$. & \hyperlink{def:operators}{Opening text}\\
$\delta\hat Z$, $\Sigma$, $\Sigma_0$ & Centered canonical phase-space vector, its symmetrized covariance, and the incident free-state covariance. Here $\delta\hat A=\hat A-\mean{\hat A}$. & Eqs.~\eqref{SM:eq:covariance_definition}, \eqref{SM:eq:input}\\
$\Sigma_{rr}$, $\Sigma_{rp}$, $\Sigma_{\pi\pi}$ & Position, mixed position--canonical-momentum, and kinetic-momentum covariance blocks. & Eqs.~\eqref{SM:eq:covariance_definition}, \eqref{SM:eq:validity_ratios}\\
$S$, $S_{\rm bulk}$ & Linear canonical transfer matrix and its uniform-dipole bulk form. The full magnet map includes entrance and exit matching. & Eqs.~\eqref{SM:eq:Sode}, \eqref{SM:eq:bulkmap}, \eqref{SM:eq:map}\\
$C_{\rm in}$, $C_{\rm out}$ & Entrance and exit matching matrices for sharp boundaries at normal incidence. & Eq.~\eqref{SM:eq:SMfaces}\\
$S_\pi$ & Maps incident canonical variables to positions and kinetic momenta. & Eq.~\eqref{SM:eq:kinetic_transfer}\\
$J$, $G$ & Canonical commutator matrix, $[\delta\hat Z_A,\delta\hat Z_B]=i\hbar J_{AB}$, and evolution generator, $\dot S=GS$. & Eqs.~\eqref{SM:eq:Sode}, \eqref{SM:eq:covariance_transport}\\
$\varepsilon_{\rm loc}$, $\mathcal H$ & Local energy $\sqrt{m^2c^4+c^2\vct\pi^2}$ and Hamiltonian for evolution in proper time $\tau$. & Eqs.~\eqref{SM:eq:SMlocal_energy}, \eqref{SM:eq:SMroot}\\
$\hat{\mathcal H}_2^{(\mathrm{FW})}$, $\hat{\mathcal H}_{\rm orb}$ & Quadratic internal Hamiltonian in FW axes and in rotating orbital axes. & Eqs.~\eqref{SM:eq:SMexpansion}, \eqref{SM:eq:Horb}\\
BMT, $\vct K_\pi$, $\vct K_a$ & Bargmann--Michel--Telegdi equation. Its generalized OAM form contains the magnetic covariance torque $\vct K_\pi$ and acceleration torque $\vct K_a$. & Eqs.~\eqref{SM:eq:SMtorques}, \eqref{SM:eq:SMcovariant}\\[3pt]
\multicolumn{3}{@{}l}{\textit{Smooth boundaries, incidence, and validity}}\\*[2pt]
$s$, $f(s)$ & Coordinate increasing into the magnet along the boundary normal, and fractional field $f(s)=H_Z(s,0)/H$ on the central plane. & Sec.~\ref{SM:sec:fringes}\\
$I_{\rm fr}$, FINT, $g$, HGAP & Fringe integral $I_{\rm fr}=\int f(1-f)\dd s$, dimensionless FINT $=I_{\rm fr}/g$, full pole gap $g$, and half-gap HGAP $=g/2$. & Eq.~\eqref{SM:eq:fringe_integral}\\
$a$, $L_{10-90}$ & Logistic-profile length $a=I_{\rm fr}$ and 10\%--90\% field-rise length $L_{10-90}=2a\ln9$. & Eq.~\eqref{SM:eq:logistic_profile}\\
$\alpha_{\rm in}$, $\beta_{\rm in}$, $\sigma_\alpha$, $\sigma_\beta$ & In-plane incidence angle from the entrance normal, elevation toward the field axis, and their rms angular spreads. & Eq.~\eqref{SM:eq:two_incidence_angles}\\
$\eta_{\rm NR}$, $\eta_a$, $\eta_{\rm fr}$ & Ratios measuring internal kinetic-momentum spread in units of $mc$, acceleration across the packet, and packet size relative to the fringe scale. & \hyperlink{def:internal_ratios}{Sec.~\ref*{SM:sec:quadratic_internal}}, Eq.~\eqref{SM:eq:spatial_validity}\\
$\lambda_C$ & Reduced Compton wavelength of the electron, $\lambda_C=\hbar/(mc)$. & \hyperlink{def:rho_compton}{Sec.~\ref*{SM:sec:entry_size_validity}}\\
$\epsilon_\perp$, $\epsilon_\parallel$ & Fractional errors in the prepared transverse and longitudinal sizes. & Eq.~\eqref{SM:eq:size_errors}\\[3pt]
\multicolumn{3}{@{}l}{\textit{Waist displacement and laboratory density}}\\*[2pt]
$R_{\perp,w}$, $R_{\parallel,w}$ & Rms sizes at the waist. Free spreading relates them to the entrance sizes $R_{\perp,\rm in}\equiv R_\perp$ and $R_{\parallel,\rm in}\equiv R_\parallel$. & Eqs.~\eqref{SM:eq:offwaist_state}, \eqref{SM:eq:offwaist_radius}\\
$d_w$, $\tau_w$ & Signed laboratory waist-to-entrance distance and proper-time interval, $\tau_w=d_w/(\gamma v)$. Positive $d_w$ describes an expanding packet. & Eq.~\eqref{SM:eq:offwaist_state}\\
$\tau_{d,\perp}$, $\tau_{d,\parallel}$, $z_{R,w}$, $z_{R,\parallel}$ & Transverse and longitudinal diffraction times in the rest frame, and laboratory Rayleigh distances for the waist state. & Eqs.~\eqref{SM:eq:psi_diffraction_times}, \eqref{SM:eq:offwaist_radius}\\
$\varrho_{\rm lab}$, $\vct j'$ & Laboratory probability density, and rest-frame probability current density per unit local proper time. & Eqs.~\eqref{SM:eq:movie_lab_density}, \eqref{SM:eq:movie_rest_current}\\
\hline
\end{tabular}\par
\par
\endgroup
\clearpage

\tableofcontents
\clearpage

\section{Incident packet and measured OAM}\label{SM:sec:preparation}
\subsection{Generalities}\label{SM:sec:incident_state}
{We specify the incident vortex state and the observables that distinguish rotation of its mean OAM from broadening of its distribution. At the magnet entrance, the packet has rms transverse radius $R_\perp$ and longitudinal length $R_\parallel$. The coordinates $x,y,z$ are measured from the packet center in its rest frame, with $y$ along the incident motion. {For the reference state, we place the waist at the entrance, where} the diagonal position--momentum covariances vanish [Eq.~\eqref{SM:eq:input}] and both rms sizes are minimal under free propagation.

Free spreading can increase a nanometer-scale transverse packet size to micrometer scales, as described by the generalized van Cittert--Zernike relation~\cite{SM:Karlovets2021}. Here $R_\perp$ and $R_\parallel$ are independent entrance-waist parameters. {Free expansion introduces position--momentum correlations, so the resulting entrance state depends on the waist location, as treated explicitly in Sec.~\ref{SM:sec:offwaist}.}

We use the normalized Laguerre--Gaussian (LG) state with radial index zero, so there are no additional \mbox{radial nodes}:
}\begin{equation}
 \begin{aligned}
 \psi_0&=\frac{[(x-i\operatorname{sgn}(\ell)z)/w]^{|\ell|}}
                 {\sqrt{\pi |\ell|!}\,w}\\
 &\quad\times\frac{\exp[-(x^2+z^2)/(2w^2)-y^2/(4R_\parallel^2)]}
 {(2\pi R_\parallel^2)^{1/4}},
 \\[-1pt]&\quad w=\frac{R_\perp}{\sqrt{|\ell|+1}}.
 \end{aligned}
 \label{SM:eq:LG}
\end{equation}
For the azimuth $\phi$ about $y$, measured from $z$ toward $x$, its angular factor is $e^{i\ell\phi}$ up to a constant phase. Hence $\hat{\mathcal L}_y=-i\hbar\partial_\phi$ acts as $\hat{\mathcal L}_y\psi_0=\ell\hbar\psi_0$.
After integration over $y$ and azimuth, the probability in a thin transverse ring is $u^{|\ell|}e^{-u}\dd u/|\ell|!$, where $u=(x^2+z^2)/w^2$. {The normalization is $\int |\psi_0|^2\,\dd^3r=1$. The radial distribution gives}
\[
\begin{aligned}
\mean{(\delta\hat x)^2+(\delta\hat z)^2}_0&=(|\ell|+1)w^2,\\
\mean{[(\delta\hat x)^2+(\delta\hat z)^2]^2}_0&=(|\ell|+1)(|\ell|+2)w^4.
\end{aligned}
\]

{We propagate canonical moments by linear phase-space transport following Bagrov et al.~\cite{SM:BagrovBelovKondrateva1994,SM:BagrovBelovTrifonov2005,SM:BagrovBelovTrifonov2006}. This method is based on the trajectory-coherent approach~{\cite{SM:BagrovBelovMaslov1989,SM:BagrovBelovTernov1983,SM:BagrovBelovTrifonov1993,SM:BagrovBelovTrifonov1996}}, developed from Maslov's complex-germ method~\cite{SM:Maslov1994}. The packet center follows a quasiclassical relativistic Lorentz-force trajectory. For nonrelativistic internal motion, we evolve the quantum state in the center's instantaneous rest frame with a Hamiltonian expanded to second order in coordinate and momentum deviations [Eq.~\eqref{SM:eq:SMexpansion}].}

At each physical crossing of a finite field step at the magnet boundary, the gauge-invariant quantities -- position and kinetic momentum, {together with their finite moments} -- are continuous. Sec.~\ref{SM:sec:boundary} derives their synchronization at a common rest-frame time [Eq.~\eqref{SM:eq:SMcrossing}], while Eq.~\eqref{SM:eq:B} converts canonical moments to kinetic ones. The mean OAM and its covariance require second and fourth moments, respectively [Eqs.~\eqref{SM:eq:mean_trace} and \eqref{SM:eq:V}]. We collect the six centered canonical operators and their symmetrized second moments in
\begin{equation}
 \begin{gathered}
 \delta\hat Z=(\delta\hat x,\delta\hat y,\delta\hat z,
              \delta\hat p_x,\delta\hat p_y,\delta\hat p_z)^T,\\
 \Sigma_{AB}=\tfrac12\mean{\{\delta\hat Z_A,\delta\hat Z_B\}},
 \qquad A,B=1,\ldots,6.
 \end{gathered}\label{SM:eq:covariance_definition}
\end{equation}
{Its position and momentum blocks are
\[
 \Sigma=\begin{pmatrix}\Sigma_{rr}&\Sigma_{rp}\\
                  \Sigma_{rp}^T&\Sigma_{pp}\end{pmatrix},\qquad
 \begin{aligned}
 (\Sigma_{rr})_{ij}&=\mean{\delta\hat r_i\delta\hat r_j},\\
 (\Sigma_{rp})_{ij}&=\tfrac12\mean{\{\delta\hat r_i,\delta\hat p_j\}},\\
 (\Sigma_{pp})_{ij}&=\mean{\delta\hat p_i\delta\hat p_j},
 \end{aligned}
\]
where $i,j=x,y,z$. The diagonal entries of $\Sigma_{rr}$ and $\Sigma_{pp}$ {give the squared rms sizes and momentum spreads, while their off-diagonal entries correlate different directions.} The mixed block $\Sigma_{rp}$ contains both expansion and rotation: its diagonal entries describe position--momentum correlations along each axis, while its antisymmetric part fixes the mean canonical OAM. For the kinetic OAM, we instead use $(\Sigma_{r\pi})_{ij}=\mean{\{\delta\hat r_i,\delta\hat\pi_j\}}/2$, obtained through Eq.~\eqref{SM:eq:B}, and
\[
 \mean{\hat{\bm{\mathcal L}}}=
 \begin{pmatrix}
 (\Sigma_{r\pi})_{yz}-(\Sigma_{r\pi})_{zy}\\
 (\Sigma_{r\pi})_{zx}-(\Sigma_{r\pi})_{xz}\\
 (\Sigma_{r\pi})_{xy}-(\Sigma_{r\pi})_{yx}
 \end{pmatrix}.
\]
For the incident field-free state, $\Sigma_{r\pi}=\Sigma_{rp}$.} Rotational symmetry makes the $x$ and $z$ variances equal. The momentum moments follow by applying $-i\hbar\nabla$ to Eq.~\eqref{SM:eq:LG}, while the OAM eigenvalue fixes the two \mbox{angular correlations}:
\begin{align}
 \Sigma_{xx}=\Sigma_{zz}&=R_\perp^2/2,&\Sigma_{yy}&=R_\parallel^2,\nonumber\\
 \Sigma_{p_xp_x}=\Sigma_{p_zp_z}&=\frac{(|\ell|+1)\hbar^2}{2w^2},&
 \Sigma_{p_yp_y}&=\frac{\hbar^2}{4R_\parallel^2},\nonumber\\
 \Sigma_{zp_x}&=\ell\hbar/2,&\Sigma_{xp_z}&=-\ell\hbar/2.
 \label{SM:eq:input}
\end{align}
Eq.~\eqref{SM:eq:input} gives the incident covariance $\Sigma_0$ before the magnet. Symmetry fixes the entries with exchanged indices, and all remaining entries vanish. {Entrance matching changes the momentum and position--momentum moments [Eq.~\eqref{SM:eq:SMfaces}]. Propagation inside the magnet starts from these matched moments, from which the mean OAM in Eq.~\eqref{SM:eq:SMcovariant} is reconstructed.}

\subsection{Relative OAM width}\label{SM:sec:preparation_guide}
{To quantify OAM rotation and broadening, we use the intrinsic kinetic OAM and its covariance,}
\begin{equation}
 \hat{\bm{\mathcal L}}=\delta\hat{\vct r}\times\delta\hat{\vct\pi},
 \qquad V_{ij}=\tfrac12\mean{\{\delta\hat{\mathcal L}_i,\delta\hat{\mathcal L}_j\}},
 \label{SM:eq:OAM_covariance}
\end{equation}
where $\delta\hat{\mathcal L}_i=\hat{\mathcal L}_i-\mean{\hat{\mathcal L}_i}$ and $i,j$ label the three spatial axes. For a nonzero mean, the relative width and tilt are
\begin{equation}
 \begin{gathered}
 \vct n=\frac{\mean{\hat{\bm{\mathcal L}}}}{|\mean{\hat{\bm{\mathcal L}}}|},\qquad
 (\Delta\mathcal L_\parallel)^2=\vct n^TV\vct n,\\
 W=\frac{\Delta\mathcal L_\parallel}{|\mean{\hat{\bm{\mathcal L}}}|},\qquad
 \alpha_L=\arccos(\vct n\cdot\vct e_t).
 \end{gathered}
 \label{SM:eq:width_angle}
\end{equation}
Here $\vct n$ is the actual mean-OAM direction and $\vct e_t$ the unit vector along the laboratory reference momentum, expressed in the comoving axes. {The subscript $\parallel$ on $\Delta\mathcal L_\parallel$ refers to the mean OAM, whereas that on $R_\parallel$ refers to the incident momentum. Thus $W$ is the rms OAM spread along $\vct n$ divided by the mean magnitude, and $\alpha_L$ is the angle between the mean OAM and the direction of motion.} Since OAM is quadratic in position and momentum, calculating $V$ requires fourth moments of these variables. An OAM spectrum measured along the outgoing tangent uses the same projection axis as $W$, up to sign, only when the mean OAM is collinear with the motion.

\subsection{Intrinsic moments and an instrument with one fixed axis}\label{SM:sec:measurement}
{To relate the intrinsic OAM in Eq.~\eqref{SM:eq:OAM_covariance} to a fixed-axis measurement, we must specify how the packet centers and observation times are combined. For a single packet after exit, all operators refer to one common time in its rest frame, with $r_t$ along the outgoing tangent. For an angular ensemble, we express every packet in the inertial rest frame of the nominal outgoing orbit. Within each packet, the operators are synchronized on the time plane through its center's exit event. These events generally have different observer times $T_\lambda$. {The ensemble average therefore combines intrinsic moments at exit, before propagation to a common observer time.}

}The packets in a mixture are labeled by $\lambda$, with weights $w_\lambda\geq0$ and $\sum_\lambda w_\lambda=1$. Their intrinsic mean OAM vectors and covariances, expressed in the same observer axes, are $\mean{\hat{\bm{\mathcal L}}}_\lambda$ and $V_\lambda$. The mixture mean is $\mean{\hat{\bm{\mathcal L}}}=\sum_\lambda w_\lambda\mean{\hat{\bm{\mathcal L}}}_\lambda$, and its total covariance is
\begin{equation}
 \begin{aligned}
 V_{\rm total}&=\sum_\lambda w_\lambda V_\lambda\\
 &\quad+\sum_\lambda w_\lambda
 \bigl(\mean{\hat{\bm{\mathcal L}}}_\lambda-\mean{\hat{\bm{\mathcal L}}}\bigr)
 \bigl(\mean{\hat{\bm{\mathcal L}}}_\lambda-\mean{\hat{\bm{\mathcal L}}}\bigr)^T.
 \end{aligned}
 \label{SM:eq:mixture}
\end{equation}
{The first term in Eq.~\eqref{SM:eq:mixture} is the mean of the individual covariances, while the second is the covariance of the individual mean OAM vectors. In the angular scans, both refer to separately centered packets, so the result excludes the extrinsic OAM of their displaced centers. We insert the combined mean and $V_{\rm total}$ into Eq.~\eqref{SM:eq:width_angle}. Averaging the individual values of $W$ would omit the spread of the mean vectors.

}For a fixed-axis measurement, we choose one position $\vct r_{\rm axis}$ and one reference momentum $\vct\pi_{\rm axis}$ at a common time in one inertial frame after leaving the magnet. For each packet with center $\vct r_c$ and mean momentum $\vct\pi_c$, we write $\vct R=\vct r_c-\vct r_{\rm axis}$ and $\Delta\vct\pi=\vct\pi_c-\vct\pi_{\rm axis}$. Here $\hat{\vct\pi}=\hat{\vct p}$ in the field-free region. The corresponding operator is
\begin{equation}
 \begin{aligned}
 \hat{\bm{\mathcal L}}_{\rm det}
 &= (\hat{\vct r}-\vct r_{\rm axis})\times
                    (\hat{\vct\pi}-\vct\pi_{\rm axis})\\
 &=\delta\hat{\vct r}\times\delta\hat{\vct\pi}
   +\vct R\times\Delta\vct\pi
   +\vct R\times\delta\hat{\vct\pi}
   +\delta\hat{\vct r}\times\Delta\vct\pi .
 \end{aligned}\label{SM:eq:detector}
\end{equation}
For ordinary OAM about a fixed axis in that frame, $\vct\pi_{\rm axis}=0$. {The last two terms} have zero mean within a centered packet, but generally nonzero variance. For a waist LG state under linear transport, inversion symmetry makes all centered third moments vanish. {The quadratic--linear cross term therefore vanishes, and the covariance of these linear terms adds directly to the intrinsic covariance.} Their contribution is positive semidefinite, whereas other incident states can also have third-moment cross terms.

{Two displaced free packets illustrate how the choice of axis changes the measured width.} Two otherwise identical waist LG packets are translated by $+d$ and $-d$ along $x$, keeping the same longitudinal mean momentum. Each has zero intrinsic variance of $\mathcal L_y$. About their common $y$ axis, however, the additional operator is $\mp d\,\delta\hat\pi_z$, and
\[
 (\Delta \mathcal L_y)^2_{\rm det}
 =d^2\mean{(\delta\hat\pi_z)^2}_0
 =\frac{d^2(|\ell|+1)^2\hbar^2}{2R_\perp^2}.
\]
For $\ell=64$, $R_\perp=3~\mu$m, and $d=1~\mu$m, the relative rms width about the common axis is $\Delta\mathcal L_{y,\rm det}/(|\ell|\hbar)=23.9\%$, although each packet has intrinsic $W=0$ [Eq.~\eqref{SM:eq:width_angle}]. This width arises from the chosen common axis, while both packets remain pure.

{Electron OAM diagnostics include spatial mode separation~\cite{SM:GrilloSpectrum2017} and proposed triangular-aperture measurements at MeV energies~\cite{SM:MaksimovDiagnostic2025}. Applying them to deformed or mixed outgoing states requires a calibrated detector response accounting for the OAM projection distribution, packet displacement, mean momentum, and acceptance. The windows in {Fig.~\Lref{fig:windows}(c) of the Paper} and Sec.~\ref{SM:sec:fixed_checks} {describe intrinsic transport. Comparison with a fixed-axis measurement must include Eq.~\eqref{SM:eq:detector}, or use a protocol conditioned on each packet's center and mean motion.} For a bunch with the parameters of Ref.~\cite{SM:JINRphoto}, the measured OAM spectrum also depends on the quantum states of the individual electrons and the distribution of their packet centers and mean momenta.}

\section{Rest-frame Hamiltonian and covariance evolution}\label{SM:sec:hamiltonian}
\label{SM:sec:moved_derivations}
\subsection{What the coordinates describe}\label{SM:sec:moved_fw}
{We describe internal motion on planes of simultaneous events in the packet center's instantaneous rest frame, orthogonal to its four-velocity. The coordinates $(\tau,\vct r)$ give the center's proper time and the displacement from it along locally nonrotating Fermi--Walker axes~\cite{SM:Roura,SM:OST}.}

{{Along the center's worldline} $X^\mu(\tau)$, the dimensionless four-velocity $u^\mu(\tau)$ and three spatial axes $e_i^\mu(\tau)$ form an orthonormal tetrad. In the inertial rest frame at a given event, their components are
\[
 \begin{aligned}
 u^\mu&=(1,0,0,0),&e_1^\mu&=(0,1,0,0),\\
 e_2^\mu&=(0,0,1,0),&e_3^\mu&=(0,0,0,1).
 \end{aligned}
\]
Their laboratory components vary with $\tau$. They satisfy $u\cdot u=1$, $u\cdot e_i=0$, $e_i\cdot e_j=-\delta_{ij}$, and the FW transport law $\dot e_i^\mu=u^\mu a'_i/c$, where $a'_i(\tau)$ are the center's proper-acceleration components. The coordinate transformation is
\[
 x^\mu(\tau,\vct r)=X^\mu(\tau)+e_i^\mu(\tau)r_i.
\]
Differentiating both the worldline and the basis gives
\[
 \begin{aligned}
 \dd x^\mu&=(c\,\dd t_{\rm lab},\dd X,\dd Y,\dd Z)\\
 &=\big(cu^\mu+\dot e_i^\mu r_i\big)\dd\tau+e_i^\mu\dd r_i\\
 &=c\left(1+\frac{\vct a'\cdot\vct r}{c^2}\right)u^\mu\dd\tau
       +e_i^\mu\dd r_i.
 \end{aligned}
\]
Thus the factor $\mathcal N(\tau,\vct r)$ comes from the changing orientation of the simultaneous planes. Contracting with the Minkowski metric and using tetrad orthogonality removes the mixed products:
\begin{equation}
 \begin{aligned}
 \dd s^2&=c^2\dd t_{\rm lab}^2-\dd X^2-\dd Y^2-\dd Z^2\\
 &=\mathcal N^2(\tau,\vct r)c^2\dd\tau^2-\dd\vct r^2,\\
 \mathcal N(\tau,\vct r)&=1+\frac{\vct a'(\tau)\cdot\vct r}{c^2}.
 \end{aligned}
 \label{SM:eq:SMmetric}
\end{equation}
For a clock at fixed $\vct r$, in the coordinate region \mbox{$\mathcal N(\tau,\vct r)>0$}, its local proper-time increment is
\[
 \dd t_{\rm loc}=\frac{\dd s}{c}=\mathcal N(\tau,\vct r)\dd\tau,
 \qquad \dd t_{\rm loc}\big|_{\vct r=0}=\dd\tau.
\]
A finite packet therefore samples different local clock rates relative to the center's common evolution time $\tau$~\cite{SM:Roura}.}

\begin{figure}[t]
 \includegraphics[width=\textwidth]{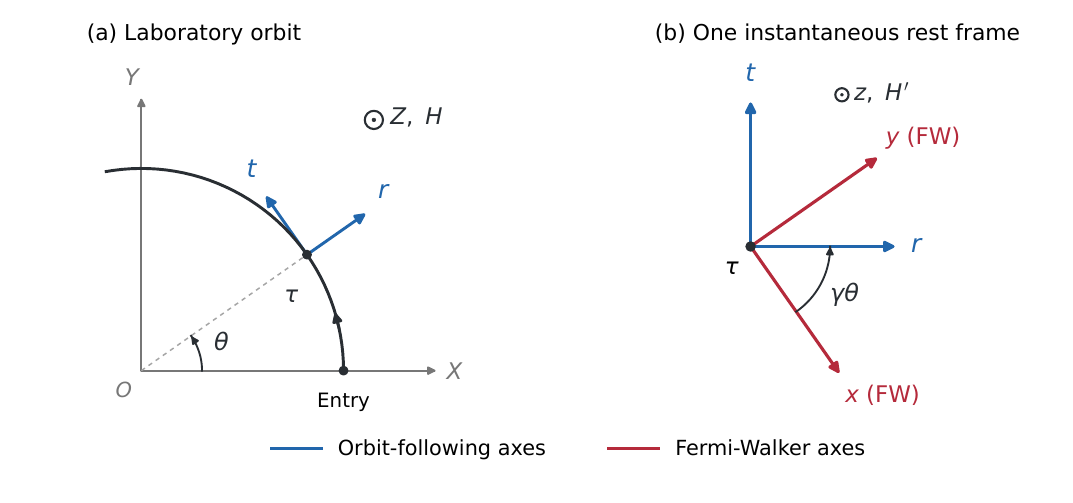}
 \caption{Laboratory, orbital, and Fermi--Walker (FW) axes. (a) The packet center follows a circular laboratory orbit about $O$, advancing through bend angle $\theta$ from entry to the event at proper time $\tau$. The fixed laboratory axes are $X,Y,Z$, while the blue axes point along the local radius $r$ and tangent $t$. The arrow on the orbit gives the direction of motion. (b) The orbital and FW axes at that event, drawn in the same instantaneous rest-frame time slice of Eq.~\eqref{SM:eq:SMmetric}. The two bases coincide at entry. Thereafter the orbital basis turns through $\gamma\theta$ relative to FW transport, where $\gamma$ is the center's Lorentz factor (Sec.~\ref{SM:sec:dipole}). The red $x,y$ labels denote FW axes, while the transfer and density plots use coordinates along $r,t,H$. Their common field direction is $z=H$. Circled dots indicate $H$ or $H'=\gamma H$ toward the reader.}
 \label{SM:fig:coordinate_frames}
\end{figure}

The FW axes have zero spatial angular velocity in the instantaneous rest frame: their derivatives $\dot e_i^\mu=u^\mu a'_i/c$ have only a time component there. By contrast, the orbital axes turn with the radius and tangent (Fig.~\ref{SM:fig:coordinate_frames}). For a circular orbit their angle relative to the FW axes is $\gamma\theta$. If $\hat U_R=\exp[-i\gamma\theta\hat L_H/\hbar]$ rotates the basis, the orbital wave function is $\psi_{\rm orb}=\hat U_R^\dagger\psi_{\rm FW}$ and
\[
 \hat{\mathcal H}_{\rm orb}=\hat U_R^\dagger\hat{\mathcal H}_2^{(\mathrm{FW})}\hat U_R
             -\gamma\dot\theta\hat L_H.
\]
The last term follows from differentiating the rotating basis. The acceleration potential instead follows from the position dependence of $\mathcal N(\tau,\vct r)$ [Eq.~\eqref{SM:eq:SMmetric}] and the transformed electric potential $\Phi'(\tau,\vct r)$, derived in Eq.~\eqref{SM:eq:SMpotential}. Eq.~\eqref{SM:eq:Horb} gives the explicit orbital Hamiltonian.

\subsection{Relativistic Hamiltonian and internal expansion}
{{The magnetic force turns the center while preserving its energy.} To describe the internal motion, we first distinguish this observable energy from the canonical Hamiltonian. For an electron, the laboratory energy is
\[
 \varepsilon_{\rm lab}=\sqrt{m^2c^4+c^2\vct\pi_{\rm lab}^2},
 \qquad \vct\pi_{\rm lab}=\vct p_{\rm lab}+e\vct A_{\rm lab}.
\]
It depends on the kinetic momentum and is gauge invariant. The Hamiltonian generating laboratory-time evolution is
\begin{equation}
 \mathcal H_{\rm lab}=\varepsilon_{\rm lab}-e\Phi_{\rm lab}.
 \label{SM:eq:SMlabroot}
\end{equation}
For a static bending magnet with no laboratory electric field, we can choose $\Phi_{\rm lab}=0$, so $\mathcal H_{\rm lab}=\varepsilon_{\rm lab}$. Expanding this energy as $mc^2+\vct\pi_{\rm lab}^2/(2m)$ would require $|\vct\pi_{\rm lab}|/(mc)\ll1$, whereas the reference momentum has $|\vct\pi_{{\rm lab},0}|/(mc)=\gamma\beta$. We instead expand about the relativistic center after changing coordinates.

In the FW frame, a local observer at fixed $\vct r$ measures the energy
\begin{equation}
 \begin{aligned}
 \varepsilon_{\rm loc}(\vct\pi)&=\sqrt{m^2c^4+c^2\vct\pi^2},\\
 \vct\pi(\tau,\vct r,\vct p)&=\vct p+e\vct A'(\tau,\vct r).
 \end{aligned}\label{SM:eq:SMlocal_energy}
\end{equation}
The clock relation in Eq.~\eqref{SM:eq:SMmetric} converts this local energy to the generator of evolution in the common time $\tau$:
\begin{equation}
 \begin{aligned}
 \mathcal H(\tau,\vct r,\vct p)
 &=\mathcal N(\tau,\vct r)\varepsilon_{\rm loc}(\vct\pi)-e\Phi'(\tau,\vct r)\\
 &=\mathcal N(\tau,\vct r)\sqrt{m^2c^4+c^2\vct\pi^2}-e\Phi'(\tau,\vct r).
 \end{aligned}
 \label{SM:eq:SMroot}
\end{equation}
This is Eq.~(4.34) of Ref.~\cite{SM:OST}, Sec.~IV B, with $V=\mathcal N$, Euclidean spatial metric, zero rotation, and charge $q=-e$. The gauge function $\chi(\tau,\vct r)$ generates $\vct A'\to\vct A'+\nabla\chi$, $\Phi'\to\Phi'-\partial_\tau\chi$, and $\vct p\to\vct p-e\nabla\chi$. Hence $\vct\pi$ and $\varepsilon_{\rm loc}$ are unchanged, while the classical Hamiltonian changes as $\mathcal H\to\mathcal H+e\partial_\tau\chi$. The Hamiltonian is therefore gauge dependent, with $\partial_\tau\chi$ taken at fixed $\vct r$. Along a trajectory, the accompanying momentum shift gives $\vct p\cdot\dot{\vct r}-\mathcal H\to\vct p\cdot\dot{\vct r}-\mathcal H-e\,\dd\chi/\dd\tau$, so the action changes only by endpoint terms. The inertial term $m\vct a'\cdot\vct r$ in Eq.~\eqref{SM:eq:SMroot} already follows from $mc^2\mathcal N$.}

\subsection{Quadratic term in the scalar potential}\label{SM:sec:potentials}
{We choose the symmetric gauge $\vct A'(\tau,\vct r)=\vct H'(\tau)\times\vct r/2$ and set $\Phi'(\tau,0)=0$. The scalar potential in Eq.~\eqref{SM:eq:SMroot} must reproduce the measured electric field in the accelerated coordinates $\tilde x^a=(c\tau,r_1,r_2,r_3)$, with $a,b=0,1,2,3$. The coordinate transformation above gives
\[
 \frac{\partial x^\mu}{\partial \tilde x^0}=\mathcal N u^\mu,
 \qquad \frac{\partial x^\mu}{\partial \tilde x^i}=e_i^\mu,
\]
and the covariant field components transform as
\[
 F^{\rm FW}_{ab}(\tilde x)=
 \frac{\partial x^\mu}{\partial \tilde x^a}
 \frac{\partial x^\nu}{\partial \tilde x^b}F^{\rm lab}_{\mu\nu}(x(\tilde x)).
\]
The fields $\vct E'(\tau),\vct H'(\tau)$ are measured along unit axes by a local observer instantaneously at rest in these coordinates. The coordinate time basis has length $\mathcal N$, whereas the spatial basis vectors already have unit length [Eq.~\eqref{SM:eq:SMmetric}]. Thus the components of the same field tensor in the coordinate basis are
\[
 F^{\rm FW}_{0i}=\frac{\mathcal N E'_i}{c},\qquad
 F^{\rm FW}_{ij}=-\epsilon_{ijk}H'_k.
\]
{The factor $\mathcal N$ converts the observer's local time to the common time $\tau$ and multiplies the electric components because they carry one time index.} With the four-potential $A^{\rm FW}_a=(\Phi'/c,-\vct A')$ and $F_{ab}=\partial_a A_b-\partial_b A_a$, where $\partial_a=\partial/\partial\tilde x^a$, this gives
\begin{equation}
 -\nabla\Phi'-\partial_\tau\vct A'=\mathcal N\vct E',
 \qquad \nabla\times\vct A'=\vct H'.
\end{equation}
For a constant laboratory field tensor, differentiating its projections on the FW basis gives
\begin{equation}
 \frac{\dd\vct H'}{\dd\tau}=-\frac{\vct a'\times\vct E'}{c^2}=0.
 \label{SM:eq:SMfield_rate}
\end{equation}
Since $m\vct a'=-e\vct E'$, $\partial_\tau\vct A'=0$ in the chosen gauge [Eq.~\eqref{SM:eq:SMfield_rate}] and $\nabla\times(\mathcal N\vct E')=0$. At fixed $\tau$, this condition makes the integral independent of the path within the coordinate region \mbox{$\mathcal N(\tau,\vct r)>0$}. We choose the straight segment $\eta\vct r$, with $\eta=0$ at the center and $\eta=1$ at the observation point $\vct r$. Using $\Phi'(\tau,0)=0$ and the line element $\vct r\,\dd\eta$, we obtain
\begin{align}
 \Phi'(\tau,\vct r)&=-\int_0^1
 \left(1+\eta\frac{\vct a'\cdot\vct r}{c^2}\right)
 (\vct E'\cdot\vct r)\dd\eta\nonumber\\
 &=-\vct E'\cdot\vct r
 -\frac{(\vct a'\cdot\vct r)(\vct E'\cdot\vct r)}{2c^2}.
 \label{SM:eq:SMpotential}
\end{align}
Differentiating at fixed $\tau$ gives
\begin{align}
 -\nabla\Phi'&=\vct E'+
 \frac{\vct a'(\vct E'\cdot\vct r)+\vct E'(\vct a'\cdot\vct r)}{2c^2}
 =\mathcal N\vct E',
 \label{SM:eq:SMpotential_gradient}
\end{align}
where the last equality uses $\vct a'\parallel\vct E'$. Eq.~\eqref{SM:eq:SMpotential} supplies the electric energy $-e\Phi'$ in the Hamiltonian~\eqref{SM:eq:SMroot}, while Eq.~\eqref{SM:eq:SMpotential_gradient} verifies that it reproduces the prescribed electric field. Combining this energy with $m\vct a'\cdot\vct r$ below cancels the linear terms and fixes the remaining quadratic acceleration potential. Eq.~\eqref{SM:eq:SMpotential} assumes a uniform laboratory field and the Lorentz-force reference orbit. For a fringe, its field gradients enter the local field and orbit equations of Sec.~\ref{SM:sec:fringes}.}

\subsection{Cancellation and quadratic approximation}\label{SM:sec:quadratic_internal}
{{We combine the electric and inertial terms} to obtain the internal Hamiltonian. Expanding Eq.~\eqref{SM:eq:SMroot} in $|\vct\pi|/(mc)$ at fixed relativistic reference motion gives}
\begin{align}
 \mathcal H-mc^2
 &=\frac{\vct\pi^2}{2m}+m\vct a'\cdot\vct r-e\Phi'\nonumber\\
 &\quad+\frac{(\vct a'\cdot\vct r)\vct\pi^2}{2mc^2}
 -\frac{\vct\pi^4}{8m^3c^2}+\cdots .
\end{align}
Using Eq.~\eqref{SM:eq:SMpotential}, the terms linear in displacement cancel: $(m\vct a'+e\vct E')\cdot\vct r=0$. The quadratic scalar energy is
\begin{equation}
 \frac e{2c^2}(\vct a'\cdot\vct r)(\vct E'\cdot\vct r)
 =-\frac m{2c^2}(\vct a'\cdot\vct r)^2.
\end{equation}
{The Lorentz-force relation thus fixes both its sign and coefficient.}

Since $\vct A'$ is linear in position, $\vct\pi$ has degree one in the internal canonical variables. {We replace the classical deviations} $\vct r,\vct\pi$ by centered operators $\delta\hat{\vct r}=\hat{\vct r}-\mean{\hat{\vct r}}$ and $\delta\hat{\vct\pi}=\hat{\vct\pi}-\mean{\hat{\vct\pi}}$. Keeping terms of total degree two {gives the Hamiltonian below, with symmetric (Weyl) ordering} of position and momentum products~{\cite{SM:Hillery}}:
\begin{align}
 \hat{\mathcal H}_2^{(\mathrm{FW})}&=\frac{(\delta\hat{\vct\pi})^2}{2m}
 -\frac m{2c^2}(\vct a'\cdot\delta\hat{\vct r})^2,\nonumber\\
 \hat{\mathcal H}-mc^2&=\hat{\mathcal H}_2^{(\mathrm{FW})}
 +\frac{\{\vct a'\cdot\delta\hat{\vct r},(\delta\hat{\vct\pi})^2\}}{4mc^2}
 -\frac{(\delta\hat{\vct\pi})^4}{8m^3c^2}+\cdots .
 \label{SM:eq:SMexpansion}
\end{align}
The two displayed corrections have degrees three and four. {\hypertarget{def:internal_ratios}{}The corresponding dimensionless expansion parameters are $\eta_{\rm NR}=\sqrt{\mean{(\delta\hat{\vct\pi})^2}}/(mc)$ and $\eta_a=\sqrt{\mean{(\vct a'\cdot\delta\hat{\vct r})^2}}/c^2$ [Eq.~\eqref{SM:eq:validity_ratios}]. Both must remain much smaller than unity throughout propagation, and the packet must be narrow enough for the spatial field expansion to be accurate.} Taking $c\to\infty$ at fixed acceleration would remove the quadratic potential as well. That is a stronger limit than nonrelativistic internal motion about a relativistic orbit.

Ref.~\cite{SM:OST}, Sec.~III, relates the spin-independent quantization used above to the Foldy--Wouthuysen transformation, including orders zero and one in $\hbar$ before dropping spin. The quadratic Hamiltonian used here generates linear Heisenberg evolution. The present model omits spin, radiation, collective effects, and higher-order quantum or gradient corrections. Spin-dependent OAM transfer has been calculated perturbatively where the field rises or falls at a solenoid entrance or exit~{\cite{SM:GatalinaBaturin2026}}.

{Although the acceleration potential has negative curvature, the full radial dynamics, including the magnetic and basis-rotation terms, is oscillatory (Sec.~\ref{SM:sec:dipole}). The potential changes the packet's internal motion and correlations, while its mean force vanishes for a \mbox{centered packet}.}

\paragraph{Packet-size conditions at entrance.}\label{SM:sec:entry_size_validity}
{The validity conditions in Eq.~\eqref{SM:eq:validity_ratios} restrict the incident LG sizes $R_\perp,R_\parallel$ [Eq.~\eqref{SM:eq:LG}].}
\hypertarget{def:rho_compton}{}The Compton wavelength is $\lambda_C=\hbar/(mc)$ and the reference bending radius is $\rho=mc\sqrt{\gamma^2-1}/(eH)$, where $H$ is the uniform laboratory field. The moments in Eq.~\eqref{SM:eq:input} give
\begin{equation}
 \frac{\mean{(\delta\hat{\vct p})^2}_0}{m^2c^2}
 =\lambda_C^2\left[\frac{(|\ell|+1)^2}{R_\perp^2}
                   +\frac{1}{4R_\parallel^2}\right].
 \label{SM:eq:SMincident_momentum}
\end{equation}
At normal entry, the proper acceleration points radially and has magnitude
$|\vct a'|=(\gamma^2-1)c^2/\rho$. Boundary matching preserves the
radial variance $R_\perp^2/2$, so
\begin{equation}
 \frac{\sqrt{\mean{(\vct a'\cdot\delta\hat{\vct r})^2}_{\rm in}}}{c^2}
 =\frac{(\gamma^2-1)R_\perp}{\sqrt2\rho}.
 \label{SM:eq:SMentry_acceleration}
\end{equation}
The same matching adds $-e\gamma H\delta\hat y$ to the radial kinetic
momentum [Eq.~\eqref{SM:eq:SMcrossing}]. Since the incident waist has
$\mean{\{\delta\hat y,\delta\hat p_x\}}_0=0$, its kinetic momentum
just \mbox{inside satisfies}
\begin{align}
 \frac{\mean{(\delta\hat{\vct\pi})^2}_{\rm in}}{m^2c^2}
 &=\lambda_C^2\left[\frac{(|\ell|+1)^2}{R_\perp^2}
                    +\frac{1}{4R_\parallel^2}\right]\nonumber\\
 &\quad+\left[\frac{\gamma\sqrt{\gamma^2-1}R_\parallel}{\rho}\right]^2.
 \label{SM:eq:SMentry_momentum}
\end{align}
The initial state and its normal-entry matching \mbox{therefore require}
\begin{equation}
 \begin{gathered}
 (|\ell|+1)\lambda_C\ll R_\perp
       \ll\frac{\sqrt2\rho}{\gamma^2-1},\\
 \frac{\lambda_C}{2}\ll R_\parallel
       \ll\frac{\rho}{\gamma\sqrt{\gamma^2-1}}.
 \end{gathered}
 \label{SM:eq:SMentry_size_bounds}
\end{equation}
The lower bounds keep the internal momentum nonrelativistic. The upper
bounds control the acceleration-size ratio $\eta_a$ and the entrance momentum ratio $\eta_{\rm NR}$, respectively [Eqs.~\eqref{SM:eq:SMentry_acceleration} and \eqref{SM:eq:SMentry_momentum}]. Subsequent evolution can impose stronger restrictions,
which we check by transporting the covariance and evaluating both ratios
in Eq.~\eqref{SM:eq:validity_ratios} along the whole trajectory. Field
variation and the vacuum-chamber aperture impose additional limits.

{For $|\ell|=64$, the lower scales in Eq.~\eqref{SM:eq:SMentry_size_bounds} are 25.1~pm for $R_\perp$ and {0.19}~pm for $R_\parallel$. In the same transverse--longitudinal order, the upper scales are $(173,115)$~mm at 1~MeV and 5~mT, and $(650,460)~\mu$m for the 25-MeV BNL $20^\circ$ magnet.} With $(R_\perp,R_\parallel)=(3,0.3)~\mu$m, the entrance ratios $(\eta_{\rm NR},\eta_a)$ [Eq.~\eqref{SM:eq:validity_ratios}] are $({8.8}\times10^{-6},{1.7}\times10^{-5})$ and $({6.5}\times10^{-4},{4.6}\times10^{-3})$, respectively. {The maxima of $\eta_{\rm NR}$ and $\eta_a$ along the trajectories are reported in Secs.~\ref{SM:sec:facilities}--\ref{SM:sec:linac}.}

\subsection{The same term from a laboratory expansion}\label{SM:sec:lab_expansion}
{To check the acceleration term independently, we expand the laboratory Hamiltonian, Eq.~\eqref{SM:eq:SMlabroot}, about the relativistic reference momentum.} Its first derivative gives the center's velocity. The quadratic terms in momentum deviations are
\[
 \frac{(\delta\pi_{{\rm lab},r})^2+(\delta\pi_{{\rm lab},H})^2}{2\gamma m}
 +\frac{(\delta\pi_{{\rm lab},t})^2}{2\gamma^3m}.
\]
Here the component subscripts $r,t,H$ denote the instantaneous radial, tangential, and field directions in the laboratory. The different longitudinal and transverse coefficients follow by differentiating $\sqrt{m^2c^4+c^2\vct\pi_{\rm lab}^2}$ twice. They account for the center's relativistic motion. Setting both coefficients to $1/(2m)$ at this stage would also make the center nonrelativistic.

For a circular orbit, express the laboratory deviations in axes turning at $eH/(\gamma m)$, using laboratory simultaneity and laboratory time. Subtracting the translation of the center and adding the rotation generator gives, in the symmetric gauge,
\begin{align}
 \hat{\mathcal H}_{2,{\rm lab}}
 &=\frac{(\delta\hat p_{{\rm lab},r}-eH\delta\hat r_{{\rm lab},t}/2)^2
            +(\delta\hat p_{{\rm lab},H})^2}{2\gamma m}\nonumber\\
 &\quad+\frac{(\delta\hat p_{{\rm lab},t}+eH\delta\hat r_{{\rm lab},r}/2)^2}{2\gamma^3m}\nonumber\\
 &\quad-\frac{eH}{\gamma m}
 (\delta\hat r_{{\rm lab},r}\delta\hat p_{{\rm lab},t}
 -\delta\hat r_{{\rm lab},t}\delta\hat p_{{\rm lab},r}).
 \label{SM:eq:labH2}
\end{align}

To compare this laboratory state with the rest-frame state, we must change the simultaneity plane. We consider a longitudinal deviation $\delta r_{{\rm lab},t}$ at laboratory time zero. After an interval $\Delta t$, its longitudinal coordinate is $\delta r_{{\rm lab},t}+v\Delta t$ to first order. The Lorentz \mbox{transformation gives}
\[
 t'=\gamma\left[\Delta t-\frac{\beta}{c}(\delta r_{{\rm lab},t}+v\Delta t)\right]
 =\frac{\Delta t}{\gamma}-\frac{\gamma\beta\delta r_{{\rm lab},t}}c.
\]
Hence the plane $t'=0$ is reached at $\Delta t=\gamma^2\beta\delta r_{{\rm lab},t}/c$. During this continuation the reference radial force $-evH$ changes the momentum by $-eH(\gamma^2-1)\delta r_{{\rm lab},t}$. Also, the longitudinal boost gives $\delta\pi_t'=\gamma(\delta\pi_{{\rm lab},t}-\beta\delta E_{\rm lab}/c)=\delta\pi_{{\rm lab},t}/\gamma$, using $\delta E_{\rm lab}=v\delta\pi_{{\rm lab},t}$. Combining these relations gives
\begin{align}
 \delta\hat r_{r}&=\delta\hat r_{{\rm lab},r},&
 \delta\hat r_{t}&=\gamma\delta\hat r_{{\rm lab},t},\nonumber\\
 \delta\hat\pi_r&=\delta\hat\pi_{{\rm lab},r}
 -eH(\gamma^2-1)\delta\hat r_{{\rm lab},t},&
 \delta\hat\pi_t&=\delta\hat\pi_{{\rm lab},t}/\gamma,
 \label{SM:eq:reslice}
\end{align}
The field-axis coordinate and momentum stay unchanged. We have obtained ordinary linear relations between trajectory deviations, which can now be promoted to operator identities. Thus the comparison of equal-time states produces a radial momentum shift as well as the longitudinal length transformation.

Next we use $\delta\hat{\vct p}=\delta\hat{\vct\pi}-e\vct A$ on the laboratory plane $t=0$ and on the rest-frame plane $t'=0$. Their symmetric-gauge potentials have $A_{{\rm lab},r}=-H\delta r_{{\rm lab},t}/2$, $A_{{\rm lab},t}=H\delta r_{{\rm lab},r}/2$, $A'_r=-\gamma H\delta r_t/2$, and $A'_t=\gamma H\delta r_r/2$. Substitution into Eq.~\eqref{SM:eq:reslice} gives the canonical transformation:
\begin{align}
 \delta\hat p_r&=\delta\hat p_{{\rm lab},r}
 -\frac{eH(\gamma^2-1)}2\delta\hat r_{{\rm lab},t},\nonumber\\
 \delta\hat p_t&=\frac{\delta\hat p_{{\rm lab},t}}\gamma
 -\frac{eH(\gamma^2-1)}{2\gamma}\delta\hat r_{{\rm lab},r}.
\end{align}
Together with the position relations it preserves $[\delta\hat r_i,\delta\hat p_j]=i\hbar\delta_{ij}$ and $[\delta\hat p_i,\delta\hat p_j]=0$, $i,j\in\{r,t,H\}$. Substitution of the inverse relations into Eq.~\eqref{SM:eq:labH2}, followed by multiplication by $\gamma$ because $\dd t=\gamma\dd\tau$ on the reference orbit, changes the evolution variable to proper time. The transformed Hamiltonian is
\begin{align*}
 \gamma\hat{\mathcal H}_{2,{\rm lab}}
 &=\frac{(\delta\hat p_r-\gamma eH\delta\hat r_t/2)^2}{2m}\\
 &\quad+\frac{(\delta\hat p_t+\gamma eH\delta\hat r_r/2)^2
 +(\delta\hat p_H)^2}{2m}\\
 &\quad-\frac{\gamma eH}{m}\hat L_H
 -\frac{(\gamma^2-1)e^2H^2}{2m}(\delta\hat r_r)^2,
\end{align*}
which is Eq.~\eqref{SM:eq:Horb}. Its last term equals $-m(\vct a'\cdot\delta\hat{\vct r})^2/(2c^2)$.
Removing the orbit-basis rotation then gives the FW Hamiltonian. Thus expanding about the relativistic laboratory orbit and changing the time slice gives the same quadratic result as transforming the exact classical Hamiltonian first. Both derivations keep the same order in internal deviations and compare all parts of the packet at the same specified time.
\subsection{Mean OAM and covariance evolution}\label{SM:sec:mean_equation}
{The rate $\dd\mean{\hat{\bm{\mathcal L}}}/\dd\tau$ of the packet's mean kinetic OAM is the mean torque about its center. Its magnetic and acceleration contributions depend on the position--momentum and position covariances [Eq.~\eqref{SM:eq:SMtorques}]. We therefore first propagate the symmetrized covariance $\Sigma$ defined in Eq.~\eqref{SM:eq:covariance_definition}.}

{For the phase-space vector $\delta\hat Z$ in Eq.~\eqref{SM:eq:covariance_definition}, we use} $J=\left(\begin{smallmatrix}0&I\\-I&0\end{smallmatrix}\right)$, with $I$ the $3\times3$ identity. Indices $A,B=1,\ldots,6$ label phase-space components, while $i,j,k=1,2,3$ label spatial components. The canonical commutators $[\delta\hat Z_A,\delta\hat Z_B]=i\hbar J_{AB}$ give the Heisenberg \mbox{equations directly}:
\begin{equation}
 \frac{\dd(\delta\hat Z)}{\dd\tau}=\frac{i}{\hbar}[\hat{\mathcal H}_2^{(\mathrm{FW})},\delta\hat Z],
 \qquad \delta\hat Z(\tau)=S(\tau)\delta\hat Z_0.
 \label{SM:eq:Sode}
\end{equation}
{Here $S(\tau)$ is the linear canonical transfer from the initial to the current state. In orbital axes, the uniform-dipole bulk matrix is given in Eq.~\eqref{SM:eq:bulkmap}, while the free-to-free transfer including both boundaries is Eq.~\eqref{SM:eq:map}. The component rotation relating the FW and orbital matrices is specified after Eq.~\eqref{SM:eq:map}.

The same $S$ governs classical deviations and Heisenberg operators because the Hamiltonian is quadratic. Applying it to the incident quantum moments propagates the packet's position and momentum uncertainties and correlations, while $SJS^T=J$ preserves the canonical commutators. {The moment hierarchy of Ref.~\cite{SM:BagrovBelovKondrateva1994}, Eq.~(12), therefore closes at each order. Section~\ref{SM:sec:wavefunction} gives the equivalent wave-function evolution.}} For the quadratic Hamiltonian, each operator factor follows the transfer $S$ in Eq.~\eqref{SM:eq:Sode} exactly, giving
\begin{equation}
 \begin{aligned}
 \Sigma(\tau)&=S(\tau)\Sigma_0S^T(\tau),\qquad SJS^T=J,\\
 \dot\Sigma(\tau)&=G(\tau)\Sigma(\tau)+\Sigma(\tau)G^T(\tau).
 \end{aligned}
 \label{SM:eq:covariance_transport}
\end{equation}
{{Here $\dot S(\tau)=G(\tau)S(\tau)$ in the bulk, with $S_{\rm bulk}(0)=I_6$. The in-field transfer from the free incident state starts from $C_{\rm in}$ [Eq.~\eqref{SM:eq:SMfaces}].} The matrix $C_H$ represents the cross product with the rest-frame field:
\[
 C_H\vct v=\vct H'\times\vct v,\qquad
 C_H=\begin{pmatrix}
 0&-H'_z&H'_y\\
 H'_z&0&-H'_x\\
 -H'_y&H'_x&0
 \end{pmatrix}.
\]
The symmetric-gauge FW generator is then}
\[
 G=\begin{pmatrix}
 eC_H/(2m)&I/m\\
 e^2C_H^2/(4m)+m\vct a'\vct a'^T/c^2&eC_H/(2m)
 \end{pmatrix}.
\]
{Eq.~\eqref{SM:eq:covariance_transport} propagates the complete covariance, including field-axis and cross-plane correlations, while preserving the canonical commutators.} Sec.~\ref{SM:sec:dipole} gives the transfer through a uniform dipole.

With the Levi-Civita symbol $\epsilon_{ijk}$, $\epsilon_{123}=1$, and Kronecker delta $\delta_{ij}$, the kinetic commutators are $[\delta\hat\pi_i,\delta\hat\pi_j]=-ie\hbar\epsilon_{ijk}H'_k$ and $[\delta\hat r_i,\delta\hat\pi_j]=i\hbar\delta_{ij}$. They give
\begin{equation}
 \dot{\delta\hat{\vct r}}=\delta\hat{\vct \pi}/m,\qquad
 \dot{\delta\hat{\vct \pi}}=\frac em\vct H'\times\delta\hat{\vct \pi}+
 \frac m{c^2}(\vct a'\cdot\delta\hat{\vct r})\vct a'.
\end{equation}
Products in the derivative of $\hat{\bm{\mathcal L}}$ are symmetrized. Equivalently, the Poisson brackets give the exact evolution of quadratic Weyl symbols under this Hamiltonian.

The position--momentum covariance is $(C_{r\pi})_{ij}=\mean{\{\delta\hat r_i,\delta\hat\pi_j\}}/2$, and its part symmetric in the spatial indices is $M_\pi=(C_{r\pi}+C_{r\pi}^T)/2$. The cross product $\delta\hat{\vct r}\times(\vct H'\times\delta\hat{\vct \pi})$ gives
\begin{equation}
 \frac{\dd\mean{\hat{\bm{\mathcal L}}}}{\dd\tau}
 =\frac em[(\tr C_{r\pi})\vct H'-C_{r\pi}^T\vct H']
 +\frac m{c^2}(\Sigma_{rr}\vct a')\times\vct a'.
\end{equation}
Because $(C_{r\pi})_{ij}-(M_\pi)_{ij}=\epsilon_{ijk}\mean{\hat{\mathcal L}_k}/2$,
$C_{r\pi}^T\vct H'=M_\pi\vct H'-\vct H'\times\mean{\hat{\bm{\mathcal L}}}/2$.
Substitution gives Eq.~(\Lref{eq:bmt}) of the Paper explicitly:
\begin{equation}
 \begin{aligned}
 \frac{\dd\mean{\hat{\bm{\mathcal L}}}}{\dd\tau}
 &=\frac e{2m}\vct H'\times\mean{\hat{\bm{\mathcal L}}}+\vct K_\pi+\vct K_a,\\
 \vct K_\pi&=\frac em[(\tr M_\pi)\vct H'-M_\pi\vct H'],\\
 \vct K_a&=\frac m{c^2}(\Sigma_{rr}\vct a')\times\vct a'.
 \end{aligned}\label{SM:eq:SMtorques}
\end{equation}
Here $(\Sigma_{rr})_{ij}=\mean{\{\delta\hat r_i,\delta\hat r_j\}}/2$. These expectation-value vectors are the torques in Eq.~(\Lref{eq:bmt}) of the Paper.

With {centered canonical OAM} $\hat{\vct L}=\delta\hat{\vct r}\times\delta\hat{\vct p}$, {the same magnetic contribution can be written as:}
\begin{align}
 \frac{(\delta\hat{\vct \pi})^2}{2m}
 &=\frac{(\delta\hat{\vct p})^2}{2m}+\frac e{2m}\vct H'\cdot\hat{\vct L}
 +\frac{e^2}{8m}(\vct H'\times\delta\hat{\vct r})^2,\nonumber\\
 \mean{\hat{\bm{\mathcal L}}}
 &=\mean{\hat{\vct L}}+\frac e2[(\tr\Sigma_{rr})\vct H'
 -\Sigma_{rr}\vct H'].\label{SM:eq:kinetic_canonical_oam}
\end{align}
The second line is the identity $\delta\hat{\vct r}\times(\vct H'\times\delta\hat{\vct r})$ averaged over the packet. This size-dependent diamagnetic contribution allows kinetic OAM to vary even in a uniform field~\cite{SM:Greenshields2014}. Differentiating the $A^2$ energy gives canonical torque $-e^2\vct H'\times(\Sigma_{rr}\vct H')/(4m)$. The full kinetic torque also includes the derivative of the kinetic--canonical OAM difference. {The gauge-invariant torque $\vct K_\pi$ therefore combines the $A^2$ contribution with the derivative of the kinetic--canonical OAM difference.} Because the final expressions involve positions, kinetic momenta, and physical fields, both torques in Eq.~\eqref{SM:eq:SMtorques} are gauge invariant. {The propagated covariance determines how these \mbox{torques evolve}.}

A four-vector equation makes the transformation between inertial observers explicit and allows comparison with relativistic mean-OAM precession~\cite{SM:Silenko2017,SM:Silenko2018,SM:Karlovets2026}. {For the covariant formulas below, we set $c=1$ and use $F^{i0}=E_i$, $F^{ij}=-\epsilon_{ijk}H_k$, and $\dot u^\mu=-eF^{\mu\nu}u_\nu/m$.} A spacelike OAM four-vector $a_\pi^\mu$ \cite{SM:Karlovets2026} satisfies $u\cdot a_\pi=0$ and has rest components $(0,\mean{\hat{\bm{\mathcal L}}})$. %
The expectation-value Pauli--Lubanski construction of Ref.~\cite{SM:Bliokh2026}, divided by the mean rest-frame energy, gives the same rest-frame intrinsic-angular-momentum representation for a free spinless packet when the same centroid is used. %

In our conventions, $F^{i\nu}a_{\pi\nu}=-(\vct H'\times\mean{\hat{\bm{\mathcal L}}})_i$ in the rest frame. The Bargmann--Michel--Telegdi (BMT) construction~{\cite{SM:BMT,SM:Karlovets2026}} combines the two field-dependent vectors $F^{\mu\nu}a_{\pi\nu}$ and $u^\mu u_\rho F^{\rho\nu}a_{\pi\nu}$. Rest-frame precession fixes the coefficient of the first to $-e/(2m)$. Differentiating $u\cdot a_\pi=0$ requires the sum of the two coefficients to be $-e/m$, so the second also equals $-e/(2m)$. The remaining torques have zero time components in the rest frame and are therefore orthogonal to $u^\mu$, {giving the covariant generalization of the BMT-like equation:}
\begin{equation}
 \dot a_\pi^\mu=-\frac e{2m}F^{\mu\nu}a_{\pi\nu}
 -\frac e{2m}u^\mu u_\rho F^{\rho\nu}a_{\pi\nu}
 +K_\pi^\mu+K_a^\mu.
 \label{SM:eq:SMcovariant}
\end{equation}
{The comoving spatial axes $e_i^\mu$ give $K_\pi^\mu=e_i^\mu K_{\pi i}$ and $K_a^\mu=e_i^\mu K_{ai}$ from Eq.~\eqref{SM:eq:SMtorques}. Both contributions are allowed by Lorentz covariance and parity: they are axial four-vectors orthogonal to $u^\mu$, as required for additional OAM torques. {Eq.~\eqref{SM:eq:SMtorques} expresses these torques through magnetic \mbox{position--momentum correlations} and the acceleration-dependent potential.}} The covariant equation already includes Thomas precession. %

The first term is orbital magnetic precession with gyromagnetic factor $g_L=1$, while the second keeps $u\cdot a_\pi=0$ as the rest frame accelerates and accounts for Thomas transport. {{The point-particle BMT equation for spin uses $g_s\simeq2$. The torques due to spatial covariances vanish in this point-particle limit.}} {For a finite packet, these torques depend on the evolving shape and \mbox{position--momentum correlations}.}

{We propagate $\Sigma$ with Eq.~\eqref{SM:eq:covariance_transport} and reconstruct $\mean{\hat{\mathcal L}_i}=\epsilon_{ijk}(C_{r\pi})_{jk}$. {The evolution described by Eq.~\eqref{SM:eq:SMcovariant} follows directly from this transfer: differentiating the reconstructed mean using Eq.~\eqref{SM:eq:covariance_transport} gives Eq.~\eqref{SM:eq:SMtorques} and its covariant form, Eq.~\eqref{SM:eq:SMcovariant}.} The generalized BMT equation is thus \textit{satisfied automatically} within our \mbox{quadratic model}.}

\subsection{Comparison with mean-OAM precession}\label{SM:sec:precession}
{Silenko \textit{et al.} derive laboratory-frame OAM precession to first order in the external fields~\cite{SM:Silenko2017,SM:Silenko2018}. The quantum equation uses canonical OAM [Eq.~(11) of Ref.~\cite{SM:Silenko2018}], whereas Eq.~\eqref{SM:eq:SMtorques} evolves kinetic OAM on instantaneous rest-frame slices. Comparing them requires the kinetic--canonical conversion and the change of simultaneity in Eq.~\eqref{SM:eq:reslice}. Their exact Foldy--Wouthuysen Hamiltonian [Eq.~(1) of Ref.~\cite{SM:Silenko2018}] contains the relativistic dynamics from which our acceleration potential follows after the internal expansion and frame transformation. Their weak-field reduction [Eq.~(3)] drops terms quadratic \mbox{in the field}, and the kinetic--canonical distinction is discussed after Eq.~(7). At fixed reference velocity, our potential $-m(\vct a'\cdot\delta\hat{\vct r})^2/(2c^2)$ [Eq.~\eqref{SM:eq:SMexpansion}] is also quadratic \mbox{in the field}.}

The mean-vector equation of Ref.~\cite{SM:Karlovets2026}, Eq.~(13) and Appendix B, includes Thomas precession, while Eq.~\eqref{SM:eq:SMcovariant} includes the additional torques of the present packet model. Its Appendix A derives the magnetic position--momentum correlations and neglects them when closing the mean equation. Here we propagate those correlations together with the acceleration potential in Eq.~\eqref{SM:eq:SMexpansion}. Thomas precession describes the rotation of successive local axes, whereas the scalar potential describes internal motion on successive accelerated \mbox{time slices}.

For a fixed bend angle, reducing the field also increases the transit time. In a uniform transverse dipole,
\[
 \frac{|\vct a'|^2}{c^2}=(\gamma^2-1)\left(\frac{eH}{m}\right)^2,
 \qquad \Delta\tau=\frac{m\theta}{eH},
\]
so $(|\vct a'|^2/c^2)(\Delta\tau)^2=(\gamma^2-1)\theta^2$ remains finite as $H$ decreases.

For the incident LG state [Eq.~\eqref{SM:eq:input}], the entrance shear [Eq.~\eqref{SM:eq:SMcrossing}] generates $\mean{\{\delta\hat r_t,\delta\hat\pi_r^+\}}/2=-eH'R_\parallel^2$. These evolving correlations contribute to the mean torque through Eq.~\eqref{SM:eq:SMtorques}, while the OAM covariance $V$ in Eq.~\eqref{SM:eq:V} further requires the fourth moments in Eq.~\eqref{SM:eq:LGfourth}.

Thomas--BMT transport and Siberian snakes control spin polarization~{\cite{SM:ManePolarized2005,SM:ManeSnakes2005}}. Silenko and Teryaev found a similar cancellation of a local field perturbation on successive passages within a mean-OAM model~{\cite{SM:SilenkoTeryaevSnake2019}}. {To test whether a finite packet preserves a narrow OAM distribution, we also propagate its spatial covariances and fourth moments [Eqs.~\eqref{SM:eq:SMtorques} and \eqref{SM:eq:V}].}

\section{Boundary matching and uniform-dipole transport}\label{SM:sec:transport}
{The Hamiltonian in Sec.~\ref{SM:sec:hamiltonian} determines evolution inside the magnet. {We obtain its entrance data} from the incident free packet and combine boundary matching with this evolution.}

\subsection{From crossing events to one time slice}\label{SM:sec:boundary}
{To construct those entrance data, we synchronize the crossing events on one rest-time slice. {We consider a packet entering normally through a sharp boundary and keep terms linear in deviations from the reference orbit.}}
Matching a free vortex packet to a longitudinal magnetic field with a sharp boundary can produce an LG state whose radius oscillates~\cite{SM:SizykhSolenoid2024}. Here the incident LG state $\psi_0$ [Eq.~\eqref{SM:eq:LG}] enters a magnetic field transverse to the center's motion, and the boundary calculation must also resolve {the different crossing times of its longitudinally separated parts.}

{Along a transmitted Lorentz trajectory, position and kinetic momentum are continuous at a finite magnetic-field step.} The Lorentz force stays finite while the crossing interval tends to zero, so its integral across the step vanishes: $\Delta\vct\pi=\lim_{\epsilon\to0}\int_{t_\times-\epsilon}^{t_\times+\epsilon}\vct F_{\rm L}\dd t=0$. {A finite field jump thus gives zero impulse in the vanishing crossing interval.} {Canonical momentum depends on the gauge, while the matching condition applies to kinetic momentum.} 

{With potentials finite and continuous at the field step, integration of the local Schr\"odinger equation across the boundary requires continuity of $\psi$ and $(-i\hbar\partial_y+eA'_y)\psi$. Here $y$ is normal to the entrance boundary. These conditions make the normal probability-current density $\operatorname{Re}[\psi^*(-i\hbar\partial_y+eA'_y)\psi]/m$ continuous at each crossing event. The exit conditions have the same form in the outgoing local axes.

To compare the free and in-field states at one rest-frame time, we continue them from their common crossing events using their respective Hamiltonians. Their equal-time phases and momentum moments can therefore differ while their boundary values agree at each crossing. See Ref.~\cite{SM:HwangLee2015} for dipole section maps and Ref.~\cite{SM:Roura}, Appendix A, for the rest-time planes.}

We obtain this continuation in the inertial rest frame at the center's entrance. At the order needed here, its Hamiltonian is $\hat{\mathcal H}=\hat{\vct\pi}^{\,2}/(2m)-e\Phi'$, where $\hat{\vct\pi}=\hat{\vct p}+e\vct A'$. This Hamiltonian describes motion in one inertial frame before subtracting the center motion. The Hamiltonian $\hat{\mathcal H}_2^{(\mathrm{FW})}$ in Eq.~\eqref{SM:eq:SMexpansion}, also Eq.~(\Lref{eq:H}) of the Paper, instead describes centered deviations in the accelerating FW frame: the uniform electric force is canceled by the center acceleration, while expanding the relativistic energy on successive rest-time slices leaves the quadratic acceleration potential derived in Sec.~\ref{SM:sec:hamiltonian}. For an operator with explicit time dependence, the Heisenberg equation contains both a commutator and a partial derivative. The commutators $[\hat r_i,\hat\pi_j]=i\hbar\delta_{ij}$ and $[\hat\pi_i,\hat\pi_j]=-ie\hbar\epsilon_{ijk}H'_k$ give
\begin{align}
 \frac{\dd\hat{\vct\pi}}{\dd t'}
 &=\frac{i}{\hbar}[\hat{\mathcal H},\hat{\vct\pi}]
   +\partial_{t'}\hat{\vct\pi}\nonumber\\
 &=-e\vct E'-\frac e{2m}
       (\hat{\vct\pi}\times\vct H'-\vct H'\times\hat{\vct\pi}).
 \label{SM:eq:SMlorentz}
\end{align}
The electric force includes $e\partial_{t'}\vct A'$, so that $\vct E'=-\nabla\Phi'-\partial_{t'}\vct A'$. For a locally uniform magnetic field, the last term reduces to $e\vct H'\times\hat{\vct\pi}/m$. This is the operator Lorentz force for charge $-e$. The constant electric force must be integrated before centering, because subtracting it at the outset would discard the effect of unequal crossing times.

In the inertial rest frame at the center's entrance, the laboratory boundary $Y=0$ moves as $y+vt'=0$, where $v$ is the center's laboratory speed. We first integrate the force along crossing trajectories labeled by the ordinary coordinate $y$. {Their crossing time $t'_\times(y)$ and momentum change $\Delta\pi_x(y)$ are ordinary functions of the trajectory label.} The transformed electric field is $E'_x=vH'$, so
\begin{align}
 t'_\times(y)&=-y/v,\nonumber\\
 \Delta\pi_x(y)&=\int_{t'_\times(y)}^{0}(-eE'_x)\dd t'=-eH'y,\nonumber\\
 \delta\hat\pi_x^+&=\delta\hat p_x^--eH'\delta\hat y.
 \label{SM:eq:SMcrossing}
\end{align}
Superscripts $-$ and $+$ denote the free incident and matched in-field states, both evaluated on the common entrance time slice. {The last line follows after subtracting the center's momentum.} For $y<0$, the same integral continues the in-field trajectory backwards from its later crossing to this reference slice. The magnetic force due to internal velocity contributes only at second order over an interval proportional to $y$. The uniform electric force disappears from the centered bulk equation, but its integral over unequal crossing times has already produced the linear momentum shear in Eq.~\eqref{SM:eq:SMcrossing}.

With $\vct A'(\delta\hat{\vct r})=\vct H'\times\delta\hat{\vct r}/2$, this yields
\begin{equation}
 \delta\hat p_x^+=\delta\hat p_x^--eH'\delta\hat y/2,\qquad
 \delta\hat p_y^+=\delta\hat p_y^--eH'\delta\hat x/2.
 \label{SM:eq:SMentry}
\end{equation}
{At this order the position operators are unchanged. The unitary matching operator can therefore be written as multiplication by a phase. The two canonical shifts in Eq.~\eqref{SM:eq:SMentry} fix its spatial derivatives:
\[
 \begin{aligned}
 \psi^+&=e^{i\chi_{\rm in}/\hbar}\psi^-,\\
 \nabla_{\delta r}\chi_{\rm in}
 &=-eH'\delta y\,\vct e_x-e\vct A'(\delta\vct r)\\
 &=-\tfrac12eH'(\delta y,\delta x,0).
 \end{aligned}
\]
Here $\delta\vct r=(\delta x,\delta y,\delta z)$ is measured from the packet center. The first term in the gradient is the kinetic shift in Eq.~\eqref{SM:eq:SMcrossing}, while the second subtracts the in-field vector potential, with zero potential in the free region. Integration fixes the phase up to an irrelevant constant:
\[
 \chi_{\rm in}=-\frac{eH'}2\delta x\delta y,\qquad
 \hat U_{\rm in}=\exp\!\left[-\frac{ieH'}{2\hbar}
                    \delta\hat x\delta\hat y\right].
\]
The same phase follows in the Schr\"odinger picture. In the inertial entrance rest frame, choose the in-field potentials $\vct A'=(0,H'\delta x,0)$ and $\Phi'=-vH'\delta x$, which give $E'_x=vH'$. At the retained order, evolution from $t'_\times=-\delta y/v$ to $t'=0$ contributes $\exp[(ie/\hbar)\int_{-\delta y/v}^{0}\Phi'\dd t']=\exp[-ieH'\delta x\delta y/\hbar]$. Transforming to the symmetric gauge adds the factor $\exp[ieH'\delta x\delta y/(2\hbar)]$, so their product is $\hat U_{\rm in}$. Writing Eq.~\eqref{SM:eq:SMentry} as $\delta\hat{\vct p}^{\,+}=\delta\hat{\vct p}^{\,-}+D\delta\hat{\vct r}^{\,-}$, together with $\delta\hat{\vct r}^{\,+}=\delta\hat{\vct r}^{\,-}$, gives the block matrices in the ordering of Eq.~\eqref{SM:eq:covariance_definition}:}
\begin{equation}
 \begin{gathered}
 C_{\rm in}=\begin{pmatrix}I&0\\D&I\end{pmatrix},\qquad
 C_{\rm out}=C_{\rm in}^{-1},\\
 D=-\frac{eH'}2\begin{pmatrix}0&1&0\\1&0&0\\0&0&0\end{pmatrix},\\
 {\Sigma^-=\Sigma_0\ \longmapsto\
 \Sigma^+=C_{\rm in}\Sigma_0C_{\rm in}^T.}
 \end{gathered}\label{SM:eq:SMfaces}
\end{equation}
{These matrices apply to normal incidence at each boundary. They act in local orbital axes, with the exit tangent along the outgoing reference momentum.} The symmetric shear block gives $C_{\rm in}JC_{\rm in}^T=J$. {The last line replaces the free-LG covariance $\Sigma_0$ of Eq.~\eqref{SM:eq:input} by the initial in-field covariance $\Sigma^+$. The bulk matrix then gives $\Sigma(\theta)=S_{\rm bulk}(\theta)\Sigma^+S_{\rm bulk}^T(\theta)$ [Eq.~\eqref{SM:eq:bulkmap}].} The kinetic variables obey a different commutator matrix on the two sides: $[\delta\hat\pi_x,\delta\hat\pi_y]=0$ outside and $-ie\hbar H'$ inside. The canonical shifts above preserve $J$ and reproduce the correct magnetic commutator after adding $e\vct A'$.

{Multiplication by $\hat U_{\rm in}$ preserves the density and position moments on the matching plane, while its phase gradient changes momentum and mixed moments. For the waist LG state in Eq.~\eqref{SM:eq:input}, Eq.~\eqref{SM:eq:SMcrossing} gives the explicit free-to-in-field changes
\[
 \begin{aligned}
 \tfrac12\mean{\{\delta\hat y,\delta\hat\pi_x\}}:&\quad 0\longmapsto-eH'R_\parallel^2,\\
 \mean{(\delta\hat\pi_x)^2}:&\quad\mean{(\delta\hat p_x)^2}_0\\
 &\quad\longmapsto\mean{(\delta\hat p_x)^2}_0+(eH')^2R_\parallel^2.
 \end{aligned}
\]
Both states refer to the common entrance time slice. The matched state $\psi^+=\hat U_{\rm in}\psi^-$ supplies \textit{the initial condition} in Eq.~\eqref{SM:eq:SMpsi}.}

For an incident LG state with zero field-axis OAM,
\begin{align}
 \mean{\hat{\mathcal L}_H}_{\rm in}
 &=\mean{\delta\hat x\delta\hat p_y-\delta\hat y(\delta\hat p_x-eH'\delta\hat y)}\nonumber\\
 &=eH'\mean{(\delta\hat y)^2}.\label{SM:eq:entrance_field_axis_mean}
\end{align}
{Eq.~\eqref{SM:eq:entrance_field_axis_mean} gives the field-axis mean OAM quoted after Eq.~(\Lref{eq:entry}) of the Paper. The same shear changes the tangential operator to $\hat{\mathcal L}_t^{(0)}-eH'\delta\hat y\delta\hat z$, where the superscript $(0)$ denotes the incident free operator before entrance matching. Since the incident LG state is an eigenstate of $\hat{\mathcal L}_t^{(0)}$ and its longitudinal and transverse densities factorize, Eqs.~\eqref{SM:eq:LG} and \eqref{SM:eq:input} give
\[
 \mean{\hat{\mathcal L}_t}_{\rm in}=\ell\hbar,\qquad
 (\Delta\mathcal L_t)^2_{\rm in}
 =(eH')^2R_\parallel^2R_\perp^2/2.
\]
Thus even an unchanged tangential mean can acquire a nonzero tangential variance. Here the projection is on the tangent, while $\Delta\mathcal L_\parallel$ in Eq.~\eqref{SM:eq:width_angle} follows the actual mean OAM.} Exit matching is inverse matching applied to the evolved state: $C_{\rm in}^{-1}S_{\rm bulk}C_{\rm in}$ in the oriented local bases. \MovieNumber{4} shows the evolution for the BNL example, while {Fig.~\Lref{fig:boundary_comparison} of the Paper} compares the two cases at a $5^\circ$ bend (Sec.~\ref{SM:sec:movie_comparison}). Oblique incidence changes the crossing times. The numerical scans keep the boundaries fixed and vary the incoming direction through the angles in Eq.~\eqref{SM:eq:two_incidence_angles}.
\subsection{Transfer through a uniform dipole}\label{SM:sec:dipole}
\label{SM:sec:moved_uniform}
To propagate the entrance-matched state $C_{\rm in}\delta\hat Z_0$ [Eq.~\eqref{SM:eq:SMfaces}], we use axes following the circular reference trajectory inside a uniform dipole. A displaced trajectory with the same momentum follows a circle of the same radius, so its radial separation oscillates as the bend advances. Motion along the field remains free. Orbit-following axes make these motions easier to see than the nonrotating FW axes.

We use the right-handed orbital axes $r,t,H$ and their position components $(r_r,r_t,r_H)$. Initially $r_{r}^{(0)}=x_0$, $r_{t}^{(0)}=y_0$, $r_{H}^{(0)}=z_0$, and the momentum components coincide in the same way. {The bend angle evolves as $\dot\theta=eH/m$ in proper time.} The basis rotation follows from the circular-orbit tetrad $u=\gamma(1,\beta\vct e_t)$, $e_r=(0,\vct e_r)$, $e_t=\gamma(\beta,\vct e_t)$, where the bold vectors are laboratory unit vectors. Differentiation gives
\[
 \dot e_r=-\gamma\beta\dot\theta\,u+\gamma\dot\theta\,e_t,
 \qquad \dot e_t=-\gamma\dot\theta\,e_r.
\]
The first term in $\dot e_r$ is its FW derivative, since $a'_r=-c\gamma\beta\dot\theta$. The remaining terms rotate the spatial basis at $\gamma\dot\theta$. Thus the axes rotate relative to FW transport at $\omega_c^\prime=\gamma eH/m=eH^\prime/m$, the rest-frame cyclotron frequency. The rotating Hamiltonian consequently contains $-\omega_c^\prime\hat L_H$, where $\hat L_H=\delta\hat r_r\delta\hat p_t-\delta\hat r_t\delta\hat p_r$ is the {centered canonical OAM} along the field. {This term accounts for the rotation of the coordinate axes used to describe the same orbit.}

With coincident bases at entry, the component rotation is
\begin{align*}
 \hat{\mathcal L}_r&=\cos(\gamma\theta)\hat{\mathcal L}_x
                  +\sin(\gamma\theta)\hat{\mathcal L}_y,\\
 \hat{\mathcal L}_t&=-\sin(\gamma\theta)\hat{\mathcal L}_x
                  +\cos(\gamma\theta)\hat{\mathcal L}_y,\\
 \hat{\mathcal L}_H&=\hat{\mathcal L}_z.
\end{align*}
Position and momentum triples transform by the same rotation. Thus
$r_r=x\cos(\gamma\theta)+y\sin(\gamma\theta)$,
$r_t=-x\sin(\gamma\theta)+y\cos(\gamma\theta)$, and $r_H=z$.
The two bases generally remain different even after a full orbit:
the accumulated relative angle is $2\pi N\gamma$.

With symmetric-gauge canonical momentum in the orbit-following frame $\delta\hat{\vct p}$, $H'=\gamma H$ and $|\vct a'|^2/c^2=(\gamma^2-1)(eH/m)^2$ yield
\begin{align}
 \hat{\mathcal H}_{\rm orb}
 &=\frac{(\delta\hat{\vct p})^2}{2m}
 -\frac{\gamma eH}{2m}(\delta\hat r_r\delta\hat p_t-\delta\hat r_t\delta\hat p_r)\nonumber\\
 &\quad+\frac{\gamma^2e^2H^2}{8m}((\delta\hat r_r)^2+(\delta\hat r_t)^2)\nonumber\\
 &\quad-\frac{(\gamma^2-1)e^2H^2}{2m}(\delta\hat r_r)^2.
 \label{SM:eq:Horb}
\end{align}
The first three terms combine symmetric-gauge kinetic energy and basis rotation. We use the same canonical variables throughout. Since $\dot\theta=eH/m$, changing from proper time to bend angle gives
\begin{equation}
 i\hbar\partial_\theta\psi
 =\hat{\mathcal K}_\theta\psi,\qquad
 \hat{\mathcal K}_\theta=\frac{m}{eH}\hat{\mathcal H}_{\rm orb}.
 \label{SM:eq:angleH}
\end{equation}
The generator has units of action. Its canonical equations are
\begin{align}
 \frac{\dd\delta\hat r_r}{\dd\theta}
 &=\frac{\delta\hat p_r}{eH}+\frac\gamma2\delta\hat r_t,&
 \frac{\dd\delta\hat p_r}{\dd\theta}
 &=\frac\gamma2\delta\hat p_t+\frac{eH}{4}\!(3\gamma^2-4)\delta\hat r_r,\nonumber\\
 \frac{\dd\delta\hat r_t}{\dd\theta}
 &=\frac{\delta\hat p_t}{eH}-\frac\gamma2\delta\hat r_r,&
 \frac{\dd\delta\hat p_t}{\dd\theta}
 &=-\frac\gamma2\delta\hat p_r-\frac{\gamma^2eH}{4}\delta\hat r_t,\nonumber\\
 \frac{\dd\delta\hat r_H}{\dd\theta}
 &=\frac{\delta\hat p_H}{eH},&
 \frac{\dd\delta\hat p_H}{\dd\theta}&=0.
 \label{SM:eq:sectorode}
\end{align}

Solving this constant-coefficient linear system gives the bulk matrix below, where
\[
 c_\theta=\cos\theta,\qquad s_\theta=\sin\theta,\qquad
 d_\theta=\gamma^2s_\theta-(\gamma^2-1)\theta.
\]
In orbital order $(\delta\hat r_r,\delta\hat r_t,\delta\hat r_H,\delta\hat p_r,\delta\hat p_t,\delta\hat p_H)^T$,
\begin{equation}
S_{\rm bulk}=\begin{pmatrix}
c_\theta+\frac{\gamma^2}{2}(1-c_\theta)&\frac\gamma2s_\theta&0&\frac{s_\theta}{eH}&\frac{\gamma(1-c_\theta)}{eH}&0\\
\frac\gamma2(d_\theta-2s_\theta)&1-\frac{\gamma^2}{2}(1-c_\theta)&0&-\frac{\gamma(1-c_\theta)}{eH}&\frac{d_\theta}{eH}&0\\
0&0&1&0&0&\frac\theta{eH}\\
eH[(\gamma^2-1)s_\theta-\frac{\gamma^2}{4}d_\theta]&\frac{\gamma eH}{4}(\gamma^2-2)(1-c_\theta)&0&c_\theta+\frac{\gamma^2}{2}(1-c_\theta)&\frac\gamma2(2s_\theta-d_\theta)&0\\
-\frac{\gamma eH}{4}(\gamma^2-2)(1-c_\theta)&-\frac{\gamma^2eH}{4}s_\theta&0&-\frac\gamma2s_\theta&1-\frac{\gamma^2}{2}(1-c_\theta)&0\\
0&0&0&0&0&1
\end{pmatrix}.
\label{SM:eq:bulkmap}
\end{equation}
Applying the entrance and exit matrices in Eq.~\eqref{SM:eq:SMfaces}, we obtain the free-to-free transfer,
\begin{equation}
 S=C_{\rm out}S_{\rm bulk}C_{\rm in}
 =\begin{pmatrix}
 c_\theta&0&0&s_\theta/(eH)&\gamma(1-c_\theta)/(eH)&0\\
 -\gamma s_\theta&1&0&-\gamma(1-c_\theta)/(eH)&d_\theta/(eH)&0\\
 0&0&1&0&0&\theta/(eH)\\
 -eHs_\theta&0&0&c_\theta&\gamma s_\theta&0\\
 0&0&0&0&1&0\\
 0&0&0&0&0&1
 \end{pmatrix}.
 \label{SM:eq:map}
\end{equation}

Thus $\delta\hat Z_{\rm out}=S\delta\hat Z_{\rm in}$ uses physical free momenta on both sides. Both matrices equal $I_6$ at $\theta=0$ and preserve $J$. Inside the field the canonical transfer is $S_{\rm bulk}C_{\rm in}$. These matrices use orbital axes. Writing $D_R(\theta)=\operatorname{diag}(R(\gamma\theta),R(\gamma\theta))$ for the component rotation above, the free-to-free FW map is $S_{\rm FW}=D_R(\theta)^TS$ for coincident entrance bases. Covariances and OAM components must be rotated with the same endpoint basis.

{The radial entry of Eq.~\eqref{SM:eq:map} has a geometric interpretation.} {Translating an incident trajectory radially at fixed momentum} gives a circle of the same radius, and its displacement projected onto the rotating radial axis is $\delta r_r=\delta r_{r}^{(0)}\cos\theta$. The acceleration term is needed to recover this dependence from the comoving Hamiltonian, as derived in Sec.~\ref{SM:sec:hamiltonian}.

{To calculate the OAM and check the approximation along the magnet, we need the kinetic momentum at each interior point.} We obtain it from the canonical transfer $S_{\rm bulk}C_{\rm in}$ by adding the vector potential:
\begin{equation}
 \begin{aligned}
 \delta\hat\pi_r&=\delta\hat p_r-\frac{\gamma eH}{2}\delta\hat r_t,\\
 \delta\hat\pi_t&=\delta\hat p_t+\frac{\gamma eH}{2}\delta\hat r_r,\\
 \delta\hat\pi_H&=\delta\hat p_H.
 \end{aligned}
 \label{SM:eq:B}
\end{equation}
{At any observation point, the kinetic transfer $S_\pi$ maps the incident canonical variables to the current positions and kinetic momenta. Inside the magnet, combining Eq.~\eqref{SM:eq:B} with $S_{\rm bulk}C_{\rm in}=C_{\rm in}S$ gives a matrix that differs from Eq.~\eqref{SM:eq:map} only in row 4:}
\begin{equation}
 (S_\pi)_{4A}=S_{4A}-\gamma eH S_{2A},\qquad A=1,\ldots,6.
 \label{SM:eq:kinetic_transfer}
\end{equation}
{All other rows are unchanged. At the exit, applying $C_{\rm out}$ gives the free-state transfer $S$ [Eqs.~\eqref{SM:eq:SMfaces} and \eqref{SM:eq:map}]. Since $\delta\hat{\vct\pi}=\delta\hat{\vct p}$ in the field-free gauge, $S_\pi=S$ there. The mean requires the propagated second moments [Eq.~\eqref{SM:eq:mean_trace}], while the variance also requires the incident LG fourth moments and quantum commutators [Eqs.~\eqref{SM:eq:LGfourth} and \eqref{SM:eq:V}].} This gives $[\delta\hat\pi_r,\delta\hat\pi_t]=-i\gamma eH\hbar$ inside and commuting momenta outside. \hypertarget{def:drift_length}{}A laboratory drift $L_{\rm drift}$ lasts proper time $L_{\rm drift}/(\gamma v)$, so its upper-right matrix block is $L_{\rm drift}I/(m\gamma v)$. Successive maps are multiplied in their order along the orbit. The drift segments modify the transfer through four dipoles, so that their combined matrix differs from the dipole matrix continued to $2\pi$.
\subsection{Boundary covariances and uncertainty}\label{SM:sec:ellipse_example}\label{SM:sec:covariance_ellipses}
\begin{figure}[t]
\centering
\includegraphics[width=0.72\textwidth]{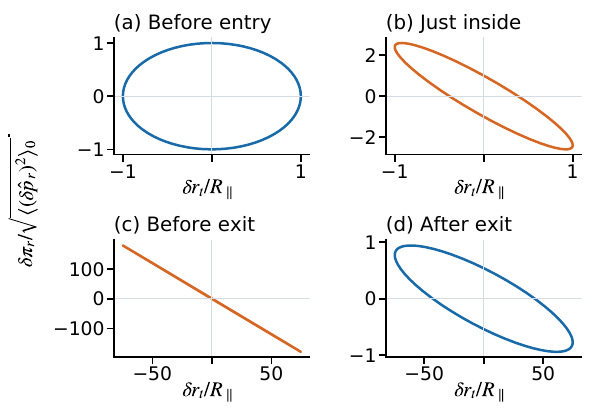}
\caption{Position--momentum correlations for the magnet in {Fig.~\Lref{fig:matching} of the Paper}:
(a) before entry, (b) just inside, (c) before exit, and
(d) after leaving the magnet. The centered longitudinal position $\delta r_t$
and radial kinetic momentum $\delta\pi_r$ are measured in the packet
rest frame. Each curve satisfies $\vct u^T\Sigma_2^{-1}\vct u=1$,
where $\vct u=(\delta r_t,\delta\pi_r)^T$ and $\Sigma_2$ is their
covariance, propagated with Eqs.~{\eqref{SM:eq:kinetic_transfer}} and \eqref{SM:eq:B}.
Eq.~\eqref{SM:eq:covariance_ellipse} gives the contour parametrization.
Axis units are the incident rms length $R_\parallel$ and radial
momentum $\sqrt{\mean{(\delta\hat p_r)^2}_0}$ of Eq.~\eqref{SM:eq:input}.
The panels use different ranges.
Entrance and exit give opposite momentum shears acting on the states
before and after the intervening evolution.
The BNL ATF example has laboratory kinetic energy $K=25$~MeV, field
$H=74.2$~mT, bend $\theta=20^\circ$, and magnetic arc length
$L_{\rm mag}=0.4$~m. The incident LG state [Eq.~\eqref{SM:eq:LG}] has
$\ell=1000$, $R_\perp=1~\mu$m, and $R_\parallel=0.3~\mu$m.}
\label{SM:fig:covariance_ellipses}
\end{figure}

{Fig.~\ref{SM:fig:covariance_ellipses} and \MovieNumber{4} show how matching changes position--momentum correlations for the BNL state of {Fig.~\Lref{fig:matching} of the Paper}. The ellipses describe $(\delta\hat r_t,\delta\hat\pi_r)$ {on a common rest-frame time slice, and \MovieNumber{4} also shows} $(\delta\hat r_r,\delta\hat\pi_H)$. We obtain their $2\times2$ covariance $\Sigma_2$ from $\Sigma$ [Eqs.~\eqref{SM:eq:covariance_definition} and \eqref{SM:eq:covariance_transport}] with the kinetic conversion in Eq.~\eqref{SM:eq:B}.}

{To compare position and momentum geometrically, we divide them by the axis units $s_1,s_2$. This gives dimensionless coordinates $\vct q=(\delta r_t/s_1,\delta\pi_r/s_2)^T$ and covariance $C_{ij}=(\Sigma_2)_{ij}/(s_i s_j)$. The figure uses the incident rms scales specified in its caption, while the movie states the units in each panel.

}The principal directions of the ellipse are the orthonormal eigenvectors $\vct e_i$ of $C$, with $C\vct e_i=\lambda_i\vct e_i$. The variance along $\vct e_i$ is $\lambda_i$, so the contour with one-rms semiaxes is
\begin{equation}
 \vct q(\phi)=\sqrt{\lambda_1}\,\vct e_1\cos\phi
          +\sqrt{\lambda_2}\,\vct e_2\sin\phi,
 \qquad 0\leq\phi<2\pi.
 \label{SM:eq:covariance_ellipse}
\end{equation}
{Here $\phi$ parametrizes the contour $\vct q^TC^{-1}\vct q=1$, or $\vct u^T\Sigma_2^{-1}\vct u=1$ in dimensional variables, with $\vct u=(\delta r_t,\delta\pi_r)^T$. The semiaxes give the rms scales of the principal combinations of observables, while the orientation shows their correlation. {The ellipses summarize second moments, while the LG probability density follows from Eq.~\eqref{SM:eq:SMpsi}.}

}For example, with $-$ denoting the incident state and $+$ the matched in-field state, the entrance shear $\delta\hat\pi_r^+=\delta\hat p_r^- -eH'\delta\hat r_t$ produces $\Sigma_{r_t\pi_r}^+=-eH'R_\parallel^2$ when the incident $r_t,p_r$ covariance vanishes. This negative covariance tilts the ellipse between Fig.~\ref{SM:fig:covariance_ellipses}(a) and (b). The initial narrow $r_r,\pi_H$ ellipse in \MovieNumber{4} instead reflects the LG correlation $\Sigma_{r_rp_H}=-\ell\hbar/2$.

The \mbox{full covariance} is also constrained by the Robertson--Schr\"odinger uncertainty relation \cite{SM:Robertson1934}. For the canonical variables and the matrix $J$ defined in Sec.~\ref{SM:sec:mean_equation},
\begin{equation}
 \Sigma+\frac{i\hbar}{2}J\succeq0.
 \label{SM:eq:matrix_uncertainty}
\end{equation}
{Here $\succeq0$ denotes positive semidefiniteness: the quadratic form is nonnegative for every complex vector.} A symplectic transfer preserves this constraint because
\begin{equation}
 \Sigma(\tau)+\frac{i\hbar}{2}J
 =S\left(\Sigma_0+\frac{i\hbar}{2}J\right)S^T,
 \qquad SJS^T=J.
\end{equation}
In dimensional variables, the ellipse has area $\pi\sqrt{\det\Sigma_2}$. For a conjugate pair $r_r,p_r$, Eq.~\eqref{SM:eq:matrix_uncertainty} gives $(\Delta r_r)^2(\Delta p_r)^2-\Sigma_{r_rp_r}^2\geq\hbar^2/4$, so this area is at least $\pi\hbar/2$. {The full matrix condition also constrains correlations among all six variables, beyond the separate position--momentum product bounds.}

For kinetic variables, the transfer $S_\pi$ of Eq.~\eqref{SM:eq:kinetic_transfer} gives $\delta\hat Z_\pi=S_\pi\delta\hat Z_0$. Both the covariance and commutator matrix transform: $\Sigma_\pi=S_\pi\Sigma_0S_\pi^T$ and $J_\pi=S_\pi JS_\pi^T$. The condition is therefore $\Sigma_\pi+i\hbar J_\pi/2\succeq0$. The displayed pairs obey $[\delta\hat r_t,\delta\hat\pi_r]=[\delta\hat r_r,\delta\hat\pi_H]=0$. {These pairs combine a position with momentum in another spatial direction. Their zero commutators allow projected ellipse areas below $\pi\hbar/2$ in dimensional variables. The full state still obeys the matrix uncertainty relation above, which couples all six variables.}

A linear canonical transformation preserves the symplectic eigenvalues $\nu_a$, the positive eigenvalues of $iJ\Sigma$. Eq.~\eqref{SM:eq:matrix_uncertainty} requires $\nu_a\geq\hbar/2$. For the incident LG packet with a longitudinal minimum-uncertainty Gaussian,
\[
 \{\nu_1,\nu_2,\nu_3\}=\{\hbar/2,\hbar/2,(2|\ell|+1)\hbar/2\}.
\]
These invariants are distinct from the ordinary eigenvalues $\lambda_i$ used to draw an individual projected ellipse. Projected areas can change as different planes become correlated. {Preservation of the symplectic eigenvalues checks the complete canonical transfer, including boundary matching. The OAM variance additionally requires fourth moments.}

\section{Mean OAM, variance, and analytic~limits}\label{SM:sec:analytic}
\subsection{Mean OAM inside and after the magnet}\label{SM:sec:mean_oam}
{We use the transfer derived in Sec.~\ref{SM:sec:transport} to find when the mean OAM returns or reverses and whether its distribution remains narrow. We first evaluate the outgoing mean kinetic OAM $\mean{\hat{\bm{\mathcal L}}}$ [Eq.~\eqref{SM:eq:OAM_covariance}] in the orbital axes $r,t,H$, using the free-to-free map in Eq.~\eqref{SM:eq:map} and the incident covariance in Eq.~\eqref{SM:eq:input}. In the field-free gauge, $\delta\hat p_i=\delta\hat\pi_i$ outside the magnet.}
Substitution of Eq.~\eqref{SM:eq:map} into $\hat{\mathcal L}_r=\delta\hat r_t\delta\hat p_H-\delta\hat r_H\delta\hat p_t$ leaves the correlation $-\gamma s_\theta\mean{\delta\hat r_{r}^{(0)}\delta\hat p_{H}^{(0)}}=\gamma\ell\hbar s_\theta/2$. Similarly,
\begin{align}
 \mean{\hat{\mathcal L}_t}_{\rm out}
 &=c_\theta(\mean{\delta\hat r_{H}^{(0)}\delta\hat p_{r}^{(0)}}-\mean{\delta\hat r_{r}^{(0)}\delta\hat p_{H}^{(0)}})
       -\theta s_\theta\mean{\delta\hat p_{H}^{(0)}\delta\hat r_{r}^{(0)}}\nonumber\\
 &=\ell\hbar(c_\theta+\theta s_\theta/2).
\end{align}
{Evaluating $\hat{\mathcal L}_H=\delta\hat r_r\delta\hat p_t-\delta\hat r_t\delta\hat p_r$ and collecting all three components gives
\begin{align}
 \mean{\hat{\mathcal L}_r}_{\rm out}&=\frac{\gamma\ell\hbar}{2}s_\theta,\nonumber\\
 \mean{\hat{\mathcal L}_t}_{\rm out}&=\ell\hbar(c_\theta+\theta s_\theta/2),\nonumber\\
 \mean{\hat{\mathcal L}_H}_{\rm out}
 &=-\gamma eH s_\theta^2\Sigma_{0,r_rr_r}+
 \frac{\gamma}{eH}(1-c_\theta)c_\theta\Sigma_{0,p_rp_r}\nonumber\\
 &\quad+\frac{\gamma}{eH}(1-c_\theta-s_\theta d_\theta)\Sigma_{0,p_tp_t}.
 \label{SM:eq:LH}
\end{align}
The first two components give Eq.~(\Lref{eq:mean}) of the Paper.}
The bulk shift in Eq.~\eqref{SM:eq:B} leaves $\hat{\mathcal L}_r$ unchanged, subtracts $\gamma eH\delta\hat r_H\delta\hat r_t$ from $\hat{\mathcal L}_t$, and adds $\gamma eH(\delta\hat r_t)^2$ to $\hat{\mathcal L}_H$. Thus
\begin{align}
 \mean{\hat{\mathcal L}_r}_{\rm bulk}&=\gamma\ell\hbar s_\theta/2,\nonumber\\
 \mean{\hat{\mathcal L}_t}_{\rm bulk}
 &=\frac{\ell\hbar}{2}
 [\gamma^2+(2-\gamma^2)c_\theta-(\gamma^2-1)\theta s_\theta],\nonumber\\
 \mean{\hat{\mathcal L}_H}_{\rm bulk}&=\mean{\hat{\mathcal L}_H}_{\rm out}+\gamma eH\Sigma_{r_tr_t},\nonumber\\
 \Sigma_{r_tr_t}&=R_\parallel^2+\gamma^2s_\theta^2\Sigma_{0,r_rr_r}\nonumber\\
 &\quad+\frac{\gamma^2(1-c_\theta)^2}{e^2H^2}\Sigma_{0,p_rp_r}
       +\frac{d_\theta^2}{e^2H^2}\Sigma_{0,p_tp_t}.
\end{align}
Here the incident variances $\Sigma_{0,r_rr_r}$, $\Sigma_{0,p_rp_r}$, and $\Sigma_{0,p_tp_t}$ are given in Eq.~\eqref{SM:eq:input}. These are local-frame components. After leaving the magnet at $\theta=\pi$, they give $\mean{\hat{\mathcal L}_t}=-\ell\hbar$ and $\mean{\hat{\mathcal L}_H}=2\gamma(\Sigma_{0,p_tp_t}-\Sigma_{0,p_rp_r})/(eH)$. Thus orbital helicity reverses sign for either sign of $\ell$, regardless of the sizes. {Choosing equal incident momentum variances,} $\Sigma_{0,p_tp_t}=\Sigma_{0,p_rp_r}$, hence $R_\parallel=R_\perp/[\sqrt2(|\ell|+1)]$, additionally cancels the field-axis mean. The full mean is then antiparallel to the outgoing motion for $\ell>0$ and parallel for $\ell<0$. After leaving the magnet at $2\pi$, all three mean OAM components regain their \mbox{incident values}, {while their variances can differ from the \mbox{incident values}.}

Changing the sign of $\ell$ reflects the incident state through $r_H=0$. The dipole field and transport equations are invariant under this reflection, while the axial vector OAM transforms as $(\mathcal L_r,\mathcal L_t,\mathcal L_H)\mapsto(-\mathcal L_r,-\mathcal L_t,\mathcal L_H)$. Consequently $W$ is unchanged and $\alpha_L\mapsto\pi-\alpha_L$. The antiparallel-alignment tolerances below therefore become parallel-alignment tolerances for negative $\ell$, with identical numerical bounds.
\subsection{Fourth moments and the OAM variance}\label{SM:sec:variance}
The mean OAM $\mean{\hat{\bm{\mathcal L}}}$ depends on averages of position times momentum. Its covariance $V_{ij}$ [Eq.~\eqref{SM:eq:OAM_covariance}] contains products of four operators, so the canonical covariance $\Sigma$ [Eq.~\eqref{SM:eq:covariance_definition}] alone is insufficient. We derive the required fourth moments of the incident LG state and then propagate them with the linear transfer $S$ in Eq.~\eqref{SM:eq:map}. The calculation accounts for the LG state's departure from a Gaussian distribution.

At $\ell=0$, Eq.~\eqref{SM:eq:LG} gives the covariance
\[
 \Sigma_{\rm G}=\operatorname{diag}\!\left(
 \frac{w^2}{2},R_\parallel^2,\frac{w^2}{2},
 \frac{\hbar^2}{2w^2},\frac{\hbar^2}{4R_\parallel^2},
 \frac{\hbar^2}{2w^2}\right).
\]
The vortex changes this to $\Sigma_0=\Sigma_{\rm G}+|\ell|Q$. The nonzero entries of the symmetric matrix $Q$ are
\[
\begin{gathered}
 {Q_{r_rr_r}=Q_{r_Hr_H}=w^2/2},\\
 Q_{p_rp_r}=Q_{p_Hp_H}=\hbar^2/(2w^2),\\
 Q_{r_Hp_r}=-Q_{r_rp_H}=\operatorname{sgn}(\ell)\hbar/2.
\end{gathered}
\]
Thus $Q$ is the additional covariance per unit of $|\ell|$ at fixed envelope widths.

{To obtain the fourth moments, we represent the LG polynomial using the dimensionless oscillator operators}
\[
 \begin{aligned}
 \hat b_r&=\frac{\delta\hat r_{r}^{(0)}/w+iw\delta\hat p_{r}^{(0)}/\hbar}{\sqrt2},\\
 \hat b_H&=\frac{\delta\hat r_{H}^{(0)}/w+iw\delta\hat p_{H}^{(0)}/\hbar}{\sqrt2}.
 \end{aligned}
\]
{For $\ell\ne0$, the circular raising operator}
$\hat b^\dagger=(\hat b_r^\dagger-i\operatorname{sgn}(\ell)\hat b_H^\dagger)/\sqrt2$
adds one factor $(x-i\operatorname{sgn}(\ell)z)/w$ to the Gaussian. The state in Eq.~\eqref{SM:eq:LG} therefore contains $|\ell|$ excitations of this circular mode, while the opposite circular mode and the longitudinal Gaussian remain in their ground states. {These oscillator operators provide a basis for the packet prepared in free space.} For $\ell=0$, the packet is the Gaussian itself.

For the occupied circular mode, $[\hat b,\hat b^\dagger]=1$ gives
\[
 e^{\xi\hat b^\dagger-\xi^*\hat b}
 =e^{-|\xi|^2/2}e^{\xi\hat b^\dagger}e^{-\xi^*\hat b},
\]
where $\xi$ is a complex dimensionless parameter. Using
$\mean{(\hat b^\dagger)^j\hat b^j}=|\ell|!/(|\ell|-j)!$
then yields
\begin{align*}
 \mean{e^{\xi\hat b^\dagger-\xi^*\hat b}}
 &=e^{-|\xi|^2/2}\sum_{j=0}^{|\ell|}
 \binom{|\ell|}{j}\frac{(-|\xi|^2)^j}{j!}\\
 &=e^{-|\xi|^2/2}L_{|\ell|}(|\xi|^2),
\end{align*}
with $L_{|\ell|}$ the Laguerre polynomial. For a six-component Fourier variable $\vct\kappa$, whose entries have the inverse units of the corresponding entries of $\delta\hat Z_0$, the occupied-mode parameter satisfies $|\xi|^2=\vct\kappa^TQ\vct\kappa/2$. Multiplying by the two unoccupied-mode factors gives
\begin{equation}
 \mean{e^{i\vct\kappa^T\delta\hat Z_0}}
 =e^{-\vct\kappa^T\Sigma_{\rm G}\vct\kappa/2}L_{|\ell|}(\vct\kappa^TQ\vct\kappa/2).
 \label{SM:eq:LGcharacteristic}
\end{equation}
Derivatives at $\vct\kappa=0$ generate moments symmetrized over all operator orders. {Expanding the logarithm to fourth order gives the expression below, where} $\mathcal O(|\vct\kappa|^6)$ collects terms of sixth and higher total order in the \mbox{Fourier variables}:
\begin{equation}
 \begin{aligned}
 \log\mean{e^{i\vct\kappa^T\delta\hat Z_0}}
 &=-\tfrac12\vct\kappa^T\Sigma_0\vct\kappa\\
 &\quad-\frac{|\ell|(|\ell|+1)}{16}(\vct\kappa^TQ\vct\kappa)^2+\mathcal O(|\vct\kappa|^6).
 \end{aligned}
\end{equation}
Equivalently, the full fourth moment is
\begin{equation}
\begin{aligned}
 &\mean{\delta\hat Z_{0A}\delta\hat Z_{0B}
             \delta\hat Z_{0C}\delta\hat Z_{0D}}_{\rm sym}\\
 &=\Sigma_{0,AB}\Sigma_{0,CD}+\Sigma_{0,AC}\Sigma_{0,BD}
   +\Sigma_{0,AD}\Sigma_{0,BC}\\
 &\quad-\frac{|\ell|(|\ell|+1)}2
 (Q_{AB}Q_{CD}+Q_{AC}Q_{BD}+Q_{AD}Q_{BC}).
\end{aligned}
\label{SM:eq:LGfourth}
\end{equation}
The subscript ``sym'' means the average over all 24 orders of the four operators. {The first three terms are the pair products for a Gaussian with covariance $\Sigma_0$. The final term accounts for the vortex state's non-Gaussian fourth moments and vanishes at $\ell=0$.}

To form the OAM operators, we use $S_\pi$, the kinetic-momentum transfer in Eq.~\eqref{SM:eq:kinetic_transfer}. Rows $(S_\pi)_j$ and $(S_\pi)_{j+3}$ express position component $j$ and its kinetic momentum, respectively, in terms of the six incident operators. The symmetric coefficient matrix of $\hat{\mathcal L}_i$ is
\begin{equation}
\begin{aligned}
 \Lambda_i=\frac12\sum_{j,k=1}^{3}\epsilon_{ijk}
 \bigl[&(S_\pi)_j^T(S_\pi)_{k+3}\\
       &+(S_\pi)_{k+3}^T(S_\pi)_j\bigr],
\end{aligned}
\label{SM:eq:OAMmatrix}
\end{equation}
where $\epsilon_{ijk}$ is defined in Sec.~\ref{SM:sec:mean_equation}. Hence
$\hat{\mathcal L}_i=(\delta\hat Z_0)^T\Lambda_i\delta\hat Z_0$
and
\begin{equation}
 \mean{\hat{\mathcal L}_i}=\tr(\Lambda_i\Sigma_0).
 \label{SM:eq:mean_trace}
\end{equation}
Contracting Eq.~\eqref{SM:eq:LGfourth} with $\Lambda_i$ and $\Lambda_j$, and subtracting $\mean{\hat{\mathcal L}_i}\mean{\hat{\mathcal L}_j}$, leaves two Gaussian pair terms and the LG correction. Converting the fully symmetrized products to the anticommutator of the two OAM operators also uses
$[\delta\hat Z_{0A},\delta\hat Z_{0B}]=i\hbar J_{AB}$
[Sec.~\ref{SM:sec:mean_equation}]. The resulting quantum-ordering term is $\hbar^2\tr(\Lambda_iJ\Lambda_jJ)/2$. Together these contributions give the OAM covariance $V_{ij}$ defined in Eq.~\eqref{SM:eq:OAM_covariance}:
\begin{align}
 V_{ij}&=2\tr(\Lambda_i\Sigma_0\Lambda_j\Sigma_0)
 +\frac{\hbar^2}{2}\tr(\Lambda_iJ\Lambda_jJ)\nonumber\\
 &\quad-\frac{|\ell|(|\ell|+1)}2
 \bigl[\tr(\Lambda_iQ)\tr(\Lambda_jQ)
       +2\tr(\Lambda_iQ\Lambda_jQ)\bigr].
 \label{SM:eq:V}
\end{align}
For the incident state, Eqs.~\eqref{SM:eq:mean_trace} and \eqref{SM:eq:V} recover $\mean{\hat{\mathcal L}_t}=\ell\hbar$ and $V_{tt}=0$. Both the LG correction and the ordering term are needed for this zero variance.

{The wave function in Eq.~\eqref{SM:eq:SMpsi}, including the vortex polynomial and boundary phases, reproduces the mean OAM and covariance in Eqs.~\eqref{SM:eq:mean_trace} and \eqref{SM:eq:V}. In the nonrelativistic laboratory limit, the same uniform-field variance follows from the exact magnetic kernel of Gallatin and McMorran~\cite{SM:Gallatin} for the same initial state. At relativistic center velocities, comparison with the comoving calculation requires transforming the state, OAM operator, and equal-time plane [Eq.~\eqref{SM:eq:reslice}], with the acceleration potential and boundary transformations included as described in Secs.~\ref{SM:sec:hamiltonian} and \ref{SM:sec:boundary}.}
\subsection{Return of the mean after complete circles}\label{SM:sec:circle}
{To test whether closing the reference orbit restores the OAM distribution, we continue the uniform-dipole solution through $N\geq1$ complete turns and apply exit matching at $\theta=2\pi N$.} The comparison is between the incident and outgoing free
states, both described in the radial, tangential, and field axes of
Eq.~\eqref{SM:eq:map}. {These orbital axes return to their initial directions after each complete turn, while the FW axes retain the relative rotation angle $2\pi N\gamma$.}

Substituting $\sin(2\pi N)=0$ and $\cos(2\pi N)=1$ into
Eq.~\eqref{SM:eq:map} gives
\begin{equation}
 \begin{aligned}
 \delta\hat r_r&=\delta\hat r_{r}^{(0)},\qquad
 \delta\hat{\vct p}=\delta\hat{\vct p}_0,\\
 \delta\hat r_t&=\delta\hat r_{t}^{(0)}-\frac{(\gamma^2-1)2\pi N}{eH}\delta\hat p_{t}^{(0)},\\
 \delta\hat r_H&=\delta\hat r_{H}^{(0)}+\frac{2\pi N}{eH}\delta\hat p_{H}^{(0)}.
 \end{aligned}
 \label{SM:eq:circlemap}
\end{equation}
The radial motion is periodic, whereas the motion along the magnetic
field remains free. Since $\dot\theta=eH/m$, the latter lasts proper time
$2\pi Nm/(eH)$ and produces the displacement in the last line.
Outside the magnet, the tangential kinetic OAM is
$\hat{\mathcal L}_t=\delta\hat r_H\delta\hat p_r-\delta\hat r_r\delta\hat p_H$.
Inserting Eq.~\eqref{SM:eq:circlemap} gives
\[
 \hat{\mathcal L}_t=\hat{\mathcal L}_{t}^{(0)}
       +\frac{2\pi N}{eH}\delta\hat p_{r}^{(0)}\delta\hat p_{H}^{(0)}.
\]
For the incident state of Eq.~\eqref{SM:eq:LG},
$\hat{\mathcal L}_{t}^{(0)}\psi_0=\ell\hbar\psi_0$.
Consequently its variance and its symmetrized covariance with the added
operator vanish. Rotational symmetry of the transverse momentum density
also gives $\mean{\delta\hat p_{r}^{(0)}\delta\hat p_{H}^{(0)}}=0$.
Taking the mean and variance therefore gives
$\mean{\hat{\mathcal L}_t}=\ell\hbar$ and the tangential variance $V_{tt}$ of Eq.~\eqref{SM:eq:OAM_covariance}:
\begin{align}
 V_{tt}(N)&=\left(\frac{2\pi N}{eH}\right)^2
            \mean{(\delta\hat p_{r}^{(0)})^2(\delta\hat p_{H}^{(0)})^2},\nonumber\\
 \mean{(\delta\hat p_{r}^{(0)})^2(\delta\hat p_{H}^{(0)})^2}
 &=\frac{(|\ell|+1)(|\ell|+2)\hbar^4}{8w^4}.
 \label{SM:eq:circlevariance}
\end{align}
For the last step, the transverse momentum magnitude is defined by $(\delta\hat p_\perp)^2=(\delta\hat p_r)^2+(\delta\hat p_H)^2$. Integration of the LG momentum density gives $\mean{(\delta\hat p_\perp)^4}=(|\ell|+1)(|\ell|+2)\hbar^4/w^4$. The angular average of $\sin^2\phi\cos^2\phi$ is $1/8$. The mean points along $t$ at a full turn, so $W=\sqrt{V_{tt}}/(|\ell|\hbar)$ for $\ell\ne0$. This proves the $N^2$ variance law for the circular continuation of the dipole solution and the specified LG state.
\subsection{Half-turn variance and mean alignment}\label{SM:sec:halfturn}
Here we derive the mean-OAM reversal condition, Eq.~(\Lref{eq:half}) of the Paper, and its half-turn width, Eq.~(\Lref{eq:halfwidth}) of the Paper. {The reversal condition equates the three incident momentum variances, $\mean{(\delta\hat p_r)^2}_0=\mean{(\delta\hat p_t)^2}_0=\mean{(\delta\hat p_H)^2}_0=\hbar^2/(4R_\parallel^2)$ [Eq.~\eqref{SM:eq:input}]. The full momentum distribution remains anisotropic for $|\ell|>0$: its transverse LG part has a vortex zero, whereas its longitudinal part is Gaussian. We use the fourth moments to find which incident sizes give a narrow OAM distribution at fixed energy, field, and bend.} {The analytic formulas below apply to the LG state in Eq.~\eqref{SM:eq:LG} {and normal incidence at sharp boundaries.} The reference examples use $K=1$~MeV and $H=5$~mT.}

At $\theta=\pi$, the transfer to the field-free region after the magnet in Eq.~\eqref{SM:eq:map} gives
\begin{align}
 \delta\hat r_r&=-\delta\hat r_{r}^{(0)}+\frac{2\gamma}{eH}\delta\hat p_{t}^{(0)}, &
 \delta\hat p_r&=-\delta\hat p_{r}^{(0)},\nonumber\\
 \delta\hat r_H&=\delta\hat r_{H}^{(0)}+\frac{\pi}{eH}\delta\hat p_{H}^{(0)}, &
 \delta\hat p_H&=\delta\hat p_{H}^{(0)}.
\end{align}
Using $\hat{\mathcal L}_t=\delta\hat r_H\delta\hat p_r-\delta\hat r_r\delta\hat p_H$, we obtain
\begin{equation}
 \hat{\mathcal L}_t=-\hat{\mathcal L}_{t}^{(0)}
 -\frac{\pi\delta\hat p_{r}^{(0)}+2\gamma\delta\hat p_{t}^{(0)}}{eH}
       \delta\hat p_{H}^{(0)}.
 \label{SM:eq:halfoperator}
\end{equation}
The incident tangential OAM is sharp, so the momentum-product term in Eq.~\eqref{SM:eq:halfoperator} is the sole source of its spread after the half turn. The incident correlations $\mean{\delta\hat p_r\delta\hat p_H}_0$ and $\mean{\delta\hat p_t\delta\hat p_H}_0$ vanish: {this term changes the variance while preserving the mean,} $\mean{\hat{\mathcal L}_t}=-\ell\hbar$, for any packet sizes. Squaring this term and averaging over the incident LG state gives
\begin{equation}
 (\Delta \mathcal L_t)^2=\frac{\hbar^4}{(eH)^2}\left[
 \frac{\pi^2(|\ell|+1)(|\ell|+2)}{8w^4}
 +\frac{\gamma^2(|\ell|+1)}{2R_\parallel^2w^2}\right].
 \label{SM:eq:halfvariance}
\end{equation}
The first term uses the fourth transverse momentum moment in Eq.~\eqref{SM:eq:circlevariance}. {The second uses the incident moments $\mean{(\delta\hat p_t)^2}_0=\hbar^2/(4R_\parallel^2)$ and $\mean{(\delta\hat p_H)^2}_0=(|\ell|+1)\hbar^2/(2w^2)$ [Eq.~\eqref{SM:eq:input}].} Both contributions are positive. {The outgoing OAM variance therefore remains positive even when the two momentum variances are matched.}

The length $R_\parallel^*=R_\perp/[\sqrt2(|\ell|+1)]$ cancels the field-axis mean. A perturbed length can be written as $R_\parallel=(1+\delta)R_\parallel^*$. The dimensionless mismatch $\delta$ measures an error in the ratio of the two sizes. For $\ell>0$, $\Delta\alpha=\pi-\alpha_L$ is the departure from antiparallel alignment. For $\ell<0$, the corresponding error is $\Delta\alpha=\alpha_L$ about parallel alignment after leaving the magnet. Using $\mean{\hat{\mathcal L}_r}=0$ and the field-axis mean in Eq.~\eqref{SM:eq:LH}, both cases give
\begin{equation}
 \tan\Delta\alpha=
 \frac{\gamma\hbar(|\ell|+1)^2}{eH R_\perp^2|\ell|}
 \left|(1+\delta)^{-2}-1\right|.
 \label{SM:eq:sizetilt}
\end{equation}
Finite angular resolution admits an interval set by Eq.~\eqref{SM:eq:sizetilt}. For example, replacing the matched length of the $\ell=10$, $R_\perp=9.6~\mu$m packet by $R_\parallel=0.6~\mu$m gives $\alpha_L=179.83^\circ$ and $W={5.8}\%$, compared with $180^\circ$ and ${5.5}\%$ at exact matching. If both sizes change, their new ratio determines $\delta$ and the new transverse radius enters the prefactor. A common rescaling leaves the mean exactly aligned.
At exact matching the projection on the mean coincides, up to sign, with $\mathcal L_t$. Substituting $R_\parallel^*$ into Eq.~\eqref{SM:eq:halfvariance} gives
\begin{equation}
 W^2=\left(\frac{\hbar}{eH R_\perp^2}\right)^2
 \frac{(|\ell|+1)^3}{\ell^2}
 \left[\gamma^2(|\ell|+1)+\frac{\pi^2(|\ell|+2)}8\right].
 \label{SM:eq:matchedwidth}
\end{equation}
For fixed $\ell,K,H$, a common increase of both sizes reduces $W$ as $R_\perp^{-2}$. The matched $\ell=64$ state reaches $W\leq10\%$ only at $R_\perp\geq16.6~\mu$m, whereas the corresponding threshold for $\ell=10$ is ${7.1}~\mu$m. For $W\leq5\%$ the thresholds are $23.4$ and $10.1~\mu$m.
{In the ideal sharp model, the width therefore approaches zero only as the matched packet sizes grow indefinitely.} {Field variation, apertures, and the conditions $\eta_{\rm NR},\eta_a\ll1$ throughout propagation [Eq.~\eqref{SM:eq:validity_ratios}] limit the physically allowed packet sizes.} This constrained half-turn problem differs from the weak-bend size-ratio minimization in Sec.~\ref{SM:sec:weaklaw}.
{A shorter half-turn magnet at $K=4.6$~MeV and $H=50$~mT has magnetic length $\pi\rho\simeq1.1$~m. For sharp boundaries, $\ell=10$, $R_\perp=10~\mu$m, and $R_\parallel=R_\perp/(11\sqrt2)\simeq0.64~\mu$m, Eq.~\eqref{SM:eq:matchedwidth} gives $W=1.6\%$. Eqs.~\eqref{SM:eq:map} and \eqref{SM:eq:kinetic_transfer} give conservative bounds $\eta_{\rm NR}<2.2\times10^{-2}$ and $\eta_a<2.2\times10^{-3}$ throughout the half turn.}
\subsection{Why weak bends admit a small OAM width}\label{SM:sec:weaklaw}
The two regions of small relative OAM width $W$ [Eq.~\eqref{SM:eq:width_angle}] in {Fig.~\Lref{fig:windows}(a,b) of the Paper} lie close to a simple relation between the initial transverse radius and longitudinal length. For the incident LG state of Eq.~\eqref{SM:eq:LG}, we derive the optimal size ratio and width in Eqs.~(\Lref{eq:shape}) and (\Lref{eq:tiltwidth}) of the Paper. The angle $\alpha_L$ denotes the tilt of the mean OAM after the magnet relative to the outgoing momentum, and it is distinct from the orbit bend $\theta$.

\subsubsection{Mean OAM after a weak bend}\label{SM:sec:weak_mean}
{A weak bend has $|\theta|\ll1$.} For $\gamma\gg1$, a bend $\theta\sim\gamma^{-1}$ turns the orbit only slightly, while the radial mean OAM in Eq.~(\Lref{eq:mean}) of the Paper is comparable to its incident tangential value. The proper flight time $\tau=m\theta/(eH)$ and field $H'=\gamma H$ give an accumulated rest-frame cyclotron angle $(eH'/m)\tau=\gamma\theta$. We therefore expand Eq.~\eqref{SM:eq:map} for small $\theta$, retaining $\gamma\theta$ of order unity. At fixed field and incident sizes,
\begin{align*}
 \gamma\sin\theta&=\gamma\theta[1+O(\theta^2)],\\
 \gamma(1-\cos\theta)&=\tfrac12\gamma\theta^2+O(\gamma\theta^4),\\
 d_\theta&=\theta-\tfrac16\gamma^2\theta^3+O(\gamma^2\theta^5).
\end{align*}
The last two coefficients are small in this scaling, and the leading transfer between the free states before and after the magnet is
\begin{equation}
\delta\hat r_{t,\rm out}=\delta\hat r_{t}^{(0)}-\gamma\theta\delta\hat r_{r}^{(0)},\qquad
\delta\hat p_{r,\rm out}=\delta\hat p_{r}^{(0)}+\gamma\theta\delta\hat p_{t}^{(0)},
\label{SM:eq:shearweak}
\end{equation}
while $\delta\hat r_r,\delta\hat r_H,\delta\hat p_t,\delta\hat p_H$ remain unchanged. Both changes are necessary: together they preserve the canonical commutators. These relations describe the net coordinate and momentum shear after entrance, propagation, and exit. The terms proportional to $\theta/(eH)$ in the full map also describe finite-time spreading. Their importance depends on the incident dimensions. We retain them in the numerical calculations at each specified finite energy and bend angle.

Immediately after leaving the magnet, the leading weak-bend state follows from the inverse coordinate map in Eq.~\eqref{SM:eq:shearweak}:
\begin{equation}
 \begin{aligned}
 \psi_{\rm out}(r_r,r_t,r_H)
 &=e^{i\gamma\theta\delta\hat r_r\delta\hat p_t/\hbar}\psi_0(r_r,r_t,r_H)\\
 &=\psi_0(r_r,r_t+\gamma\theta r_r,r_H).
 \end{aligned}
 \label{SM:eq:weakwavefunction}
\end{equation}
The unitary operator acts on the incident wave function $\psi_0$ of Eq.~\eqref{SM:eq:LG}. The coordinate arguments are measured from the center in its rest frame just outside the magnet, along the outgoing radial, tangential, and field axes. The shear preserves volume, and therefore the normalization.
The real envelope correlates $r_r$ and $r_t$, while the vortex polynomial
vanishes on the longitudinal line $r_r=r_H=0$.
On a loop around that line, oriented about the positive $r_t$ axis, the
phase changes by $2\pi\ell$. This is the phase singularity of the limiting
state immediately after leaving the magnet. At finite bend angle or after a
subsequent drift, its zeros and phase winding can be calculated from
the full complex wave function in Eq.~\eqref{SM:eq:SMpsi}.

For either sign of nonzero $\ell$, inserting Eq.~\eqref{SM:eq:shearweak} into
$\delta\hat{\vct r}\times\delta\hat{\vct p}$ and using the incident
correlations in Eq.~\eqref{SM:eq:input} gives the mean OAM after leaving the magnet:
\begin{equation}
\mean{\hat{\bm{\mathcal L}}}_{\rm out}=\ell\hbar(\gamma\theta/2,1,0).
 \label{SM:eq:weakmean}
\end{equation}
The coordinate axes are the radial, tangential, and field directions at the magnet exit. {For the positive bend angles considered here, $\alpha_L=\arctan(\gamma\theta/2)$ for $\ell>0$, whereas} $\alpha_L=\pi-\arctan(\gamma\theta/2)$ for $\ell<0$. In the following projection and variance formulas, positions and momenta are incident centered operators, with the labels $(0)$ and $\delta$ suppressed. Direct substitution into \mbox{$\hat{\vct r}\times\hat{\vct p}$ gives}
\begin{equation}
\begin{aligned}
\hat{\mathcal L}_{\vct n}&=\frac{\operatorname{sgn}(\ell)}{\sqrt{1+\gamma^2\theta^2/4}}
\bigg[\hat{\mathcal L}_{t}^{(0)}\\
&\quad+\frac{\gamma\theta}{2}(\hat r_t\hat p_H+\hat r_H\hat p_t)
 -\frac{\gamma^2\theta^2}{2}\hat r_r\hat p_H\bigg].
\end{aligned}
\label{SM:eq:weakprojection}
\end{equation}
The factor $\operatorname{sgn}(\ell)$ makes the projection axis follow the mean OAM for either sign, but it drops out of the variance. The incident state is an eigenstate of $\hat{\mathcal L}_{t}^{(0)}$. Its covariance with either added operator therefore vanishes. The covariance between the two added operators also vanishes, since one is odd and the other even under simultaneous reversal of $r_t$ and $p_t$ in the longitudinal Gaussian.

\subsubsection{Minimizing the width by varying the packet sizes}
{With the leading mean direction fixed by the weak bend, we vary the incident sizes to minimize $W$ [Eq.~\eqref{SM:eq:width_angle}]. The LG moments give}
\begin{align}
\frac{\big[\Delta(\hat r_t\hat p_H+\hat r_H\hat p_t)\big]^2}{\hbar^2}
 &=\frac{(|\ell|+1)^2R_\parallel^2}{2R_\perp^2}
 +\frac{R_\perp^2}{8R_\parallel^2}+\frac12,\label{SM:eq:weakmixed}\\
\frac{\big[\Delta(\hat r_r\hat p_H)\big]^2}{\hbar^2}
 &=\frac{\ell^2+|\ell|+2}{8}.\label{SM:eq:weakxp}
\end{align}
The first two terms in Eq.~\eqref{SM:eq:weakmixed} are respectively $R_\parallel^2\mean{(\delta\hat p_H)^2}_0/\hbar^2$ and $R_\perp^2\mean{(\delta\hat p_t)^2}_0/(2\hbar^2)$. The constant $1/2$ arises from the ordered cross terms, using $\mean{\hat r_t\hat p_t}=\mean{\hat r_H\hat p_H}=i\hbar/2$ at the waist. Retaining this quantum term is necessary even though the transfer matrix itself is classical.

At fixed $R_\perp$, the first contribution grows as $R_\parallel^2$, whereas the second decreases as $R_\parallel^{-2}$ because $\mean{(\delta\hat p_t)^2}_0=\hbar^2/(4R_\parallel^2)$. Their product is independent of $R_\parallel$, so their sum is smallest when they are equal:
\begin{equation}
\begin{gathered}
R_\parallel^2\mean{(\delta\hat p_H)^2}_0=\frac{R_\perp^2}{2}\mean{(\delta\hat p_t)^2}_0,\\
R_\perp/R_\parallel=\sqrt{2(|\ell|+1)}.
\label{SM:eq:weakbalance}
\end{gathered}
\end{equation}
{The resulting ratio differs from the equal-momentum condition at a sharp half turn [Sec.~\ref{SM:sec:halfturn}],} $R_\perp/R_\parallel=\sqrt2(|\ell|+1)$.

Combining Eqs.~\eqref{SM:eq:weakprojection}--\eqref{SM:eq:weakxp} and dividing by $|\mean{\hat{\bm{\mathcal L}}}|^2=\ell^2\hbar^2(1+\gamma^2\theta^2/4)$ gives
\begin{equation}
W^2=\frac{\gamma^2\theta^2\big[\Delta(\hat r_t\hat p_H+\hat r_H\hat p_t)\big]^2
 +\gamma^4\theta^4\big[\Delta(\hat r_r\hat p_H)\big]^2}
 {4\ell^2\hbar^2(1+\gamma^2\theta^2/4)^2}.
\end{equation}
At the optimum, $\big[\Delta(\hat r_t\hat p_H+\hat r_H\hat p_t)\big]^2/\hbar^2=(|\ell|+2)/2$. Using $\sin^2\alpha_L=\gamma^2\theta^2/(4+\gamma^2\theta^2)$ then yields
\begin{equation}
W_{\min}^2=\tfrac12\sin^4\alpha_L+
\frac{|\ell|+2}{2\ell^2}\sin^2\alpha_L.
\label{SM:eq:weakwidth}
\end{equation}
For negative $\ell$, the mean reverses relative to the corresponding positive-$\ell$ state. Eq.~\eqref{SM:eq:weakwidth} is unchanged, since it contains $|\ell|$, $\ell^2$, and even powers of $\sin\alpha_L$.

{Eq.~\eqref{SM:eq:weakbalance} fixes the aspect ratio while leaving the absolute size free.} {Terms omitted from the weak-bend approximation, together with fringe corrections, select packet sizes of a few micrometers in the finite-angle calculations.} At $|\ell|=64$ and $\gamma\theta=0.87$, Eq.~\eqref{SM:eq:weakwidth} gives $W_{\min}=11.8\%$, close to the IMP and BNL results calculated with finite bend angles, fringes, and incidence-angle spread. At large $|\ell|$, {the remaining term} {$W_{\min}\simeq\sin^2\alpha_L/\sqrt2$ sets the minimum over the two incident sizes within this \mbox{reduced model}.}

The filled circles in {Fig.~\Lref{fig:mechanism}(b) of the Paper} compare the leading weak-bend expression
with the transfer at the finite angle $\theta=5^\circ$, $H=5$~mT, and
$\ell=64$. With the magnetic length $l_H=\sqrt{\hbar/(eH)}$, we keep
$R_\perp=\sqrt{65}\,l_H\simeq{2.9}~\mu$m and
$R_\parallel=l_H/\sqrt2\simeq{0.26}~\mu$m. The Lorentz factors satisfy $\gamma=2\tan\alpha_L/\theta$ for the leading tilts $10^\circ$, $40^\circ$, and $60^\circ$, giving $\gamma\simeq{4}$, $19.2$, and $39.7$. Using Eq.~\eqref{SM:eq:map}, including both sharp boundaries, changes these tilts by less than $0.01^\circ$ and gives $W={4.2}\%$, $30.1\%$, and $53.9\%$, respectively.
These points use the specified pure incident state and abrupt boundaries perpendicular to the reference orbit.
The single open circle near $23.6^\circ$ shows the IMP calculation
with smooth fringes and incidence-angle averaging from {Fig.~\Lref{fig:windows}(a) of the Paper}.
Along these orbits, the ratios in Eq.~\eqref{SM:eq:validity_ratios} obey
$\eta_{\rm NR}<8.4\times10^{-4}$ and $\eta_a<2.5\times10^{-4}$.

{The weak-bend formulas describe the leading shear at fixed $\gamma\theta$, while the finite-angle calculations include spreading and fringe correlations. Increasing $\gamma$ at fixed field and packet sizes eventually violates the entrance bounds in Sec.~\ref{SM:sec:entry_size_validity}, as illustrated by the size scans in Sec.~{\ref{SM:sec:size_scan}}. The limiting state in Eq.~\eqref{SM:eq:weakwavefunction} {also shows how the mean OAM and the phase singularity can acquire different tilts.}

}\subsection{Example: a 1-MeV electron in a $20^\circ$ bend}\label{SM:sec:worked_example}
We illustrate the full finite-angle transfer, including finite-time spreading, for a single dipole. The incident state has $\ell=64$, $R_\perp=3~\mu$m, and $R_\parallel=0.3~\mu$m, and its covariance is given by Eq.~\eqref{SM:eq:input}. The laboratory energy $K=1$~MeV gives $\gamma\simeq{3}$. At $H=5$~mT, the magnetic coefficient is $eH\simeq7.6\hbar/\mu\mathrm m^2$.

At $\theta=20^\circ$, Eq.~\eqref{SM:eq:map} expresses the outgoing position and momentum operators in terms of the incident ones. We form the OAM matrices with Eq.~\eqref{SM:eq:OAMmatrix}, calculate the OAM mean and covariance with Eqs.~\eqref{SM:eq:mean_trace} and \eqref{SM:eq:V}, and project along the mean using Eq.~\eqref{SM:eq:width_angle}:
\[
 W=22.2\%,\qquad \alpha_L=27.3^\circ.
\]
At zero bend, the same calculation recovers $\mean{\hat{\bm{\mathcal L}}}=(0,64,0)\hbar$ and zero tangential variance. For a point inside the magnet, the kinetic-momentum conversion in Eq.~\eqref{SM:eq:B} is required before forming the OAM matrices. Replacing the incident LG state by a Gaussian with the same covariance would already produce an incorrect nonzero tangential variance before entry.

{\section{Wave function, probability density, and OAM projections}\label{SM:sec:wavefunction_spectrum}}

{The OAM moments calculated in Secs.~\ref{SM:sec:transport} and \ref{SM:sec:analytic} describe the mean and spread of the intrinsic kinetic OAM. To determine the packet's spatial shape, phase, and probabilities of individual OAM projections, we now propagate the full incident LG state [Eq.~\eqref{SM:eq:LG}]. We use the trajectory-coherent-state construction of Bagrov et al.~{\cite{SM:BagrovBelovMaslov1989,SM:BagrovBelovTrifonov1996,SM:BagrovBelovTrifonov2005,SM:BagrovBelovTrifonov2006}} with the same spinless quadratic Hamiltonian and entrance and exit matching. The resulting wave function also reproduces the moments obtained from the covariance matrix.}

\subsection{The full wave function inside the magnet and after exit}\label{SM:sec:wavefunction}

{We first evolve the incident amplitude and phase on the simultaneous planes of the packet's instantaneous rest frame. The centered orbital coordinates are $\delta\vct r=(\delta r_r,\delta r_t,\delta r_H)$, and the subscript $0$ denotes their entrance values. The orbital and Fermi--Walker (FW) axes coincide there, so $(\delta r_{r,0},\delta r_{t,0},\delta r_{H,0})=(\delta x_0,\delta y_0,\delta z_0)$. In these coordinates, the incident state of Eq.~\eqref{SM:eq:LG} is}

\begin{equation}
 \begin{aligned}
 \psi_0(\delta\vct r_0)&=\mathcal N_0
 \left(\frac{{\delta r_{r,0}}-i\operatorname{sgn}(\ell){\delta r_{H,0}}}{w}\right)^{|\ell|}
 e^{-\delta\vct r_0^T\sigma_0^{-2}\delta\vct r_0/2},\\
 \sigma_0&=\operatorname{diag}\big(w,\sqrt2R_\parallel,w\big),\\
 \mathcal N_0&=[\pi w^2|\ell|!\sqrt{2\pi}R_\parallel]^{-1/2}.
 \end{aligned}\label{SM:eq:psi_initial}
\end{equation}
Here $w=R_\perp/\sqrt{|\ell|+1}$ and $\int|\psi_0|^2\dd^3\delta r_0=1$. {The Gaussian widths in $\sigma_0$ follow Ref.~\cite{SM:KarlovetsSerbo2020}. The polynomial supplies the phase winding and also contributes to $R_\perp$ for $\ell\ne0$.} These sizes and the corresponding momentum widths determine the transverse and longitudinal diffraction times~\cite{SM:Karlovets2021},
\begin{equation}
 \tau_{d,\perp}=\frac{mw^2}{\hbar}
 =\frac{mR_\perp^2}{\hbar(|\ell|+1)},\qquad
 \tau_{d,\parallel}=\frac{2mR_\parallel^2}{\hbar}.
 \label{SM:eq:psi_diffraction_times}
\end{equation}
They are the rest-frame times over which free spreading increases each rms size by $\sqrt2$. {{At the entrance waist, the quadratic spatial phase of the Gaussian \mbox{envelope vanishes}.}}

{The matched entrance state $\psi(0^+)=\hat U_{\rm in}\psi_0$ is the initial condition for the Schr\"odinger equation. Entrance matching synchronizes the packet on a common rest-frame time plane and adds the phase derived in Sec.~\ref{SM:sec:boundary} from Eq.~\eqref{SM:eq:SMentry}. Inside the magnet, the state evolves with $\hat{\mathcal H}_{\rm orb}$ [Eq.~\eqref{SM:eq:Horb}]: kinetic energy causes spreading, the magnetic field couples radial and tangential motion, and the acceleration potential changes the radial evolution. At exit, we apply the inverse matching phase in the local exit axes. For a uniform magnet with sharp boundaries perpendicular to the reference orbit,}
\begin{equation}
 \begin{aligned}
 \psi_{\rm in\,field}(\tau)
 &=e^{-i\tau\hat{\mathcal H}_{\rm orb}/\hbar}
   e^{-ieH'{\delta r_{r,0}}{\delta r_{t,0}}/(2\hbar)}\psi_0,\\
 \psi_{\rm out}(\delta\vct r)
 &=e^{ieH'\delta r_r\delta r_t/(2\hbar)}
   \psi_{\rm in\,field}(\delta\vct r,\tau_{\rm exit}),\\
 \psi_{\rm drift}(\Delta\tau)
 &=e^{-i\Delta\tau(\delta\hat{\vct p})^2/(2m\hbar)}\psi_{\rm out},
 \end{aligned}\label{SM:eq:psi_stages}
\end{equation}
where $H'=\gamma H$, $\theta=eH\tau/m$, and $\Delta\tau$ is the proper time since leaving the magnet.

{The canonical transport used for the moments also determines the propagation phase in Eq.~\eqref{SM:eq:psi_stages}. The required classical action for the internal motion is
\begin{equation}
 \mathcal S_{\rm cl}
 =\int_0^\tau\!\big(\delta\vct p\cdot\delta\dot{\vct r}-\mathcal H\big)\,\dd\tau'
 +\sum_{\text{crossed boundaries}}\chi.
 \label{SM:eq:psi_classical_action}
\end{equation}
The integral follows the matched classical trajectory. Here $\mathcal H$ is the classical Hamiltonian of Eq.~\eqref{SM:eq:Horb} inside the magnet and $(\delta\vct p)^2/(2m)$ during a free drift. The sum adds $\chi_{\rm in}=-eH'{\delta r_{r,0}}{\delta r_{t,0}}/2$ at entrance and, after exit, $\chi_{\rm out}=eH'\delta r_r\delta r_t/2$ evaluated on the exit plane [Eq.~\eqref{SM:eq:psi_stages}]. Its endpoint derivatives are $\delta\vct p=\nabla_{\delta r}\mathcal S_{\rm cl}$ and $\delta\vct p_0=-\nabla_{\delta r_0}\mathcal S_{\rm cl}$ at fixed times. For the quadratic Hamiltonian retained in Eq.~\eqref{SM:eq:SMexpansion}, this action gives the exact Gaussian propagator~\cite{SM:Littlejohn1986,SM:Folland1989,SM:deGosson2011}. The corresponding canonical transfer is}
\begin{equation}
 S_\psi=\begin{cases}
 S_{\rm bulk}(\theta)C_{\rm in},
   &\text{inside the magnet},\\
 S(\theta_{\rm exit}),
   &\text{just after exit},\\
 \begin{pmatrix}
 I&\Delta\tau I/m\\
 0&I
 \end{pmatrix}
 S(\theta_{\rm exit}),
   &\text{after a free drift}.
 \end{cases}
 \label{SM:eq:psi_transfer}
\end{equation}

The matrices $C_{\rm in}$, $S_{\rm bulk}$, and $S$ are given in Eqs.~\eqref{SM:eq:SMfaces}, \eqref{SM:eq:bulkmap}, and \eqref{SM:eq:map}, respectively, and $I$ is the $3\times3$ identity matrix. {The boundary phases in Eq.~\eqref{SM:eq:psi_stages} enter through $C_{\rm in}$ and, after exit, through $C_{\rm out}$ in $S=C_{\rm out}S_{\rm bulk}C_{\rm in}$.} {We divide this matrix into position and \mbox{momentum blocks},}
\begin{equation}
 S_\psi=\begin{pmatrix}S_{rr}&S_{rp}\\S_{pr}&S_{pp}\end{pmatrix},\qquad
 \delta\hat{\vct r}=S_{rr}\delta\hat{\vct r}_0+S_{rp}\delta\hat{\vct p}_0,
 \label{SM:eq:psi_blocks}
\end{equation}
in the order $(r_r,r_t,r_H,p_r,p_t,p_H)$. {The $3\times3$ blocks of $S_\psi$ give the dependence of final coordinates and momenta on their initial values~\cite{SM:BagrovBelovTrifonov2005}. For example, $S_{rp}=\partial\delta\vct r/\partial\delta\vct p_0$ at fixed $\delta\vct r_0$. The first index labels the final variable and the second the initial one. Thus $S_{rr},S_{pp}$ are dimensionless, $S_{rp}$ has units of length/momentum, and $S_{pr}$ of momentum/length.}

{To obtain the propagator, we first express the action in terms of the initial and final positions. For invertible $S_{rp}$, Eq.~\eqref{SM:eq:psi_blocks} and the symplectic identity $S_\psi JS_\psi^T=J$ give the endpoint momenta,
\[
 \begin{aligned}
 \delta\vct p_0&=S_{rp}^{-1}(\delta\vct r-S_{rr}\delta\vct r_0),\\
 \delta\vct p&=S_{pp}S_{rp}^{-1}\delta\vct r-(S_{rp}^{-1})^T\delta\vct r_0.
 \end{aligned}
\]
Integrating the endpoint relations gives the action below, including the boundary phases already contained in $S_\psi$. Substitution into the Schr\"odinger equation determines the propagator's prefactor up to a constant, which is fixed by unitarity:
\[
 \int_{\mathbb R^3} U_\psi(\delta\vct r,\delta\vct r_0)
 U_\psi^*(\delta\vct r',\delta\vct r_0)\,\dd^3\delta r_0
 =\delta^{(3)}(\delta\vct r-\delta\vct r'),
\]
Both propagators are evaluated at the same final time. The prefactor therefore has modulus $(2\pi\hbar)^{-3/2}|\det S_{rp}|^{-1/2}$. The absolute Van Vleck determinant $|\det S_{rp}|^{-1}$ measures the focusing of neighboring trajectories: it is the Jacobian from final position to initial momentum at fixed initial position and endpoint times. Combining this prefactor with the action gives~\cite{SM:BagrovBelovTrifonov2005}
\begin{equation}
 \begin{aligned}
 U_\psi(\delta\vct r,\delta\vct r_0)
 &=\frac{e^{i\mathcal S_{\rm cl}(\delta\vct r,\delta\vct r_0)/\hbar}}
 {\sqrt{\det(2\pi i\hbar S_{rp})}},\\
 \mathcal S_{\rm cl}&=\frac12\delta\vct r^TS_{pp}S_{rp}^{-1}\delta\vct r
 -\delta\vct r_0^TS_{rp}^{-1}\delta\vct r\\
 &\quad+\frac12\delta\vct r_0^TS_{rp}^{-1}S_{rr}\delta\vct r_0,\\
 \psi(\delta\vct r,\tau)&=\int_{\mathbb R^3}U_\psi(\delta\vct r,\delta\vct r_0)
 \psi_0(\delta\vct r_0)\,\dd^3\delta r_0.
 \end{aligned}\label{SM:eq:psi_propagator}
\end{equation}
For a free interval, $S_{rr}=S_{pp}=I$ and $S_{rp}=\Delta\tau I/m$, so Eq.~\eqref{SM:eq:psi_propagator} reduces to
\[
 U_\psi=\left(\frac{m}{2\pi i\hbar\Delta\tau}\right)^{3/2}
 \exp\!\left[\frac{im|\delta\vct r-\delta\vct r_0|^2}{2\hbar\Delta\tau}\right],
\]
which tends to $\delta^{(3)}(\delta\vct r-\delta\vct r_0)$ as $\Delta\tau\to0^+$. This fixes the initial square-root phase, which is continued along the evolution. At a classical focus, including $\theta=2\pi N$, $S_{rp}$ is singular and the coordinate kernel has a distributional limit. The Gaussian integral below gives a regular finite packet through the focus.}

{To propagate the LG polynomial in Eq.~\eqref{SM:eq:psi_initial}, we introduce an auxiliary dimensionless source $J_s$ and replace the polynomial with $\exp[J_s(\delta r_{r,0}-i\operatorname{sgn}(\ell)\delta r_{H,0})/w]$~\cite{SM:BagrovBelovTrifonov1996,SM:BagrovBelovTrifonov2005}. Taking $|\ell|$ derivatives with respect to $J_s$ at $J_s=0$ restores the polynomial. Completing the square in the three incident coordinates then gives}
\begin{equation}
 \begin{aligned}
 &\mathcal N_0\int_{\mathbb R^3} U_\psi
 e^{-\delta\vct r_0^T\sigma_0^{-2}\delta\vct r_0/2
 +J_s({\delta r_{r,0}}-i\operatorname{sgn}(\ell){\delta r_{H,0}})/w}\,\dd^3\delta r_0\\
 &\qquad=\frac{\mathcal N_0}{\sqrt{\det \mathcal D_\psi}}
 e^{-\delta\vct r^T\sigma_\psi^{-2}\delta\vct r/2
 +J_s\vct d_\psi^T\delta\vct r+J_s^2b_\psi/2}.
 \end{aligned}\label{SM:eq:psi_gaussian_integral}
\end{equation}
{Combining the propagator's quadratic action with the incident Gaussian gives the spreading matrix $\mathcal D_\psi$ and the complex inverse squared width $\sigma_\psi^{-2}$~\cite{SM:BagrovBelovTernov1983,SM:BagrovBelovTrifonov1993,SM:BagrovBelovTrifonov2006},}
\begin{equation}
 \begin{aligned}
 \mathcal D_\psi&=S_{rr}+i\hbar S_{rp}\sigma_0^{-2},\\
 \sigma_\psi^{-2}&=-\frac{i}{\hbar}(S_{pr}+i\hbar S_{pp}\sigma_0^{-2})\mathcal D_\psi^{-1}.
 \end{aligned}\label{SM:eq:psi_gaussian_coefficients}
\end{equation}
{The dimensionless matrix $\mathcal D_\psi$ fixes the normalization and overall phase. The real part of $\sigma_\psi^{-2}$ determines the Gaussian envelope, while its imaginary part gives the quadratic spatial phase. For free propagation from the incident waist, these coefficients give}
\begin{equation}
 \begin{aligned}
 (\mathcal D_\psi)_{ii}&=1+i\Delta\tau/\tau_{d,i},\\
 (\sigma_\psi^{-2})_{ii}&=\frac{m}{\hbar\tau_{d,i}}
                \frac{1}{1+i\Delta\tau/\tau_{d,i}},\\
 R_i(\Delta\tau)&=R_i(0)\sqrt{1+(\Delta\tau/\tau_{d,i})^2},
 \end{aligned}
 \label{SM:eq:psi_free_spreading}
\end{equation}
where $i=r,t,H$, $\tau_{d,r}=\tau_{d,H}=\tau_{d,\perp}$, $\tau_{d,t}=\tau_{d,\parallel}$, and $R_i$ is the rms size in that direction. {The laboratory Rayleigh length is $z_{R,i}=\gamma v\tau_{d,i}$.} The transverse and longitudinal Gouy phases are $-(|\ell|+1)\arctan(\Delta\tau/\tau_{d,\perp})$ and $-\tfrac12\arctan(\Delta\tau/\tau_{d,\parallel})$, respectively. {Inside the magnet, radial and tangential motion become coupled, producing the off-diagonal elements of $\sigma_\psi^{-2}$.}

{For finite incident widths, $\det\mathcal D_\psi$ stays nonzero by Property~10.10 of Ref.~\cite{SM:BagrovBelovTrifonov2005}. Its Eq.~(10.16) gives $\operatorname{Re}\sigma_\psi^{-2}=(\mathcal D_\psi^\dagger)^{-1}\sigma_0^{-2}\mathcal D_\psi^{-1}>0$, preserving a normalizable Gaussian envelope.} {Diffraction thus keeps the packet finite at a classical focus.}

{The terms linear and quadratic in $J_s$ determine the evolving vortex polynomial. Completing the square in Eq.~\eqref{SM:eq:psi_gaussian_integral} gives its coefficients, the inverse-length vector $\vct d_\psi$ and the dimensionless scalar $b_\psi$:}
\begin{equation}
 \begin{aligned}
 \vct d_\psi&=\frac1w(\mathcal D_\psi^{-1})^T
 \begin{pmatrix}1\\0\\-i\operatorname{sgn}(\ell)\end{pmatrix},\\
 b_\psi&=\frac{i\hbar}{w^2}\Big[(\mathcal D_\psi^{-1}S_{rp})_{rr}
 -(\mathcal D_\psi^{-1}S_{rp})_{HH}\\
 &\hspace{34mm}-2i\operatorname{sgn}(\ell)(\mathcal D_\psi^{-1}S_{rp})_{rH}\Big].
 \end{aligned}\label{SM:eq:psi_polynomial_coefficients}
\end{equation}
{Differentiating the generating Gaussian in Eq.~\eqref{SM:eq:psi_gaussian_integral} $|\ell|$ times with respect to $J_s$ at $J_s=0$ gives the full wave function for the canonical transfer matrix $S_\psi$ in Eq.~\eqref{SM:eq:psi_transfer}:}
\begin{equation}
 \begin{aligned}
 \psi(\delta\vct r,\tau)=&\,\frac{\mathcal N_0}{\sqrt{\det \mathcal D_\psi}}
 e^{-\delta\vct r^T\sigma_\psi^{-2}\delta\vct r/2}\,|\ell|!\\
 &\times\sum_{j=0}^{\lfloor|\ell|/2\rfloor}
 \frac{b_\psi^j(\vct d_\psi^T\delta\vct r)^{|\ell|-2j}}
 {2^j j!(|\ell|-2j)!}.
 \end{aligned}\label{SM:eq:SMpsi}
\end{equation}
{Before entrance matching, the canonical transfer matrix is $S_\psi=I_6$, so Eq.~\eqref{SM:eq:SMpsi} recovers the incident LG state in Eq.~\eqref{SM:eq:psi_initial}. The magnet changes the spatial phase and relative spreading in different directions, so the propagated state generally departs from a single LG mode.}

{\textit{Connection with the Bagrov--Maslov method.---}The trajectory-coherent construction propagates a packet around its center's classical trajectory using linearized Hamiltonian equations~\cite{SM:BagrovBelovMaslov1989,SM:BagrovBelovTernov1983,SM:BagrovBelovTrifonov1993}. We apply it to our quadratic rest-frame Hamiltonian [Eq.~\eqref{SM:eq:SMexpansion}], including boundary matching. Quantum propagation is exact within this quadratic approximation.

Complex coordinate and momentum variations form the $3\times3$ matrices $C$ and $B$. Before entrance matching, the incident Gaussian gives $C_0=I$ and $B_0=i\hbar\sigma_0^{-2}$. The same canonical matrix $S_\psi$ that transports the \mbox{moments gives}
\[
 \begin{pmatrix}C\\B\end{pmatrix}
 =S_\psi\begin{pmatrix}I\\i\hbar\sigma_0^{-2}\end{pmatrix}.
\]
Thus $C=\mathcal D_\psi$ and $Q=BC^{-1}=i\hbar\sigma_\psi^{-2}$ [Eq.~\eqref{SM:eq:psi_gaussian_coefficients}]. The real part of $Q$ fixes the quadratic spatial phase, while its positive imaginary part determines the decaying Gaussian envelope. These variations around the real trajectory span Maslov's complex germ~\cite{SM:Maslov1994,SM:BagrovBelovTrifonov1996}.

Between the magnet boundaries, differentiating $Q=BC^{-1}$ using the linearized Hamiltonian equations gives the Riccati equation [Ref.~\cite{SM:BagrovBelovTrifonov2005}, Eq.~(9.12)]
\[
 \dot Q+H_{rr}+H_{rp}Q+QH_{pr}+QH_{pp}Q=0.
\]
Here $\dot Q=\dd Q/\dd\tau$, and $H_{ab}$ contains second derivatives of the classical Hamiltonian with respect to coordinates and canonical momenta, $a,b\in\{r,p\}$. The acceleration potential contributes $-m\vct a'\vct a'^T/c^2$ to $H_{rr}$ in FW axes. Thus linear transport determines $Q$, which satisfies this equation automatically. Boundary matching enters $S_\psi$ through $C_{\rm in}$ and $C_{\rm out}$ [Eq.~\eqref{SM:eq:psi_transfer}].

For $\ell=0$, the Gaussian is the vacuum state of Ref.~\cite{SM:BagrovBelovTrifonov2005}, Eq.~(11.5): its lowering operators annihilate this state. The covariance $\Sigma_{rr}=\tfrac12\operatorname{Re}(C\sigma_0^2C^\dagger)$ gives the same transport of coordinate fluctuations about the classical trajectory as Ref.~\cite{SM:BagrovBelovMaslov1989} (Russian original, p.~90), with our incident widths [Eq.~\eqref{SM:eq:covariance_transport}].

Combining the two transverse raising operators with relative phase factor $-i\operatorname{sgn}\ell$ and applying this combination $|\ell|$ times produces the LG polynomial and its phase winding {[Eq.~\eqref{SM:eq:psi_initial} and Ref.~\cite{SM:BagrovBelovTrifonov2005}, Eq.~(11.6)]}. Our $|\ell|$ derivatives with respect to $J_s$ at $J_s=0$ implement this construction [Eqs.~\eqref{SM:eq:psi_gaussian_integral} and \eqref{SM:eq:SMpsi}]. The Gaussian envelope and vortex polynomial evolve with the same $S_\psi$, while the kinetic OAM can change.

For comparison, the momentum--position order in Ref.~\cite{SM:BagrovBelovTrifonov2005}, Eq.~(10.36), gives
\[
 \lambda_1=S_{rr}^T,\quad \lambda_2=-S_{pr}^T,\quad
 \lambda_3=-S_{rp}^T,\quad \lambda_4=S_{pp}^T.
\]
Its Eq.~(14.20) gives our propagator, Eq.~\eqref{SM:eq:psi_propagator}, with the same Van Vleck determinant, after separating the center's action and translation phase. In its generating function, Eq.~(12.12), set $D_0=\hbar\sigma_0^{-2}$, $C=\mathcal D_\psi$, and
\[
 \vct\alpha=-\frac{iJ_s}{\sqrt2}(1,0,-i\operatorname{sgn}\ell)^T,
\]
then remove the coherent-state factor $e^{-|\vct\alpha|^2/2}$ and choose normalization $\mathcal N_0$. This reproduces both source terms in Eq.~\eqref{SM:eq:psi_gaussian_integral}. As a limiting check, the uniform-field bulk transfer in FW axes reduces, for nonrelativistic center motion, to its Example~10.3, Eqs.~(10.42)--(10.43), after converting Gaussian units to SI and taking charge $-e$.}

We use the intrinsic kinetic-OAM operator $\hat{\bm{\mathcal L}}$ in Eq.~\eqref{SM:eq:OAM_covariance}, with $\delta\hat{\vct\pi}=-i\hbar\nabla_{\delta r}+e\vct A$ inside and $-i\hbar\nabla_{\delta r}$ outside. The exit phase in Eq.~\eqref{SM:eq:psi_stages} and the disappearance of the vector potential together change the kinetic OAM, even though the matched densities coincide. For $i,j\in\{r,t,H\}$, we obtain the mean and symmetrized covariance by integration:
\begin{align}
 \mean{\hat{\mathcal L}_i}&=\int\psi^*\hat{\mathcal L}_i\psi\,\dd^3\delta r,\nonumber\\
 V_{ij}&=\operatorname{Re}\int
 [ (\hat{\mathcal L}_i-\mean{\hat{\mathcal L}_i})\psi]^*
 [ (\hat{\mathcal L}_j-\mean{\hat{\mathcal L}_j})\psi]\,\dd^3\delta r.
 \label{SM:eq:SMpsiMom}
\end{align}
{In Eq.~\eqref{SM:eq:SMpsiMom}, the spatial derivatives in $\delta\hat{\vct\pi}$ act on both the Gaussian and the vortex polynomial.} {These integrals reproduce Eqs.~\eqref{SM:eq:mean_trace} and \eqref{SM:eq:V}.}

For a smooth fringe, we integrate the reference Lorentz trajectory and its linearized deviations as described in Sec.~\ref{SM:sec:fringes}, Eq.~\eqref{SM:eq:fringe_transfer}. Converting the resulting transfer to canonical variables on the entrance and final rest-time slices gives $S_\psi$. Substitution into Eqs.~\eqref{SM:eq:psi_gaussian_coefficients}--\eqref{SM:eq:SMpsi} then gives the propagated wave function.

\textit{Phase singularities.---}A vortex line is a zero of $\psi$ around which its phase winds by an integer multiple of $2\pi$. A transported state can have a narrow intrinsic-OAM distribution [Eq.~\eqref{SM:eq:width_angle}] even when its mean OAM is tilted and the incident multiply charged vortex has split~\cite{SM:BliokhNori2012}. {Thus small $W$ can recur in a state whose spatial structure differs from the incident LG mode.}

The unequal phase evolution and spreading in the bending plane and along the field act analogously to optical astigmatism~{\cite{SM:Dennis2006}}. {For $b_\psi\ne0$, the polynomial in Eq.~\eqref{SM:eq:SMpsi} is a Hermite polynomial of a complex argument. If $\operatorname{Re}\vct d_\psi\times\operatorname{Im}\vct d_\psi\ne0$, each simple zero imposes two independent real conditions on the position $\delta\vct r$, defining a vortex line.}

For the $\ell=64$ example in Fig.~\ref{SM:fig:packet_story} (\MovieNumber{2}), the polynomial in Eq.~\eqref{SM:eq:SMpsi} changes from one zero of multiplicity 64 at $b_\psi=0$ to 64 simple zeros at $b_\psi\ne0$. They define parallel vortex lines along $\operatorname{Re}\vct d_\psi\times\operatorname{Im}\vct d_\psi$, generally distinct from the mean-OAM direction. The exit phase preserves these zeros. Splitting is therefore compatible with the outgoing $W\simeq12\%$. Subsequent free propagation preserves the OAM distribution while the zeros move.

\subsection{OAM projection after exit}\label{SM:sec:projection}\label{SM:sec:projection_spectrum}
{The outgoing state determines the probabilities of individual OAM projections. To calculate them, we resolve the outgoing state $\psi_{\rm out}$ [Eq.~\eqref{SM:eq:SMpsi} with $S_\psi=S(\theta_{\rm exit})$] into eigenstates of $\hat{\mathcal L}_t=-i\hbar(\delta r_H\partial_{\delta r_r}-\delta r_r\partial_{\delta r_H})$. The coordinates are measured from the packet center on the exit rest-time plane. The probability $P_k$ of measuring $k\hbar$, $k\in\mathbb Z$, includes all radial and longitudinal states.}

{The orbital helicity $\hat{\mathcal L}_t/\hbar$~\cite{SM:IvanovSerbo2011,SM:IvanovHelicity2012} is defined along the outgoing tangent, fixed by the laboratory mean kinetic momentum and carried into the packet rest frame. For a tilted mean OAM, this axis differs from the mean-OAM axis $\vct n$ used in $W$ [Eq.~\eqref{SM:eq:width_angle}].}

{We expand the outgoing state in orthonormal eigenstates of $\hat{\mathcal L}_t$. Rotation through an angle $\phi$ multiplies every component with projection $k\hbar$ by $e^{-ik\phi}$. Orthogonality therefore makes the overlap with the rotated state a sum of these phases weighted by the probabilities $P_k$. Fourier integration isolates each probability:}
\begin{align}
 \chi(\phi)&=\int\psi_{\rm out}^*(\delta\vct r)
 \psi_{\rm out}(R_t^{-1}(\phi)\delta\vct r)\dd^3\delta r
          =\sum_kP_ke^{-ik\phi},\nonumber\\
 P_k&=\frac1{2\pi}\int_0^{2\pi}e^{ik\phi}\chi(\phi)\dd\phi .
 \label{SM:eq:projection}
\end{align}
{Here $R_t(\phi)$ rotates the coordinates through $\phi$ about the outgoing tangent, $\chi$ is the complex overlap, and $|\chi|^2$ is the overlap probability. Normalization gives $\chi(0)=\sum_kP_k=1$. For the state in Eq.~\eqref{SM:eq:SMpsi}, we integrate each polynomial term with its Gaussian weight analytically and evaluate the remaining integral over $\phi$ numerically.}

\begin{figure}[!t]
 \includegraphics[width=\textwidth]{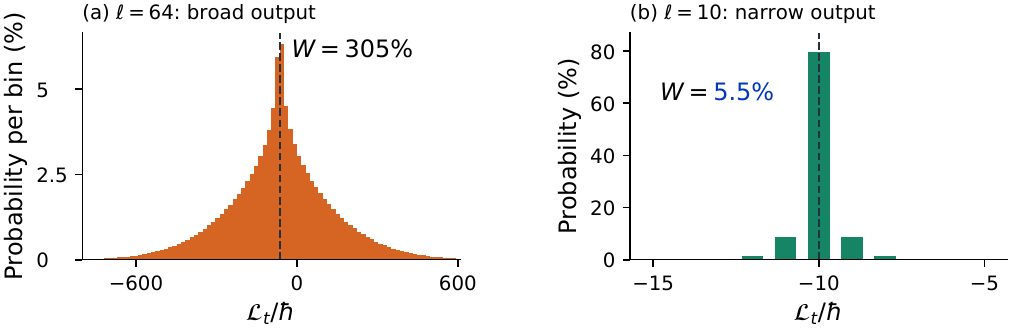}
 \caption{Probabilities $P_k$ of measuring the {tangential intrinsic kinetic OAM projection}
 $\mathcal L_t=k\hbar$ after leaving the magnet, calculated from the complex wave
 function with Eq.~\eqref{SM:eq:projection}. The integer $k$ labels OAM
 eigenvalues along the outgoing momentum, and all radial and longitudinal
 states are included.
 (a) Incident $\ell=64$, $R_\perp=3~\mu$m: each bar sums $P_k$ over
 16 adjacent integers.
 (b) Incident $\ell=10$, $R_\perp=9.6~\mu$m: each bar gives one $P_k$.
 Dashed lines mark the mean projections, $-64\hbar$ and $-10\hbar$,
 with probabilities $P_{-64}={0.7\%}$ and $P_{-10}=79.6\%$.
 Both packets are incident LG states of Eq.~\eqref{SM:eq:LG}, with
 rms length $R_\parallel=R_\perp/[\sqrt2(|\ell|+1)]$.
 The kinetic energy is $K=1$~MeV, the field is $H=5$~mT, and the bend is
 $\theta=180^\circ$.}
 \label{SM:fig:projection}
\end{figure}

After leaving a $180^\circ$ bend, the two states below have mean $(0,-\ell,0)\hbar$. Both satisfy the incident momentum-width matching condition $R_\parallel=R_\perp/[\sqrt2(|\ell|+1)]$ [Eq.~(\Lref{eq:half}) of the Paper]. The laboratory energy is $K=1$~MeV and the field is $H=5$~mT:
\begin{center}
\begin{tabular}{rrrr}
$\ell$&$R_\perp$ ($\mu$m)&$W$ (\%)&$P_{-\ell}$ (\%)\\\hline
64&3&305&{0.7}\\
10&9.6&{5.5}&79.6
\end{tabular}\par
\end{center}
{The probability $P_{-\ell}$ is the fraction of measurements yielding the reversed projection $-\ell\hbar$. For $\ell=10$ and $R_\perp=9.6~\mu$m, the mean OAM reverses exactly while $P_{-\ell}=79.6\%$ [Fig.~\ref{SM:fig:projection}].}

The probabilities are normalized and reproduce the operator mean and variance. For the incident eigenstate, Eq.~\eqref{SM:eq:projection} gives $\chi=e^{-i\ell\phi}$ and $P_k=\delta_{k\ell}$, as expected.

\subsection{Density and OAM through a weak dipole}\label{SM:sec:weak_density}
\begin{figure}[!t]
\centering
\includegraphics[width=\textwidth]{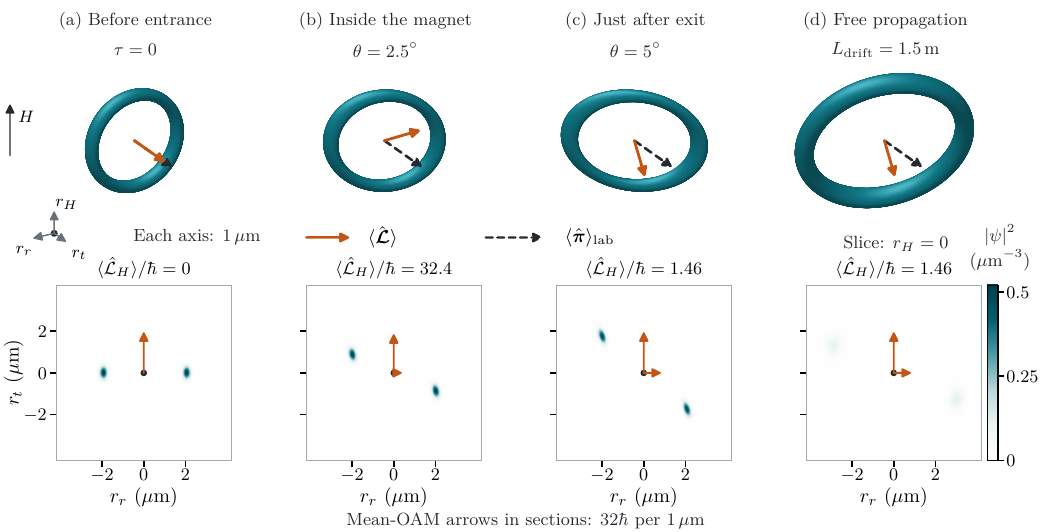}
\caption{{Probability density from Eq.~\eqref{SM:eq:SMpsi} at four proper times $\tau$.
Surfaces: 20\% of the instantaneous density maximum, with illumination shading.
Lower row: $|\psi(r_r,r_t,0;\tau)|^2$ on a common color scale in $\mu$m$^{-3}$.
The centered instantaneous rest-frame coordinates $r_r,r_t,r_H$ are radial,
tangential, and along the field. Each row has a common spatial scale, and each triad axis represents $1~\mu$m.
Upper solid and dashed arrows give only the directions of $\mean{\hat{\bm{\mathcal L}}}$
and $\mean{\hat{\vct\pi}}_{\rm lab}$. Lower orange arrows give
$\mean{\hat{\mathcal L}_r}$ and $\mean{\hat{\mathcal L}_t}$ with calibration $32\hbar$ per $1~\mu$m.
Numbers give $\mean{\hat{\mathcal L}_H}/\hbar$.
Parameters: $K=4.6$~MeV~\cite{SM:IMP2024}, $H=5$~mT, a $5^\circ$ bend with sharp boundaries,
$\ell=64$, $R_\perp=2~\mu$m, and $R_\parallel=0.2~\mu$m.
In (d), the exit axes remain fixed during a laboratory drift $L_{\rm drift}=\gamma v\Delta\tau=1.5$~m,
with $\Delta\tau=500$~ps. \MovieNumber{2} gives the intervening states and the clocks $\tau$ and $t=\gamma\tau$.}}
\label{SM:fig:packet_story}
\end{figure}

\begin{figure}[!htbp]
 \centering
 \includegraphics[width=0.82\columnwidth]{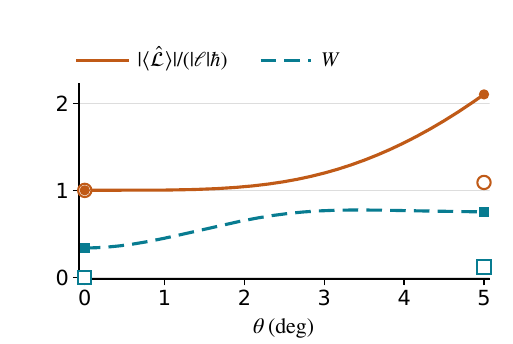}
 \caption{Evolution of $|\mean{\hat{\bm{\mathcal L}}}|/(|\ell|\hbar)$
 and the relative rms width $W$ for the packet in {Fig.~\ref{SM:fig:packet_story}}.
 The width is taken along the mean OAM, as in
 Eq.~\eqref{SM:eq:width_angle}, and both curves follow from
 Eq.~\eqref{SM:eq:SMpsiMom}.
 The curves show the in-field state over $0\leq\theta\leq5^\circ$.
 Filled markers give its boundary values, and open markers the
 corresponding free states. After leaving the magnet,
 $|\mean{\hat{\bm{\mathcal L}}}_{\rm out}|=69.8\hbar$ and
 $W={12\%}$. The magnet has $H=5$~mT, and the incident electron has $K=4.6$~MeV,
 $\ell=64$, $R_\perp=2~\mu$m, and $R_\parallel=0.2~\mu$m.}
 \label{SM:fig:weak_oam_trace}
\end{figure}

{Figs.~\ref{SM:fig:packet_story} and \ref{SM:fig:weak_oam_trace} and \MovieNumber{2} follow the $\ell=64$ packet through a $5^\circ$ bend at common instantaneous rest-frame time. After exit, its density spreads while the mean OAM and $W$ remain constant. Sec.~\ref{SM:sec:supplemental_movies} collects all movie descriptions, including the $90^\circ$ laboratory-time examples in Movies~\MovieIndex{1} and \MovieIndex{3}.}

\Needspace{8\baselineskip}
\subsection{Internal evolution and boundary matching}
\label{SM:sec:movie_comparison}
{The entrance and exit phases are inverse operators, yet the packet generally leaves in a different state because it evolves between the boundaries. {Fig.~\Lref{fig:boundary_comparison} of the Paper}, separates these effects for the $5^\circ$ packet in Fig.~\ref{SM:fig:packet_story}. One calculation holds the internal state fixed between the boundaries, while the other evolves it with Eq.~\eqref{SM:eq:Horb}. Both apply the same entrance and exit matrices $C_{\rm in,out}$ [Eq.~\eqref{SM:eq:SMfaces}].}

Writing $\hat U_{\rm bulk}$ for the internal evolution operator
from the entrance to the exit, the outgoing state is
\begin{equation}
 \begin{aligned}
 \psi_{\rm out}&=\hat U_{\rm out}\hat U_{\rm bulk}
                       \hat U_{\rm in}\psi_0,\\
 \hat U_{\rm in}&=e^{-ieH'\delta\hat r_r\delta\hat r_t/(2\hbar)},
 \qquad \hat U_{\rm out}=\hat U_{\rm in}^{-1}.
 \end{aligned}\label{SM:eq:movie_two_modes}
\end{equation}
The entrance and exit operators act in their respective local boundary
axes. With $\hat U_{\rm bulk}=\hat I$, where $\hat I$ is the identity
operator, the two phases cancel and $\psi_{\rm out}=\psi_0$.
This equality restores the internal state relative to the outgoing
center and its local axes, although the center's mean momentum has turned
through $5^\circ$. Evolution between the boundaries uses the transfer in {Eq.~\eqref{SM:eq:bulkmap}}, with the mean and covariance obtained from
Eq.~\eqref{SM:eq:SMpsiMom}.

{The density sections $|\psi(\delta r_r,\delta r_t,0;\tau)|^2$ in {Fig.~\Lref{fig:boundary_comparison} of the Paper} use common spatial and color scales. To show the momentum change accompanying the boundary phase, arrows give the local radial kinetic momentum,}
\begin{equation}
 \begin{aligned}
 \pi_r^{\rm loc}(\delta\vct r,\tau)
 &=\frac{\operatorname{Re}[\psi^*{\delta\hat\pi_r}\psi]}{|\psi|^2}\\
 &=\hbar\operatorname{Im}\!\left(\frac{\partial_{{\delta r_r}}\psi}{\psi}\right)
       +eA_r.
 \end{aligned}\label{SM:eq:local_kinetic_momentum}
\end{equation}

Here the derivative acts on the state in Eq.~\eqref{SM:eq:SMpsi}, and
$A_r=-H'{\delta r_t}/2$ inside the uniform magnet and zero in vacuum.
{At $\times1$, the arrows represent $\pi_r^{\rm loc}$ on a scale of $100\hbar/\mu$m per $1~\mu$m of the spatial axes, and the factors printed in the panels give their magnification.} Motion of the density in orbital coordinates also includes the rotation of the axes described in Sec.~\ref{SM:sec:dipole}. 

{Because the packet front ($r_t>0$) enters the field earlier, matching at a common rest-frame time gives the local radial kinetic-momentum shift $-eH'r_t$ [Sec.~\ref{SM:sec:boundary}].} Since
$\hat{\mathcal L}_H=\delta\hat r_r\delta\hat\pi_t-
\delta\hat r_t\delta\hat\pi_r$, the resulting change of the
field-axis mean OAM is $eH'R_\parallel^2={3}\hbar$.
{At exit, the opposite shift $eH'r_t$ removes}
\begin{equation}
 eH'\mean{(\delta\hat r_t)^2}_{\rm exit}
 \label{SM:eq:exit_H_removal}
\end{equation}
{from the field-axis mean OAM. The average is taken in the state that reaches the exit, so cancellation depends on its evolution inside the magnet. Holding that state fixed removes exactly the induced $3\hbar$ and restores the incident state.}

With internal evolution, $\mean{(\delta\hat r_t)^2}$ grows from
$0.04~\mu$m$^2$ at entry to ${1.6}~\mu$m$^2$ at exit.
Just before exit the mean is
$(27.9,51.8,121.03)\hbar$ in radial, tangential,
and field directions. Exit matching then gives
\begin{equation}
 \mean{\hat{\bm{\mathcal L}}}_{\rm out}
 \simeq(27.9,64,1.46)\hbar.
 \label{SM:eq:movie_exit_mean}
\end{equation}
{Exit matching removes $119.57\hbar$ along the field. The remaining correlations $\mean{\delta\hat r_t\delta\hat\pi_r}_{\rm out}=-1.43\hbar$ and $\mean{\delta\hat r_r\delta\hat\pi_t}_{\rm out}=0.0313\hbar$ give their difference $\mean{\hat{\mathcal L}_H}_{\rm out}=1.46\hbar$. The final mean has magnitude $69.8\hbar$, $\alpha_L=23.6^\circ$, and $W=12\%$.}

\subsection{Acceleration potential in a uniform dipole}
\label{SM:sec:movie_acceleration}
\begin{figure}[tp]
 \centering\includegraphics[width=.73\textwidth]{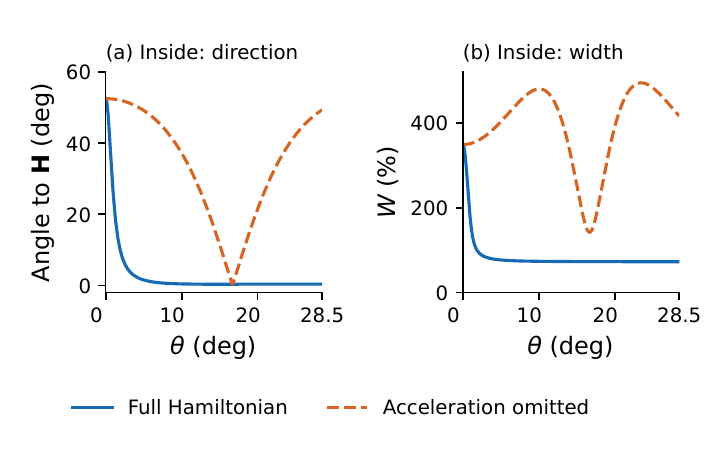}
 \caption{Effect of the acceleration potential for identical entrance data. Solid blue retains the full quadratic Hamiltonian in Eq.~\eqref{SM:eq:SMexpansion}, whereas dashed orange omits only the term $-m(\vct a'\cdot\delta\hat{\vct r})^2/(2c^2)$. Both evolutions follow the linear Heisenberg equations [Eq.~\eqref{SM:eq:Sode}], with kinetic conversion [Eq.~\eqref{SM:eq:B}] and OAM moments [Eqs.~\eqref{SM:eq:mean_trace} and \eqref{SM:eq:V}]. (a) In-field angle $\arccos[\mean{\hat{\mathcal L}_H}/|\mean{\hat{\bm{\mathcal L}}}|]$ of the mean OAM $\mean{\hat{\bm{\mathcal L}}}$ to $+\vct H$. (b) In-field relative rms width $W$ along the mean [Eq.~\eqref{SM:eq:width_angle}]. Parameters are $\ell=64$, $R_\perp=3~\mu$m, $R_\parallel=0.3~\mu$m, $K=5$~MeV, $H=33.3$~mT, $\rho=0.55$~m, and final bend $28.5^\circ$.}
 \label{SM:fig:acceleration}
\end{figure}

{The acceleration potential changes the correlations accumulated between entrance and exit. To quantify its contribution, Fig.~\ref{SM:fig:acceleration} compares the full Hamiltonian in Eq.~\eqref{SM:eq:Horb} with evolution in which only this potential is omitted.} The incident electron has $\ell=64$, $R_\perp=3~\mu$m, $R_\parallel=0.3~\mu$m, and $K=5$~MeV, and the magnet has $\rho=0.55$~m and a $28.5^\circ$ bend. Both calculations use the same reference orbit, field, incident state, and entrance matching. Applying the same exit transformation then gives the two outgoing states.

The entrance shear in Eq.~\eqref{SM:eq:SMcrossing} gives
\[
 \tfrac12\mean{\{\delta\hat r_t,\delta\hat\pi_r\}}=-eH'R_\parallel^2,
 \qquad \mean{\hat{\mathcal L}_H}_{\rm in}=eH'R_\parallel^2,
\]
{because $[\delta\hat r_t,\delta\hat\pi_r]=0$ and the other initial cross moment vanishes. {Both calculations therefore start} with $\mean{\hat{\mathcal L}_H}=49.1\hbar$, at $52.5^\circ$ to the field. Just before exit, retaining the acceleration potential brings the mean within $0.358^\circ$ of the field, with $\mean{\hat{\mathcal L}_H}=6.46\times10^4\hbar$ and $W=72.5\%$ [Eq.~\eqref{SM:eq:width_angle}]. Omitting it gives an angle of $49.4^\circ$, $\mean{\hat{\mathcal L}_H}=49.3\hbar$, and $W=416\%$. The potential changes the position--momentum correlations and $\Sigma_{r_tr_t}$ entering $\delta\hat\pi_r=\delta\hat p_r-\gamma eH\delta\hat r_t/2$ [Eq.~\eqref{SM:eq:B}].}

{After exit, retaining and omitting the potential gives $|\mean{\hat{\bm{\mathcal L}}}|=580\hbar$ and $1550\hbar$, respectively, with tilts $\alpha_L=83.7^\circ$ and $93.6^\circ$. The corresponding rms spreads are $\Delta\mathcal L_\parallel=441\hbar$ and $1160\hbar$. Thus similar relative widths, $W=76\%$ and $74.8\%$, can accompany substantially different mean OAM and absolute spread. }

With the acceleration potential retained, $\max\eta_{\rm NR}={2.3}\times10^{-3}$ satisfies the internal-momentum condition in Eq.~\eqref{SM:eq:validity_ratios}. {The comparison with earlier precession approximations} is given in Sec.~\ref{SM:sec:precession}. The geometrical test in Sec.~\ref{SM:sec:dipole} establishes why the potential is required. {The corresponding in-field one-turn comparison is shown in \MovieNumber{5} (Sec.~\ref{SM:sec:movie_mirrorcle}).}

\section{Smooth magnet edges and angular windows}\label{SM:sec:fringes}
{{We extend the analysis of outgoing OAM in Sec.~\ref{SM:sec:analytic} to smooth edges, where the packet front and rear experience different fields.}} The coordinate $s$ increases inward along a fixed normal to the boundary, and $f(s)=H_Z(s,0)/H$ is the fractional field on the central plane. With $\Theta(s)$ the Heaviside step function, the effective edge satisfies $\int[f(s)-\Theta(s)]\dd s=0$. In the MAD-X convention {\cite{SM:MADX}},
\begin{equation}
 \mathrm{FINT}=\frac1g\int f(1-f)\dd s,\quad
 I_{\rm fr}=g\,\mathrm{FINT},\quad g=2\,\mathrm{HGAP}.\label{SM:eq:fringe_integral}
\end{equation}
FINT is dimensionless, HGAP is the magnetic half-gap, and $g$ is the full pole gap. Hwang and Lee use $K_2$ for FINT in their Eq.~(30), hence $I_{\rm fr}=gK_2$~\cite{SM:HwangLee2015}. This profile-dependent integral differs from the pole gap, vacuum aperture, and percentage rise length.

For the logistic profile,
\begin{equation}
 \begin{gathered}
 f(s)=\frac{1}{1+e^{-s/a}},\qquad a>0,\\
 I_{\rm fr}=a,\qquad L_{10-90}=2a\ln9.
 \end{gathered}
 \label{SM:eq:logistic_profile}
\end{equation}
Here $a$ sets the profile length and $L_{10-90}$ spans 10\%--90\% of $H$. Since $\dd f/\dd s=f(1-f)/a$, Eq.~\eqref{SM:eq:fringe_integral} gives $I_{\rm fr}=a\int_0^1\dd f=a$. Inverting the profile, $s=a\ln[f/(1-f)]$, gives the rise length: $I_{\rm fr}=1$~mm corresponds to $L_{10-90}={4.4}$~mm.

With $Z$ along the field and $H_s$ normal to the boundary, a local vacuum field is
\begin{equation}
 H_Z=H\operatorname{Re} f(s+iZ),\qquad H_s=H\operatorname{Im} f(s+iZ).
 \label{SM:eq:fringe_field}
\end{equation}
The Cauchy--Riemann relations give $\partial_sH_s+\partial_ZH_Z=0$ and $\partial_sH_Z-\partial_ZH_s=0$, so this two-dimensional field is locally divergence-free and curl-free. The logistic profile has its nearest complex poles at $s=0$, $Z=\pm\pi I_{\rm fr}$, so the analytic strip is $|Z|<\pi I_{\rm fr}$. The spatial expansion therefore requires a packet small on the local fringe scale, in addition to small internal momentum.

{In Eqs.~\eqref{SM:eq:fringe_orbit} and \eqref{SM:eq:fringe_transfer}, dots denote derivatives with respect to laboratory time $t$. The center's position $\vct R$}, kinetic momentum $\vct\pi_{\rm lab}$ [Eq.~\eqref{SM:eq:SMlabroot}], and velocity $\vct v$ obey
\begin{align}
 \dot{\vct R}&=\frac{\vct\pi_{\rm lab}}{m\gamma},\qquad
 \dot{\vct\pi}_{\rm lab}=-e\vct v\times\vct H(\vct R),\nonumber\\
 \gamma&=\sqrt{1+\vct\pi_{\rm lab}^2/(m^2c^2)}=1+\frac{K}{mc^2},\qquad \vct v=\dot{\vct R}.
 \label{SM:eq:fringe_orbit}
\end{align}
The laboratory frame is convenient because the field is static. Linearizing the Lorentz equations about each packet center gives
\begin{align}
 \delta\dot{\hat{\vct R}}&=\delta\hat{\vct v},\qquad
 \delta\hat{\vct v}=\frac{1}{m\gamma}\left[I_3-\frac{\vct v\vct v^T}{c^2}\right]\delta\hat{\vct\pi}_{\rm lab},\nonumber\\
 \delta\dot{\hat{\vct\pi}}_{\rm lab}&=-e\left[\delta\hat{\vct v}\times\vct H
 +\vct v\times(\delta\hat{\vct R}\cdot\nabla)\vct H\right].
 \label{SM:eq:fringe_transfer}
\end{align}
{The terms describe velocity-dependent forces and field variation across the packet. Integrating these equations transports the position and momentum operators and their quantum moments.}

{Each center is followed to a plane beyond the exit fringe along its normal, where the residual field is negligible. Internal deviations on this laboratory-time slice are continued to the simultaneous plane through the exit event in the nominal outgoing rest frame. The packets reach this plane at different observer times [Sec.~\ref{SM:sec:measurement}]. We combine their separately centered moments [Eqs.~\eqref{SM:eq:mean_trace} and \eqref{SM:eq:V}] using Eq.~\eqref{SM:eq:mixture}, then calculate $W$ and $\alpha_L$ from Eq.~\eqref{SM:eq:width_angle}.}

To compare all packets at a common observer time $T_*$, each must additionally evolve freely from its exit time $T_\lambda$:
\[
\begin{aligned}
 \delta\hat{\vct r}_\lambda(T_*)&=\delta\hat{\vct r}_{e,\lambda}
 +(T_*-T_\lambda)\bigl[\vct v(\hat{\vct\pi})
 -\mean{\vct v(\hat{\vct\pi})}\bigr],\\
 \vct v(\vct\pi)&=\frac{c^2\vct\pi}{\sqrt{m^2c^4+c^2\vct\pi^2}}.
\end{aligned}
\]
Here $e$ labels the exit event, all quantities use the same observer, and the field-free momentum $\hat{\vct\pi}=\hat{\vct p}$ remains constant. Linear transport expands this velocity about each packet's mean momentum. {A packet whose mean momentum differs from the reference value moves relative to the observer,} so its centered intrinsic OAM can change during free evolution, consistently with conserved OAM about a fixed origin.

{A planar sensitivity check ($\beta_{\rm in}=0$) sets the extrapolated incoming waist-center events at the entrance origin to the same laboratory time, takes $T=0$ at nominal exit, and propagates every sampled packet to $T_*=1$~ns. At the IMP and BNL weak-bend minima and the 1-, 5-, and 10-MeV near-half-turn minima with $I_{\rm fr}=1$~mm and 0.3-mrad incidence spread, the changes in $W$ are below $2\times10^{-6}$. The displayed widths therefore remain approximately $12\%,12\%,3.1\%,7.3\%,14\%$. {The sensitivity depends on the prepared state:} at the BNL point $(R_\perp,R_\parallel)\simeq(0.112,0.001)~\mu$m, $W$ rises from {2.2} to {2.4}. {The exit-time check covers all sampled trajectories, which have exited by $T_*$.} The figures and tables use the exit-event convention.

}The quadratic internal expansion requires small momentum and acceleration-size ratios throughout \mbox{the trajectory}:
\begin{equation}
 \eta_{\rm NR}=\frac{\sqrt{\tr\Sigma_{\pi\pi}}}{mc},\qquad
 \eta_a=\frac{\sqrt{\vct a'{}^T\Sigma_{rr}\vct a'}}{c^2}.
 \label{SM:eq:validity_ratios}
\end{equation}
Here $\Sigma_{rr}$ and $\Sigma_{\pi\pi}$ are the position and kinetic-momentum blocks of $\Sigma_\pi=S_\pi\Sigma_0S_\pi^T$ [Eqs.~\eqref{SM:eq:covariance_transport} and \eqref{SM:eq:kinetic_transfer}], evaluated on the instantaneous rest-frame slice. The proper acceleration $\vct a'$ is defined in Sec.~\ref{SM:sec:moved_fw}. For a uniform field the blocks follow explicitly from Eqs.~{\eqref{SM:eq:kinetic_transfer}} and \eqref{SM:eq:B}. For a smooth edge, the spatial field expansion is also monitored with
\begin{equation}
 \begin{gathered}
 \sigma_n^2=\mean{(\delta\hat s)^2}_{\rm lab},\qquad
 \sigma_Z^2=\mean{(\delta\hat Z)^2}_{\rm lab},\\
 \eta_{\rm fr}=\frac{\sqrt{\sigma_n^2+\sigma_Z^2}}{I_{\rm fr}},
 \qquad f(1-f)>10^{-4}.
 \end{gathered}
 \label{SM:eq:spatial_validity}
\end{equation}
The scalars $\delta\hat s=\vct n_{\rm b}\cdot\delta\hat{\vct R}$ and $\delta\hat Z$ are laboratory position deviations along the edge normal $\vct n_{\rm b}$ and field axis. {Here $\delta\hat Z$ is the scalar field-axis displacement, whereas the phase-space vector in Eq.~\eqref{SM:eq:covariance_definition} has six components.} These two widths measure whether the packet sees an approximately linear field. Their laboratory variances follow from Eq.~\eqref{SM:eq:fringe_transfer}, while the condition on $f$ selects the region with an appreciable gradient. Each reported maximum includes sampled points along every trajectory and, for an ensemble, all incidence nodes. {The size scans require all three maxima below 0.1 to select small expansion scales.}

The laboratory angle $\alpha_{\rm in}$ in {Fig.~\Lref{fig:matching} of the Paper} describes the incident momentum projected onto the bending plane $(X,Y)$, measured from the fixed entrance normal. {We also include its elevation} $\beta_{\rm in}$ toward the field axis $Z$. With laboratory momentum magnitude $\pi_{\rm ref}=mc\sqrt{\gamma^2-1}$ and nominal entrance normal $+Y$,
\begin{equation}
 \frac{\vct\pi_{{\rm lab},{\rm in}}}{\pi_{\rm ref}}
 =(-\sin\alpha_{\rm in}\cos\beta_{\rm in},\,
 \cos\alpha_{\rm in}\cos\beta_{\rm in},\,\sin\beta_{\rm in}).
 \label{SM:eq:two_incidence_angles}
\end{equation}
The incoming LG axis follows this momentum. We draw the two angles independently from zero-mean Gaussians with rms widths $\sigma_\alpha=\sigma_\beta=0.3$~mrad, follow each center through the same fixed boundaries and full field in Eq.~\eqref{SM:eq:fringe_field}, and combine its centered OAM moments with Eq.~\eqref{SM:eq:mixture}. Averaging the individual values of $W$ would miss the variation of their means. The OAM tilt $\alpha_L$ remains the distinct angle in Eq.~\eqref{SM:eq:width_angle}.

For the common weak-bend state $(R_\perp,R_\parallel)=(2,0.2)~\mu$m, the variances $(\Delta\mathcal L_\parallel)^2$ without incidence spread are $70.5\hbar^2$ for IMP and $68.5\hbar^2$ for BNL. Averaging over $\alpha_{\rm in}$ alone changes them by $-7.6\times10^{-4}\%$ and $-0.009\%$, whereas averaging over $\beta_{\rm in}$ alone increases them by $0.011\%$ and $0.29\%$. Adding $\beta_{\rm in}$ {to the in-plane ensemble} also increases the variance by $0.011\%$ and $0.29\%$. The elevation has the larger effect in these weak bends because it displaces the center along the field and changes the edge focusing. At 1~MeV, 5~mT, $I_{\rm fr}=5$~mm, $\ell=10$, $(R_\perp,R_\parallel)=(10,1)~\mu$m, and $\theta\simeq179.38^\circ$, the variance without incidence spread is $(\Delta\mathcal L_\parallel)^2={0.12}\hbar^2$. In-plane averaging increases it by $24\%$, while elevation averaging alone adds $0.32\%$. The in-plane change is about 74 times larger, and averaging in both planes gives $(\Delta\mathcal L_\parallel)^2={0.15}\hbar^2$.

For this nearly 3-m half turn, $\sigma_\beta=0.3$~mrad produces about 0.9~mm rms displacement along $Z$. The largest sampled elevation angle is {1.6}~mrad, and at zero in-plane incidence its trajectory reaches $|Z|\simeq4.7$~mm for $I_{\rm fr}=0.5$ or 1~mm, beyond the respective bounds $\pi I_{\rm fr}\simeq1.6$ or 3.1~mm of the local field model. Thus the restriction concerns the chosen off-plane field description. For planar incidence the centers remain at $Z=0$: with 0.3-mrad in-plane spread, the same 1-MeV packet gives minima $W={3.1}\%$ and ${3.1}\%$ at $\theta\simeq179.91^\circ$ and $179.85^\circ$, respectively. {Fig.~\Lref{fig:windows}(c) of the Paper} uses $I_{\rm fr}=5$~mm, whose 15.7-mm bound contains all sampled incidence trajectories.

{Sharp and smooth profiles are compared at fixed effective dipole length $L_{\rm mag}$. Along the nominal planar orbit, $s_{\rm orb}$ denotes path length, distinct from the coordinate normal to the boundary $s$ [Eq.~\eqref{SM:eq:fringe_integral}]. The momentum $\pi_{\rm ref}=mc\sqrt{\gamma^2-1}$ gives $\rho=\pi_{\rm ref}/(eH)$ and $\dd\theta/\dd s_{\rm orb}=eH_Z(s_{\rm orb},0)/\pi_{\rm ref}$, hence}
\begin{equation}
 L_{\rm mag}\equiv\frac{1}{H}\int H_Z(s_{\rm orb},0)\dd s_{\rm orb}
 =\rho\theta.
 \label{SM:eq:effective_length}
\end{equation}
For a sharp uniform dipole $L_{\rm mag}$ is its orbit length, whereas for a smooth profile it is the field-weighted path length including both transitions. An oblique trajectory can have a different integrated field from the nominal one. %

\subsection{Incidence angles and the sharp-boundary limit}\label{SM:sec:incidence}
{To connect with the boundary matching of Sec.~\ref{SM:sec:boundary}, we take {the limit of sharp boundaries} at fixed boundary orientation, field integral $\int H_Z\dd s_{\rm orb}=HL_{\rm mag}$ [Eq.~\eqref{SM:eq:effective_length}], and incidence with $|\cos\alpha_{\rm in,out}|$ bounded away from zero.} {At normal incidence, the linear entrance and exit maps approach $C_{\rm in,out}$ in Eq.~\eqref{SM:eq:SMfaces}. {At oblique incidence, the crossing-time construction uses the actual normal to the boundary to obtain the corresponding matching matrices.}} Position and kinetic momentum remain continuous at each crossing, while the map compares their values on common-time slices.

The fringe path length and crossing-time spread increase by $1/|\cos\alpha|$, so allowed incidence depends on the fringe, packet dimensions, and orbit radius.

The small-packet condition in Eq.~\eqref{SM:eq:spatial_validity} gives {$\sigma_n\ll I_{\rm fr}$, and hence $\sigma_n\ll L_{10-90}$ for the logistic profile}. The fringe length therefore sets the crossing scale, and the local-curvature condition reduces to
\begin{equation}
 |\cos\alpha|\gg\frac{L_{10-90}}{\rho}.
 \label{SM:eq:local_crossing}
\end{equation}
Here $\alpha$ is the full angle between momentum and the normal to the boundary. This condition limits the orbit curvature during crossing, whereas the preceding small-packet condition limits the field variation across the packet. At $I_{\rm fr}=5$~mm, the right-hand sides are about $0.0065$, $0.0016$, and $0.023$ for the IMP, BNL, and 1-MeV half-turn examples. Their nominal cosines are near unity, exceeding these scales by factors of about 150, 630, and 43.

\subsection{Width of an angular recovery window}\label{SM:sec:window}
{At fixed state, energy, field, and fringe profile, the recovery angle $\theta_*$ minimizes $W$ [Eq.~\eqref{SM:eq:width_angle}]. Its angular tolerance follows by expanding about a \mbox{nondegenerate minimum}:}
\[
W^2(\theta)=W^2(\theta_*)+\frac12\left.\frac{\dd^2W^2}{\dd\theta^2}\right|_{\theta_*}(\theta-\theta_*)^2+O((\theta-\theta_*)^3),
\]
with $\left.\dd^2W^2/\dd\theta^2\right|_{\theta_*}>0$. Solving the quadratic approximation for $W<20\%$ gives the full interval
\begin{equation}
\Delta\theta\simeq2\sqrt{\frac{2[(0.2)^2-W^2(\theta_*)]}{\left.\dd^2W^2/\dd\theta^2\right|_{\theta_*}}}.
\label{SM:eq:window}
\end{equation} The interval exists only for $W(\theta_*)<20\%$. Higher terms can make it asymmetric.

\begingroup

The numerical half-turn example and its size and angle tolerances are collected in Sec.~\ref{SM:sec:recovery}. For the common weak-bend states used above, the sampled two-angle maxima $(\eta_{\rm NR},\eta_a,\eta_{\rm fr})$ are bounded by $(3.7\times10^{-5},4.2\times10^{-5},3\times10^{-4})$ for IMP and $(2.1\times10^{-4},2.6\times10^{-4},2.9\times10^{-4})$ for BNL.
\endgroup

\begingroup
\section{Incident packet away from its~waist}\label{SM:sec:offwaist}
Moving the LG waist [Eq.~\eqref{SM:eq:LG}] changes the density and phase at the magnet entrance. We first keep the source-waist sizes fixed, then isolate the phase dependence by comparing identical entrance densities. The outgoing mean kinetic OAM and relative width $W$ use Eq.~\eqref{SM:eq:width_angle}.

\begin{figure}[t]
 \centering\includegraphics[width=\textwidth]{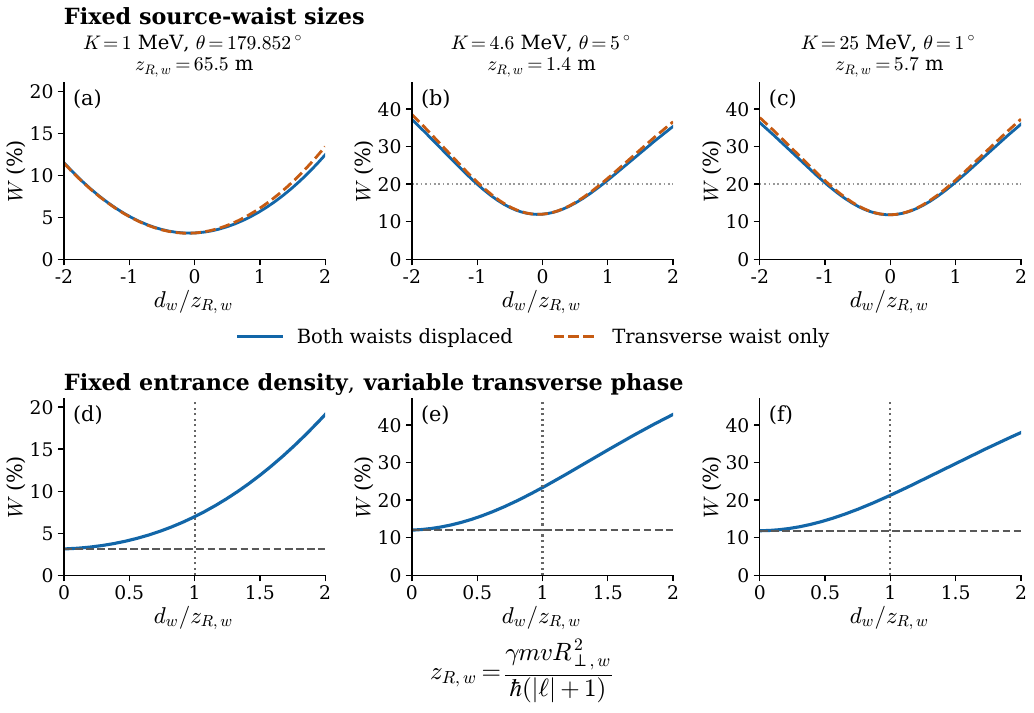}
 \caption{Outgoing relative OAM width $W$ [Eq.~\eqref{SM:eq:width_angle}] for two preparations. {$R_{\perp,w}$ {is the rms radius at the transverse waist, while $R_{\perp,\rm in}\equiv R_\perp$ is its value at the magnet entrance} [Eq.~\eqref{SM:eq:offwaist_radius}].} (a--c) Fixed source-waist sizes: both waists move together (solid blue), or only the transverse waist moves while the longitudinal waist stays at entrance (dashed orange). Positive $d_w$ gives an expanding packet, and negative $d_w$ a converging one. (d--f) Identical entrance LG densities and different transverse quadratic phases: the transverse source waist and its distance upstream vary together, while the longitudinal waist stays at entrance. Here $z_{R,w}$ is recalculated for each source waist [Eq.~\eqref{SM:eq:offwaist_radius}]. Horizontal dashed lines: waist at entrance with the same density. Vertical dotted lines: maximum physical $d_w$. {The common parameters $(K,H,\theta,\ell,I_{\rm fr})$ are (a,d) $(1~\mathrm{MeV},5~\mathrm{mT},179.852^\circ,10,1~\mathrm{mm})$, (b,e) $(4.6~\mathrm{MeV},5~\mathrm{mT},5^\circ,64,5~\mathrm{mm})$, and (c,f) $(25~\mathrm{MeV},6.19~\mathrm{mT},1^\circ,64,5~\mathrm{mm})$. The fixed size pairs are $(R_{\perp,w},R_{\parallel,w})$ in (a--c) and $(R_{\perp,\rm in},R_{\parallel,\rm in})$ in (d--f): $(10,1)$, $(1.9,0.17)$, and $(1.7,0.15)~\mu\mathrm m$ from left to right.} Magnet settings and Gaussian incidence averaging, $\sigma_\alpha=0.3$~mrad and $\beta_{\rm in}=0$ [Eq.~\eqref{SM:eq:two_incidence_angles}], are unchanged between rows. {At $d_w=0$, the size pairs coincide with the 1-MeV smooth-reversal and weak IMP/BNL preparations in Table~\ref{SM:tab:favorable}. The incidence average here remains planar.} Horizontal dotted lines in (b,c): $W=20\%$.}
 \label{SM:fig:offwaist}
\end{figure}

\subsection{Fixed source-waist sizes}
The laboratory distance $d_w$ from the waist to the effective entrance plane is positive for an expanding packet and negative for a converging one. We first place the transverse and longitudinal waists in the same plane, with rest-frame rms sizes $R_{\perp,w}$ and $R_{\parallel,w}$. The corresponding LG state $\psi_w$ is Eq.~\eqref{SM:eq:LG}, with $w_w=R_{\perp,w}/\sqrt{|\ell|+1}$. Free evolution to the entrance gives
\begin{equation}
 \tau_w=\frac{d_w}{\gamma v},\qquad
 \psi_{\rm in}=\exp\!\left[-\frac{i\tau_w(\delta\hat{\vct p})^2}{2m\hbar}\right]\psi_w.
 \label{SM:eq:offwaist_state}
\end{equation}
In the ordering of Eq.~\eqref{SM:eq:covariance_definition}, the same evolution gives
\begin{equation}
 S_{\rm free}(d_w)=
 \begin{pmatrix}I_3&\tau_w I_3/m\\0&I_3\end{pmatrix},\qquad
 \Sigma_{\rm in}=S_{\rm free}\Sigma_wS_{\rm free}^{T},
 \label{SM:eq:offwaist_transfer}
\end{equation}
where $I_3$ is the identity matrix and $\Sigma_w$ is Eq.~\eqref{SM:eq:input} at the waist. Using $SS_{\rm free}$ in Eqs.~\eqref{SM:eq:mean_trace} and \eqref{SM:eq:V}, and $S_\psi S_{\rm free}$ in Eq.~\eqref{SM:eq:SMpsi}, propagates the waist moments and wave function through the magnet.

Free propagation leaves all momentum moments unchanged but gives
\begin{equation}
 \frac12\mean{\{\delta\hat r_i,\delta\hat p_i\}}_{\rm in}
 =\frac{\tau_w}{m}\mean{(\delta\hat p_i)^2}_w,
 \qquad i=r,t,H.
 \label{SM:eq:offwaist_correlation}
\end{equation}
The transverse radius and longitudinal length at entrance are
\begin{equation}
 \begin{aligned}
 R_{\perp,\rm in}&=R_{\perp,w}\sqrt{1+(d_w/z_{R,w})^2},\\
 R_{\parallel,\rm in}&=R_{\parallel,w}\sqrt{1+(d_w/z_{R,\parallel})^2},\\
 z_{R,w}&=\frac{\gamma mvR_{\perp,w}^2}{\hbar(|\ell|+1)},\qquad
 z_{R,\parallel}=\frac{2\gamma mvR_{\parallel,w}^2}{\hbar}.
 \end{aligned}\label{SM:eq:offwaist_radius}
\end{equation}
{The entrance radius $R_{\perp,\rm in}$ {is $R_\perp$ of Sec.~\ref{SM:sec:preparation} and equals the waist radius} $R_{\perp,w}$ when $d_w=0$.} Here $z_{R,w}$ and $z_{R,\parallel}$ are the laboratory Rayleigh distances associated with Eq.~\eqref{SM:eq:psi_diffraction_times}. {The solid blue curves in Fig.~\ref{SM:fig:offwaist}(a--c) include simultaneous transverse and longitudinal spreading. For comparison, the dashed orange curves move only the transverse waist, with the longitudinal waist at entrance.} In this case, we use $(\tau_w/m)\operatorname{diag}(1,0,1)$ in the upper-right block of Eq.~\eqref{SM:eq:offwaist_transfer}, with $R_{\parallel,\rm in}=R_{\parallel,w}$ and $\Sigma_{{\rm in},r_tp_t}=0$.

For abrupt boundaries perpendicular to the reference orbit, Eq.~\eqref{SM:eq:map} gives \emph{the same outgoing radial and tangential means} as Eq.~(\Lref{eq:mean}) of the Paper. The field-axis mean follows Eq.~\eqref{SM:eq:LH} with the propagated entrance variances and the additional correlation term
\begin{equation}
 \Delta\mean{\hat{\mathcal L}_H}_{\rm corr}
 =\gamma s_\theta\big[(2c_\theta-1)\Sigma_{{\rm in},r_rp_r}
                         -\Sigma_{{\rm in},r_tp_t}\big].
 \label{SM:eq:offwaist_mean}
\end{equation}
Here $s_\theta=\sin\theta$ and $c_\theta=\cos\theta$. The full change combines this term with the changed position variances in Eq.~\eqref{SM:eq:LH}. Both the projection axis and the variance entering $W$ generally depend on $d_w$.

At $\theta=\pi$, the outgoing tangential OAM in Eq.~\eqref{SM:eq:halfoperator} contains only the incident $\hat{\mathcal L}_t$ and momentum products. Under the preceding drift, {$\delta\hat r_i\to\delta\hat r_i+\tau_w\delta\hat p_i/m$, and the two added terms} in $\hat{\mathcal L}_t=\delta\hat r_H\delta\hat p_r-\delta\hat r_r\delta\hat p_H$ cancel. The momenta also stay constant. Hence both $\mean{\hat{\mathcal L}_t}=-\ell\hbar$ and its variance [Eq.~\eqref{SM:eq:halfvariance}] are independent of $d_w$ at fixed waist state. The radial mean is zero and the field-axis mean is $2\gamma(\Sigma_{w,p_tp_t}-\Sigma_{w,p_rp_r})/(eH)$. Equal momentum variances therefore keep the mean collinear with the outgoing momentum and give $W(d_w)=W(0)$, with Eq.~\eqref{SM:eq:matchedwidth} evaluated at the waist. This conserved condition corresponds to $R_{\parallel,w}=R_{\perp,w}/[\sqrt2(|\ell|+1)]$. The ratio of the \emph{entrance sizes} generally changes according to Eq.~\eqref{SM:eq:offwaist_radius}. The same argument applies when only the transverse waist is moved, because both transverse coordinates undergo the same drift.

At fixed source-waist sizes, smooth edges and incidence averaging make $W$ sensitive to displacement [Fig.~\ref{SM:fig:offwaist}(a--c)]. For the near-half turn, $z_{R,w}=65.5$~m and a 1-m displacement changes $W$ by less than $0.01$ percentage point from $3.1\%$. The corresponding changes are $12\%$ to $17.1\%$ at 4.6~MeV and $11.8\%$ to $12.2\%$ at 25~MeV. Large Rayleigh distances thus explain why placing the waist at entrance works well when $|d_w|\ll z_{R,w}$. Moving only the transverse waist changes the plotted $W$ by at most 1.4 percentage points relative to moving both.

\subsection{Packets with identical entrance densities}
To isolate the phase dependence, we keep the full entrance LG density, $\ell$, and the longitudinal waist fixed. With $\zeta=d_w/z_{R,w}$, Eq.~\eqref{SM:eq:offwaist_radius} gives
\[
 R_{\perp,w}=\frac{R_{\perp,\rm in}}{\sqrt{1+\zeta^2}},\qquad
 d_w=\frac{\gamma mvR_{\perp,\rm in}^2}{\hbar(|\ell|+1)}
      \frac{\zeta}{1+\zeta^2}.
\]
Thus $d_w$ is maximal at $\zeta=1$. Reciprocal values $0<\zeta<1$ and $1/\zeta>1$ give the same distance but different source waists. Beyond $\zeta=1$, the physical distance decreases as the transverse phase curvature increases.

Up to a constant phase, the entrance states are
\[
 \psi_{\rm in}^{(\zeta)}=
 \exp\!\left[\frac{i(|\ell|+1)\zeta(\delta r_r^2+\delta r_H^2)}
                    {2R_{\perp,\rm in}^2}\right]\psi_{\rm in}^{(0)}.
\]
Their densities coincide, while their position--momentum correlations and momentum moments differ. In Fig.~\ref{SM:fig:offwaist}(d--f), increasing $\zeta$ from 0 to 2 raises $W$ from $3.1\%$ to $19.1\%$, from $12\%$ to $42.8\%$, and from $11.8\%$ to $38\%$, respectively. {The outgoing OAM distribution therefore depends on both the entrance density and the position--momentum correlations.}

Across both rows and all incidence nodes, the sampled expansion ratios satisfy $\max\eta_{\rm NR}<4\times10^{-4}$, $\max\eta_a<5\times10^{-4}$, and $\max\eta_{\rm fr}<0.017$ [Eqs.~\eqref{SM:eq:validity_ratios} and \eqref{SM:eq:spatial_validity}]. Free propagation and the explicit phase above give the same second and fourth moments.

For the 4.6-MeV magnet, a much smaller transverse waist, $R_{\perp,w}=10$~nm, reaches the same $R_{\perp,\rm in}=1.9~\mu$m after $d_w=7.5$~mm. With the longitudinal waist still at entrance, $R_\parallel=0.17~\mu$m, it gives $W=72.3\%$ and a mean almost along $H$. Its nominal-path maxima are $\eta_{\rm NR}\simeq0.003$, $\eta_a\simeq0.0014$, and $\eta_{\rm fr}\simeq0.012$.
\endgroup

\begingroup

\section{Accelerator geometries, prepared states, and size dependence}\label{SM:sec:facilities}
We compare published magnet geometries and proposed test dipoles by asking three questions: \textit{(i)} what state enters the magnet, \textit{(ii)} what OAM distribution leaves it, and \textit{(iii)} whether the approximation remains applicable throughout the path. Table~{\ref{SM:tab:facilities}} collects the geometries and their status. The relative rms width $W$ and mean-direction angle $\alpha_L$ are defined in Eq.~\eqref{SM:eq:width_angle}. {A small $W$ means a narrow projection \textit{along the outgoing mean OAM}.}

The reference momentum fixes the field $H$ and magnetic length $L_{\rm mag}$:
\begin{equation}
 \pi_{\rm ref}c=\sqrt{K(K+2mc^2)},\quad
 H=\frac{\pi_{\rm ref}}{e\rho},\quad L_{\rm mag}=\rho\theta.\label{SM:eq:rigidity}
\end{equation}
Every quoted packet size is a single-electron rest-frame rms size at the chosen entrance, unless a waist is explicitly displaced. {The bunch dimensions additionally reflect the distribution of packet centers.} For example, the JINR 5-MeV design gives a calculated 59-keV rms energy spread, bunch sizes $0.72/0.66$~mm, and horizontal/vertical angular spreads $0.3/0.4$~mrad~\cite{SM:JINRphoto}. The supplied horizontal, vertical, and longitudinal axes correspond to our entrance axes $(x,z,y)$.

Unless specified otherwise, the comparisons below use a single centered packet or a Gaussian in-plane incidence ensemble, with the out-of-plane incidence angle $\beta_{\rm in}=0$. The weak-bend size maps in Fig.~\Lref{fig:windows}(a,b) of the Paper include independent $0.3$-mrad rms spreads in \mbox{both planes}.

\begin{table}[tp]

\caption{Magnet geometries used in the comparisons. {Fields follow from Eq.~\eqref{SM:eq:rigidity}, and $L_{\rm mag}=\rho\theta$ is the effective orbit length.} ``Chosen'' denotes a calculation at a selected energy, ``continuation'' a setting outside the cited channel range, and ``proposed'' an additional magnet. {The last column gives outcomes for the specified incident packets and magnet settings.} {Unless specified otherwise, the reference trajectory crosses the boundaries at normal incidence, with the stated model fringes.}}
\label{SM:tab:facilities}\label{SM:tab:size_geometries}
\setlength{\tabcolsep}{3pt}\renewcommand{\arraystretch}{1.16}
\begin{ruledtabular}
\begin{tabular}{p{.135\textwidth}rrrrp{.44\textwidth}}
Facility / model & $K$ (MeV)&$\rho$ (m)&$H$ (mT)&$\theta$ (deg)&Status, incident state, and calculation\\
JINR photo-injector &1.9 / 5&0.55&14.3 / 33.3&28.5&{Commissioning simulation / design,} $L_{\rm mag}=0.274$~m~\cite{SM:JINRcurrent,SM:JINRphoto}. Fixed $(\ell,R_\perp,R_\parallel)=(64,3,0.3)$, sizes in $\mu$m: $W=69.9/76\%$, $\alpha_L=62.5/83.7^\circ$ (Table~{\ref{SM:tab:finite_bends}}).\\
JINR LINAC &20 / 30&0.532&129 / 191&35&20~MeV is a continuation below the 25--60-MeV {channel range, while 30~MeV lies within it}~\cite{SM:JINRmagnets}. Selected submicrometer states pass the sampled bounds, with $W\simeq72$--$73\%$ (Sec.~{\ref{SM:sec:size_scan}}).\\
 &60 / 70&0.532&379 / 442&35&{Channel endpoint / magnet-design calculation,} $L_{\rm mag}=0.325$~m, gap 41~mm. {For the fixed $(64,3,0.3)$ state, the internal-momentum ratio exceeds the allowed range along the path (Sec.~\ref{SM:sec:linac}), placing the predicted $W$ outside the approximation's controlled regime.}\\
BNL ATF dipole &25&1.15&74.2&20&Published 0.4-m component at a chosen energy~\cite{SM:BNL,SM:BNLcurrent}. {Fixed-state result: Table~{\ref{SM:tab:finite_bends}}. A different} $\ell=1000$ packet gives $W=70.7\%$ (Table~\ref{SM:tab:favorable}).\\
BNL weak spectrometer &25&13.8&6.2&1&Chosen setting of a legacy 0.24-m, $0$--$8^\circ$ spectrometer{~\cite{SM:BNL}, with availability requiring confirmation}~\cite{SM:BNLcurrent}. Optimized $\ell=64$ packet: $W=11.9\%$, $\alpha_L=23.5^\circ$.\\
IMP test dipole &4.6&3.4&5&2 / 5&Proposed addition at a demonstrated energy{~\cite{SM:IMP2024}, with} $L_{\rm mag}=0.118/0.296$~m. Fixed $(64,3,0.3)$ packet: $W\simeq3/12.5\%$, {$\alpha_L=9.9/23.6^\circ$. Optimized states:} Table~\ref{SM:tab:favorable}.\\
MESA MARC1 &30&0.7&145&45&{Published arc geometry~\cite{SM:MESA},} $L_{\rm mag}=0.55$~m, gap 40~mm. Selected states and size searches remain broad (Tables~\ref{SM:tab:favorable} and {\ref{SM:tab:size_additional}}).\\
ELBE-inspired bend &12&0.42&99.6&21&{Chosen energy, with the published} 0.15-m length interpreted as separation of parallel boundaries~\cite{SM:ELBEchicane}. $\ell=64$ size minimum: $W=71.4\%$ (Table~{\ref{SM:tab:size_additional}}).\\
Proposed finite bend &100&5&67.1&15&$L_{\rm mag}=1.3$~m. Selected $\ell=10$ packet: $W=86.9\%$, {$\alpha_L=88.8^\circ$. All states in the restricted $\ell=64$ size map lie outside the applicability selection.}\\
\end{tabular}\par
\end{ruledtabular}
\end{table}

The JINR photo-injector parameters combine the 1.9-MeV commissioning simulation and the 5-MeV design supplied by M.~A.~Nozdrin~\cite{SM:JINRcurrent,SM:JINRphoto}. {A 24-MeV LINAC test beam has also been documented~\cite{SM:JINR}. Section~\ref{SM:sec:linac} assesses the applicability of the model to the fixed 60--70-MeV packet.} At BNL, typical current user energies are 50--75~MeV, so the chosen 25-MeV settings and effective boundary orientations require facility consultation~\cite{SM:BNLcurrent}. The IMP source uses a 60-kV gridded thermionic gun {and a $60^\circ$ diagnostic dipole. Near 4.3~MeV} its measured bunch energy distribution has about 0.2~MeV full width at half maximum~\cite{SM:IMP2024,SM:IMPdiagnostics}. This spread is separate from the longitudinal quantum uncertainty of the assumed pure packet.

MESA is commissioning in 2026, with 105-MeV energy-recovery and 155-MeV external-beam design modes{~\cite{SM:MESAcurrent}. The published} MARC dipoles operate at 30, 55, 80, 105, and 130~MeV, with quoted fields 143--619~mT~\cite{SM:MESA}. PERLE's 250-MeV recirculator design has six $30^\circ$ dipoles per $180^\circ$ arc, lengths 0.33 or 0.66~m, radii 0.63 or 1.26~m, and quoted fields 0.45--0.91~T{~\cite{SM:PERLE}. First photogun beam} was reported in May 2026~\cite{SM:PERLEcurrent}. IOTA has tracked a single electron at 150~MeV/$c${~\cite{SM:IOTA}. Its four} $30^\circ$ and four $60^\circ$ dipoles have nominal field 0.7~T and inferred radius about 0.72~m~\cite{SM:IOTAfacility}. The tested main-dipole states fail the internal-momentum criterion (Sec.~{\ref{SM:sec:size_scan}}). DELTA's single-electron operation at 1.5~GeV~\cite{SM:DELTA} is a higher-energy prospect requiring radiation as well as lattice transport.

An experimental comparison requires repeated preparation of the same centered state and measurement of the incident and outgoing OAM distributions with a calibrated axis and acceptance. {Beam imaging and centroid tracking characterize the spatial profile and mean trajectory, while $W$ requires an OAM-sensitive diagnostic.}

\subsection{Fixed packets and variation of the preparation}
The finite-angle results in Table~{\ref{SM:tab:finite_bends}} use the same entrance state, $\ell=64$, $R_\perp=3~\mu$m, $R_\parallel=0.3~\mu$m. Keeping the state fixed separates changes caused by the magnet from changes caused by preparation.
\begin{table}[tp]

\caption{{Fixed-state comparison after bends with sharp boundaries:} $\ell=64$, $R_\perp=3~\mu$m, $R_\parallel=0.3~\mu$m, {at normal incidence.} OAM means and rms spreads {are in units of $\hbar$, while $W$ and $\alpha_L$ use} Eq.~\eqref{SM:eq:width_angle}. The internal-momentum maximum includes entrance matching and the sampled magnetic path.}\label{SM:tab:finite_bends}
\begin{ruledtabular}
\begin{tabular}{lrrr}
 & Photo-injector & Photo-injector & BNL ATF\\
$K$ (MeV)&1.9&5&25\\
$\theta$ (deg)&28.5&28.5&20\\
$H$ (mT)&14.3&33.3&74.2\\
$|\mean{\hat{\bm{\mathcal L}}}|/\hbar$&138&580&3000\\
$\Delta\mathcal L_\parallel/\hbar$&96.7&441&2420\\
$W$ (\%)&69.9&76&80.7\\
$\alpha_L$ (deg)&62.5&83.7&88.8\\
$\eta_{\rm NR}^{\max}$&${1.8}\!\times\!10^{-4}$&${2.3}\!\times\!10^{-3}$&${7.9}\!\times\!10^{-2}$
\end{tabular}\par
\end{ruledtabular}
\end{table}
The incident rms sizes $R_\perp,R_\parallel$ set the momentum uncertainties [Eq.~\eqref{SM:eq:input}] and entrance shear [Eq.~\eqref{SM:eq:SMcrossing}]. At fixed magnet geometries in Fig.~\ref{SM:fig:facilitywidths}, we vary both rms dimensions independently on the grid
\[
 \begin{gathered}
 \ell\in\{10,64,100,1000\},\qquad
 R_\perp\in\{0.3,1,3,10,30\}~\mu\mathrm m,\\
 R_\parallel\in\{0.03,0.1,0.3,1,3\}~\mu\mathrm m.
 \end{gathered}
\]
Each state is a radial-index-zero LG state propagated through the same magnet. Keeping only the states whose ratios $\eta_{\rm NR},\eta_a$ in Eq.~\eqref{SM:eq:validity_ratios} remain below 0.1 along the trajectory {gives {the smallest widths for sharp boundaries listed below:}}
\begin{center}
\begin{tabular}{lrrr}
 & JINR, 1.9 MeV & JINR, 5 MeV & BNL, 25 MeV\\
$W_{\min}$ (\%) & 47.3 & 64.6 & 70.7\\
$\ell$ &64&1000&1000\\
$R_\perp$ ($\mu$m)&1&3&1\\
$R_\parallel$ ($\mu$m)&0.3&0.3&0.3
\end{tabular}\par
\end{center}
Fig.~\ref{SM:fig:facilitywidths} gives the smallest admitted $W$ at each sampled $\ell$. Every sampled minimum exceeds $20\%$.

For the original $\ell=64$, $R_\perp=3~\mu$m, $R_\parallel=0.3~\mu$m packet, the same three geometries with equal logistic fringe integrals $I_{\rm fr}$ [Eq.~\eqref{SM:eq:fringe_integral}] give
\begin{center}
\begin{tabular}{lrrr}
$I_{\rm fr}$ (mm)&1.9 MeV&5 MeV&25 MeV\\
\hline
0 (sharp)&69.9&76&80.7\\
1&70.1&76&80.7\\
5&70.7&76.2&80.7\\
15&72.5&76.7&80.8
\end{tabular}\par
\end{center}
All entries are $100W$. Changing the fringe at fixed state shifts these broad widths by at most about 2.6 percentage points.

For the 5-MeV photo-injector at $I_{\rm fr}=5$~mm, thirteen equally spaced in-plane incidence angles from $-3$ to $+3$~mrad change $W$ by less than 0.1 percentage point around $76.2\%$. These are separately centered pure packets. The BNL fixed-state momentum maximum, 0.079, lies near the 0.1 selection threshold. This $\ell=64$ packet differs from the $\ell=1000$, $(R_\perp,R_\parallel)=(1,0.3)~\mu$m example in Fig.~\Lref{fig:matching} of the Paper.
\begin{figure}[tbp]
\centering
\includegraphics[width=0.82\columnwidth]{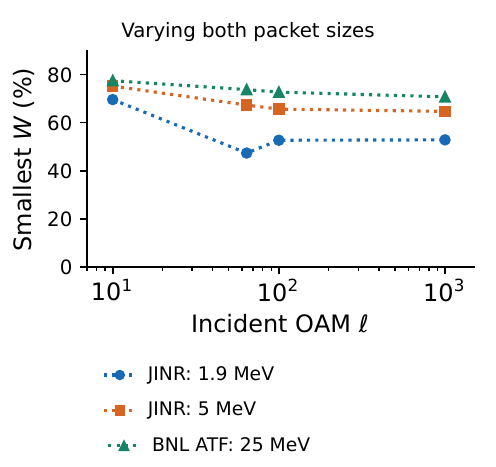}
\caption{{Smallest $W$ for sharp boundaries at each tested} $\ell$ for the JINR 1.9- and 5-MeV and BNL 25-MeV bends in Table~{\ref{SM:tab:facilities}}. Both incident sizes vary independently on the grid stated in the text. The sampled path ratios $\eta_{\rm NR},\eta_a$ must remain below 0.1. {These minima refer to the stated grid and \mbox{incident-state family}.}}
\label{SM:fig:facilitywidths}
\end{figure}

\Needspace{8\baselineskip}
\subsection{Representative outcomes: weak bends, reversal, and larger bends}\label{SM:sec:favorable}\label{SM:sec:highenergy}
Table~\ref{SM:tab:favorable} distinguishes three ways the mean can evolve: \textit{(i)} a tilted mean after a weak bend, \textit{(ii)} reversal near a half turn, and \textit{(iii)} return in a uniform circular reference. It also shows broad finite-bend examples. {The narrow-distribution criterion is $W\leq20\%$. Reversal or return additionally requires} the mean within $3^\circ$ of the corresponding direction. The half-turn preparation and its tolerances are developed together in Sec.~\ref{SM:sec:fixed_checks}.

In the proposed half-turn family, $\rho=0.949$~m is fixed: $K=1$, 4.6, and 5~MeV require $H=5$, 17.9, and 19.3~mT and an arc length about 3~m. {Thus the IMP-energy and JINR-energy rows specify reversal in additional magnets with the stated half-turn geometry.} The uniform circular reference has a 6-m orbit at 1~MeV and 5~mT. The sharp and circular rows are single packets. Smooth half turns and broad finite bends {use a 0.3-mrad rms in-plane ensemble, while the weak optimized rows average} in \mbox{both planes}.
\begin{table}[tp]

\caption{Representative outgoing OAM distributions. In (a), packet entries are $(\ell,R_\perp,R_\parallel)$ {with sizes in $\mu$m. Mean magnitudes} and rms spreads along the mean are in $\hbar$. The listed angles are nominal settings, with tolerances given separately in Table~{\ref{SM:tab:recovery}}. In (b), minima and independent size-error ranges refer to the two-angle weak-bend maps of Fig.~\Lref{fig:windows}(a,b) of the Paper. In (c), expansion ratios are maxima over the sampled path and every incidence node for the three broad examples in (a).}
\label{SM:tab:favorable}\label{SM:tab:weak_minima}\label{SM:tab:highenergy_checks}
\setlength{\tabcolsep}{3pt}
\begin{ruledtabular}
\begin{tabular}{lrrlrrrr}
\multicolumn{8}{l}{{(a) Specified incident states, with geometries in} Table~{\ref{SM:tab:facilities}} and the text}\\
Regime / model&$K$ (MeV)&$\theta$ (deg)&$(\ell,R_\perp,R_\parallel)$&$|\mean{\hat{\bm{\mathcal L}}}|$&$\Delta\mathcal L_\parallel$&$\alpha_L$ (deg)&$W$ (\%)\\
Sharp reversal&1&180&$(10,9.6,0.617)$&10&0.55&180&5.5\\
Smooth reversal&1&179.85&$(10,10,1)$&10&0.31&178.4&3.1\\
 &$4.6$&179.88&$(10,10,1)$&10&0.69&178.4&6.9\\
 &5&179.88&$(10,10,1)$&10&0.74&178.4&7.4\\
Uniform circle&1&360&$(10,10,1)$&10&0.37&0&3.7\\
Weak IMP&4.6&2&$(64,3,0.3)$&65&1.9&9.9&3\\
 &4.6&5&$(64,3,0.3)$&69.8&8.7&23.6&12.5\\
Broad BNL&25&20&$(1000,1,0.3)$&$1.5\times10^4$&$1.1\times10^4$&86.2&70.7\\
Broad MESA&30&45&$(64,0.3,1)$&1700&1340&87.9&78.7\\
Broad proposed&100&15&$(10,0.1,1)$&468&407&88.8&86.9\\
\end{tabular}\par
\medskip
\begin{tabular}{llrrrr}
\multicolumn{6}{l}{(b) Weak bends: $\ell=64$, $I_{\rm fr}=5$~mm, $\sigma_\alpha=\sigma_\beta=0.3$~mrad}\\
Model &\shortstack{Central $(R_\perp,R_\parallel)$\\($\mu$m)}&$|\mean{\hat{\bm{\mathcal L}}}|/\hbar$&\shortstack{$\alpha_L$\\(deg)}&\shortstack{Minimum $W$\\(\%)}&\shortstack{$W$ at independent\\$\pm20\%$ (\%)}\\
IMP, $5^\circ$&$(1.9,0.17)$&69.8&23.6&12&12--12.3\\
BNL, $1^\circ$&$(1.7,0.15)$&69.8&23.5&11.9&11.9--12\\
\multicolumn{6}{l}{IMP wider rectangle: $R_\perp=1$--$5~\mu$m, $R_\parallel=0.1$--$0.3~\mu$m gives $W=12$--$19.6\%$.}\\
\end{tabular}\par
\medskip
\begin{tabular}{lrrr}
\multicolumn{4}{l}{(c) Applicability of the broad finite-bend examples in (a)}\\
Model&$\max\eta_{\rm NR}$&$\max\eta_a$&$\max\eta_{\rm fr}$\\
BNL&0.044&0.0046&$5.2\times10^{-4}$\\
MESA&0.072&0.0025&$1.4\times10^{-4}$\\
Proposed&0.048&0.0016&$4.4\times10^{-5}$\\
\end{tabular}\par
\end{ruledtabular}
\end{table}

The smooth half-turn and single-packet weak IMP rows in Table~\ref{SM:tab:favorable}(a) {use $I_{\rm fr}=1$~mm, while the broad rows use} 5~mm. The sharp-reversal longitudinal size {is rounded in the table, while exact matching uses} Eq.~(\Lref{eq:half}) of the Paper. For the weak IMP packet, $I_{\rm fr}=1,5,10$~mm gives $W=3,3,3.1\%$ at $2^\circ$ and $12.5,12.5,12.6\%$ at $5^\circ$. The sharp-path maxima are $\eta_{\rm NR}<5.6\times10^{-5}$ and $\eta_a<6.2\times10^{-5}$. {Thus this tilted narrow distribution arises in the weak-bend regime.}

For a separate MESA check, $\ell=100$ and $(R_\perp,R_\parallel)=(0.3,1)~\mu$m give $W=75.1\%$ {for sharp boundaries} and $75.2$--$75.7\%$ for $I_{\rm fr}=10,20,30$~mm. The corresponding path maxima satisfy $\eta_{\rm NR}<0.093$, $\eta_a<0.0037$, and $\eta_{\rm fr}<1.1\times10^{-4}$. The three broad rows in Table~\ref{SM:tab:favorable}(a) {use different incident states, so their widths reflect both energy and preparation.} Their Gaussian incidence average uses seven-node Gauss--Hermite quadrature over $\pm1.125$~mrad and the separately centered exit-event moments of Eq.~\eqref{SM:eq:mixture}.
A further survey used $|\ell|=10,64,100,1000$, $R_\perp=0.1,0.3,1,3,10~\mu$m, and $R_\parallel=0.003,0.01,0.03,0.1,0.3,1~\mu$m. Besides BNL and MESA above, it included the compact 60-MeV, $35^\circ$, $\rho=0.532$-m bend and proposed $15^\circ$ bends at 60, 100, and 155~MeV with $\rho=5$~m, plus 155~MeV at $\rho=10$~m. With $I_{\rm fr}=5$~mm and zero incidence spread, no grid point with all three sampled ratios below 0.1 had $W<20\%$.
\endgroup

\begingroup

\subsection{Dependence on the incident packet size}
\label{SM:sec:size_scan}

\label{SM:sec:size_preparations}
\begingroup

\begin{figure}[tp]

\includegraphics[width=\textwidth]{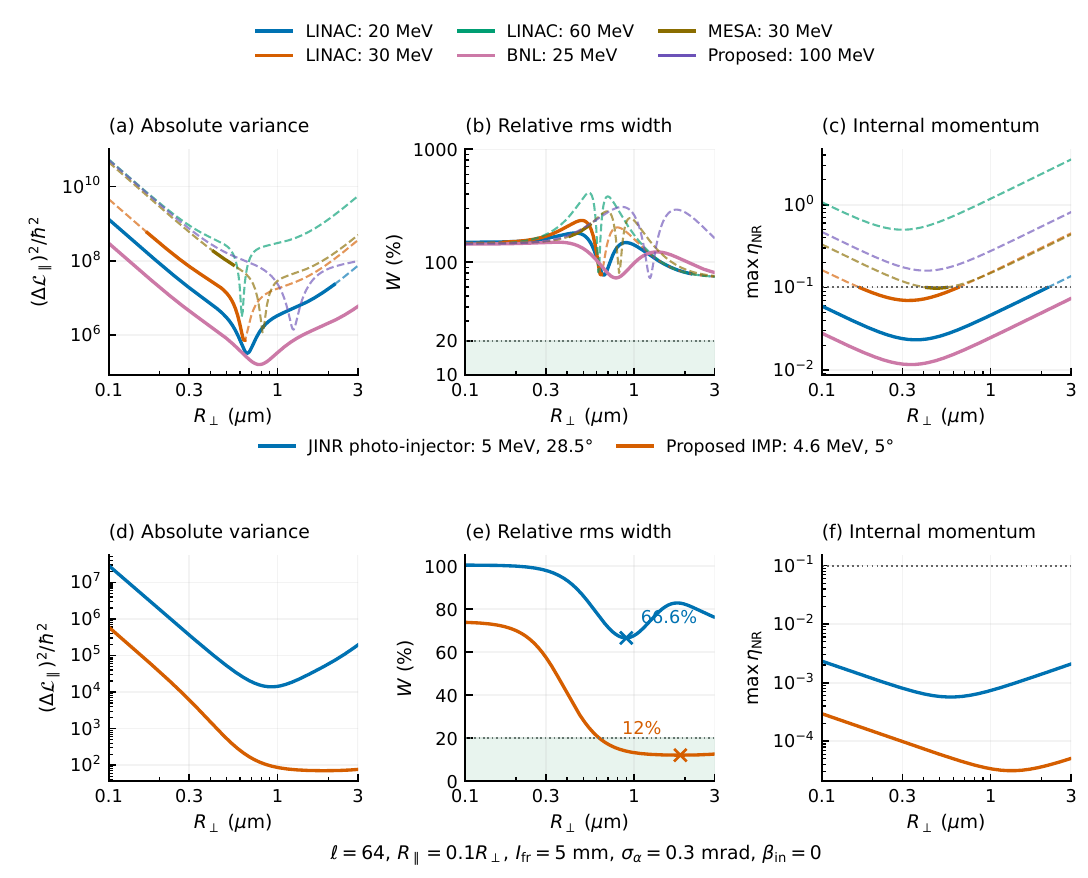}
\caption{One-dimensional size comparisons for $\ell=64$ and $R_\parallel=0.1R_\perp$. The columns give absolute OAM variance along the mean, relative rms width $W$, and the largest sampled internal-momentum ratio. {(a--c) LINAC, BNL, MESA, and the proposed 100-MeV bend. (d--f) The 5-MeV} JINR photo-injector and weak 4.6-MeV IMP bend. All use the geometries in Table~{\ref{SM:tab:facilities}}, $I_{\rm fr}=5$~mm, $\sigma_\alpha=0.3$~mrad, and $\beta_{\rm in}=0$. {Solid curves pass all three sampled 0.1 thresholds, while dashed curves continue} the quadratic model outside this selection. All low-energy points pass. {The width panels use the same observable in percent. The absolute variance is shown} to distinguish broadening from a change in the mean magnitude. Shading marks $W<20\%$, and crosses in (e) mark minima on the plotted size line.}
\label{SM:fig:size_fixed_aspect}\label{SM:fig:size_low}
\end{figure}
\begin{figure}[tp]

\includegraphics[width=\textwidth]{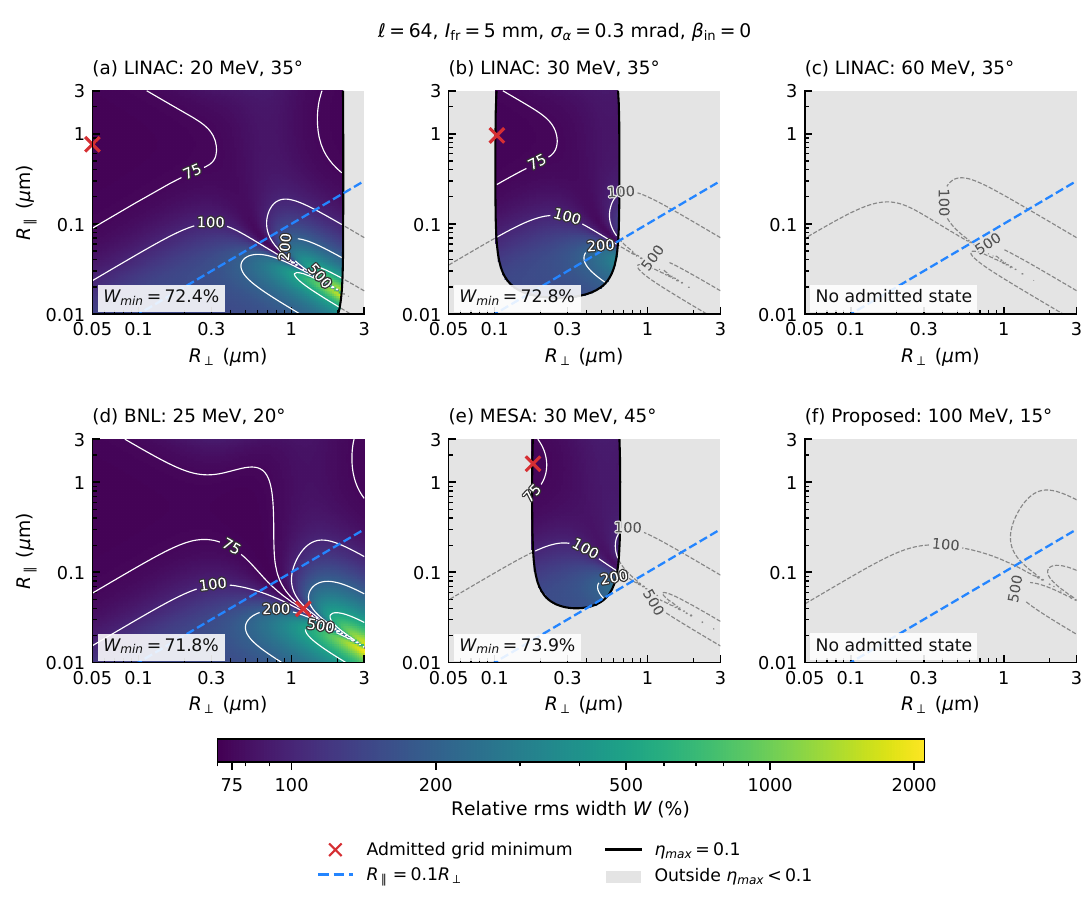}
\caption{Relative rms OAM width $W$ as both incident sizes vary independently, for $\ell=64$, $I_{\rm fr}=5$~mm, $\sigma_\alpha=0.3$~mrad, and $\beta_{\rm in}=0$. Geometry is listed in Table~{\ref{SM:tab:facilities}}. Color and labeled white contours give $100W$, with levels at 75, 100, 200, and 500\%. Gray dashed contours are labeled at 100 and 500\%. Crosses mark the smallest $W$ among admitted points of the $121\times101$ size grid. The black boundary is $\eta_{\max}=0.1$, where $\eta_{\max}$ is the largest of the three expansion ratios over the sampled trajectory and incidence nodes. Gray regions lie outside this selection, where dashed contours show the formal continuation of the quadratic model. Panels (c,f) lie entirely outside the selected domain of applicability. The dashed blue line is the section $R_\parallel=0.1R_\perp$ in Fig.~\ref{SM:fig:size_fixed_aspect}(a--c).}
\label{SM:fig:size_surfaces}
\end{figure}
\endgroup

{To minimize $W$ [Eq.~\eqref{SM:eq:width_angle}] in {Fig.~\Lref{fig:windows}(a,b) of the Paper}, we vary both rms sizes independently at $\ell=64$ [Eq.~\eqref{SM:eq:input}]. Each pair $(R_\perp,R_\parallel)$ specifies a new incident state with zero quadratic \mbox{spatial phase}. This preparation differs from Lorentz contraction, which gives $R_{\perp,\rm lab}=R_\perp$ and $R_{\parallel,\rm lab}\simeq R_\parallel/\gamma$ at equal laboratory time {for a packet with a narrow momentum spread}. Preparing the waist requires control of both the longitudinal size and \mbox{spatial phase}.}

{For each size pair, Eqs.~\eqref{SM:eq:mean_trace}, \eqref{SM:eq:V}, and \eqref{SM:eq:mixture} give the outgoing mean OAM and its covariance matrix. {Since the projection axis and normalization in $W$ vary with the packet sizes, the minima of $W$ and of the absolute variance can occur at different sizes.}}

{The IMP and BNL weak-bend models} use {nominal boundaries perpendicular to the reference orbit}, logistic fringe integrals $I_{\rm fr}=5$~mm [Eq.~\eqref{SM:eq:fringe_integral}], and independent Gaussian incidence spreads $\sigma_\alpha=\sigma_\beta=0.3$~mrad in the two planes defined by Eq.~\eqref{SM:eq:two_incidence_angles}. The two size ranges are $R_\perp=0.01$--$30~\mu$m and $R_\parallel=0.001$--$30~\mu$m. All displayed points satisfy $\max\eta_{\rm NR},\max\eta_a,\max\eta_{\rm fr}<0.1$ [Eqs.~\eqref{SM:eq:validity_ratios} and \eqref{SM:eq:spatial_validity}]. Neither map contains a sampled or numerically refined $W<10\%$ region.

\subsubsection{The two weak bends in {Fig.~\Lref{fig:windows}(a,b) of the Paper}}\label{SM:sec:size_weak}
We vary both sizes for the two weak geometries in Table~{\ref{SM:tab:facilities}}, {with normal incidence of the reference trajectory and the stated model fringes.} The table distinguishes the proposed IMP element from the chosen legacy BNL setting.

The optimized states in Table~\ref{SM:tab:favorable}(b) have small orbit bends, but the accumulated rest-frame phases are comparable: $\gamma\theta=0.873$ and 0.871. Their narrow projections therefore accompany a resolved change in mean direction.

{{The regions of small width are curved, so allowed size pairs must lie inside the joint region in the two-dimensional map.}} In the proposed IMP magnet, the simpler rectangle $R_\perp=1$--$5~\mu$m, $R_\parallel=0.1$--$0.3~\mu$m remains below $W=20\%$.

{The common sizes $(R_\perp,R_\parallel)=(2,1)~\mu$m give $W=17.5/17.4\%$ for IMP/BNL, whereas equal $2~\mu$m sizes give $28.7\%$ in both. The independent $\pm70\%$ variations in {Fig.~\Lref{fig:windows}(a,b) of the Paper}, $R_\perp=0.6$--$3.4~\mu$m and $R_\parallel=0.06$--$0.34~\mu$m, give $W=12$--$27.5\%$ and $11.9$--$19.1\%$. Both maxima occur at $(0.6,0.34)~\mu$m, away from the preferred aspect ratio, and only BNL stays below \mbox{$20\%$ throughout}.}
\endgroup
\subsubsection{Why the minima occur near these sizes}
{Eq.~\eqref{SM:eq:weakbalance} fixes the preferred size ratio, while the numerical minima also select an absolute scale. To distinguish these dependences, we expand at fixed $\gamma$:}
\begin{equation}
\begin{aligned}
 W^2=\frac{\theta^2}{\ell^2}\Bigg\{&
 \left(\frac{eH}{\hbar}\right)^2
 \frac{(|\ell|+2)R_\perp^4}{8(|\ell|+1)}\\
 &+\frac{\gamma^2}{4}\left[
 \frac{(|\ell|+1)^2R_\parallel^2}{2R_\perp^2}
 +\frac{R_\perp^2}{8R_\parallel^2}+\frac12\right]\Bigg\}
 +o(\theta^2).
\end{aligned}
\label{SM:eq:size_small_bend}
\end{equation}
This follows by expanding Eqs.~\eqref{SM:eq:map}, \eqref{SM:eq:V}, and \eqref{SM:eq:width_angle}, including the mean direction. The two terms depending on $R_\parallel$ reflect its coordinate extent and momentum uncertainty. Their derivative vanishes at
\begin{equation}
 \frac{(|\ell|+1)^2R_\parallel}{R_\perp^2}
 -\frac{R_\perp^2}{4R_\parallel^3}=0,
 \qquad
 R_\parallel=\frac{R_\perp}{\sqrt{2(|\ell|+1)}}.
 \label{SM:eq:weak_aspect_ratio}
\end{equation}
The positive second derivative gives the same ratio as Eq.~\eqref{SM:eq:weakbalance}. It makes the Gaussian envelope spherical, while the vortex factor keeps the complete density anisotropic. This ratio differs from the equal-momentum condition that fixes the mean direction at a sharp half turn [Eq.~(\Lref{eq:half}) of the Paper].

At $\ell=64$, the estimate $R_\perp/R_\parallel=11.4$ agrees with 11.3 and 11.4 at the full numerical minima. It explains the orientation of their low-width regions, although $\gamma\theta\simeq0.87$ and the smooth fields require the complete transfer. {Eqs.~\eqref{SM:eq:weakbalance} and \eqref{SM:eq:weak_aspect_ratio} give the preferred ratio of transverse to longitudinal size. The finite-angle calculation determines the sizes themselves by accounting for both momentum uncertainty and the momentum variation induced by the magnetic field.}

The natural magnetic length is $l_H=\sqrt{\hbar/(eH)}$. With the vortex factor included, an order-of-magnitude scale is $R_\perp\sim\sqrt{|\ell|+1}\,l_H$, which gives {2.9} and {2.6}~$\mu$m for the two fields in {Fig.~\Lref{fig:windows}(a,b) of the Paper}. The finite-bend calculation multiplies these estimates by about 0.64. This factor depends on the energy, angle, boundaries, and prepared correlations. A check with geometrically similar IMP magnets at 1.25, 5, and 20~mT moves the minimizing radius $R_\perp$ from {3.8} to {1.9} and 0.94~$\mu$m while leaving $W\simeq12\%$. This comparison rescales the radius, effective dipole length, and fringe length together.

\begingroup

\subsubsection{Why other examined bends remain broad}
{Table~{\ref{SM:tab:size_additional}} extends the optimization to larger bends at $\ell=64$. {We use normal incidence of the reference trajectory except for the ELBE-inspired first bend,} where we interpret the documented 0.15-m length as the separation of parallel effective boundaries~\cite{SM:ELBEchicane}. With the quoted $21^\circ$ bend, this gives $\rho=0.15/\sin21^\circ=0.42$~m, and we choose $K=12$~MeV.}

\begin{table}[tp]

\caption{Additional independent-size minima at $\ell=64$, with $\sigma_\alpha=0.3$~mrad and $\beta_{\rm in}=0$. Each size spans 1~nm--10~$\mu$m for JINR, LINAC, and BNL, or 1~nm--30~$\mu$m for MESA and ELBE. {Geometry is specified in Table~{\ref{SM:tab:facilities}}. All three} sampled expansion ratios must be below 0.1, with the stricter $\max\eta_{\rm NR}\leq0.08$ used for MESA. The entries are smallest found widths in this stated family and domain.}
\label{SM:tab:size_additional}
\begin{ruledtabular}
\begin{tabular}{lrrrr}
Model & $K$ (MeV)&$\theta$ (deg)&$I_{\rm fr}$ (mm)& Smallest found $W$ (\%)\\
JINR photo-injector &1&28.5&5&29.1\\
JINR photo-injector &1.9&28.5&5&44.7\\
JINR photo-injector &3&28.5&5&56.5\\
JINR photo-injector &5&28.5&5&66\\
LINAC channel geometry &5, 10, 20, 30&35&5&69.9--73\\
BNL dipole &25&20&5&71.7\\
MESA MARC1 &30&45&20&74.2\\
ELBE-inspired magnet &12&21&5&71.4
\end{tabular}\par
\end{ruledtabular}
\end{table}

For the photo-injector, $\rho=0.55$~m and $\theta=28.5^\circ$ are retained while the field changes with energy. 5~MeV is the design maximum, whereas 1.9~MeV is discussed in the published simulation {\cite{SM:JINRphoto,SM:JINRcurrent}}. The LINAC continuation holds the channel-2 geometry $\rho=0.532$~m and $\theta=35^\circ$ fixed {\cite{SM:JINRmagnets}}. Points below the cited channel range are model continuations. {In the tests at 40--200~MeV, the internal-momentum ratio exceeds the selection threshold across the tested sizes, placing $W$ outside the controlled regime of the model.} The main IOTA dipoles at 100 and 150~MeV/$c$ likewise fail this criterion in the examined family, leaving $W$ undetermined.

For MESA, allowing $\max\eta_{\rm NR}$ to approach 0.1 {reduces the smallest width found to} $W\simeq73.2\%$ on the selection boundary. {The minimum under the stricter condition in Table~}{\ref{SM:tab:size_additional}} has $(R_\perp,R_\parallel)=({0.19},{1.6})~\mu$m. ELBE has an interior minimum at $({0.45},0.27)~\mu$m with $\eta_{\rm NR}={0.0036}$. Neither case reaches $W<20\%$ within the tested sizes and fringes.

These finite bends accumulate substantial correlations: the radial mean contains $\gamma\sin\theta/2$ [Eq.~(\Lref{eq:mean}) of the Paper], and the longitudinal transfer contains $\gamma^2(\theta-\sin\theta)$ [Eq.~\eqref{SM:eq:map}]. {Adjusting two waist sizes balances some contributions, while the full transfer retains its shears and position--momentum correlations.} {Section~\ref{SM:sec:offwaist} treats position--momentum correlations generated by free propagation when the packet waist is displaced from the entrance field boundary.} Half-turn recovery uses the distinct cancellation of Sec.~\ref{SM:sec:halfturn}.
\endgroup
\paragraph{Locating the minima and checking applicability.}
For a fixed magnet, Eqs.~\eqref{SM:eq:mean_trace} and \eqref{SM:eq:V} give continuous functions of the two sizes. The canonical covariance $\Sigma=S\Sigma_0S^T$ [Eq.~\eqref{SM:eq:covariance_transport}] is linear in $R_\perp^2$, $R_\parallel^2$, $R_\perp^{-2}$, $R_\parallel^{-2}$, and a constant, whereas the OAM covariance $V$ is quadratic in these five quantities. We evaluate these expressions directly and locate candidate minima by independent local and global searches from multiple \mbox{initial points}.

For each sampled trajectory point, any one of the three expansion ratios has the size dependence
\begin{equation}
 \eta^2=c_1 R_\perp^2+c_2 R_\parallel^2+c_3 R_\perp^{-2}
       +c_4 R_\parallel^{-2}+c_5.
 \label{SM:eq:size_box_bound}
\end{equation}
Here $\eta$ denotes $\eta_{\rm NR}$, $\eta_a$, or $\eta_{\rm fr}$ of Eqs.~\eqref{SM:eq:validity_ratios} and \eqref{SM:eq:spatial_validity}. The only varied quantities are the two sizes $(R_\perp,R_\parallel)$. Substituting Eq.~\eqref{SM:eq:input} into $\Sigma=S\Sigma_0S^T$, with Eq.~\eqref{SM:eq:B} for kinetic momentum, fixes all five coefficients at each sampled position for the chosen magnet and $\ell$. Thus the coefficients of $R_j^2$ and $R_j^{-2}$ stay fixed together as $R_j$ changes. Since $c_1,\ldots,c_4\geq0$, the largest rms expansion ratio over two independent size intervals occurs at one of their four endpoint pairs. Checking these four pairs bounds the rms ratios throughout that size range at each sampled trajectory point.

\begingroup

{Minimizing with both incidence spreads $\sigma_\alpha=\sigma_\beta=0.3$~mrad also finds no $W<20\%$ region for the eleven settings in Table~{\ref{SM:tab:size_additional}}. The comparisons below use planar ensembles, $\beta_{\rm in}=0$, and include the distinct \mbox{weak IMP model}.}

\subsubsection{Restricted-domain comparisons and fixed aspect ratio}
{Restricting the size domain tests how the minima depend on the available preparation. Fig.~\ref{SM:fig:size_fixed_aspect} uses $0.1\leq R_\perp\leq3~\mu$m at fixed aspect ratio. Fig.~\ref{SM:fig:size_surfaces} varies both sizes over $0.05\leq R_\perp\leq3~\mu$m and $0.01\leq R_\parallel\leq3~\mu$m. These restricted ranges give grid minima that can differ from Table~{\ref{SM:tab:size_additional}}. Along $R_\parallel=0.1R_\perp$, the admitted $R_\perp$ intervals are $0.1$--$2.2~\mu$m for LINAC at 20~MeV, $0.17$--$0.64~\mu$m at 30~MeV, $0.1$--$3~\mu$m for BNL, and $0.41$--$0.55~\mu$m for MESA. Neither the 60-MeV LINAC nor the proposed 100-MeV bend passes: their smallest sampled $\max\eta_{\rm NR}$ are 0.5 and 0.16 [Eq.~\eqref{SM:eq:validity_ratios}], or 0.3 and 0.11 when both sizes vary independently.}

{Independent variation in Fig.~\ref{SM:fig:size_surfaces} gives admitted minima $W\simeq72$--$74\%$ for LINAC at 20 and 30~MeV, BNL, and MESA. For the 5-MeV photo-injector it gives $66\%$, compared with $66.6\%$ at fixed aspect ratio [Fig.~{\ref{SM:fig:size_low}}(d--f)], while the weak IMP bend retains $12\%$. Table~\ref{SM:tab:favorable} uses different states, allowing $\ell$ and the size ratio to vary.}
\endgroup

\begingroup

\section{Half-turn recovery, tolerances, and repeated bends}\label{SM:sec:fixed_checks}
\subsection{Preparation and reversal at sharp boundaries}
{{We first distinguish reversal of the mean from recovery of a narrow OAM distribution, then examine smooth edges, angular acceptance, and preparation errors.} The angle scan (Fig.~{\ref{SM:fig:half_extended}}) and ring (Fig.~{\ref{SM:fig:ringturns_extended}}) start from the LG state $\psi_0$ [Eq.~\eqref{SM:eq:LG}] with $\ell=64$, $R_\perp=3~\mu$m, and $R_\parallel=0.3~\mu$m. It has $\mean{\hat{\mathcal L}_t}=64\hbar$ and $V_{tt}=0$ [Eq.~\eqref{SM:eq:OAM_covariance}]. {Each scan point starts from the same incident state, while the ring propagates that state through successive bends.} We use the observables in Eq.~\eqref{SM:eq:width_angle} and planar incidence $\alpha_{\rm in}$ [Eq.~\eqref{SM:eq:two_incidence_angles}], with $\beta_{\rm in}=0$.}

For the $K=1$~MeV, $H=5$~mT angle scan, the orbit radius is $\rho=0.949$~m. The magnet length changes as $\rho\theta$, with normal entrance and exit at every angle. At $\theta=180^\circ$, the components in the outgoing radial, tangential, and field directions $(r,t,H)$ are
\[
 \begin{gathered}\mean{\hat{\bm{\mathcal L}}}/\hbar=(0,-64,-181),\\
 W={77.5\%},\quad \alpha_L=110^\circ.\end{gathered}
\]
The solid curves retain $R_\parallel=0.3~\mu$m, whereas the dashed curves change it to the matched value $R_\parallel=R_\perp/[\sqrt2(|\ell|+1)]\simeq33$~nm from Eq.~(\Lref{eq:half}) of the Paper. In the same $(r,t,H)$ axes, Eq.~\eqref{SM:eq:LH} at $\theta=\pi$ gives
\[
 \mean{\hat{\bm{\mathcal L}}}_{\rm out}
 =\left(0,-\ell\hbar,\frac{2\gamma}{eH}
 [\Sigma_{0,p_tp_t}-\Sigma_{0,p_rp_r}]\right).
\]
Thus, the mean OAM becomes antiparallel to the outgoing momentum for $\ell>0$ only when the incident momentum variances match. The state shown by the dashed curves then has $\alpha_L=180^\circ$ but $W\simeq305\%$, with its spectrum in Fig.~\ref{SM:fig:projection}(a) and size tolerance in Sec.~{\ref{SM:sec:sizetolerance}}.

Its nonmonotonic relative width also reflects the changing mean magnitude: between $176^\circ$ and $178.5^\circ$, $\Delta\mathcal L_\parallel$ falls from $785\hbar$ to $573\hbar$, while $|\mean{\hat{\bm{\mathcal L}}}|$ falls from $158\hbar$ to $83.9\hbar$, so $W$ rises. At $180^\circ$ the tangential rms spread is $\Delta\mathcal L_t=195\hbar$. The largest sampled ratios [Eq.~\eqref{SM:eq:validity_ratios}] in the original angle scan are $\eta_{\rm NR}={1.1}\times10^{-4}$ and $\eta_a={2}\times10^{-5}$.

\begin{figure}[!t]

 \centering
 \includegraphics[trim=0 0 171bp 0,clip,width=.70\textwidth]{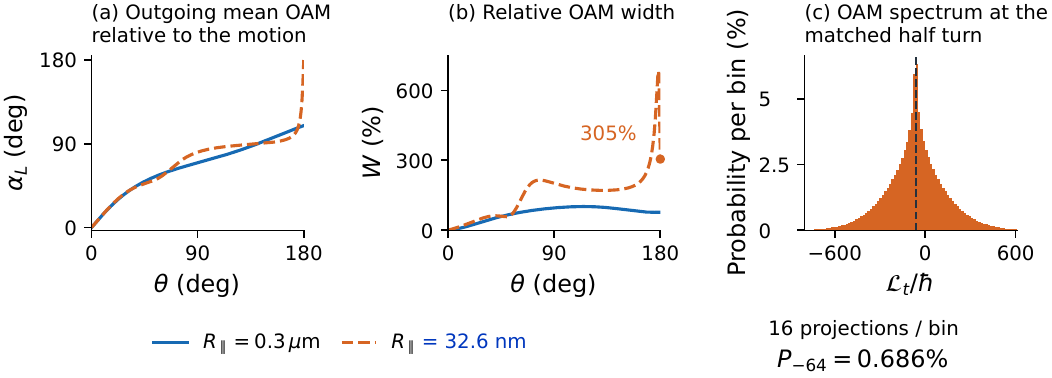}
 \caption{Mean direction (a) and relative rms width (b) {after bends with sharp boundaries} at $K=1$~MeV and $H=5$~mT. Both packets have $\ell=64$ and $R_\perp=3~\mu$m. {Solid curves use $R_\parallel=0.3~\mu$m, while dashed curves use} the equal-momentum value $R_\parallel\simeq32.6$~nm, Eq.~(\Lref{eq:half}) of the Paper. Each angle uses a magnet of length $\rho\theta$ {at normal incidence.} {At $180^\circ$ the matched mean reverses exactly within a broad distribution, $W=305\%$.}}
 \label{SM:fig:half_extended}
\end{figure}
\label{SM:sec:fixed_recovery}
{We compare half-turn reversal with the weak-bend tilted states of Sec.~\ref{SM:sec:analytic}.} For $K=1$~MeV, $H=5$~mT, $\ell=10$, and $R_\perp=9.6~\mu$m, the matching condition in Eq.~(\Lref{eq:half}) of the Paper fixes $R_\parallel=R_\perp/[\sqrt2(|\ell|+1)]\simeq0.62~\mu$m. These sizes give a narrow outgoing OAM distribution, $\mean{\hat{\bm{\mathcal L}}}=(0,-10,0)\hbar$ and $W=5.5\%$ at $180^\circ$. Its connected $W\leq20\%$ interval extends 7.9~mrad below $180^\circ$, with $\max\eta_{\rm NR}={1.6}\times10^{-4}$. Sections~\ref{SM:sec:recovery} and {\ref{SM:sec:sizetolerance}} {compare the angular and size tolerances, while Fig.~\ref{SM:fig:sizetolerance} isolates} the fringe effect at fixed bend.

{A weak bend means $|\theta|\ll1$, while} $\gamma\theta$ can remain of order unity [Sec.~\ref{SM:sec:weak_mean}]. At a small finite angle, the fixed state of Fig.~{\ref{SM:fig:half_extended}} has $\alpha_L=14.5^\circ$ and $W={8.6}\%$ at $10^\circ$. Such a tilted mean with a relatively narrow distribution motivates a spatiotemporal-vortex analysis~\cite{SM:BliokhNori2012}, while its phase singularities follow from the full complex state in Eq.~\eqref{SM:eq:SMpsi}.

For the narrow sharp state, imposing both $W\leq20\%$ and direction error below $3^\circ$ gives $179.548^\circ\leq\theta\leq180.399^\circ$. The uniform-circle state $(\ell,R_\perp,R_\parallel)=(10,10,1)$, with sizes in $\mu$m, instead has the return interval $359.555^\circ\leq\theta\leq360.359^\circ$. Its formal free-exit moments obey $\mean{\hat{\mathcal L}_t}=10\hbar$, $\Delta\mathcal L_t=0.37N\hbar$, {and $W=3.7N\%$, so the width reaches} $18.5\%$ after five turns even though the mean returns. The segmented ring in Sec.~{\ref{SM:sec:repeated_bends}} has a different transfer.
\subsection{Smooth edges at a fixed bend}
The angular averages in this subsection and Sec.~{\ref{SM:sec:matched_search}} set $\beta_{\rm in}=0$.
For $\ell=10$ and $(R_\perp,R_\parallel)\simeq(9.6,0.617)~\mu$m, the sizes matched to sharp boundaries give $W={5.5}\%$ at the sharp half turn. Keeping both nominal boundaries at $180^\circ$ while adding logistic edges gives
\begin{center}
\begin{tabular}{rrrr}
&\multicolumn{3}{c}{$W$ (\%) for the stated $\sigma_\alpha$}\\
$I_{\rm fr}$ (mm)&0&0.3 mrad&1 mrad\\
\hline
0.2&{5.9}&6.1&{7.9}\\
0.5&6.8&{7}&{8.6}\\
1&{8.8}&{9}&10.2\\
2&13.7&13.8&14.6\\
5&29.6&29.7&30.1
\end{tabular}\par
\end{center}
Each angular average uses the same fixed magnet boundaries and a Gaussian in-plane spread, with $\beta_{\rm in}=0$. At every listed fringe length, increasing the angular spread \textit{broadens the OAM distribution}. The fringe itself has the larger effect as its length grows. Independent ten-percent size changes affect this state less than a 1--2-mm fringe integral at fixed bend.

\begin{figure}[tp]

\includegraphics[width=\textwidth,trim=0 3bp 0 0,clip]{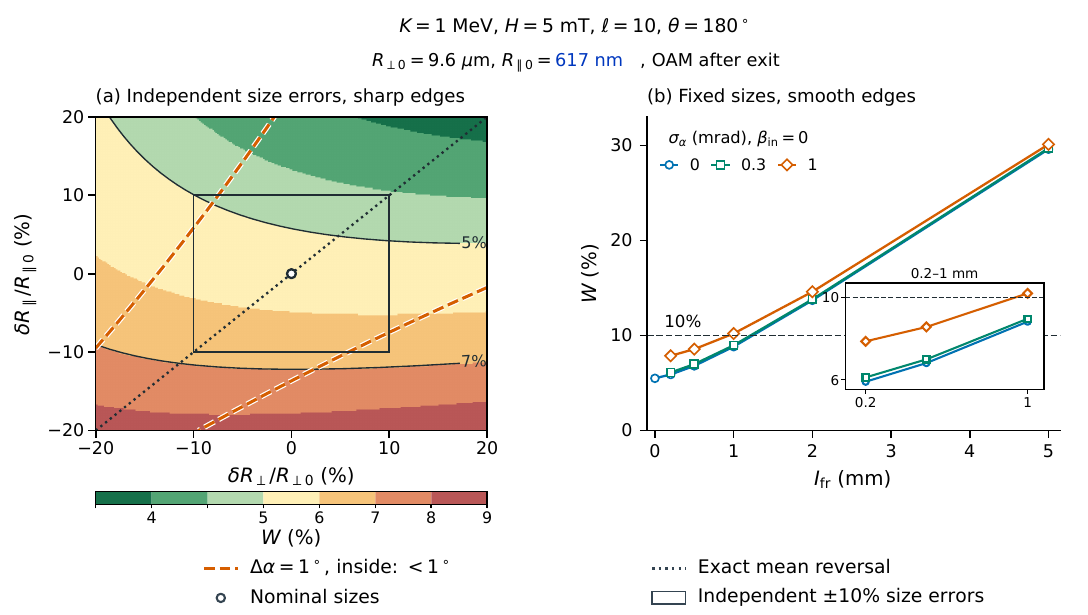}
\caption{Size and fringe sensitivity at fixed $180^\circ$ bend: $K=1$~MeV, $H=5$~mT, $\ell=10$, $R_{\perp0}=9.6~\mu$m, and $R_{\parallel0}=R_{\perp0}/[\sqrt2(|\ell|+1)]\simeq0.62~\mu$m. The subscript $0$ labels the nominal entrance sizes. (a) Independent fractional size errors, {with sharp boundaries and normal incidence.} {Color gives $W$ along the actual mean, and orange dashed curves bound} a direction error below $1^\circ$. The diagonal retains exact reversal, and the entire panel has $W<10\%$. (b) Nominal sizes held fixed while both logistic fringe integrals vary together. Symbols show the tabulated planar incidence ensembles, {$\beta_{\rm in}=0$, and lines guide the eye.} The inset uses the same axes and units, and horizontal dashed lines mark $W=10\%$. This fixed-angle comparison differs from the retuned-angle windows in Fig.~{\ref{SM:fig:windows_extended}}.}
\label{SM:fig:sizetolerance}
\end{figure}

A sharper fringe can dominate otherwise modest preparation errors. At $K=3$~MeV, $H=0.1$~T, $\ell=100$, $(R_\perp,R_\parallel)=(10,2)~\mu$m, and $\theta=180^\circ$, a logistic 1-mm 10--90\% rise ($I_{\rm fr}\simeq0.23$~mm) raises $W$ from $1.01\%$ to $21.9\%$. The orbit is recomputed with the geometry and state fixed. Independent $\pm20\%$ size changes give $W=0.62$--$2\%$ {for sharp boundaries} and $14.8$--$31\%$ for the smooth profile. At nominal sizes, independent $\pm1\%$ changes of energy or field {give $21.5$--$22.3\%$, while varying the rise} from 0.8 to 1.2~mm gives $17.7$--$26.1\%$. Individual incidences $-0.3,0,+0.3$~mrad give $18.7,21.9,25.1\%$, respectively. The size box has sampled $(\eta_{\rm NR},\eta_a,\eta_{\rm fr})<(0.024,0.0034,0.038)$. Over the wider domain $R_\perp=1$--$20~\mu$m, $R_\parallel=0.5$--$5~\mu$m, the sharp and smooth ranges become $0.21$--$111\%$ and $9$--$110\%$, with the widest smooth range bounded by $(0.039,0.0057,0.063)$.
\subsection{Recovery with fringes and incidence spread}\label{SM:sec:recovery}
{{We test whether adjusting the bend angle compensates for smooth edges in the presence of incidence spread.}} For the incident state $\ell=10$, $(R_\perp,R_\parallel)=(10,1)~\mu$m, we vary the design bend over $177^\circ$--$181^\circ$ at fixed radius $\rho\simeq0.949$~m. {The reference trajectory crosses both boundaries at normal incidence, and rigidity sets the field at each energy.} With the fringe integral $I_{\rm fr}$ defined in Eq.~\eqref{SM:eq:fringe_integral} and no incidence spread, the 1-MeV minimum is $W={2.6}\%$ at $\theta-180^\circ\simeq-{2.6}$~mrad for $I_{\rm fr}=1$~mm. Increasing the integral to 5 or 10~mm and adjusting only the bend gives minima of ${3.5}\%$ and ${5.9}\%$.

{Here $\sigma_\alpha$ is the rms spread of packet-center incidence angles, with a Gaussian distribution proportional to $\exp[-\alpha_{\rm in}^2/(2\sigma_\alpha^2)]$ in the bending plane ($\beta_{\rm in}=0$). We combine separately centered moments using Eq.~\eqref{SM:eq:mixture} and the exit-event convention of Sec.~\ref{SM:sec:measurement}. Section~\ref{SM:sec:fringes} {gives the limitation imposed by the fringe-field model. The two-angle result is stated below.}} Table~{\ref{SM:tab:recovery}} separates intervals that constrain both width and direction from intervals that constrain width alone. {Angle refinement can shift the minima from the nominal settings in Table~\ref{SM:tab:favorable}.} For the selected 1-, 4.6-, and 5-MeV settings, the sampled $\eta_{\rm NR}$ maxima are $1.6\times10^{-4}$, $7.4\times10^{-3}$, and $9.2\times10^{-3}$, with $\eta_a<8.7\times10^{-4}$ and $\eta_{\rm fr}<7.2\times10^{-3}$. {Changing $\ell$ from $+10$ to $-10$ preserves $W$ and turns the mean from near reversal to about $1.6^\circ$ from the outgoing momentum.}

\begin{table}[tp]

\caption{Angular and incidence tolerances for the fixed state $\ell=10$, $(R_\perp,R_\parallel)=(10,1)~\mu$m in the proposed $\rho=0.949$-m half-turn family. {The reference trajectory crosses both boundaries at normal incidence, and Gaussian incidence averages are planar} ($\beta_{\rm in}=0$). (a) At $I_{\rm fr}=1$~mm and $\sigma_\alpha=0.3$~mrad, the interval constrains both $W$ and the error $180^\circ-\alpha_L$. {Sizes stay fixed for that interval, while the final two columns vary} each size independently by $\pm10\%$ at the listed setting. (b) The bounds constrain $W$ alone, with $\delta\theta=\theta-180^\circ$ in mrad and endpoints rounded inward. {A dash marks an unattained bound. Each interval varies the design angle of the magnet at the stated incidence spread.}}
\label{SM:tab:recovery}\label{SM:tab:favorable_windows}
\setlength{\tabcolsep}{3.5pt}
\begin{ruledtabular}
\textit{{(a) Width and mean direction, with size sensitivity at the selected setting}}\par\smallskip
\begin{tabular}{rrrrrr}
\shortstack{$K$\\(MeV)}&\shortstack{Design setting\\(deg)}&\shortstack{$W\leq20\%$, error $\leq3^\circ$:\\$\theta$ (deg)}&\shortstack{Width\\(mrad)}&\shortstack{$W$ range\\(\%)}&\shortstack{Max. error\\(deg)}\\
1&179.85&$[179.457,180.256]$&14&{2.9}--{3.6}&{2.4}\\
4.6&179.88&$[179.771,179.983]$&{3.7}&{6}--{8.1}&{2.4}\\
5&179.88&$[179.78,179.974]$&3.4&{6.4}--{8.6}&{2.4}
\end{tabular}\par
\medskip\textit{(b) Width-only intervals at fixed incident sizes}\par\smallskip
\begin{tabular}{rrrrll}
$K$ (MeV)&$I_{\rm fr}$ (mm)&$\sigma_\alpha$ (mrad)&Min. $W$ (\%)&$W\leq5\%$: $\delta\theta$ (mrad)&$W\leq20\%$: $\delta\theta$ (mrad)\\
1&1&0&{2.6}&$[-4.08,-1.07]$&$[-9.49,4.5]$\\
1&1&0.3&{3.1}&$[-3.96,-1.21]$&$[-9.47,4.46]$\\
1&1&1&6.2&---&$[-9.21,4.2]$\\
5&1&0&3.4&$[-2.49,-1.82]$&$[-3.94,-0.35]$\\
5&1&0.3&{7.3}&---&$[-3.83,-0.46]$\\
5&1&1&21.9&---&---\\
5&5&0.3&28.4&---&---\\
10&1&0&{7.1}&---&$[-3,-1.26]$\\
10&1&0.3&14.2&---&$[-2.77,-1.47]$
\end{tabular}\par
\end{ruledtabular}
\end{table}

The 1-MeV intervals illustrate the tradeoff between width and field tolerance. At 0.3-mrad incidence spread, the common $10\%$ interval for $I_{\rm fr}=0.8$, 1, and 1.2~mm has width 5.87~mrad and extends approximately from $\delta\theta=-5.5$ to $+0.37$~mrad [Fig.~{\ref{SM:fig:windows_extended}}(a)]. The common $20\%$ interval is 13.1~mrad. The stricter $5\%$ interval has width 2.76~mrad for the 1-mm profile and 1.92~mrad for all three, approximately $\delta\theta=-3.54$ to $-1.62$~mrad. At their common midpoint, independent $\pm1\%$ changes of both sizes and the nominal values give $W=3.1$--$3.39\%$ across the profiles. The mean is about $1.6^\circ$ from reversal. For the planar ensemble, the wider fringe range 0.5, 1, and 5~mm leaves a common $20\%$ interval of 4.43~mrad.

At 5~MeV and 0.3-mrad incidence spread, the $10\%$ interval has width 1.24~mrad and extends approximately from $\delta\theta=-2.78$ to $-1.54$~mrad. A 1-mrad spread or 5-mm fringe removes even the $20\%$ interval in this scan. At 10~MeV the minimum is $14.2\%$ and the $20\%$ interval is 1.31~mrad. Adding an independent Gaussian packet-center energy spread of 59~keV, the source-design value in Sec.~\ref{SM:sec:facilities}, to the 0.3-mrad incidence ensemble gives a 5-MeV minimum $W=7.32\%$ with the same exit-event convention. {This tests the proposed half-turn geometry using the energy spread of the source designed for the $28.5^\circ$ photo-injector magnet.}

{The sampled maxima of $\eta_{\rm NR}$ [Eq.~\eqref{SM:eq:validity_ratios}] at 1, 5, and 10~MeV are ${1.6}\times10^{-4}$, ${9.2}\times10^{-3}$, and ${6.5}\times10^{-2}$, and $\eta_a<3.2\times10^{-3}$ and $\eta_{\rm fr}\simeq7.2\times10^{-3}$.} The 10-MeV minimum occurs near $179.878^\circ$. Formal minimization at 20 and 30~MeV gives $\max\eta_{\rm NR}=0.48$ and 1.6. At 25, 60, and 100~MeV in the same compact family, $\max\eta_{\rm NR}$ reaches 0.93, 12.4, and 56.7, {placing the prediction of $W$ outside the controlled regime of the Hamiltonian in Eq.~\eqref{SM:eq:SMexpansion}.} Other high-energy geometries can remain within the approximation: with $\ell=1000$, $(R_\perp,R_\parallel)=(6.8,{0.15})~\mu$m, a 25-MeV, $0.5^\circ$ bend in the 0.24-m BNL spectrometer model gives $W\simeq{3.2}\%$, while proposed 5-mT bends at 24 and 60~MeV, through $0.5^\circ$ and $0.25^\circ$, give ${3}\%$ and ${4.5}\%$. Each value covers nine individually centered cases with $I_{\rm fr}=1$, 5, and 10~mm and incidence $0,\pm1$~mrad. These are individual-packet widths in the specified \mbox{weak elements}.

\paragraph{Extension to two incidence planes.}
For the 1-MeV packet $(\ell,R_\perp,R_\parallel)=(10,10,1)$, with sizes in $\mu$m, Fig.~\Lref{fig:windows}(c) of the Paper uses $I_{\rm fr}=5$~mm and $\sigma_\alpha=\sigma_\beta=0.3$~mrad. Its minimum is $W=3.8\%$ at $\theta_*\simeq179.378^\circ$. Expressed as $\pi-\theta$, the calculated intervals are 3.96--17.6~mrad for $W<20\%$ and 7.63--14~mrad for $W<10\%$, with widths 13.6 and 6.4~mrad. Adding the elevation spread shifts the planar interval toward larger bend angles by about 0.06~mrad and barely changes its width. At the fixed optimum angle, independent ranges $R_\perp=10$--$16~\mu$m and $R_\parallel=0.7$--$1.3~\mu$m give $W=3.4$--$7.1\%$, with direction error below $2.9^\circ$. The nominal-state sampled maxima satisfy $(\eta_{\rm NR},\eta_a,\eta_{\rm fr})<(1.7\times10^{-4},5.8\times10^{-5},1.5\times10^{-3})$. {The 5-mm profile contains all sampled incidence trajectories within the local field model and supports this calculation with incidence spread in \mbox{both planes}.}
\begin{figure}[tp]

 \includegraphics[width=\textwidth]{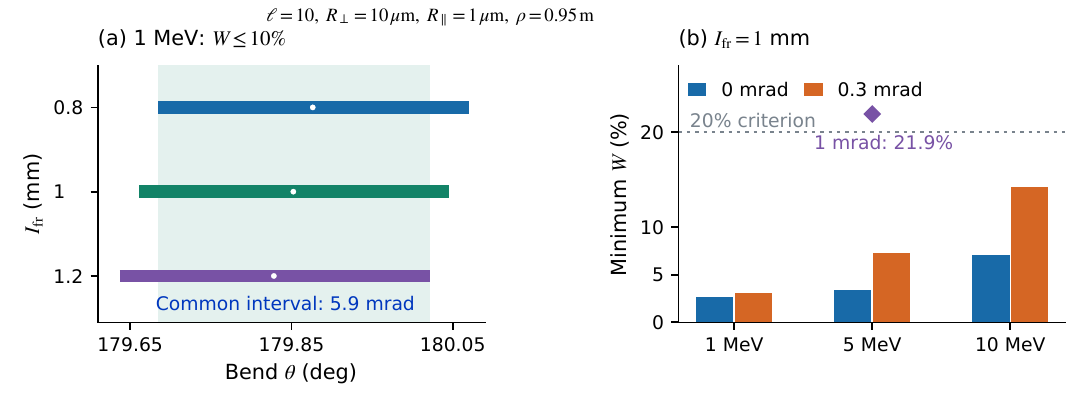}
 \caption{Angular windows after retuning a smooth near-half-turn magnet for $\ell=10$, $(R_\perp,R_\parallel)=(10,1)~\mu$m and fixed $\rho=0.949$~m. {The reference trajectory crosses both boundaries at normal incidence, and the incidence ensemble is planar.} (a) At 1~MeV and $\sigma_\alpha=0.3$~mrad, segments give $W\leq10\%$ {for the stated fringe integrals, dots mark minima, and green shading marks their common interval.} (b) Minimum $W$ under a design-angle scan, for $I_{\rm fr}=1$~mm and $\sigma_\alpha=0$ or 0.3~mrad. The diamond is the 5-MeV result at 1-mrad spread. The field changes with energy to keep the radius fixed. These windows hold the incident sizes fixed, unlike the tolerance map in Fig.~\ref{SM:fig:sizetolerance}.}
 \label{SM:fig:windows_extended}
\end{figure}

\subsection{Sensitivity to wave-packet preparation}\label{SM:sec:sizetolerance}
{We test the sensitivity of half-turn alignment and $W$ [Eq.~\eqref{SM:eq:width_angle}] to size errors at fixed $180^\circ$ bend. The first state in Table~{\ref{SM:tab:sizetolerance}} needs about 0.1-nm longitudinal precision for $1^\circ$ alignment, although its matched distribution remains broad. The second admits $R_\parallel=0.53$--$0.76~\mu$m for $1^\circ$, with $W=4$--$7.2\%$, and 611--623~nm for 1~mrad, with $W=5.4$--$5.6\%$. {The direction bound sets both intervals, which remain unchanged when the width criterion is relaxed from $W<10\%$ to $20\%$.} Displaced waists and their quadratic phases are treated in Eqs.~\eqref{SM:eq:offwaist_state}--\eqref{SM:eq:offwaist_correlation}.}

Fig.~\ref{SM:fig:sizetolerance}(a) varies both sizes independently by $\pm20\%$, giving $W={3.8}$--${8.9}\%$ and a largest mean-direction error of ${2.5}^\circ$. Within the central $\pm10\%$ range, these values are ${4.5}$--${6.8}\%$ and ${1.2}^\circ$, respectively. Away from matching we project the full covariance along the actual mean, whereas Eq.~\eqref{SM:eq:halfvariance} alone gives the tangential variance.

\paragraph{Matching the sizes with smooth edges.}\label{SM:sec:matched_search}
We impose Eq.~(\Lref{eq:half}) of the Paper exactly and search $R_\perp=0.1$--60~$\mu$m and $\theta=179^\circ$--$180.5^\circ$ at 1~MeV, 5~mT, and 0.3-mrad rms incidence spread. For $I_{\rm fr}=1$~mm, the numerical minima lie near $(R_\perp,\theta)=(13~\mu\mathrm m,179.87^\circ)$ for $\ell=10$ and $(32~\mu\mathrm m,179.87^\circ)$ for $\ell=64$, with $W\simeq{4.1}\%$ and ${3.7}\%$. Fixing $\theta=180^\circ$ and varying only the matched radius instead gives ${9}\%$ and ${8.3}\%$. At 5-mm fringes the best bends move to about $179.39^\circ$, giving $4.8\%$ and $4.3\%$, compared with $21.1\%$ and $19.6\%$ at $180^\circ$. A finite optimal scale now balances inverse-size momentum terms against the residual coordinate-dependent shear.

The tolerance calculation holds the rounded bend angle, field, energy, fringe profile, and incidence distribution fixed. The nominal transverse radius is $R_{\perp0}=13~\mu$m for $\ell=10$ or $32~\mu$m for $\ell=64$. The corresponding longitudinal length is $R_{\parallel0}=R_{\perp0}/[\sqrt2(|\ell|+1)]$, approximately 0.84 or 0.35~$\mu$m. The subscript $0$ labels this nominal setting. The actual entrance sizes $R_\perp,R_\parallel$ vary with fractional errors $\epsilon_\perp,\epsilon_\parallel$, whose common allowed magnitude is $d$:
\begin{equation}
 \begin{gathered} R_\perp=(1+\epsilon_\perp)R_{\perp0},\qquad
 R_\parallel=(1+\epsilon_\parallel)R_{\parallel0},\\
 |\epsilon_\perp|,|\epsilon_\parallel|\leq d.\end{gathered}
 \label{SM:eq:size_errors}
\end{equation}

{A common size rescaling, $\epsilon_\perp=\epsilon_\parallel$, preserves Eq.~(\Lref{eq:half}) of the Paper. At 1-mm fringes, common $\pm20\%$ changes give $W=4.1$--$5\%$ for $\ell=10$ and $3.7$--$4.5\%$ for $\ell=64$, with the mean within $0.18^\circ$ of reversal. Thus {the matching condition for sharp boundaries} identifies a finite smooth-edge neighborhood with small residual tilt. All listed independent size ranges have sampled $\eta_{\rm NR},\eta_a,\eta_{\rm fr}<0.03$ [Eqs.~\eqref{SM:eq:validity_ratios} and \eqref{SM:eq:spatial_validity}], bounded by the four endpoint pairs in Eq.~\eqref{SM:eq:size_box_bound}.}

\begin{table}[tp]

\caption{Size tolerances near reversal at $K=1$~MeV and $H=5$~mT. (a) Sharp $180^\circ$ bend, normal incidence, and fixed $R_\perp$: the intervals vary only $R_\parallel$, with fractional error $R_\parallel/R_\parallel^*-1$ and $R_\parallel^*=R_\perp/[\sqrt2(|\ell|+1)]$. {Direction bounds are rounded inward, and the final column gives $W$ at exact matching.} (b) Smooth edges and $\sigma_\alpha=0.3$~mrad, $\beta_{\rm in}=0$: both sizes vary independently as in Eq.~\eqref{SM:eq:size_errors}. The central radii are 13 and 32~$\mu$m for $\ell=10$ and 64, with longitudinal lengths set by the same matching relation. Bend, field, energy, and fringe profile stay fixed. Width ranges and maximum errors $180^\circ-\alpha_L$ are rounded. All smooth size boxes pass the sampled applicability bounds.}
\label{SM:tab:sizetolerance}\label{SM:tab:matched_neighborhoods}
\setlength{\tabcolsep}{3pt}
\begin{ruledtabular}
\textit{(a) Sharp boundary: longitudinal-size error at fixed transverse radius}\par\smallskip
\begin{tabular}{rrrrrrr}
$\ell$ & $R_\perp$ ($\mu$m)&$R_\parallel^*$ (nm)&$\Delta\alpha\leq1^\circ$&$\Delta\alpha\leq0.1^\circ$&$\Delta\alpha\leq1$ mrad&$W$ (\%)\\
64&3&32.6&$[-0.3,\,+0.3]\%$&$[-0.03,\,+0.03]\%$&$[-0.017,\,+0.017]\%$&305\\
10&9.6&617&$[-13,\,+23]\%$&$[-1.6,\,+1.7]\%$&$[-0.96,\,+0.99]\%$&{5.5}
\end{tabular}\par
\medskip\textit{(b) Smooth boundaries: independent errors in both sizes}\par\smallskip
\begin{tabular}{rrrrrrrr}
$\ell$ &\shortstack{$I_{\rm fr}$\\(mm)}&\shortstack{$\theta$\\(deg)}&\shortstack{Central $W$\\(\%)}&\shortstack{$W$ at\\$d=0.1$ (\%)}&\shortstack{$W$ at\\$d=0.2$ (\%)}&\shortstack{Max. error\\($d=0.1$)}&\shortstack{Max. error\\($d=0.2$)}\\
10&1&179.87&{4.1}&3.5--{5.1}&3.2--{6.5}&$0.69^\circ$&${1.4}^\circ$\\
64&1&179.87&{3.7}&{3.1}--{4.5}&{2.8}--{5.8}&$0.62^\circ$&${1.3}^\circ$\\
10&5&179.39&4.8&{4.2}--{5.6}&{3.9}--{6.8}&${1.7}^\circ$&${2.2}^\circ$\\
64&5&179.39&{4.3}&{3.7}--{5.1}&{3.4}--{6.1}&${1.6}^\circ$&${2.1}^\circ$
\end{tabular}\par
\end{ruledtabular}
\end{table}

The distinct planar near-reversal packet $(\ell,R_\perp,R_\parallel)=(10,13,1)$, with sizes in $\mu$m, gives $W=3.6\%$ and a mean $0.53^\circ$ from reversal at 1~MeV, 5~mT, $I_{\rm fr}=1$~mm, $\theta=179.87^\circ$, and 0.3-mrad incidence spread. The rectangle $R_\perp=10$--$16~\mu$m, $R_\parallel=0.7$--$1.3~\mu$m retains $W=2.8$--$6.3\%$ and direction error below $2.1^\circ$. The wider ranges $R_\perp=1$--$26~\mu$m, $R_\parallel=0.3$--$3~\mu$m allow $W$ up to $136\%$ and error up to $78^\circ$, although their sampled expansion ratios remain below $(6\times10^{-4},2.2\times10^{-4},0.027)$.

For the $(10,10,1)$ planar packet, independent size factors 0.9, 1, and 1.1 give $W\simeq3$--$4\%$ at $179.852^\circ$ and $I_{\rm fr}=0.8,1,1.2$~mm. At 5~MeV, $179.877^\circ$, and $I_{\rm fr}=1$~mm, they give $6.4$--$8.6\%$. Both use 0.3-mrad incidence spread. These fixed-angle size checks complement the fixed-size angular intervals in Table~{\ref{SM:tab:recovery}}.
\paragraph{Wider variation of packet sizes and OAM.}\label{SM:sec:wide_preparation_scan}
{We vary the preparation at fixed magnet settings over integer $\ell=1,\ldots,100$ and $200,500,1000$, with independent size factors $0.5,0.6,\ldots,1.5$: $R_\perp=5$--$15~\mu$m and $R_\parallel=0.5$--$1.5~\mu$m. The offsets are $\delta\theta\simeq-2.6$~mrad at 1~MeV, with $I_{\rm fr}=0.8,1,1.2$~mm, and $-2.2$~mrad at 5~MeV, with $I_{\rm fr}=1$~mm. Both use the preceding 0.3-mrad planar ensemble. At $\ell=10$, the sampled widths are $W=2.6$--$12.9\%$ and $5.5$--$22.2\%$, respectively. At $\ell=64$ they are $3.8$--$47.9\%$ and $3.6$--$95.5\%$, whereas the smallest $\ell=1000$ widths exceed $50\%$. Thus narrow distributions occupy only part of the tested family.}

{For every tested $\ell$, Eq.~\eqref{SM:eq:size_box_bound} bounds each size rectangle at its four corners along the sampled incidence trajectories. The resulting maxima satisfy $(\eta_{\rm NR},\eta_a,\eta_{\rm fr})<(9.1\times10^{-4},1.6\times10^{-4},0.088)$ at 1~MeV across all three fringes and $(0.015,0.0013,0.019)$ at 5~MeV [Eqs.~\eqref{SM:eq:validity_ratios} and \eqref{SM:eq:spatial_validity}], so every tested state passes the sampled \mbox{0.1 criteria}.}
\endgroup
\subsection{Successive transport through four dipoles}\label{SM:sec:ring}\label{SM:sec:repeated_bends}
{To test recurrence, we propagate the initial 1-MeV packet ($\ell=64$, $R_\perp=3~\mu$m, $R_\parallel=0.3~\mu$m) {through four 5-mT dipoles and drifts, carrying the evolving state through every element.}} The matrix $M$ transfers the state through one $90^\circ$ dipole, resolved in its local radial, tangential, and field axes. For abrupt boundaries perpendicular to the reference orbit, it is Eq.~\eqref{SM:eq:map} at $\theta=\pi/2$, while smooth edges use Eq.~\eqref{SM:eq:fringe_transfer} and the geometry specified below. The matrix $D$ describes the following drift. With $I_3$ the three-dimensional identity, $v$ the reference speed, and {$N$ the number of complete turns}, {the transfer matrix for sharp boundaries} is
\begin{equation}
 \begin{gathered}
 D=\begin{pmatrix}I_3&\dfrac{L_{\rm drift}}{\gamma mv}I_3\\0&I_3\end{pmatrix},\qquad
 S_N=(DM)^{4N}.
 \end{gathered}
 \label{SM:eq:ring_transfer}
\end{equation}
{For smooth dipoles, $M$ includes the field tails up to negligible field. The free length in $D$ is the remaining distance between these endpoints, avoiding overlap of the two propagation intervals.}

\begin{figure}[!t]
 \centering
 \includegraphics[width=0.85\textwidth]{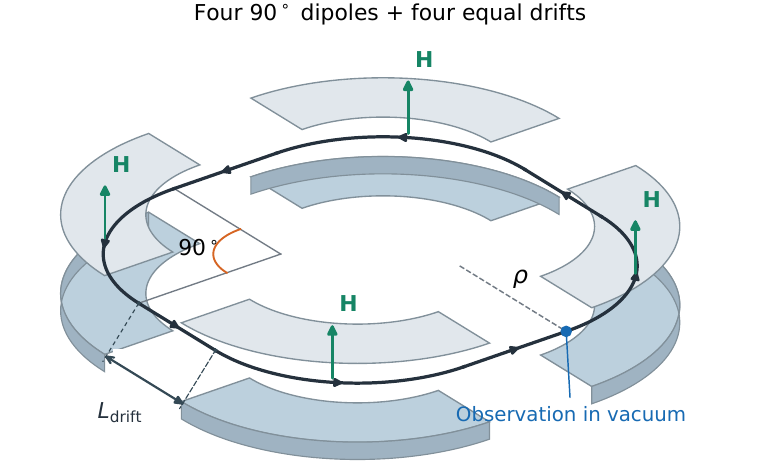}
 \caption{{Reference geometry with sharp boundaries:} four $90^\circ$ dipoles joined by four equal field-free drifts of length $L_{\rm drift}$. The matrices $M$ and $D$ describe one dipole and the following drift. {Their product $DM$ gives one cell, and $(DM)^4$ gives one complete turn [Eq.~\eqref{SM:eq:ring_transfer}]. The packet is observed in vacuum before the first entrance after each turn}, where the reference orbit closes.}
 \label{SM:fig:ring}
\end{figure}

{For sharp boundaries,} a drift of proper duration $\tau_d=L_{\rm drift}/(\gamma v)$ gives $\delta\hat{\vct r}\mapsto\delta\hat{\vct r}+\tau_d\delta\hat{\vct p}/m$, with unchanged momentum. Hence the free intrinsic OAM is conserved in the packet rest frame while the coordinate and mixed covariances change. At the next dipole the changed covariance $\Sigma$ enters the torques $\vct K_\pi,\vct K_a$ [Eq.~\eqref{SM:eq:SMtorques}] and relative width $W$ [Eq.~\eqref{SM:eq:width_angle}]. This accounts for the nonmonotonic dependence on drift length in Table~{\ref{SM:tab:fixedrings}}.

\begingroup

\begin{table}[tp]

\caption{{Relative rms OAM width $W$ [Eq.~\eqref{SM:eq:width_angle}] in the four-dipole model for $K=1$~MeV, $H=5$~mT, $\ell=64$, $R_\perp=3~\mu$m, and $R_\parallel=0.3~\mu$m. The subscripts 1 and 8 denote complete turns at the vacuum observation point in Fig.~{\ref{SM:fig:ring}}. The quoted $\eta_{\rm NR}^{\max}$ [Eq.~\eqref{SM:eq:validity_ratios}] {covers eight turns with sharp boundaries} where the expansion remains applicable. Smooth results use logistic $I_{\rm fr}=10$~mm and a reference bend of $90^\circ$ per magnet. Dashes denote short gaps requiring a calculation with overlapping fringe fields.}}
\label{SM:tab:fixedrings}
\label{SM:tab:ringfringes}
{
\setlength{\tabcolsep}{4pt}
\begin{ruledtabular}
\begin{tabular}{rrrrr}
&\multicolumn{3}{c}{Sharp boundaries}&Smooth edges\\
$L_{\rm drift}$ (m)&$W_1$ (\%)&$W_8$ (\%)&$\eta_{\rm NR}^{\max}$&$W_1$ (\%)\\
0.01&204&328&$7.4\times10^{-4}$&---\\
0.1&135&78.7&$7.2\times10^{-4}$&---\\
0.5&75.7&86.4&$7\times10^{-4}$&73.9\\
1&76.6&80.8&$7\times10^{-4}$&80.2\\
3&87.6&\multicolumn{2}{c}{outside range}&87.4
\end{tabular}\par
\end{ruledtabular}
}
\end{table}
\endgroup

The analytic circular reference has $V_{tt}\propto N^2$ [Eq.~\eqref{SM:eq:circlevariance}]. The four-dipole sequence instead has its own period matrix. The oscillations are described by the shifted radial coordinate $\hat q=\delta\hat r_r-\gamma\delta\hat p_t/(eH)$ and momentum $\delta\hat p_r$. Setting $\theta=\pi/2$ in Eq.~\eqref{SM:eq:map} gives $\hat q\mapsto\delta\hat p_r/(eH)$ and $\delta\hat p_r\mapsto-eH\hat q$. The following drift adds $L_{\rm drift}\delta\hat p_r/(\gamma mv)$ to $\hat q$. Using $\rho=\gamma mv/(eH)$ therefore gives the cell matrix acting on $(\hat q,\delta\hat p_r)$:
\begin{equation}
 \begin{pmatrix}-L_{\rm drift}/\rho&1/(eH)\\-eH&0\end{pmatrix}.
 \label{SM:eq:ring_cell}
\end{equation}
{Its eigenvalues have unit modulus for $0<L_{\rm drift}<2\rho$.} They can then be written $e^{\pm i\mu}$, with $\cos\mu=-L_{\rm drift}/(2\rho)$. Thus $\mu$ is the phase advance of the internal radial oscillation per cell, distinct from the centroid bend and the wave-function phase. Beyond $2\rho$, drift expansion overcomes radial focusing by the bends, and {one eigenvalue has modulus greater than unity, giving growing radial motion}. At $L_{\rm drift}=0.1$ and $1$~m the phases are $93^\circ$ and $122^\circ$. Modulo $360^\circ$, four cells give $12.1^\circ$ and $127^\circ$ per turn, respectively, even though the centroid closes every turn. The next magnet therefore reinforces or partially cancels correlations produced earlier.

The subscript $N$ denotes the operators at the observation point
after $N$ complete turns. One magnet followed by one drift shifts
$\delta\hat r_t-\gamma\delta\hat p_r/(eH)$ by a constant multiple of
$\delta\hat p_t$. Repeating this shift $4N$ times gives
\begin{align}
 \delta\hat r_{t,N}-\frac{\gamma}{eH}\delta\hat p_{r,N}
 &=\delta\hat r_{t}^{(0)}-\frac{\gamma}{eH}\delta\hat p_{r}^{(0)}\nonumber\\
 &\quad+\frac{N[4L_{\rm drift}/\rho-2\pi(\gamma^2-1)]}{eH}
              \delta\hat p_{t}^{(0)},\nonumber\\
 \delta\hat r_{H,N}&=\delta\hat r_{H}^{(0)}+
       \frac{N(2\pi+4L_{\rm drift}/\rho)}{eH}\delta\hat p_{H}^{(0)}.
 \label{SM:eq:ring_secular_shifts}
\end{align}
The momenta $\delta\hat p_t$ and $\delta\hat p_H$ remain constant.
For $0<L_{\rm drift}<2\rho$, the radial pair in
Eq.~\eqref{SM:eq:ring_cell} oscillates and stays bounded, while these
two coordinate shifts accumulate linearly. Each OAM component
therefore contains a bounded oscillatory part and a part linear in
$N$ with an oscillating coefficient. Its variance has a quadratic
upper envelope, with oscillations absent from the exact
uniform-circle law. The variance along the changing mean has the
same upper envelope, but division by the changing mean magnitude can
produce additional peaks in $W^2$.

If $4N$ times the cell phase is an integer multiple of $2\pi$, the
radial pair returns exactly and the mean LG OAM again points along
the tangent. Substituting
Eq.~\eqref{SM:eq:ring_secular_shifts} into
$\hat{\mathcal L}_t=\delta\hat r_H\delta\hat p_r-
\delta\hat r_r\delta\hat p_H$ gives
\begin{equation}
 V_{tt}(N)=
 \left[\frac{N(2\pi+4L_{\rm drift}/\rho)}{eH}\right]^2
 \mean{(\delta\hat p_{r}^{(0)})^2(\delta\hat p_{H}^{(0)})^2}.
 \label{SM:eq:ring_return_variance}
\end{equation}
Here $V_{tt}$ is the tangential OAM variance defined in
Eq.~\eqref{SM:eq:V}, and the initial fourth moment is evaluated in
Eq.~\eqref{SM:eq:circlevariance}. The $N^2$ law holds on these
{turn numbers satisfying this return condition}, with the extra drift time changing its
coefficient. {The zero-drift limit recovers Eq.~\eqref{SM:eq:circlevariance}.}

\begin{figure}[tp]
 \includegraphics[width=\textwidth]{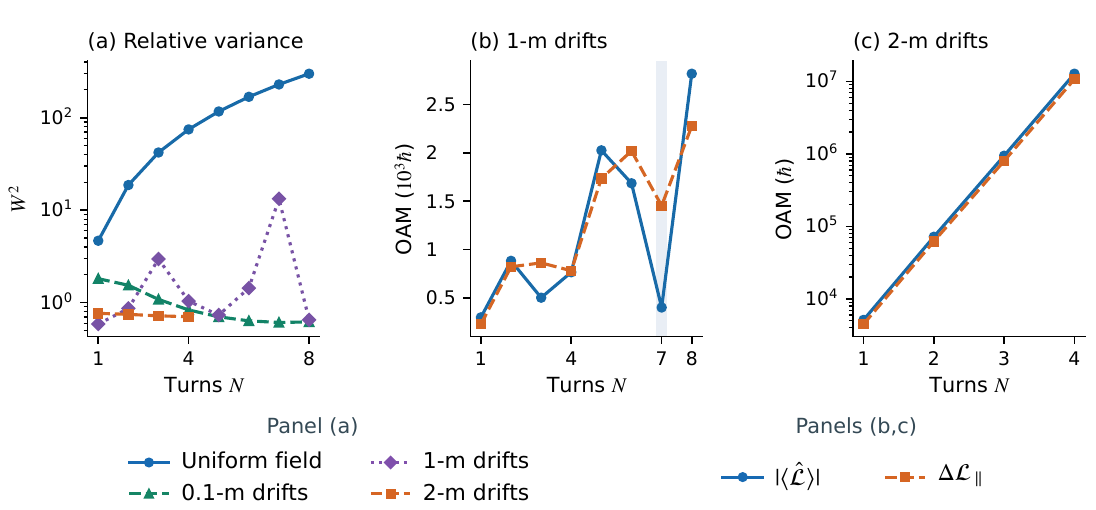}
 \caption{OAM at the fixed observation point after $N$ complete turns. The four-dipole model has four $90^\circ$ bends and equal drifts of length $L_{\rm drift}$, with $K=1$~MeV, $H=5$~mT, $\ell=64$, $R_\perp=3~\mu$m, and $R_\parallel=0.3~\mu$m. {Moments for sharp boundaries follow} from Eqs.~\eqref{SM:eq:ring_transfer}, \eqref{SM:eq:mean_trace}, and \eqref{SM:eq:V}, with the width and projection axis of Eq.~\eqref{SM:eq:width_angle}. (a) Relative variance $W^2$ {along the mean OAM at each turn} for 0.1-, 1-, and 2-m drifts and the uniform-circle reference. In the uniform circle, formal matching to free motion gives a mean magnitude $|\ell|\hbar$ after each turn and the $N^2$ variance in Eq.~\eqref{SM:eq:circlevariance}. (b,c) Mean magnitude and rms spread along the mean for 1- and 2-m drifts. Lines connect observations at integer turns. The shaded strip in (b) marks $N=7$, where $W^2$ is largest over these eight observations. Panel (c) includes four checked turns. By $N=4$, $\max\eta_{\rm NR}={0.017}$ [Eq.~\eqref{SM:eq:validity_ratios}], and the largest longitudinal rms size is $\sqrt{\Sigma_{r_t r_t}}={2.2}$~mm and the largest three-dimensional rms radius is $\sqrt{\tr\Sigma_{rr}}={2.5}$~mm. {$R_\parallel$ and $R_\perp$ denote the entrance sizes [Eq.~\eqref{SM:eq:input}]. During propagation their longitudinal and transverse counterparts are $\sqrt{\Sigma_{r_t r_t}}$ and $\sqrt{\Sigma_{r_r r_r}+\Sigma_{r_H r_H}}$. The quoted $\sqrt{\tr\Sigma_{rr}}$ includes all three directions and initially equals $\sqrt{R_\perp^2+R_\parallel^2}$, where $\Sigma_{rr}$ is the position block of Eq.~\eqref{SM:eq:covariance_definition}.} These maxima include the full four-turn path.}
 \label{SM:fig:ringturns_extended}
\end{figure}

{At 0.1-m drift the radial phase changes slowly over eight turns, with $W^2$ rising slightly from 0.61 at turn 7 to 0.619 at turn 8, whereas at 1~m successive turns sample substantially different correlations [Fig.~{\ref{SM:fig:ringturns_extended}}(a)].} Since $W^2=(\Delta\mathcal L_\parallel)^2/|\mean{\hat{\bm{\mathcal L}}}|^2$, {its peak can occur at a different turn from a peak of the absolute variance.} For the 1-m drift, the calculation gives
\begin{center}
\begin{tabular}{rrrr}
$N$&$|\mean{\hat{\bm{\mathcal L}}}|/\hbar$&$\Delta\mathcal L_\parallel/\hbar$&$W^2$\\
\hline
6&1680&2020&{1.4}\\
7&399&1450&13.3\\
8&2820&2280&{0.65}
\end{tabular}\par
\end{center}
{Between turns 6 and 7 the mean magnitude falls faster than the rms spread, increasing their ratio [Fig.~{\ref{SM:fig:ringturns_extended}}(b)]. The projection follows the mean at each turn.} The sampled maximum remains $\max\eta_{\rm NR}={7}\times10^{-4}$ [Eq.~\eqref{SM:eq:validity_ratios}], so this {nonmonotonic behavior} follows from the radial-oscillation phase and the changing mean magnitude within the quadratic model.

For this radius, the radial stability boundary is $2\rho=1.9$~m. The 2-m example in Fig.~{\ref{SM:fig:ringturns_extended}}(c) lies just beyond this boundary. Its radial growth factor is {3.7} per turn. The same incident state gives
\begin{center}
\begin{tabular}{rrrr}
$N$&$|\mean{\hat{\bm{\mathcal L}}}|/\hbar$&$\Delta\mathcal L_\parallel/\hbar$&$W$ {(\%)}\\
\hline
1&5140&4500&{87.5}\\
2&${7.2}\times10^4$&${6.2}\times10^4$&{86.4}\\
3&${9.4}\times10^{5}$&$8\times10^{5}$&{84.9}\\
4&${1.3}\times10^{7}$&${1.1}\times10^{7}$&{83.9}\\
\end{tabular}\par
\end{center}
{Both moments grow rapidly while their ratio decreases. Through turn 4, $\max\eta_a=0.0076$ [Eq.~\eqref{SM:eq:validity_ratios}], while Fig.~{\ref{SM:fig:ringturns_extended}}(c) gives the momentum and size maxima. The 3-m drift lies farther outside the radial stability range, which additional focusing can alter.}

{In a stable focusing lattice whose one-turn map is rotationally symmetric in normalized phase-space variables, Ref.~\cite{SM:BaturinTransport2026} constructs a periodic Lewis--Floquet OAM invariant. {This invariant and the intrinsic kinetic OAM defined in Eq.~\eqref{SM:eq:OAM_covariance} are generally different observables. For a state with a conserved invariant distribution [Ref.~\cite{SM:BaturinTransport2026}, Eq.~(27)], the kinetic-OAM width follows from a separate moment calculation.} A matched transverse Floquet mode returns after each turn up to a global phase [Ref.~\cite{SM:BaturinTransport2026}, Supplemental Material, Eq.~(96)]. {Its longitudinal kinetic-OAM distribution therefore recurs at the same observation point, with a width set by the \mbox{chosen mode}.}}

{{Our four-dipole reference with sharp boundaries} {has free vertical spreading [Eq.~\eqref{SM:eq:ring_secular_shifts}], which prevents a matched periodic transverse mode. An appropriately focused lattice can instead preserve a narrow kinetic-OAM distribution for suitably prepared states.} For the same quadratic Hamiltonian, both approaches describe the same unitary transport, with the covariance evolving as $\Sigma=S\Sigma_0S^T$ [Eq.~\eqref{SM:eq:covariance_transport}].}

\paragraph{Smooth fringes.}
{The smooth comparison in Table~{\ref{SM:tab:fixedrings}} uses logistic edges with $I_{\rm fr}=10$~mm [Eq.~\eqref{SM:eq:fringe_integral}] and the off-plane field component $H_s$ of Eq.~\eqref{SM:eq:fringe_field}. The effective boundary positions are adjusted to keep each reference bend at $90^\circ$ and close the orbit. Here $L_{\rm drift}$ is the nominal boundary separation along the outgoing tangent. Each tail extends to $24I_{\rm fr}$, where the relative field is below $4\times10^{-11}$, so the remaining free step has length $L_{\rm drift}-48I_{\rm fr}$. For $L_{\rm drift}=0.5$~m, the first-turn path gives $\max\eta_{\rm NR}=1.1\times10^{-4}$ and $\max\eta_a=2.5\times10^{-5}$ [Eq.~\eqref{SM:eq:validity_ratios}], with the largest coordinate rms size about $19~\mu$m, well below $I_{\rm fr}$. At 0.01 and 0.1~m the neighbouring fringe fields overlap. These two gaps require transport in the combined field and specification of the preparation and observation point, so their smooth-edge entries remain unevaluated.}

\begingroup

\paragraph{Absolute broadening and the limit of the model.}
At $L_{\rm drift}=3$~m the cell eigenvalues are approximately $-2.8$ and $-0.36$, so the growing radial amplitude is multiplied by about 62 per turn. Since $\delta\hat r_t-\gamma\delta\hat p_r/(eH)$ grows only linearly, the leading field-axis OAM is $-\gamma(\delta\hat p_r)^2/(eH)$. Thus the quadratic map gives
\[
 |\mean{\hat{\bm{\mathcal L}}}|,\ \Delta\mathcal L_\parallel\propto|\lambda|^{8N},
 \qquad (\Delta\mathcal L_\parallel)^2\propto|\lambda|^{16N},
\]
with $|\lambda|\simeq2.8$. A nearly constant ratio $W$ can therefore conceal rapid absolute broadening. For turns 1 and 2, the formal mean magnitudes are $8.1\times10^4\hbar$ and $3\times10^8\hbar$, and rms spreads $7.1\times10^4\hbar$ and $2.6\times10^8\hbar$: $W$ changes only from $87.6\%$ to $86.2\%$ while the variance grows by $1.3\times10^7$.

The first two momentum maxima are $1.4\times10^{-3}$ and 0.083, and the second-turn radial rms size is 10.3~mm. The third turn reaches $(\eta_{\rm NR},\eta_a)=(5.2,2)$ and is excluded. With 10-mm smooth fringes, the formal second-turn radial size is likewise 9.3~mm, comparable to the fringe scale, despite $\eta_{\rm NR,out}=0.011$. At a 10-m drift, the first-turn quartic mean correction is 0.15\% {of the retained mean kinetic energy, while the cubic correction} has zero mean by parity but 4.2\% rms magnitude, evaluated on the quadratically evolved state. The formal second-turn maxima, $(693,250)$, invalidate that expansion.
\endgroup
\begingroup

\paragraph{Focusing and intrinsic OAM.}\label{SM:sec:ring_focusing}
{To assess focusing in a ring, we compare two ideal thin lenses.}

For an axisymmetric thin lens between field-free regions, the effective focusing potential is $\hat V(\tau)=\kappa(\tau)[(\delta\hat r_r)^2+(\delta\hat r_H)^2]/2$. Neglecting displacement during the short transit makes the interaction a function of the fixed position operators. Its values at different times commute, so integrating $i\hbar\partial_\tau\psi=\hat V(\tau)\psi$ gives $\hat U=\exp[-i\int\hat V(\tau)\dd\tau/\hbar]$, or
\begin{equation}
 \hat U=\exp[-i\mathcal K_\perp((\delta\hat r_r)^2+(\delta\hat r_H)^2)/(2\hbar)],\label{SM:eq:axisymmetric_lens}
\end{equation}
where $\mathcal K_\perp=\int\kappa(\tau)\dd\tau$ has units of momentum per length. Applying $-i\hbar\nabla$ to the acquired phase gives the momentum kicks $\delta\hat p_r\mapsto\delta\hat p_r-\mathcal K_\perp\delta\hat r_r$ and $\delta\hat p_H\mapsto\delta\hat p_H-\mathcal K_\perp\delta\hat r_H$. Their contributions to $\hat{\mathcal L}_t=\delta\hat r_H\delta\hat p_r-\delta\hat r_r\delta\hat p_H$ cancel. Hence $[\hat U,\hat{\mathcal L}_t]=0$, and the complete tangential-OAM distribution is preserved while the radial size and phase change, consistent with axial symmetry [Ref.~\cite{SM:Karlovets2021}, Sec.~4.3].

A quadrupole focuses one transverse direction and defocuses the other [Ref.~\cite{SM:BaturinTransport2026}, Eq.~(41)]. Its effective potential is $\hat V_Q(\tau)=\kappa_q(\tau)[(\delta\hat r_r)^2-(\delta\hat r_H)^2]/2$. The same phase integration as in Eq.~\eqref{SM:eq:axisymmetric_lens} gives $\hat U_Q=\exp[-i\mathcal K_q((\delta\hat r_r)^2-(\delta\hat r_H)^2)/(2\hbar)]$, with $\mathcal K_q=\int\kappa_q(\tau)\dd\tau$. Substituting the opposite kicks $-\mathcal K_q\delta\hat r_r$ and $+\mathcal K_q\delta\hat r_H$ into $\hat{\mathcal L}_t$ gives
\[
 \hat U_Q^\dagger\hat{\mathcal L}_t\hat U_Q-\hat{\mathcal L}_t
 =-2\mathcal K_q\delta\hat r_r\delta\hat r_H.
\]
These relations compare the field-free states immediately before and after the lens, where canonical and kinetic momenta coincide. For the incident LG state, Eq.~\eqref{SM:eq:LG}, $\mean{\delta\hat r_r\delta\hat r_H}=0$, so the outgoing mean remains $\mean{\hat{\mathcal L}_t}_{\rm out}=\ell\hbar$. Its radial moments nevertheless give $\mean{\delta\hat r_r^2\delta\hat r_H^2}=R_\perp^4(|\ell|+2)/[8(|\ell|+1)]$ and
\[
 \Delta\mathcal L_t=|\mathcal K_q|R_\perp^2
 \sqrt{\frac{|\ell|+2}{2(|\ell|+1)}}.
\]
Thus $W=\Delta\mathcal L_t/(|\ell|\hbar)$ as in Eq.~\eqref{SM:eq:width_angle}. A thin magnetic quadrupole of gradient $G$ and length $L$ has $|\mathcal K_q|=|eGL|$. At $\ell=10$ and $R_\perp=10~\mu$m, an uncompensated $|GL|\simeq4.5~\mu$T gives $W=5\%$. Successive focusing elements can compensate these correlations, so lattice tolerances require the complete sequence.
\endgroup

\begingroup

\section{LINAC: magnet parameters and~applicability}\label{SM:sec:linac}
{{We examine whether the internal nonrelativistic approximation applies to the LINAC parallel-transfer geometry, using the conditions of Sec.~\ref{SM:sec:hamiltonian}.} For the incident state specified below, the internal momentum spread exceeds $mc$ before exit, although matching reduces it again [Fig.~{\ref{SM:fig:linacvalidity}} and Table~{\ref{SM:tab:linacformal}}]. {These examples therefore lie outside the internal nonrelativistic regime of Eq.~\eqref{SM:eq:Horb}} and are excluded from the controlled width estimates in the Paper.}
\subsection{Magnet geometry and the incident state}
Seroshtanov \textit{et al.}~{\cite{SM:JINRmagnets}} describe LINAC channel~2 for 25--60~MeV. We consider the 60-MeV endpoint and the 70-MeV magnet-design calculation, with bend magnitude $\theta=35^\circ$, radius $\rho=0.532$~m, and full pole gap $g=41$~mm [Figs.~2--3 and pp.~766--770 of Ref.~{\cite{SM:JINRmagnets}}]. The 335.5-mm mechanical yoke dimension differs from the orbit \mbox{arc length}.

With the normal entrance and exit specified on p.~770 of Ref.~{\cite{SM:JINRmagnets}}, we use planar effective boundaries and a uniform field. The rigidity relation, Eq.~\eqref{SM:eq:rigidity}, gives the effective dipole length $L_{\rm mag}$ and uniform fields $H$:
\begin{align}
 L_{\rm mag}&=\rho\theta={32.5~\mathrm{cm}},\\
 H_{60}&={379~\mathrm{mT}},\qquad
 H_{70}={442~\mathrm{mT}}.
\end{align}
The subscripts on $H$ give $K$ in MeV. The orbit-integrated fields are $\int H_Z\dd s_{\rm orb}=HL_{\rm mag}={0.12}$ and ${0.14}$~T\,m, respectively [Eq.~\eqref{SM:eq:effective_length}], where $s_{\rm orb}$ is the distance along the reference orbit used in that equation, distinct from the coordinate normal to the boundary $s$ in Eq.~\eqref{SM:eq:fringe_integral}. The incident LG state $\psi_0$ [Eq.~\eqref{SM:eq:LG}] has $\ell=64$, $R_\perp=3~\mu$m, and $R_\parallel=0.3~\mu$m, with covariance $\Sigma_0$ from Eq.~\eqref{SM:eq:input}.

\begin{figure}[tbp]
 \centering
 \includegraphics[width=0.82\columnwidth]{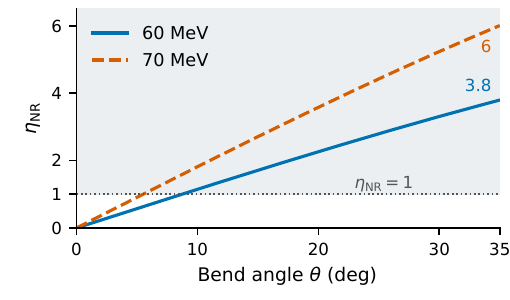}
 \caption{Internal-momentum ratio $\eta_{\rm NR}$ [Eq.~\eqref{SM:eq:validity_ratios}] along the LINAC model dipole. At $K=60$ and 70~MeV it exceeds unity before the $35^\circ$ exit, despite the small values after leaving the magnet in Table~{\ref{SM:tab:linacformal}}. The bend radius is $\rho=0.532$~m, with abrupt boundaries perpendicular to the reference orbit and incident state $\ell=64$, $R_\perp=3~\mu$m, $R_\parallel=0.3~\mu$m. Shading identifies $\eta_{\rm NR}\geq1$, where the internal nonrelativistic expansion is already uncontrolled. The expansion requires $\eta_{\rm NR}\ll1$ throughout the bend. The curves are calculated with the quadratic Hamiltonian of Eq.~\eqref{SM:eq:Horb}.}
 \label{SM:fig:linacvalidity}
\end{figure}

\subsection{Inside the magnet and after exit}
{Exit matching, $C_{\rm out}=C_{\rm in}^{-1}$ [Eq.~\eqref{SM:eq:SMfaces}], removes the position-dependent part of kinetic momentum in Eq.~\eqref{SM:eq:B}. To test validity throughout the bend, we track the internal-momentum and acceleration-size ratios [Eq.~\eqref{SM:eq:validity_ratios}]:}
\begin{equation}
 \eta_{\rm NR}(\theta)=\frac{\sqrt{\tr\Sigma_{\pi\pi}(\theta)}}{mc},\qquad
 \eta_a(\theta)=\frac{\sqrt{\mean{(\vct a'\cdot\delta\hat{\vct r})^2}}}{c^2}.\label{SM:eq:linac_validity}
\end{equation}
Here $\Sigma_{\pi\pi}$ is the kinetic-momentum covariance block obtained from the transfer $S_\pi$ [Eq.~\eqref{SM:eq:kinetic_transfer}]. For both energies, $\eta_{\rm NR}$ exceeds unity well before the exit (Fig.~{\ref{SM:fig:linacvalidity}}). Table~{\ref{SM:tab:linacformal}} gives the crossing angles, path maxima, and field-free endpoints. 
\begin{table}[htbp]

\caption{Applicability diagnostics for the fixed LINAC packet $\ell=64$, $(R_\perp,R_\parallel)=(3,0.3)~\mu$m, with sharp boundaries perpendicular to the reference orbit and a $35^\circ$ bend. {``Out'' means the field-free endpoint, while maxima cover the sampled path. The internal-path violation places the outgoing OAM width outside the controlled regime even when the exit ratios are small.}}
\label{SM:tab:linacformal}
\begin{ruledtabular}
\begin{tabular}{lrr}
$K$ (MeV)&60&70\\
$\eta_{\rm NR,out}$&$2.7\times10^{-4}$&$3.2\times10^{-4}$\\
$\max\eta_{\rm NR}$&3.8&6\\
$\max\eta_a$&0.056&0.076\\
Angle where $\eta_{\rm NR}=1$ (deg)&8.7&5.5\\
\end{tabular}\par
\end{ruledtabular}
\end{table}
At fixed packet dimensions, field, and effective boundaries, we replace the steps by logistic profiles [Eq.~\eqref{SM:eq:logistic_profile}] with fringe integrals $I_{\rm fr}=\int f(1-f)\dd s=1,5,10$~mm [Eq.~\eqref{SM:eq:fringe_integral}]. {These lengths characterize each transition, whereas $\int H_Z\dd s_{\rm orb}$ characterizes the complete bend.} The smooth transfer [Eq.~\eqref{SM:eq:fringe_transfer}] gives $\max\eta_{\rm NR}={3.7},{3.5},{3.3}$ at 60~MeV and $5.9,{5.6},{5.3}$ at 70~MeV. The formal outgoing relative OAM widths remain $W\simeq74\%$ [Eq.~\eqref{SM:eq:width_angle}], although every profile violates $\eta_{\rm NR}\ll1$.

Varying the incident positions and momenta slightly in the exact laboratory Lorentz trajectories reproduces the linear transfer $S$ [Eq.~\eqref{SM:eq:map}] for this geometry, with the boundary synchronization of Sec.~\ref{SM:sec:boundary}. This agreement verifies the classical transfer.

{To find states for which this single-dipole calculation remains valid, Sec.~{\ref{SM:sec:size_scan}} varies the incident sizes at the same radius and $35^\circ$ bend. Submicrometer packets at 20 and 30~MeV can satisfy $\max\eta_{\rm NR},\max\eta_a,\max\eta_{\rm fr}<0.1$ [Eqs.~\eqref{SM:eq:validity_ratios} and \eqref{SM:eq:spatial_validity}]. They differ from the fixed 3-$\mu$m packet above. The 20-MeV case extrapolates the geometry below the published 25--60-MeV channel range.}
\endgroup
\subsection{Two opposite bends}
{We ask whether the LINAC parallel-transfer geometry restores the internal state as well as the momentum direction. Our model uses two identical $+35^\circ$ and $-35^\circ$ {dipoles at normal incidence, separated by a drift} $L_{\rm drift}$. The outgoing reference orbit is parallel to the incident orbit and laterally displaced. Components retain the first magnet's field axis. Let $S_+$ be Eq.~\eqref{SM:eq:map} at $\theta=35^\circ$. Reversing the radial coordinate and its canonical momentum gives}
\[
 P=\operatorname{diag}(-1,1,1,-1,1,1),\qquad S_-=P S_+P.
\]
{This reflection reverses the field and bend, with positive travel time. The drift advances positions by $L_{\rm drift}\delta\hat{\vct p}/(\gamma mv)$ and leaves momenta unchanged. With $I_3$ the three-dimensional identity, the pair transfer is}
\begin{equation}
 \begin{gathered}
 D(L_{\rm drift})=\begin{pmatrix}I_3&\dfrac{L_{\rm drift}}{\gamma mv}I_3\\0&I_3\end{pmatrix},\\
 S_{\rm pair}(L_{\rm drift})=S_-D(L_{\rm drift})S_+.
 \end{gathered}\label{SM:eq:opposite_bends}
\end{equation}
{Even at $L_{\rm drift}=0$, multiplication gives, with $c_\theta,s_\theta,d_\theta$ as in Eq.~\eqref{SM:eq:map},}

{
\[
S_{\rm pair}(0)=
\begin{pmatrix}
\cos2\theta&0&0&\dfrac{\sin2\theta}{eH}&\dfrac{2\gamma c_\theta(1-c_\theta)}{eH}&0\\[3pt]
-2\gamma s_\theta(1-c_\theta)&1&0&\dfrac{2\gamma c_\theta(1-c_\theta)}{eH}&\dfrac{2[d_\theta+\gamma^2s_\theta(1-c_\theta)]}{eH}&0\\[3pt]
0&0&1&0&0&\dfrac{2\theta}{eH}\\[3pt]
-eH\sin2\theta&0&0&\cos2\theta&-2\gamma s_\theta(1-c_\theta)&0\\
0&0&0&0&1&0\\
0&0&0&0&0&1
\end{pmatrix}.
\]
}

{At $\theta=35^\circ$, the radial diagonal entries are $\cos70^\circ\ne1$, and motion along the field adds $2\theta\delta\hat p_H/(eH)$ to $\delta\hat r_H$. {{\textit{Internal evolution continues through the reversed bend.} The resulting mean OAM $\mean{\hat{\bm{\mathcal L}}}$ and covariance $V$ generally differ from their incident values [Eqs.~\eqref{SM:eq:mean_trace} and \eqref{SM:eq:V}].}}}

\section{Summary}\label{SM:sec:summary}
{Following the trajectory-coherent approach~{\cite{SM:BagrovBelovMaslov1989,SM:BagrovBelovTernov1983,SM:BagrovBelovTrifonov1993,SM:BagrovBelovTrifonov1996}}, we have developed {a method to calculate the quantum transport of vortex electrons through bending magnets}. {The center follows a quasiclassical relativistic trajectory, while the internal quantum state evolves under a quadratic Hamiltonian in its instantaneous rest frame.} {We have obtained the wave function and the kinetic-OAM distribution both inside the magnet and outside it.}}

\begingroup

For a proposed 1-MeV, 5-mT half-turn magnet ($L_{\rm mag}\simeq3$~m), $\ell=10$, $I_{\rm fr}=5$~mm, and $\sigma_\alpha=\sigma_\beta=0.3$~mrad, retuning to the angle $\theta\simeq179.38^\circ$ compensates the fringes. At this angle, independent entrance rms sizes $R_\perp=10$--$16~\mu$m and $R_\parallel=0.7$--$1.3~\mu$m in the packet rest frame give relative rms OAM widths $W=3.4$--$7.1\%$ [Eq.~\eqref{SM:eq:width_angle}], with the mean within $2.9^\circ$ of reversal. At the fixed packet sizes, $(R_\perp,R_\parallel)=(10,1)~\mu$m, the $W<10\%$ angular interval has a width of 6.4~mrad (Sec.~\ref{SM:sec:recovery}). We conclude that experimental recovery stays demanding, requiring a dedicated magnet, measured fringe fields, and coherent LG preparation.

For instance, the weak-bend models for IMP (4.6~MeV, $5^\circ$, 0.30~m) and BNL ATF (25~MeV, $1^\circ$, 0.24~m) give $W\simeq12\%$ and the mean-OAM tilt $\alpha_L\simeq24^\circ$ for $\ell=64$ and the same fringe and incidence parameters. Their optimal entrance sizes are $(1.9,0.17)$ and $(1.7,0.15)~\mu$m, respectively. Independent $\pm20\%$ size errors keep $W\leq12.3\%$ [Table~\ref{SM:tab:favorable}(b)], corresponding to longitudinal tolerances of $\pm34$ and $\pm30$~nm in the packet rest frame. Achieving this precision and coherent preparation experimentally seems challenging, though not impossible. %
\endgroup

\clearpage\section{Supplemental movies}\label{SM:sec:supplemental_movies}
\newcommand{\SupplementalMoviePanel}[3]{%
  \hypertarget{movie:#2}{}\includegraphics[width=\textwidth]{#1}%
  \par\smallskip\noindent\href{run:anc/#2}{\texttt{\detokenize{#2}}}%
}

\subsection*{Supplemental movie guide}\label{SM:sec:movie_guide}
{\begin{itemize}
 \item \MovieLink{Movie1_Gaussian_lab.mp4}{Movie~1}: Gaussian laboratory density, $\ell=0$ (Sec.~\ref{SM:sec:lab_density_movies}).
 \item \MovieLink{Movie2_Vortex_comoving.mp4}{Movie~2}: rest-frame density, phase, and OAM, $\ell=64$ (Sec.~\ref{SM:sec:packet_visualization}).
 \item \MovieLink{Movie3_Vortex_lab.mp4}{Movie~3}: vortex laboratory density, $\ell=10$ (Sec.~\ref{SM:sec:lab_vortex_movie}).
 \item \MovieLink{Movie4_Correlations.mp4}{Movie~4}: position--momentum correlations in the BNL dipole (Fig.~\ref{SM:fig:covariance_ellipses}).
 \item \MovieLink{Movie5_Ring.mp4}{Movie~5}: OAM over one turn with and without the acceleration potential (Sec.~\ref{SM:sec:movie_mirrorcle}).
\end{itemize}}
The counters give physical time, with $t_{\rm lab}=\gamma\tau$ at the packet center. Playback rates vary between stages.

{\subsection{Gaussian packet in the laboratory: Movie~1}
\label{SM:sec:lab_density_movies}}

\MovieNumber{1} uses $K=4.6$~MeV, $H=5$~mT, and a $90^\circ$ bend. Its entrance-waist sizes $R_\perp=2~\mu$m and $R_\parallel=1~\mu$m are specified in the instantaneous rest frame. At the waist, the laboratory longitudinal rms size is approximately $R_\parallel/\gamma=0.1~\mu$m, while either transverse coordinate has rms size $R_\perp/\sqrt2=1.4~\mu$m. During the $17.9$~ns laboratory transit, the density ellipse in the orbit plane turns relative to the local tangent and radius, while spreading along the field is weak. Movies~\MovieIndex{1}--\MovieIndex{3} use the quadratic Hamiltonian. {For the $90^\circ$ examples in Movies~\MovieIndex{1} and \MovieIndex{3}, the local expansion ratios [Eq.~\eqref{SM:eq:validity_ratios}] remain small.}

\begin{center}\SupplementalMoviePanel{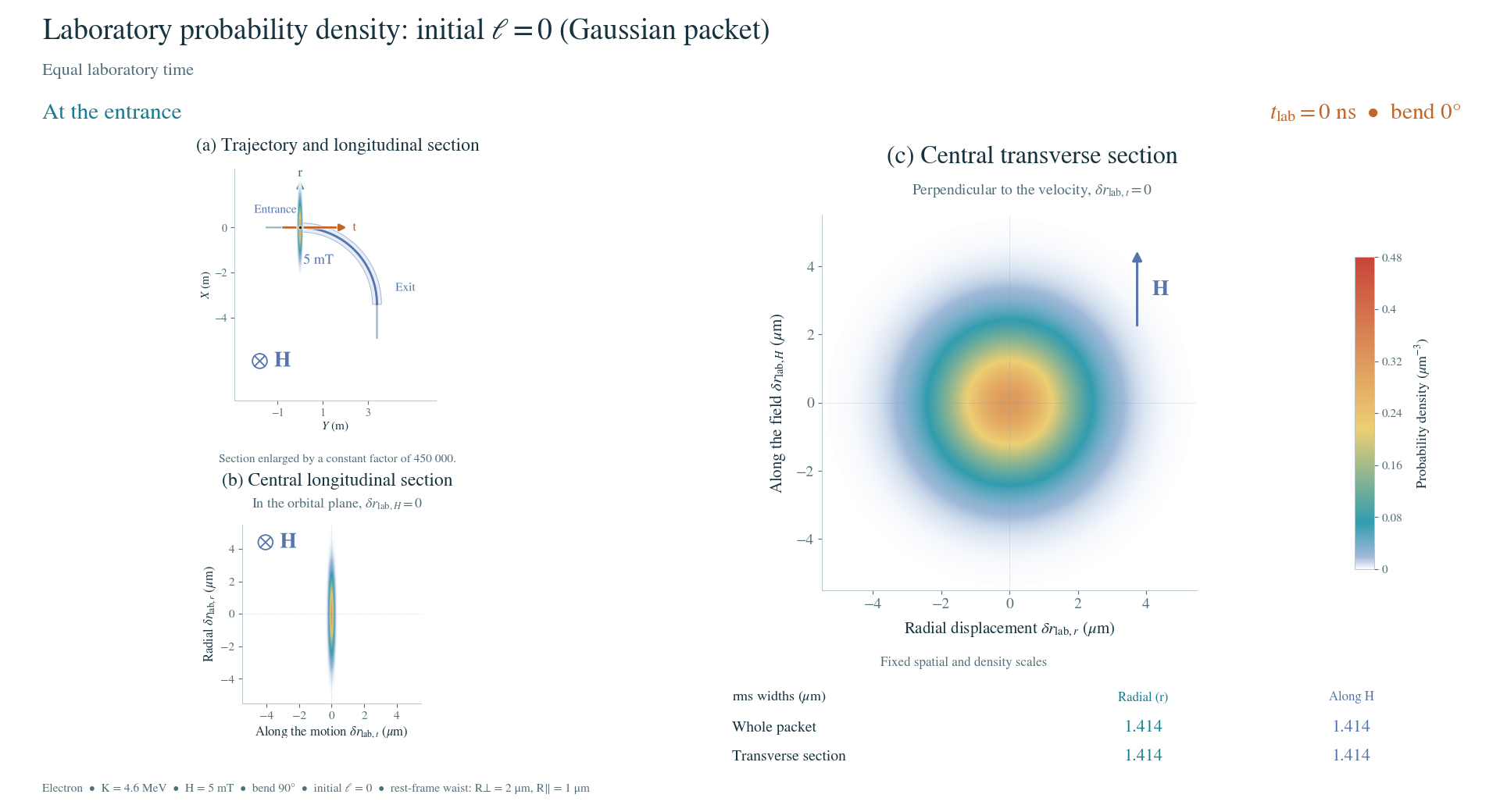}{Movie1_Gaussian_lab.mp4}{Movie 1: Gaussian packet at equal laboratory time}\end{center}

{A laboratory observer compares all points at one laboratory time $t_{\rm lab}$, whereas Eq.~\eqref{SM:eq:SMpsi} describes simultaneous points in the instantaneous rest frame. Constructing the laboratory density therefore requires both the change of simultaneity and the transformation of probability density and current density.

We first find the probability current density from the quadratic Hamiltonian. Multiplying the Schr\"odinger equation by $\psi^*$ and subtracting its complex conjugate cancels the real acceleration potential. The kinetic term gives, in FW coordinates,
\[
 \partial_\tau|\psi|^2
 =-\nabla\cdot\frac{\operatorname{Re}[\psi^*\delta\hat{\vct\pi}\psi]}{m}.
\]
The expression under the divergence is the probability flux per unit center proper time $\tau$. A clock at fixed FW position measures $\dd t_{\rm loc}=\mathcal N\dd\tau$, with $\mathcal N=1+\vct a'\cdot\delta\vct r/c^2$ from the metric in Eq.~\eqref{SM:eq:SMmetric}. Dividing the flux by this clock-rate factor gives the current density per unit local proper time,
\begin{equation}
 j'_i=\frac{\operatorname{Re}[\psi^*\delta\hat\pi_i\psi]}{m\mathcal N}.
 \label{SM:eq:movie_rest_current}
\end{equation}
Thus probability conservation reads $\partial_\tau|\psi|^2+\nabla\cdot(\mathcal N\vct j')=0$. The spatial derivatives in $\delta\hat\pi_i$ in Eq.~\eqref{SM:eq:movie_rest_current} act on the complete state in Eq.~\eqref{SM:eq:SMpsi}, including its polynomial. In orbital coordinates, the axes rotate through $\gamma\theta$ about the field and move at $\gamma\dot\theta\vct e_H\times\delta\vct r$ per unit $\tau$. Subtracting this velocity times $|\psi|^2$ gives the additional coordinate flux associated with the rotating basis in Eq.~\eqref{SM:eq:Horb}. The current density in Eq.~\eqref{SM:eq:movie_rest_current} describes the effective probability flow at this order.

Next we identify which rest-frame event corresponds to each point of a laboratory image. The coordinate transformation in Sec.~\ref{SM:sec:moved_fw} is $x^\mu_{\rm lab}=X^\mu(\tau)+e_i^\mu(\tau)\delta r_i$, where $X^\mu$ is the reference worldline and $e_i^\mu$ are its spatial basis vectors. Taking its scalar products with the dimensionless four-velocity $u^\mu$ and with $e_i^\mu$ gives
\begin{equation}
 \begin{aligned}
 u_\mu(\tau)[x^\mu_{\rm lab}-X^\mu(\tau)]&=0,\\
 \delta r_i&=-e_{i\mu}(\tau)[x^\mu_{\rm lab}-X^\mu(\tau)].
 \end{aligned}
 \label{SM:eq:movie_lab_events}
\end{equation}
We fix $x^\mu_{\rm lab}=(ct_{\rm lab},\vct r_{\rm lab})$, solve the first line for $\tau$, and use the second to find the position on that rest-time plane. This follows from $u\cdot e_i=0$ and $e_i\cdot e_j=-\delta_{ij}$. The spatial basis is chosen along $r,t,H$, as in Eq.~\eqref{SM:eq:SMpsi}. For our constant-speed orbit, the time transformation is $t_{\rm lab}=\gamma(\tau+\beta\delta r_t/c)$, with $\beta=v/c$. Points at different longitudinal positions therefore require different values of $\tau$ even within one laboratory image.

In the local rest frame, the effective four-current has time component $c|\psi|^2$ and spatial components $j'_i$. Expanding this vector on the orthonormal basis gives the first line below. A boost along the tangent has $u^0=\gamma$, $e_t^0=\gamma\beta$, and $e_r^0=e_H^0=0$, which gives its laboratory time component and hence the laboratory probability density:
\begin{equation}
 \begin{aligned}
 J^\mu&=c|\psi|^2u^\mu+\sum_{i=r,t,H}e_i^\mu j'_i,\\
 \varrho_{\rm lab}(\vct r_{\rm lab},t_{\rm lab})
 &=\frac{J^0}{c}
 =\gamma\left[|\psi(\delta\vct r,\tau)|^2
       +\frac{\beta j'_t(\delta\vct r,\tau)}{c}\right].
 \end{aligned}
 \label{SM:eq:movie_lab_density}
\end{equation}
Here $\tau$ and $\delta\vct r$ are obtained separately for each laboratory point from Eq.~\eqref{SM:eq:movie_lab_events}. Conservation of this effective current gives $\int\varrho_{\rm lab}\dd^3r_{\rm lab}=1$. Near a simple vortex zero, $|\psi|^2$ vanishes quadratically whereas the current correction is linear in displacement. The truncated expression can therefore give negative $\varrho_{\rm lab}$ at distances below $\hbar/(mc)$, outside its accuracy as a positive relativistic density and far below the movie resolution. When internal flow and evolution during the simultaneity offset are negligible, Eq.~\eqref{SM:eq:movie_lab_density} reduces to the familiar longitudinal contraction and the density increases by $\gamma$. Movies~\MovieIndex{1} and \MovieIndex{3} include the current term and evaluate the wave function at the different rest-frame times required by a common $t_{\rm lab}$.

The position $\vct r_{\rm lab}$ labels points in the rest frame of the magnet. Eq.~\eqref{SM:eq:movie_lab_events} relates these positions to the instantaneous rest-time plane on which $\delta\hat{\vct r}$ acts. The resulting $\tau$ selects free, in-field, or outgoing evolution in Eq.~\eqref{SM:eq:psi_stages}. Propagating backwards from the entrance waist describes the incident packet converging toward that waist. These movies use sharp boundaries, while Sec.~\ref{SM:sec:fringes} treats a spatially resolved fringe.

\MovieNumber{1} shows central longitudinal and transverse sections of the laboratory density, with an enlarged longitudinal section on the reference orbit. Whole-packet rms widths use the three-dimensional density, while section widths use the normalized two-dimensional sections. }

\clearpage

{\subsection{Vortex packet in the instantaneous rest frame: Movie~2}
\label{SM:sec:density_notes}\label{SM:sec:packet_visualization}}

{{Fig.~\ref{SM:fig:packet_story}} shows the density evolution, while \MovieNumber{2} also resolves the phase and its zeros, using Eq.~\eqref{SM:eq:SMpsi} for $\ell=64$, $(R_\perp,R_\parallel)=(2,0.2)~\mu$m, and $K=4.6$~MeV through a $5^\circ$ bend in $H=5$~mT. The energy follows the IMP demonstration~\cite{SM:IMP2024}, while the magnet and packet sizes are chosen here. The rms transverse size is $R_\perp=w\sqrt{65}$, with $w\simeq0.25~\mu$m [Eq.~\eqref{SM:eq:LG}]. {Fig.~\Lref{fig:windows}(a) of the Paper} additionally includes the fringe integral $I_{\rm fr}=5$~mm and incidence spreads $\sigma_\alpha=\sigma_\beta=0.3$~mrad [Eqs.~\eqref{SM:eq:fringe_integral} and \eqref{SM:eq:two_incidence_angles}].}

\begin{center}\SupplementalMoviePanel{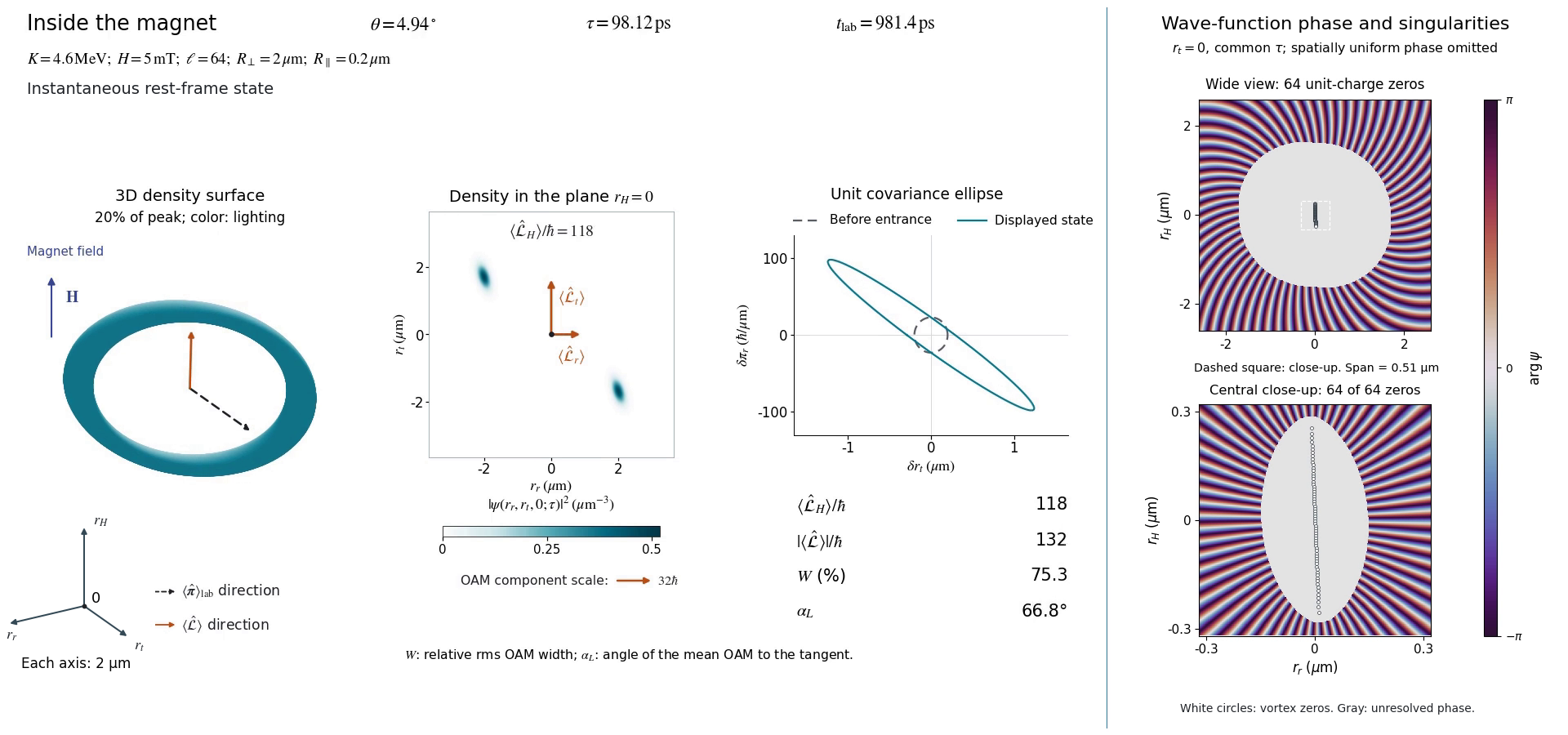}{Movie2_Vortex_comoving.mp4}{Movie 2: Vortex density, phase, and OAM through a weak dipole}\end{center}

{\MovieNumber{2} uses a common instantaneous rest-frame time $\tau$, with centered coordinates $(r_r,r_t,r_H)=(\delta r_r,\delta r_t,\delta r_H)$. The exit axes remain fixed during the drift, and $t_{\rm lab}=\gamma\tau$ refers to the center. Movies~\MovieIndex{1} and \MovieIndex{3} instead use a common laboratory time.}

{Inside the magnet, $\theta=eH\tau/m$, so the $5^\circ$ bend takes $\tau_{\rm mag}\simeq99.2$~ps of proper time ($t_{\rm mag}\simeq993$~ps in the laboratory). To illustrate diffraction after exit, we choose a free drift of $\Delta\tau=500$~ps, corresponding to the laboratory distance}
\begin{equation}
 L_{\rm drift}=\gamma v\Delta\tau={1.5}~\mathrm m,
 \qquad v=c\sqrt{1-\gamma^{-2}},
 \label{SM:eq:movie_times}
\end{equation}
where $v$ is the center's laboratory speed.

{The surfaces in \MovieNumber{2} and {Fig.~\ref{SM:fig:packet_story}} mark 20\% of the density maximum at the displayed time, with color showing illumination:}
\begin{equation}
 |\psi(\delta\vct r,\tau)|^2
 =0.2\max_{\delta\vct r'}|\psi(\delta\vct r',\tau)|^2.
 \label{SM:eq:density_surface}
\end{equation}
{The density section at $r_H=0$ uses a fixed color scale. \MovieNumber{2} also shows the unit covariance ellipse of $(\delta r_t,\delta\pi_r)$ [Eq.~\eqref{SM:eq:covariance_ellipse}], {which summarizes second moments. The enclosed probability depends on the full state.}}

{The phase plots show $\arg\psi(\delta r_r,0,\delta r_H;\tau)$ up to a spatially uniform phase. White circles locate analytic vortex-line intersections with $r_t=0$, including those in the low-density core. {Each marker locates the intersection of a distinct vortex line with the plotted plane. The direction of each line follows from its three-dimensional geometry.} Phase colors depend on gauge, while the zeros are gauge invariant.}

{To display the matching at each boundary separately from propagation, \MovieNumber{2} holds $\tau$ fixed while changing the boundary phase. At entrance it uses $\psi_\lambda=\exp[-i\lambda eH' r_rr_t/(2\hbar)]\psi_0$, with $0\leq\lambda\leq1$. The same $\lambda$ multiplies $\vct A'$, and hence $H'$, in the kinetic-momentum operator used to evaluate OAM. At exit, the inverse phase grows with $\lambda$ while the vector potential decreases as $(1-\lambda)\vct A'$. } 

{The mean kinetic OAM $\mean{\hat{\bm{\mathcal L}}}$ and its covariance $V$ follow from Eq.~\eqref{SM:eq:SMpsiMom}, giving the tilt $\alpha_L$ and relative width $W$ through Eq.~\eqref{SM:eq:width_angle}.} Fig.~\ref{SM:fig:weak_oam_trace} shows the corresponding mean-OAM
magnitude and relative width through the magnet. Once the packet
is free, both remain constant in its rest frame while its density spreads.

\par\medskip

{\subsection{Vortex packet in the laboratory: Movie~3}
\label{SM:sec:lab_vortex_movie}}

{\MovieNumber{3} allows a direct comparison with the Gaussian packet in \MovieNumber{1}: we change the initial OAM to $\ell=10$ and keep the same energy, field, bend, and physical waist sizes. Both use Eq.~\eqref{SM:eq:movie_lab_density}, identical spatial scales, and the fixed range $0\leq\varrho_{\rm lab}\leq0.48~\mu\mathrm m^{-3}$. The incident vortex packet has a ring-shaped transverse section. At fixed $R_\perp$, its larger momentum spread shortens the transverse diffraction time by a factor of eleven [Eq.~\eqref{SM:eq:psi_diffraction_times}]. Its whole-packet rms size along the field grows by $15\%$ during the bend, compared with $0.13\%$ for $\ell=0$. }

\begin{center}\SupplementalMoviePanel{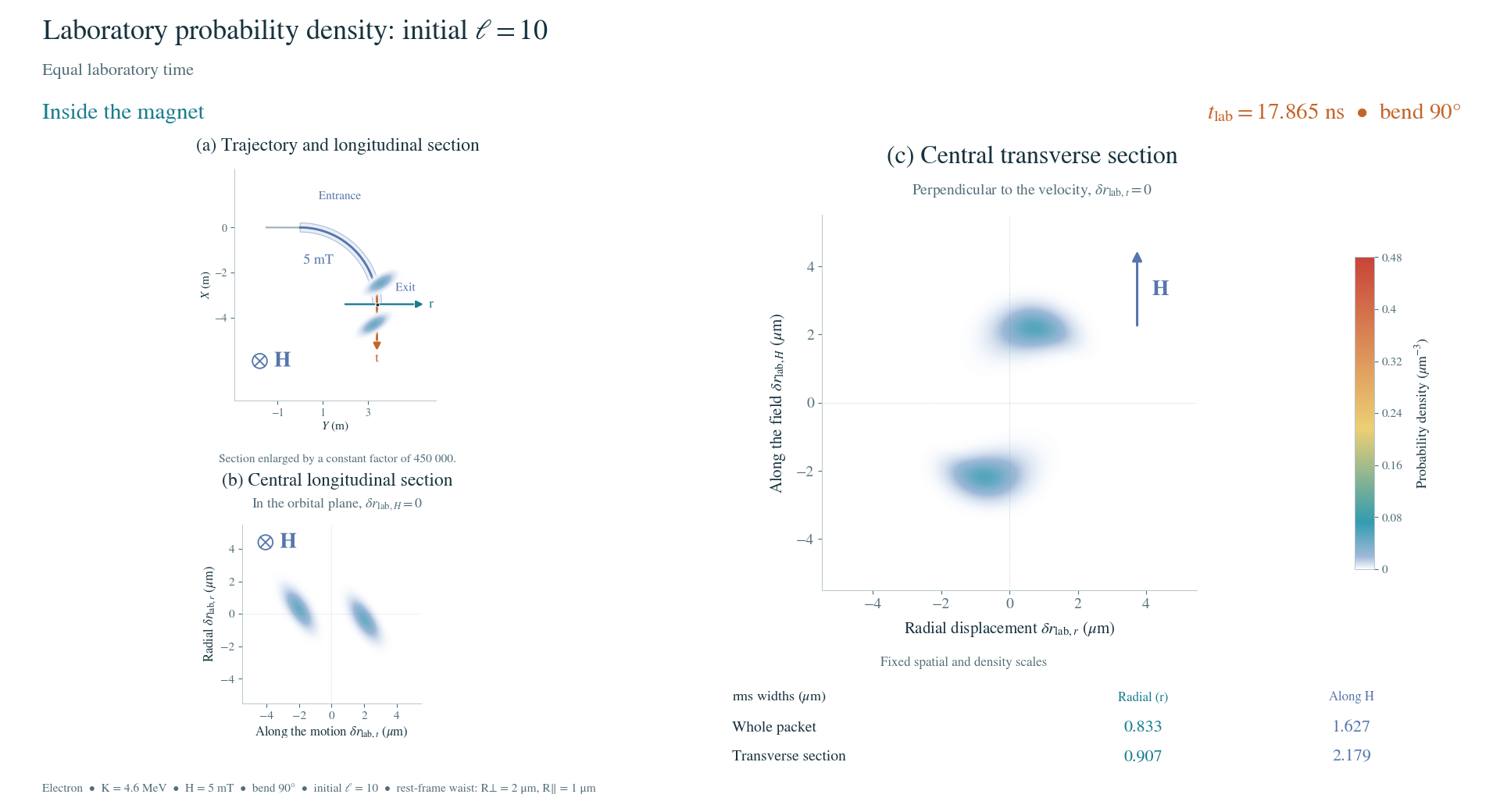}{Movie3_Vortex_lab.mp4}{Movie 3: Vortex packet at equal laboratory time}\end{center}

\clearpage

\subsection{Position--momentum correlations: Movie~4}\label{SM:sec:movie_correlations}

The incident LG packet has $\ell=1000$, $K=25$~MeV, $R_\perp=1~\mu$m, and $R_\parallel=0.3~\mu$m. It enters a $20^\circ$ dipole normally, with $H=74.2$~mT and $L_{\rm mag}=0.4$~m.

\begin{center}\SupplementalMoviePanel{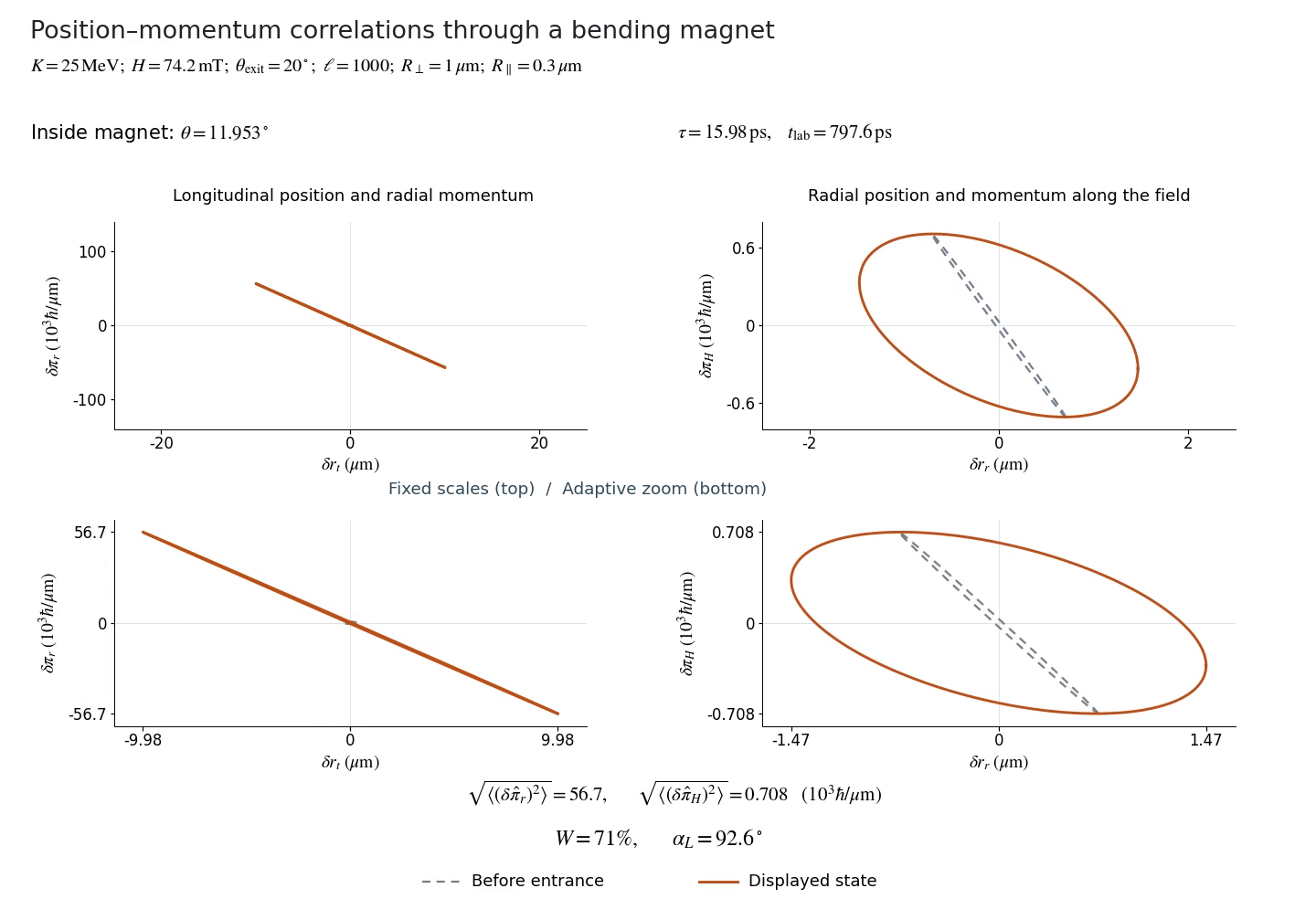}{Movie4_Correlations.mp4}{Movie 4: Position--momentum correlations through a 20-degree dipole}\end{center}

Unit covariance ellipses [Eq.~\eqref{SM:eq:covariance_ellipse}] compare the incident and evolving states. The upper row uses fixed scales and the lower row follows the same contours with adaptive zoom. At each boundary, $\lambda$ resolves matching at fixed center proper time. After exit, $W=70.7\%$ and $\alpha_L=86.2^\circ$ (Fig.~\ref{SM:fig:covariance_ellipses}).

\clearpage

\subsection{One orbit in a uniform magnetic field: Movie~{5}}
\label{SM:sec:movie_mirrorcle}
{\MovieNumber{5} compares the two Hamiltonians of Sec.~\ref{SM:sec:movie_acceleration} over one complete turn in a uniform field, using $K=6$~MeV and $\rho=15$~cm from the MIRRORCLE-6X parameters in Table~1 of Ref.~\cite{SM:MIRRORCLE2006}.} We take a spatially uniform field $H=144$~mT, fixed by relativistic rigidity [Eq.~\eqref{SM:eq:rigidity}], and the incident LG state of Eq.~\eqref{SM:eq:LG} with $\ell=64$, $R_\perp=1~\mu$m, and $R_\parallel=0.2~\mu$m. The blue curves use $\hat{\mathcal H}_{\rm orb}$ in Eq.~\eqref{SM:eq:Horb}, whereas the orange curves omit its acceleration potential $-m(\vct a'\cdot\delta\hat{\vct r})^2/(2c^2)$. Both use the same uniform field, entrance matrix $C_{\rm in}$ [Eq.~\eqref{SM:eq:SMfaces}], kinetic conversion [Eq.~\eqref{SM:eq:B}], and OAM moments [Eqs.~\eqref{SM:eq:mean_trace} and \eqref{SM:eq:V}].

\begin{center}\SupplementalMoviePanel{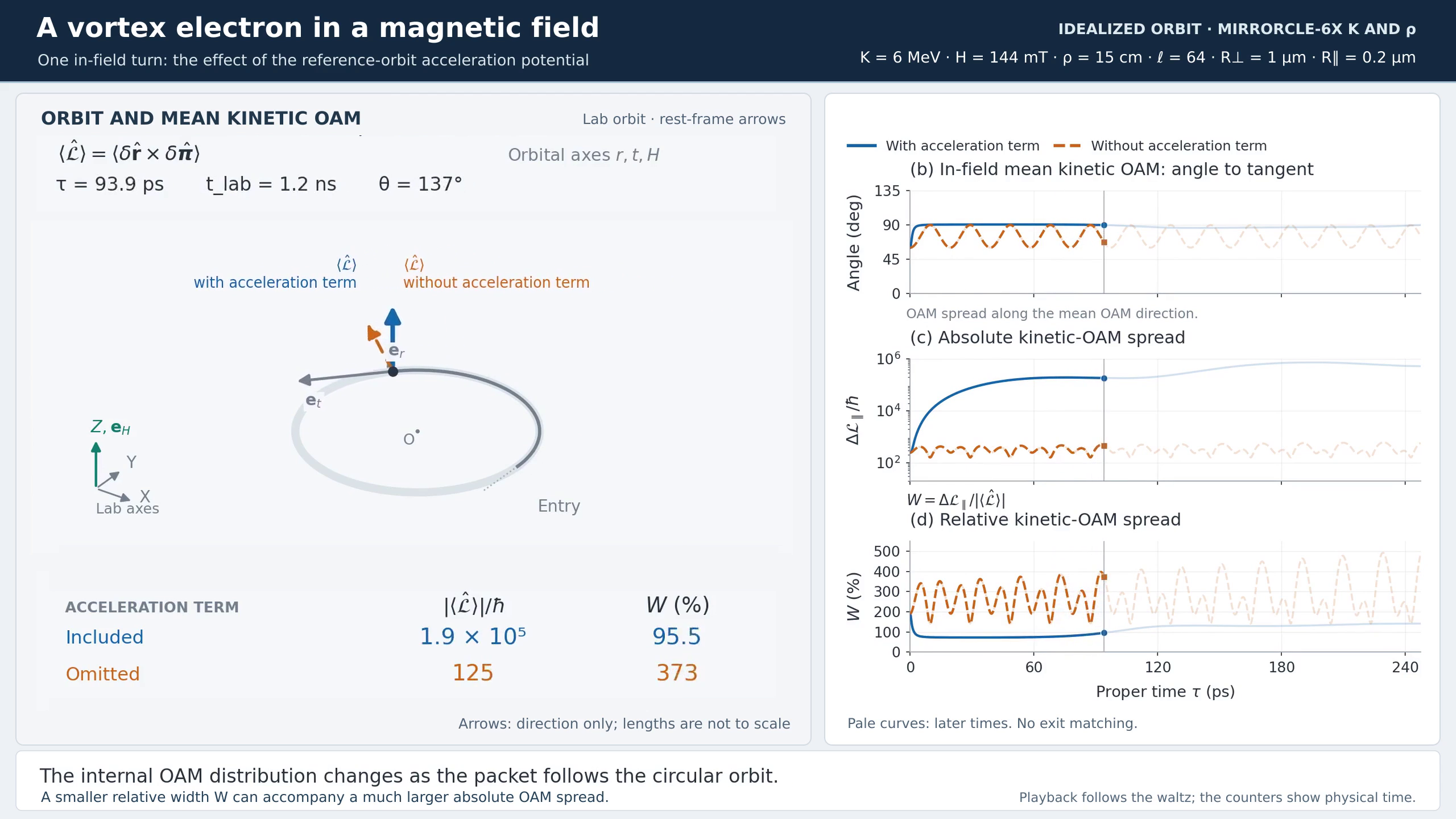}{Movie5_Ring.mp4}{Movie 5: OAM over one orbit in a uniform magnetic field}\end{center}

Music: J.~Strauss~II, \textit{The Blue Danube}, {Op.~314, performed by the U.S.~Marine Band, conducted by} A.~F.~Schoepper (public-domain recording).

{At fixed center time, the opening sequence interpolates the entrance shear $\Delta\pi_r=-\lambda eH'r_t$, $0\leq\lambda\leq1$. It takes the rest-frame mean from $(0,64,0)\hbar$ to $(0,64,112)\hbar$ in $(r,t,H)$ axes, with $W=191\%$ [Eq.~\eqref{SM:eq:width_angle}]. {The subsequent turn describes motion within the field.} The free-to-free calculation in Sec.~\ref{SM:sec:circle} includes the exit. After one turn, $\tau=247$~ps or $t_{\rm lab}=3.2$~ns, the full Hamiltonian gives $W\simeq141\%$ and $\Delta\mathcal L_\parallel\simeq5.3\times10^5\hbar$, versus $484\%$ and $585\hbar$ without the acceleration potential. The smaller relative width accompanies a larger mean and absolute spread.}

{For evolution with $\hat{\mathcal H}_{\rm orb}$, the maximum expansion ratios $\max\eta_{\rm NR}\simeq0.015$ and $\max\eta_a\simeq8.2\times10^{-4}$ [Eq.~\eqref{SM:eq:validity_ratios}] support the internal nonrelativistic expansion along this model orbit. The uniform field isolates the acceleration contribution.}

\end{document}